\documentclass[twoside,12pt]{thesis}

\usepackage{floatflt,bm,psfrag,layout,enumitem}
\usepackage{amsmath}
\usepackage{graphicx,amssymb,rotate}
\usepackage{mathrsfs}
\usepackage{color}
\usepackage{graphicx}
\usepackage{subfigure}
\usepackage{multirow}
\usepackage{float}
\usepackage{pstricks}
\usepackage[toc,page]{appendix}
\usepackage{pifont}
\usepackage[utf8]{inputenc}
\usepackage{tabls}

\usepackage{makeidx}

\usepackage[noadjust]{cite}
\newcommand{\cmark}{\text{\ding{51}}}
\newcommand{\xmark}{\text{\ding{55}}}
\usepackage[bottom]{footmisc}
\usepackage[bookmarks=true,bookmarksnumbered=true,%
pdftitle={Thesis},%
pdfauthor={Najimuddin Khan (IIT Indore)}]{hyperref}
\renewcommand\baselinestretch{1.5}
           
\renewcommand{\floatpagefraction}{0.8}
\renewcommand\bibname{References}
\newcommand{\begp}{\begingroup}
\newcommand{\eegp}{\endgroup}

\newcommand{\beq}{\begin{equation}}
\newcommand{\eeq}{\end{equation}}
\newcommand{\bea}{\begin{eqnarray}}
\newcommand{\eea}{\end{eqnarray}}
\newcommand{\Z}{\mathbb{Z}}
\def\sv{\left\langle\sigma v\right\rangle}
\def\mc{\mathcal}
\def\nn{\nonumber}
\def\Eqn#1{Eq.\ (\ref{#1})}
\def\ra{\rightarrow}
\def\mpl{M_{\rm Pl}}
\def\vector#1#2{\left( \begin{array}{c}#1\\ #2\end{array}\right)}

\newcommand{\MS}{\overline{\mbox{\sc ms}}}

\def\eg{ {\em e.g.\ }}
\def\etc{ {\em etc.\ }}
\def\ie{ {\em i.e.,\ }}
\def\viz{ {\em viz.\ }}
\def\etal{\!{\em et al.\ }}
\def\mpl{M_{\rm Pl}}
\def\hw{\qquad \boxed{\mathbf {HW}}}

\usepackage[compact]{titlesec}
\usepackage{setspace}
\titlespacing*{\section}{0pt}{3ex}{0ex}
\titlespacing*{\subsection}{0pt}{3ex}{0ex}
\titlespacing*{\subsubsection}{0pt}{3ex}{0ex}

\begin{document}



\graphicspath{{FirstPage/}}
\pagestyle{plain}
\pagenumbering{roman}\setcounter{page}{1}
\addcontentsline{toc}{section}{Title}

\begin{center}
\vspace*{-2 cm}
{\Large \textbf{Exploring Extensions of the Scalar Sector of the Standard Model}}\\

\vfill
{\large\textbf{\textsf{Ph.D. Thesis}}}\\
\vspace*{0cm}

\vfill\vspace*{0cm}
{\Large\textbf{by}}\\
\vspace*{0.5cm}
{\Large\textbf{Najimuddin Khan}}\\
\vspace*{1cm}

\includegraphics[width=5cm,clip]{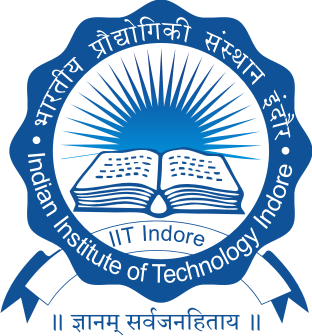}\\
\vspace*{2cm}

{\large\textbf{Discipline of Physics, Indian Institute of Technology Indore,
\\
 Khandwa Road, Simrol, Indore - 453552, India}}\\
\vspace*{1cm}

\end{center}

\begin{center}
\newpage
\null
\newpage
\end{center}

\begin{center}
\vspace*{-2.0cm}
{\Large\textbf{Exploring Extensions of the Scalar Sector of the Standard Model}}\\

\vfill
{\large\textbf{\textsf{A THESIS}}}\\
\vspace*{0cm}

\vfill
{\large\it{Submitted in partial fulfillment of the \\requirements for the award of the degree}}\\
\vspace*{0.2cm}
{\Large\textbf{of}}\\
\vspace*{0.5cm}

{\large\textbf{\textsf{DOCTOR OF PHILOSOPHY}}}\\
\vspace*{0cm}

\vfill\vspace*{0cm}
{\Large\textbf{by}}\\
\vspace*{0.5cm}
{\Large\textbf{Najimuddin Khan}}\\
\vspace*{1cm}

\includegraphics[width=5cm,clip]{./Plots/IITimage}\\
\vspace*{2cm}

{\large\textbf{Discipline of Physics, Indian Institute of Technology Indore,\\
 Khandwa Road, Simrol, Indore - 453552, India}}\\
\vspace*{1cm}

\end{center}

\newpage
\null
\newpage

\vspace*{5cm}
{\Huge\textbf{ \hspace{4cm} Dedicated}}\\
        
      {\Huge\textbf{  \hspace{5.8cm} to}}\\
 
  {\Huge\textbf{ \hspace{3.8cm} My Family}}

\newpage
\null
\newpage
\begin{center}
{\Huge\textbf{Acknowledgements}}\\

\phantomsection
\addcontentsline{toc}{section}{Acknowledgements}
\printindex

\end{center}
\vspace{28pt}
First and foremost, I gratefully acknowledge the continual guidance and support of my adviser, Dr. Subhendu Rakshit. His dedication to research and pursuit of physics has been an invaluable source of inspiration and encouragement to me.
I would also like to thank my PSPC committee members, Dr. Manavendra N Mahato, Dr. Antony Vijesh for serving as my committee members even at hardship.

I would like to thank Prof. Amitava Raychaudhuri, Prof. Amitava Datta, Prof. Biswarup Mukhopadhyaya, Dr. Avirup Shaw, Avinanda Chaudhuri, Nabanita Ganguly for their wonderful collaboration. You supported me greatly and were always willing to help me.

A special thanks to my family. Words cannot express how grateful I am to my {\bf MAA-ABBA} for all of the sacrifices that you’ve made on my behalf. Your prayer for me was what sustained me thus far. I would also like to thank all of my relatives who supported, and incented me to strive towards my goal. 

``There are some people in life that make you laugh a little louder, smile a little bigger and live just a little bit better"....FRIENDS. My jolly friends Minarul Islam, Lalsha Khan, Imran Ali. Thanks dude for helping to live.
I would like to thank my JAJABOR friend at IIT Indore Mr. Sudeep Ghosh. You are the best person at IIT Indore who always fight with me and gave a lot of love to me.  It is also unforgettable moments with VAI-BON: Sidharta Karmakar and Sujata.

\newpage
\null
\newpage
\begin{center}
{\Huge\textbf{Abstract}}\\
\phantomsection
\addcontentsline{toc}{section}{Abstract}
\printindex
\end{center}
\vspace{28pt}


The Standard Model is one of the most successful theories which describes strong, weak and electromagnetic forces and interactions between the elementary particles. The scalar boson has been found at the Large Hadron Collider (LHC) on 4th July 2012 that has confirmed the Higgs mechanism. Although, the properties of this scalar are consistent with the Higgs as predicted by the standard model (SM), the experimental data still allow an extension of the scalar sector. The standard model fails to explain few physical phenomena in Nature, for example, the presence of dark matter, existence of massive neutrino, the observed matter-antimatter asymmetry, inflation of the Universe, etc. These phenomena indicate the presence of new physics beyond the standard model. In this thesis, various extensions of the scalar sector have been considered comprising of different ${SU(2)_L}$ multiplets. The main purpose for writing this thesis is to explore properties of these new scalars using the weak vector boson scattering processes and from the (meta)stability of the scalar potential.

If the extended scalar sector participates in the electroweak symmetry breaking then these extra scalars need to couple with the known standard model particles. In this work, it has been shown that the vector boson scattering involving scalar boson exchanges provide a complimentary way to direct search methods to probe into the scalar sector. As different extended scalar sectors have similar types of scalar fields, e.g., an extra $CP$-even Higgs, charged Higgs etc., these new physics models can give rise to the similar types of experimental signatures. In this work, it has been shown that it is possible to distinguish between such models at various vector boson scattering processes by looking at the resonances. Also, the shapes of the resonances can provide further insight to the relevant parameter space of these models.

As in the SM, the electroweak vacuum is metastable, it is important to explore if an extended scalar has an answer in its reserve. As the scalar weakly interacting massive scalar particles protected by $Z_2$ symmetry can serve as viable dark matter candidates, it is interesting to explore if they help prolong the lifetime of the Universe. The effective Higgs potential gets modified in the presence of these new extra scalars, improving the stability of electroweak vacuum. Such an exercise has been undertaken in various kinds of extended scalar sectors. In order to show the explicit dependence of the electroweak stability on different parameters of these extended sectors, various kinds of phase diagrams have been presented. Graphical demonstrations have been provided to illustrate how the confidence level, at which stability of electroweak vacuum is excluded, depends on such new physics parameters. This study will help to estimate the lifetime of the electroweak vacuum, especially if it still remains in the metastable state in the extended scalar sectors.

\null
\newpage
\begin{center}
{\Huge\textbf{List of Publications}}\\
\phantomsection
\addcontentsline{toc}{section}{List of Publications}
\printindex
\end{center}
\vspace{28pt}

\begin{flushleft}

A. \underline{Published}:\\

\end{flushleft}

\begin{flushleft}

B. \underline{Under review}:\\
\end{flushleft}

\begin{flushleft}
{\bf N.B}: Entries A1, A2, B2 and B3 are parts of my thesis.
\end{flushleft}

\newpage
\null
\newpage

 \begin{singlespace}
{\hypersetup{linkbordercolor=white}
\tableofcontents
}
{
\hypersetup{linkbordercolor = white}
\listoffigures
}
{
\hypersetup{linkbordercolor = white}
\listoftables
}
\end{singlespace}
 \graphicspath{{ABb/}}

\newpage
{\Huge\textbf{List of abbreviations}}\\
\pagestyle{plain}
\phantomsection
\addcontentsline{toc}{section}{List of abbreviations}
\printindex
\vspace{0pt}
\begp
\allowdisplaybreaks
\bea
\rm SM \hspace{3cm} && \rm Standard~Model\nn\\
\rm BSM \hspace{3cm} && \rm  Beyond~ Standard~Model  \nn\\
\rm VEV \hspace{3cm} && \rm   Vacuum~expectation~value    \nn\\
\rm EW \hspace{3cm} && \rm   Electroweak   \nn\\
\rm EWSB \hspace{3cm} && \rm Electroweak~symmetry~breaking     \nn\\
\rm EWPT \hspace{3cm} && \rm Electroweak~precision~test  \nn\\
\rm LHC \hspace{3cm} && \rm  Large~Hadron~Collider    \nn\\
\rm  ATLAS \hspace{3cm} && \rm   A~Toroidal~LHC~Apparatus    \nn\\
\rm  CMS \hspace{3cm} && \rm   Compact~Muon~Solenoid    \nn\\
\rm LEP \hspace{3cm} && \rm  Large~Electron~Positron~Collider   \nn\\
\rm  CMBR \hspace{3cm} && \rm  Cosmic~Microwave~Background~Radiation \nn\\
\rm WIMP \hspace{3cm} && \rm Weakly~interacting~massive~particles     \nn\\
\rm DM \hspace{3cm} && \rm  Dark~Matter   \nn\\
S\rm \text{-}matrix \hspace{3cm} && \rm   Scattering\text{-}matrix   \nn\\
\rm WMAP \hspace{3cm} && \rm Wilkinson~Microwave~Anisotropy~Probe      \nn\\
{\rm SM+}S \hspace{3cm} && \rm  Standard~Model~with~a~real~singlet~scalar     \nn\\
\rm 2HDM \hspace{3cm} && \rm  Two~Higgs~doublet~model     \nn\\
\rm HTM \hspace{3cm} && \rm   Higgs~triplet~model   \nn\\
\rm IDM \hspace{3cm} && \rm   Inert~doublet~model    \nn\\
\rm  ID\hspace{3cm} && \rm    Inert~doublet   \nn\\
\rm  ITM\hspace{3cm} && \rm    Inert~triplet~model    \nn\\
\rm IT \hspace{3cm} && \rm   Inert~triplet    \nn\\
\rm VBS  \hspace{3cm} && \rm  Vector~boson~scattering     \nn\\
\rm RG  \hspace{3cm} && \rm  Renormalization~group     \nn\\
\rm RGE  \hspace{3cm} && \rm Renormalization~group~equation     \nn\\
\rm  NNLO\hspace{3cm} && \rm  next\textbf{-}to\textbf{-}next\textbf{-}to~leading~order \nn\\
\rm  VC\hspace{3cm} && \rm Veltman’s~ condition\nn
\eea
\eegp

\chapter{Introduction}
\label{chap:Intro}
\linespread{0.1}
\graphicspath{{Chapter1/}}
\pagestyle{headings}
\noindent\rule{15cm}{1.5pt} 
\pagenumbering{arabic}
The unification of electric and magnetic fields in classical electromagnetism was realized by James Clerk Maxwell in 1864.  Similarly, the Standard Model (SM) electroweak theory unifies the electromagnetic and weak forces. This was first proposed by Sheldon Glashow in 1961. Abdus Salam and Steven Weinberg revised the Glashow's electroweak theory by having the masses for $W^\pm$ and $Z$ bosons. The SM describes what matter is made of and how it holds together. The basic ideas are: all matter is made of elementary particles, and these particles interact with each other by exchanging other particles associated with the fundamental forces. 
The SM Lagrangian is designed to respect certain mathematical symmetries. The equations of motion derived from this Lagrangian have enabled physicists to make predictions about various observables which have been successfully tested in particle physics laboratories.
Nearly every quantity that has been measured in particle physics laboratories over the past five decades up to energy of TeV falls right on the predicted value, within experimental error margins. This has made the SM of particle physics one of the best tested and established fundamental theory of the Nature. In the SM, the Higgs mechanism is believed to give rise to the masses of all the elementary particles through spontaneous symmetry breaking of the gauge symmetry. Discovery of a Higgs-like scalar boson at the Large Hadron Collider (LHC) on 4th July 2012~\cite{Aad:2012tfa, Chatrchyan:2012ufa}, confirms that the Higgs mechanism is responsible for electroweak symmetry breaking.
 
Although the SM is one of the most successful theories, it is unable to explain various experimental observations, e.g., it fails to explain non-zero neutrino masses, baryon-antibaryon asymmetry in Nature, mysterious nature of dark matter and dark energy. Moreover the SM does not incorporate the theory of gravitation. Also it is plagued with its own theoretical problems such as the hierarchy problem related to the mass of the Higgs, mass hierarchy and mixing patterns in leptonic and quark sectors etc. The mechanism by which the SM particles get their respective masses, and the deficiencies of the SM from the theoretical as well as experimental points of view, will be discussed in the following section.
\section{The Standard Model}
The standard model is the theory used to describe the fundamental forces (except gravity) and the interactions between fundamental particles. It can explain the three fundamental forces namely the strong, weak and electromagnetic interactions in terms of local gauge symmetries ${ SU(3)_C}$, ${ SU(2)_L}$, and ${ U(1)}_Y$ respectively. The strong interaction between quarks and gluons is governed by ${ SU(3)_C}$. The group ${ SU(2)_L} \otimes { U(1)}_Y$ is responsible for electroweak interactions. The weak force is mediated by the massive $W^\pm$ and $Z$ bosons, whereas a massless vector gauge boson, the photon, mediate the electromagnetic force between electrically charged particles. In the SM, there are three different gauge coupling constants $g_1,~g_2$, and $g_3$, corresponding to these groups which are instrumental in determining the strength of the forces. The gauge coupling constant $g_3$ of group ${ SU(3)_C}$ is large, e.g., $g_3=1.17$ at an energy scale $M_t=173$ GeV. The electromagnetic force is small compared to the strong force due to smallness of $g_1$ and $g_2$, e.g., $g_1=0.36$ and $g_2=0.67$ at energy scale $M_t$. The weak force is even smaller than the electromagnetic force as it is suppressed by a massive gauge boson propagator.
\begin{figure}[ht]
\begin{center}
{
\includegraphics[width=4in,height=2.8in, angle=0]{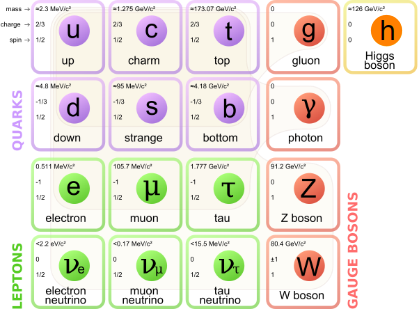}}
\caption{ \textit{The particle spectrum of the SM of elementary particles~\cite{SMImage}.}}
\label{fig:SMparticle}
\end{center}
\end{figure}

In the SM, there are six quarks and six leptons and their antiparticles. Each quark comes with three colors namely red, green, and blue. The SM also includes gauge bosons such as photon, $W^\pm$, $Z$, and gluons and one neutral Higgs field (see Fig.~\ref{fig:SMparticle}). Higgs is the only particle in the SM which is responsible for the masses of all particles.

In the chiral representation, the four-component Dirac fermions ($\psi$) can be split into two separate pieces for the left-handed and right-handed fermions:
\begingroup
\allowdisplaybreaks \beq
\psi_L= \frac{1-\gamma_5}{2}\psi \qquad {\rm and} \qquad \psi_R= \frac{1+\gamma_5}{2}\psi\nn.
\eeq
\endgroup\par
A left-handed $up$-$type$ quark and a left-handed $down$-$type$ quark together form a quark doublet under ${ SU(2)_L}$. Similarly, a left-handed charged lepton and the corresponding left-handed neutrino forms a doublet. Right-handed fermions are singlets under the same group.

The left-handed quark doublet $Q_L$ and lepton doublet $\chi_L$ are denoted as,
\begingroup
\allowdisplaybreaks \beq
Q_L \equiv \begin{pmatrix}
u_{L} \\  d_{L}
\end{pmatrix} \,
\qquad {\rm and} \qquad
\chi_L \equiv \begin{pmatrix}
\nu_{lL} \\  l_{L}
\end{pmatrix}  ,
\label{lepquarlL}
\eeq
\endgroup\par
Right-handed quark singlet $q_R$ and lepton singlet $l_R$ are given by,
\begingroup
\allowdisplaybreaks 
\bea
 q_R=u_{R},~d_{R};~~&{\rm and }&~~ l_{R},\label{lepquarlR}
\eea
\endgroup\par
where $u$ represents the $up$-$type$ quarks of the three generations $u,c,t$; and $d$ stands for the $down$-$type$ quarks $d,s,b$. The charged leptons are denoted by $l = e,\mu,\tau$ with the corresponding left-handed neutrinos $\nu_l = \nu_e,\nu_\mu,\nu_\tau$. Right-handed neutrinos are not included in the SM. 

Also, the SM complex scalar transform as a doublet under ${ SU(2)_L}$ and is given by,
\begp
\allowdisplaybreaks \beq
\phi= \begin{pmatrix}
\phi^+ \\  \phi^0
\end{pmatrix}\nn,
\eeq
\eegp\par
where $\phi^+=\frac{\phi_1+i\phi_2}{\sqrt{2}}$ is the complex charged scalar field and $\phi^0=\frac{\phi_3+ i \phi_4}{\sqrt{2}}$ is the neutral complex scalar field.

In the following, the electroweak part of the SM Lagrangian symmetric under ${ SU(2)_L}\otimes{ U(1)}_Y$ will be presented, which determines electroweak interactions and masses of the particles.
\begin{table}
\begin{center}
{\setlength\tablinesep{3pt}
\setlength\tabcolsep{7pt}
\begin{tabular}{|c|c|c|c|c|c|}
\hline
Field &~~~${ SU(3)_C}$ &~~ ${ SU(2)_L}$ ~& $T^3$ & $U(1)_Y$ & $Q=T^3+\frac{Y}{2}$ \\
\hline
$Q_L=\begin{pmatrix}
u_L \\[2pt]  d_L
\end{pmatrix}$ & {\bf 3} & {\bf 2} & $\begin{pmatrix}
 \frac{1}{2} \\[2pt] -\frac{1}{2} 
\end{pmatrix}$ & $\frac{1}{3}$ & $\begin{pmatrix}
\frac{2}{3} \\[2pt] -\frac{1}{3} 
\end{pmatrix}$ \\
$u_R$ & {\bf 3} & {\bf 1} & 0 & $\frac{4}{3}$ & $\frac{2}{3}$ \\
$d_R$ & {\bf 3} & {\bf 1}& 0 & $-\frac{2}{3}$ & $-\frac{1}{3}$ \\
$L_L=\begin{pmatrix}
\nu_L \\[2pt]  l_L
\end{pmatrix}$ & {\bf 1} & {\bf 2}& $\begin{pmatrix}
 \frac{1}{2} \\[2pt] -\frac{1}{2}
\end{pmatrix}$ & $-1$ & $\begin{pmatrix}
 0\\[2pt]-1  
\end{pmatrix}$ \\
$l_R$ & {\bf 1} & {\bf 1}& 0 & $-2$ & $-1$ \\ 
\hline
$\phi=\begin{pmatrix}
\phi^+ \\  \phi^0
\end{pmatrix}$  & {\bf 1} & {\bf 2} & $\begin{pmatrix}
\frac{1}{2} \\[2pt]  -\frac{1}{2}
\end{pmatrix}$ &  $1$ & $\begin{pmatrix}
1 \\[2pt] 0  
\end{pmatrix}$ \\
\hline
\end{tabular}}
\end{center}
\caption{\em Charges of the SM fermions and scalars. $C$ is the color charge under ${ SU(3)_C}$ group, $T^3$ is the third component of weak isospin of ${ SU(2)_L}$ group, $Y$ is the hypercharge quantum number of ${U(1)_Y}$ group and $Q$ is the electric charge quantum number. \label{table:charges}}
\end{table}
It is given by,
\begingroup
\allowdisplaybreaks
\beq
\mathcal{L}_{SM} = \mathcal{L}_{fermions} + \mathcal{L}_{gauge} + \mathcal{L}_{Higgs} + \mathcal{L}_{Yukawa}.\nn
\eeq
\endgroup\par
The kinetic terms of the fermions and their interactions with gauge bosons can be written as,  
\begingroup
\allowdisplaybreaks \beq
\mathcal{L}_{fermions}= i \bar{Q}_L~\gamma^{\mu} D^L_\mu~ Q_L+i \bar{\chi}_L~\gamma^{\mu} D^L_\mu ~\chi_L  + i\bar{q}_R~\gamma^{\mu} D^R_\mu~ q_R+ i \bar{l}_R~\gamma^{\mu}D^R_\mu ~l_R,
\label{fermionew}
\eeq
\endgroup\par
where the covariant derivative of fermion doublet with left chirality is defined as,
\begingroup
\allowdisplaybreaks \beq
D^L_\mu=\left(\partial_{\mu}+i g_2 T^a W^a_{\mu}+i g_1 \frac{Y}{2} B_{\mu} \right),
\label{coVderi}
\eeq
\endgroup\par
and the covariant derivative of singlet fermion with right chirality is given by,
\begingroup
\allowdisplaybreaks \beq
D^R_\mu=\left(\partial_{\mu}+i g_1 \frac{Y}{2} B_{\mu}\right). 
\eeq
\endgroup\par
The second and third terms of the equation~\ref{coVderi} are related to ${ SU(2)_L}$ and ${ U(1)}_Y$ gauge transformations respectively. $W^{a}_\mu$ ($a$=1,2,3) are the ${ SU(2)_L}$ gauge bosons, corresponding to three generators of ${ SU(2)_L}$ group and $B_\mu$ is the ${ U(1)}_Y$ gauge boson. In the SM, the generators of ${ SU(2)_L}$ are $2\times2$ matrices $ T^a=\frac{1}{2} \tau^a$, where the $\tau^a$, are the Pauli spin matrices,
\begingroup
\allowdisplaybreaks \beq
\tau^1= \begin{pmatrix}
0 & 1\\ 1& 0
\end{pmatrix},
\qquad
\tau^2= \begin{pmatrix}
0 & -i\\ i& 0
\end{pmatrix},
\qquad
\tau^3= \begin{pmatrix}
1 & 0\\ 0& -1
\end{pmatrix}.\nn
\eeq
\endgroup\par
$Y$ is the weak hypercharge operator, generator of ${ U(1)}_Y$ group. The hypercharge operator, defined as a linear combination of the electromagnetic charge operator $Q$ and the third generator $T^3=\frac{\tau^3}{2}$ of ${ SU(2)_L}$, is given by,
\begingroup
\allowdisplaybreaks \beq
Y=2(Q-T^3). \nn
\eeq
\endgroup\par
The third component of weak isospin, $T_3$, hypercharge quantum numbers $Y$ and the electric charge $Q$  of left-handed and right-handed fermions and scalar fields are summarized in Table~\ref{table:charges}.

The gauge part of the Lagrangian contains the kinetic term and interaction term of the gauge bosons and can be written as,
\begingroup
\allowdisplaybreaks
\beq
\mathcal{L}_{gauge}= -\frac{1}{4} {W}_{\mu\nu}^a { W}^{a,\mu\nu}- \frac{1}{4} B_{\mu\nu} B^{\mu\nu}.\nn
\eeq
\endgroup\par
The field strength tensors are defined as, 
\begingroup
\allowdisplaybreaks
\bea
{B}_{\mu\nu}&=&\partial_\mu B_\nu - \partial_\nu B_\mu  ,\nn\\
{ W}_{\mu\nu}^a&=&\partial_\mu W_\nu^a - \partial_\nu W_\mu^a - g_2 \epsilon^{abc} W_{\mu}^b W_{\nu}^c\nn, 
\eea
\endgroup\par
where $\epsilon^{abc}$ is structure constant of $SU(2)_L$ group such that $[T^a,T^b]=i\epsilon^{abc} T^c$.

In the SM, the gauge symmetry prevents us from adding explicit mass terms for gauge bosons. As a result, in the limit of exact symmetry, all gauge bosons are massless.
To incorporate the massive $W^\pm$ and $Z$ bosons into the SM, the Higgs mechanism has been developed to circumvent this constraint on the mass. In this mechanism, the masses of all particles (except neutrinos) are obtained through the spontaneous breaking of the $SU(2)_L \otimes U(1)_Y$ gauge symmetry at the electroweak scale.

The part of the Lagrangian, which gives rise to the masses of the gauge bosons and the Higgs and also to the interaction between the Higgs and the gauge bosons, is given by,
\begingroup
\allowdisplaybreaks \bea
\mathcal{L}_{Higgs} &=& (D^{L,\mu} \phi)^\dagger (D^L_\mu \phi) - V(\phi),
\label{LagHiggs}
\eea
\endgroup\par
where $V(\phi)$ is the SM Higgs potential, and is given by,
\begingroup
\allowdisplaybreaks \bea
 V(\phi) &=& m^2 \phi^\dagger \phi + \lambda (\phi^\dagger \phi)^2,\label{SMSSBpot}\\
 {\rm with,}~~~
 \phi &\equiv& \vector{\phi^+}{\phi^0}= \vector{\frac{\phi_1+i\phi_2}{\sqrt{2}}}{\frac{\phi_3+ i \phi_4}{\sqrt{2}}},\label{PotScalefield}
\eea
\endgroup\par
The electroweak symmetry breaking and how the particles get masses will be discussed in the following.

In the SM the electroweak symmetry breaking (EWSB) is realized by the so-called Higgs mechanism proposed by Robert Brout,  Fran\c{c}ois Englert and 
Peter Higgs~\cite{Higgs:1964pj, Higgs:1966ev, Englert:1964et, Higgs:1964ia}. In this mechanism, the real component $\phi_3$ of the neutral complex scalar of the electroweak doublet acquires a non-vanishing vacuum expectation value (VEV) leading to EWSB. As a result, the gauge group ${SU(2)_L \otimes U(1)_Y}$ is broken down to $U(1)_{EM}$, the symmetry group that corresponds to electromagnetism.

In the SM, the Higgs potential $V(\phi)$, which is responsible for spontaneous symmetry breaking, is given in eqn.~\ref{SMSSBpot},  
\begingroup
\allowdisplaybreaks 
\beq
V(\phi) = m^2 \phi^\dagger \phi + \lambda (\phi^\dagger \phi)^2\nn.
\eeq
\eegp\par
For $\lambda<0$, the potential goes to $-\infty$, i.e., it gets unbounded from below at very high field values. So $\lambda$ is taken to be positive. If $m^2>0$, the minimum of the potential is found at $|\phi|=\sqrt{\left\langle 0| \phi^\dagger \phi |0\right\rangle} = 0$, where $|0\rangle$ represents the ground state. However, the minimum occurs at $|\phi|=\sqrt{\left\langle 0| \phi^\dagger \phi |0\right\rangle} = \sqrt{-\frac{m^2}{2\lambda}}=\frac{v}{\sqrt{2}}$ for $m^2<0$ and $\lambda>0$. In the former case, the symmetry is unbroken while in the latter case symmetry is apparently broken.
\begin{figure}[ht]
\begin{center}
{
\includegraphics[width=2.8in,height=2.in, angle=0]{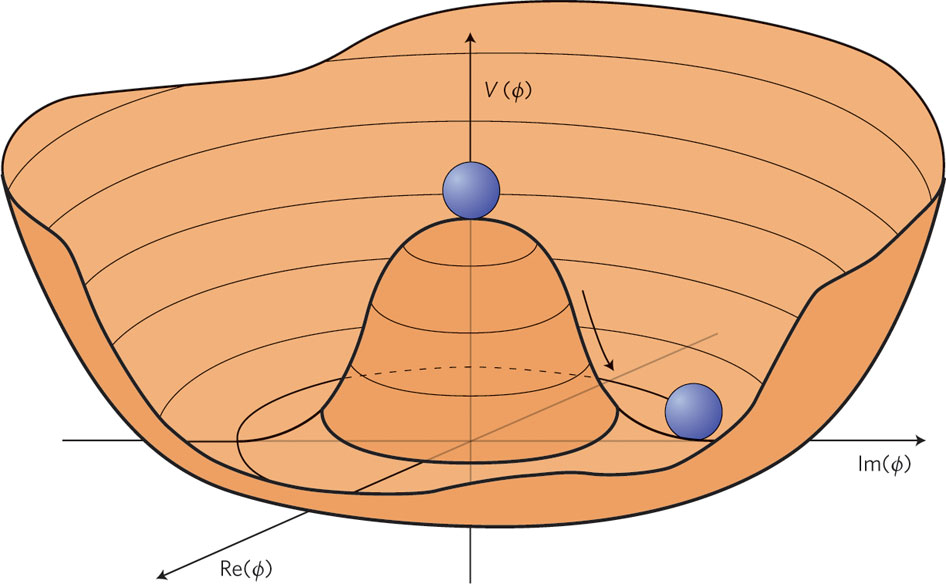}}
\caption{ \textit{Schematic diagram of the Higgs potential. The potential is symmetric about the vertical axis but at the minimum the symmetry is broken. The circular line indicates the remaining U(1) symmetry.}}
\label{fig:SMPotSSB}
\end{center}
\end{figure}

The fields $\phi_1$, $\phi_2$ and $\phi_4$ in eqn.~\ref{PotScalefield} are not physical fields and do not receive any VEV. They can be removed by a particular gauge choice, known as unitary gauge choice. These unphysical fields, known as the Goldstone bosons, are `eaten' by the massless $W^\pm$ and the $Z$ bosons, so that they get massive. In this gauge, the scalar field can be written as,
\begingroup
\allowdisplaybreaks \beq
\phi= \begin{pmatrix}
0 \\  \frac{h+v}{\sqrt{2}}
\end{pmatrix} 
,
\label{SSBphi2}
\eeq
\endgroup\par
where $\phi_3=h+v$, $h$ is the physical Higgs boson. Using the eqns.~\ref{SMSSBpot}, and~\ref{SSBphi2}, the eqn.~\ref{LagHiggs} can be written explicitly as,
\begingroup
\allowdisplaybreaks \bea
\mathcal{L}_{Higgs}&=& \frac{1}{2} \partial_\mu h \partial^\mu h + (h+v)^2 \left( \frac{g_2^2}{4} W_{\mu}^+ W^{\mu-} + \frac{g_1^2 + g_2^2}{8} Z_\mu Z^\mu  \right) \nn\\
&& -\frac{1}{4} \lambda (h+v)^2 ( (h + v)^2 - 2  v^2) + ...
\label{MassWZh}
\eea
\endgroup\par
The charged $W^\pm$ bosons are defined as $W_{\mu}^\pm = \frac{W_{\mu}^1 \mp i W_{\mu}^2}{\sqrt{2}}$. The $Z$ boson and photon are orthogonal combinations of $W_\mu^3$ and $B_\mu$: $Z_\mu= c_W W_\mu^3 - s_W B_\mu$ and $A_\mu= s_W W_\mu^3 + c_W B_\mu$.
$c_W \equiv \cos\theta_W$ and $s_W \equiv \sin\theta_W$, where $\theta_W$ is the Weinberg angle. It can be expressed in terms of the gauge coupling constants as,
\begingroup
\allowdisplaybreaks \beq
\theta_W = \cos^{-1}\left(\frac{g_2}{\sqrt{g_1^2 + g_2^2}}\right).
\label{ThetaWein}
\eeq
\endgroup\par
One can express the electric charge in terms of the gauge coupling constants $g_1$ and $g_2$ as, $e=\frac{g_1 g_2}{\sqrt{g_1^2 +g_2^2}}$, which determines the strength of the electromagnetic interaction. 

The mass terms for $W^\pm$ and $Z$ bosons as well as for the Higgs boson $h$ from eqn.~\ref{MassWZh} can be identified as,
\begingroup
\allowdisplaybreaks \bea
M_W^2 &=& \frac{1}{4} g_2^2 v^2,\nn\\
M_Z^2 &=& \frac{1}{4} (g_1^2 + g_2^2) v^2,\nn\\
M_h^2 &=& 2 \lambda v^2,\nn
\eea
\endgroup\par
where the photon $A_\mu$ remains massless after symmetry breaking, i.e., the vacuum breaks the original symmetry  $ SU(2)_L \times U(1)_Y$ in such a way that only the ${U(1)_{EM}}$ survives as the residual symmetry with a conserved charge $Q=T_3+\frac{Y}{2}$.

Like the gauge bosons, explicitly adding the quark or lepton mass terms ($-m \bar{\psi}\psi$) in eqn.~\ref{fermionew} violates the $SU(2)_L$ gauge symmetry. To retain the gauge invariance, the fermion mass needs to be introduced into the SM after EWSB through the Yukawa interaction between the Higgs field and the fermions fields. The Yukawa part of the SM Lagrangian is given by,
\begingroup
\allowdisplaybreaks \bea
\mathcal{L}_{Yukawa} &=& - \sum_{i,j=1}^3 ( y^{ij}_d\bar{Q}_{iL}~\phi~ d_{jR} + y^{ij}_u \bar{Q}_{iL}~\tilde{\phi} ~u_{jR} + y^{ij}_l\bar{\chi}_{iL}~\phi~ l_{jR})+ h.c.,
\label{LagYuk}
\eea
\endgroup\par
where $\phi$ is the SM complex scalar doublet and $\tilde{\phi}=i \tau_2 \phi^*$, $y_{d=d,s,b}$ are the Yukawa couplings for $down$-$type$ quarks, $y_{u=u,c,t}$ for $up$-$type$ quarks and $y_{l=e,\mu,\tau}$ for charged leptons. $Q_L$ and $\chi_L$ represent the left-handed quark and lepton doublet, $u_R,d_R$ are right-handed singlet quarks and $l_{R}$ are the right-handed singlet charged leptons. Here $i,j$ are the generation indices.

Using eqns.~\ref{lepquarlL},~\ref{lepquarlR} and~\ref{SSBphi2} and substituting them into the eqn.~\ref{LagYuk}, the Yukawa Lagrangian can be written as,
\begingroup
\allowdisplaybreaks \bea
\mathcal{L}_{Yukawa} &=&  \frac{1}{\sqrt{2}}(h+v) ( y^{ij}_d \bar{d}_{iL} d_{jR}+ y^{ij}_u \bar{u}_{iL} u_{jR} +y^{ij}_l \bar{l}_{iL} l_{jR} ) + h.c.
\label{Yukmass}\nn
\eea
\endgroup\par
The mass matrix of the fermions in flavor basis are then given by,
\begingroup
\allowdisplaybreaks \beq
m^{ij}_f = \frac{1}{\sqrt{2}} y^{ij}_f v, \qquad {\rm with},~~~ f=d,u~{\rm and }~l,\nn
\eeq
\endgroup\par
which can be diagonalized via biunitary transformations with $3\times3$ unitary
matrices $V_L^f$ and $V_R^f$ as,
\begingroup
\allowdisplaybreaks \beq
m^{diag}_f = \frac{1}{\sqrt{2}} (V^f_L)_{ki} y^{ij}_f (V^f_R)^\dagger_{jk} v = \frac{1}{\sqrt{2}} y^{diag}_f v \label{femmas},\nn
\eeq
\endgroup\par
and the mass eigenstates of quarks and leptons are defined as,
\begingroup
\allowdisplaybreaks \bea
d_{iL}^\prime &=& (V^d_L)_{ij} d_{jL},~ ~ d^\prime_{iR} = (V^d_R)_{ij} d_{jR},\label{unit3}\\
u^\prime_{iL} &=& (V^u_L)_{ij} u_{jL}, ~~ u^\prime_{iR} = (V^u_R)_{ij} u_{jR},\label{unit2}\\
l^\prime_{iL} &=& (V^l_L)_{ij} l_{jL}, ~~ l^\prime_{iR} = (V^l_R)_{ij} l_{jR}\label{unit1}.
\eea
\endgroup\par
It the absence of right-handed neutrinos in the leptonic sector, the flavor and mass bases are the same for the leptons in the SM. But for the quark sector, both bases are different. One can get from eqns.~\ref{unit3} and~\ref{unit2}, that the quarks of different flavors mix to form the mass eigenstates which take part in flavor changing charged current interactions. Using eqn.~\ref{fermionew}, these interactions can be written as the following Lagrangian,
\begingroup
\allowdisplaybreaks \bea
-\frac{g_2}{\sqrt{2}} (\bar{u^\prime}_L ~ \bar{c^\prime}_L~ \bar{t^\prime}_L)\gamma^\mu W^+_\mu V_{CKM}(d^\prime_L~ s^\prime_L~ b^\prime_L)^T + h.c.,\nn
\eea
\endgroup\par
where $V_{CKM}=V^{u}_{L}V^{d\dagger}_{L}$, stands for the Cabibbo-Kobayashi-Maskawa matrix (CKM)~\cite{Cabibbo:Kobayashi1,Cabibbo:Kobayashi2}. It is parameterized by three mixing angles and a phase and is given by~\cite{Agashe:2014kda},
\begingroup
\allowdisplaybreaks \beq
V_{CKM}=\resizebox{0.7\textwidth}{!}{$\begin{pmatrix}
|V_{ud}|=  0.97425 \pm 0.00022 &  |V_{us}|= 0.2253 \pm 0.0008 & |V_{ub}|= 0.00413 \pm 0.00049 \\
|V_{cd}|=0.225 \pm 0.008 & |V_{cs}|= 0.986 \pm 0.016 & |V_{cb}|=0.0411 \pm 0.0013 \\
|V_{td}|=0.0084 \pm 0.0006 & |V_{ts}|=0.0400 \pm 0.0027 & |V_{tb}| =1.021 \pm 0.032
\end{pmatrix}$}.
\label{VCKM}
\eeq
\endgroup\par
However, the SM does not predict any flavor changing neutral current (FCNC) at the tree level.
\section{Shortcomings of the SM}
Although the SM has been successfully tested at the permill level at LEP, there exist various kinds of theoretical inconsistencies as well as intriguing experimental observations that call for the introduction of new physics beyond the SM. A few of these shortcomings will be briefly discussed in the following subsections.
\subsection{The Higgs mass hierarchy problem}
The mass of the Higgs boson is $125.7\pm0.3$ GeV, as measured at the LHC. The SM cannot predict the Higgs mass, but relates it as $m_h^2=2\lambda v^2$. If the radiative corrections are included for the Higgs mass, then the following problem occurs. The one-loop radiative correction to the Higgs boson mass arising from its self-interaction and couplings with gauge boson and fermions is given by,
\begingroup
\allowdisplaybreaks \bea
\delta m_h^2 &=& \frac{\Lambda_{cut}^2}{16 \pi^2 v^2 }(3 m_h^2 + 3 m_Z^2 + 6 m_W^2 - 12 m_t^2)+ {\rm logarithmic~divergences} \nn\\&&\hspace{0cm}+  {\rm finite~terms}.\nn
\eea
\endgroup\par
If the cut-off scale is set as $ \Lambda_{cut}= \mpl$ then the Higgs mass scale is naturally pushed to the Planck scale. This is known as hierarchy problem for the Higgs mass. In order to bring down $m_h$ to the observed scale $m_{EW}$, a severe fine-tuning between the quadratic radiative corrections and the bare mass (at least one part in $10^{34}$) is required. This level of fine-tuning is deemed unnatural. The new physics beyond the SM, for example, supersymmetry, is needed to solve this problem. In supersymmetric theories, the radiative correction to the Higgs mass arising from SM gauge bosons and fermions are canceled with the correction terms generated by their supersymmetric degrees of freedom. However, so far no experiment could confirm the presence of these superpartners of the known particles.
\subsection{Flavor Problems}
Several unresolved issues are associated with the fermionic sector, termed as flavor problems. There is no explanation of why the SM includes only three generations of fermions. In the SM, by construction, the neutrinos are massless as it does not include right-handed neutrinos. But from the neutrino oscillation experiments we got convinced that at least two neutrinos have non-zero mass. These experiments give information about the mass squared differences between neutrino eigenstates, although the individual values of the masses are not known. From cosmological measurements, it has been shown that the sum of the three neutrino masses is less than $0.1$ eV~\cite{Agashe:2014kda}. As neutrino masses are tiny compared to other fermion masses, it is believed that the mechanism behind neutrino mass generation is different from the other fermions which obtain mass from the Higgs mechanism. The most popular natural explanation of small neutrino masses is via the see-saw mechanism. There are broadly three classes of such models namely type-I, type-II, and type-III see-saw models requiring involvements of right-handed neutrinos, a ${ SU(2)_L}$ triplet scalar with hypercharge $Y=2$ or ${ SU(2)_L}$ hyperchargeless triplet fermions respectively~\cite{Abada:2007ux}. Neutrinos can also get mass from non-seesaw mechanisms namely the Zee mechanism~\cite{Zee:1980ai}, supersymmetry with $R$-parity violation etc.

Fermions have masses in a range of eleven orders of magnitude, from the neutrino mass $\lesssim$ 0.1 eV, up to the top quark mass $\sim$ 173 GeV. A new dynamical mechanism, e.g., a theory with horizontal symmetries is needed to explain the fermion mass hierarchy. The SM does not have an answer to the origin of the structure of the Yukawa matrices and therefore to that of the CKM matrix of eqn.~\ref{VCKM}. Also, as in the SM the neutrinos are massless, the neutrino flavor mixing remains unexplained. The mixing angles of the quarks are small ($\theta_{12}\sim13^\circ$, $\theta_{23}\sim2.4^\circ$ and $\theta_{13}\sim0.2^\circ$)~\cite{Agashe:2014kda}, whereas the mixing angles in the neutrino sector are $\sim32^\circ$, $\sim45^\circ$ and $\sim9^\circ$ respectively~\cite{Agashe:2014kda}. There is no underlying mechanism in the SM to reproduce the observed fermion masses and mixing pattern.
\subsection{Stability of the electroweak vacuum}
With the revelation of the Higgs on 2012 at LHC, the existence of all SM particles have been confirmed and the values of all the parameters in the Lagrangian are known at the electroweak scale. These data indicate that if the validity of the SM is extended to $\mpl$, a second, deeper minimum is located near the Planck scale such that, the electroweak vacuum is metastable, i.e., the transition lifetime of the electroweak vacuum to the deeper minimum is finite $\tau \sim 10^{300}$ years~\cite{Degrassi:2012ry, Buttazzo:2013uya,Khan:2014kba}.

As it was shown in Refs.~\cite{Masina:2011aa,Isidori:2007vm,Bezrukov:2007ep,Kamada:2012se} that if the electroweak vacuum is metastable then Higgs cannot play the role of inflaton. Extra new degrees of freedom are needed with the SM to explain the inflation of the Universe. However, in this thesis, the issues related to inflation will not be addressed. New physics beyond the SM can alter the Higgs potential so that it enhances its stability.
\subsection{Gravity}
Massive particles experience gravitational interactions. Gravity is expected to dominate over all other fundamental forces at energy scale $\gtrsim \mpl$. However, construction of models based on quantum gravity theories is outside the scope of the SM.
\subsection{Gauge Unification}
In the electromagnetism, $E$ and $B$ are unified, as with a single electric charge one can explain any electromagnetic phenomenon. Hence from a theoretical standpoint, a theory which can illustrate unification of all the fundamental forces at higher energies would be an appealing one. The gauge coupling constants $g_1$, $g_2$ and $g_3$ corresponding to the three gauge groups in the SM, namely ${ U(1)}_Y$, ${ SU(2)_L}$ and ${ SU(3)_C}$, assume different values at the electroweak scale. Assuming SM to be valid all the way to the Planck scale, one can evolve these coupling constants using RG equations. Given the accuracy involved in measured values of the SM parameters at the electroweak scale, $g_1$, $g_2$ and $g_3$ do not merge to a single coupling constant at any energy before $\mpl$, which is desirable to attain gauge unification. One can take refuge to grand unified theories based on a single gauge group like ${ SU(5)}$ or ${ SO(10)}$ to realize this dream.
\subsection{Dark matter}
Various kinds of astrophysical observations, e.g., anomalies in the galactic rotation curves, gravitational lensing effects in Bullet cluster etc., have indicated the existence of dark matter in the Universe. Dark matter interacts gravitationally. It has no electric charge, so we cannot observe it through its interactions with photons. The satellite based experiments such as Wilkinson Microwave Anisotropy Probe (WMAP)~\cite{Bennett:2012zja} and Planck~\cite{Ade:2013zuv} have measured the Cosmic Microwave Background Radiation (CMBR) of the Universe with unprecedented accuracy and suggest that the Universe consists of about 4$\%$ ordinary matter, 27$\%$ dark matter and the rest 69$\%$ is a mysterious unknown energy called dark energy which is thought to be the cause of accelerated expansion of the Universe. The SM fails to provide a viable dark matter candidate. To explain the observed presence of the dark matter, the presence of new physics beyond the SM is required. A popular scheme in this regard is to introduce `weakly interacting massive particles' (WIMP) protected by a discrete symmetry that ensures stability of these particles. The issues related to the dark matter in various extended scalar sectors will be discussed in Chapter~\ref{chap:MetaSMextended}.
\subsection{Baryon-antibaryon asymmetry}
Astrophysical evidences have indicated that our galaxy and its neighborhood are predominantly made of matter. The asymmetry is defined as the ratio of the difference of the baryon and anti-baryon number densities to the photon number densities in the Universe. According to recent data from WMAP\cite{Bennett:2012zja} the asymmetry is given by,
\begingroup
\allowdisplaybreaks \bea
\eta= (6.19\pm0.14)\times 10^{-10}.\nn
\eea
\endgroup\par
To produce this baryon asymmetry, three ingredients are necessary as outlined by Sakharov conditions~\cite{Sakharov:1967dj}, such as baryon number violation, C and CP violation, and departure from thermal equilibrium.

Although the SM has the required ingredients, it cannot produce large enough baryon-to-photon ratio ($\eta$) as observed by the experiments. The new physics beyond the SM is  required as one needs additional sources of $CP$ violation. Several mechanisms have been proposed to explain the baryon asymmetry, e.g., electroweak baryogenesis, leptogenesis and GUT baryogenesis etc. 
\subsection{Recent LHC data and scope for an extended scalar sector}
\label{Recendatanewphysics}
LHC-I, with 20\, fb$^{-1}$ data in its $\sqrt s=8$ TeV run, has confirmed discovery of the Higgs boson of mass $M_h\sim125$ GeV. The Higgs signal strengths are shown in the Table~\ref{table1LHC} as measured from ATLAS and CMS collaborations at the LHC. The measured properties of this Higgs boson are consistent with the minimal choice of the scalar sector as in the SM. But the data still comfortably allow an extended scalar sector, which, in turn, can accommodate a more elaborate mechanism for EWSB. It is also noted that an extended scalar sector may not always participate in the electroweak symmetry breaking, e.g., as in the case of an extended inert scalar sector protected by a discrete symmetry.
\begin{table}
\begin{center}
  {
    \begin{tabular}{ | c | c | c |}
         \hline
     Signal strength, $\mu_{\chi \chi}=\frac{\Gamma_{exp}(h\ra \chi \chi)}{\Gamma_{SM}(h\ra \chi \chi)}$ &  ${\rm ATLAS}$ & ${\rm CMS}$  \\
     \hline
     $\mu_ {\bar{\tau} \tau}$ & ~$1.41\pm0.40$~ & ~$0.89\pm0.3$~   \\
     \hline
     $\mu_ {W W^*}$  & $1.23\pm0.22$ & $0.91\pm0.23$  \\
     \hline
     $\mu_ {Z Z^*}$  & $1.5\pm0.35$ & $1.05\pm0.31$ \\
     \hline
     $\mu_ {\gamma\gamma}$  & $1.15\pm0.27$ & $1.12\pm0.23$ \\
     \hline
     \end{tabular}}
     \end{center}
    \caption{ \textit{ The observed signal strengths and uncertainties for different Higgs boson decay channels~{\rm\cite{CMS:2015kwa}} for mass, $M_h\sim125$ GeV. The data have been taken for proton-proton collision of center-of-mass energies, $\sqrt{s}=7,~8$ TeV with luminosity, 20 $fb^{-1}$. $W^*,~Z^*$ decay to fermions. $\Gamma_{exp}(h\ra \chi \chi)$ is the decay width as observed in the experiments and $\Gamma_{SM}(h\ra \chi \chi)$ is the same predicted by the SM.} }
    \label{table1LHC}
\end{table}

Recently the ATLAS and CMS collaborations have analyzed the $\sqrt{s}=13$ TeV data and announced the possible presence of a scalar resonance around 750 GeV~\cite{newgamresonance}. The significances of the signals are $3.9\sigma$ and $3.4\sigma$ in the respective experiments. This indication has fuelled speculations about an extended scalar sector.

In the following section, the particle content of the extended scalar sectors and the constraints on these new models will be discussed.
\section{Extended scalar sector}
In this thesis, various kinds of extended scalar sectors containing an additional ${ SU(2)_L}$ singlet, doublet or triplet with hypercharges $Y=0,2$, have been considered along with the SM doublet. The main purpose for writing this thesis is to explore characteristics of these new scalars using the weak vector boson scattering processes and from the (meta)stability of the scalar potential.

Each extended scalar sector can consist of different ${ SU(2)_L}$ multiplets with different isospin, hypercharge, etc. So the structure of the scalar potential is changes in different extended scalar sectors. The SM doublet $\phi$, given in eqn.~\ref{PotScalefield}, with isospin $I=1/2$ and hypercharge $Y=1$, consists of two singly-charged, a neutral $CP$-even and a neutral $CP$-odd scalar fields respectively. Depending on the isospin $I_i$ and hypercharge $Y_i$, other scalar multiplets contain different numbers of scalar fields with different charges. For example, a real triplet scalar with hypercharge $Y=0$ consists of a pair of singly-charged fields and a $CP$-even neutral scalar field, whereas a complex triplet scalar with hypercharge $Y=2$, has an extra doubly-charged fields, and a $CP$-odd neutral scalar field. After the electroweak symmetry breaking, the fields of the SM scalar doublet mix with the new fields of extra scalar sector. In the extended scalar sector, one of the linear combinations of the $CP$-even states is identified as the observed Higgs, whereas the other combinations become new physical $CP$-even scalars. Similarly, one of the combinations of singly-charged states forms a charged Goldstone boson. This is eaten by the electroweak charged gauge boson $W$ which becomes massive.
The other combinations become new singly-charged physical scalars. Also, a combination of $CP$-odd scalar states becomes the component of the $Z$ boson and the other combinations of these become physical pseudoscalars.
But the fields with electric charge quantum number more than one, remain as physical scalar fields.
These scalar fields can have direct couplings with the SM particles, or these may get generated after the electroweak symmetry breaking.
If the extra scalar fields are $odd$ under a discrete $Z_2$ symmetry, then the SM particles cannot couple with $odd$ number of the new scalar fields.
This symmetry prevents the extra scalars to acquire VEVs, and hence these extra scalars do not mix with the fields of the SM scalar doublet.
In this case, the electroweak symmetry breaking is fully driven by the SM Higgs doublet. The charged and $CP$-odd components of the SM doublet remain as Goldstone bosons and $CP$-even neutral component is identified with the observed Higgs.
\subsection{Constraints on the extended scalar sectors}
From experimental and theoretical considerations such as the electroweak precision experiments, absolute vacuum stability of the scalar potential and unitarity of the scattering matrix, one can put constraints on parameter space of the extended scalar sectors. In the following sections, some of these constraints are discussed.
\subsubsection{The electroweak precision constraints}\label{EWPTcons}
The electroweak precision constraints on the physics parameter spaces come from the precise measurement of the neutral current and charged interaction at the Large Electron–Positron Collider~\cite{ALEPH:2005ab} and the Stanford Linear Collider~\cite{Abe:2000dq}, etc. The relative strength of the neutral and charge currents denoted by $\rho$ can be expressed as,
\begingroup
\allowdisplaybreaks \beq
\rho = \frac{M_W^2}{M_Z^2 \cos \theta_W^2}.
\label{rhoparam}\nn
\eeq
\endgroup\par
In the SM, at tree-level, the value of the $\rho$ parameter is equal to 1. If an extra scalar multiplet is added along with the SM scalar doublet, then the $\rho$ parameter may be modified at the tree-level. These modifications are different for different types of multiplets. If the extra scalar is odd under any discrete symmetry under which the SM particles are even, then $\rho$ parameter remains the same as in the SM. Note that, it will differ when radiative corrections are included into the vector boson self-energies. Let us now concentrate at tree-level only.

It was shown in the Ref.~\cite{Lee:1972gj}, that the $\rho$ parameter can be easily calculable for any number of scalar multiplets present in the model. Let the vacuum expectation values of scalars are $v_i$, where $i$ is the number of multiplets, with the ${ SU(2)}_L$ isospin $I_i$ and hypercharge $Y_i$. Masses of the gauge bosons corresponding to the neutral and charge current processes are expressed as,
\begingroup
\allowdisplaybreaks \bea
M_Z^2 &=&  \frac{1}{4} (g_1^2 + g_2^2) \sum_i |Y_i|^2 v_i^2, \nn\\
M_{W^\pm}^2 &=&  \frac{1}{8} g_2^2 \sum_i \lbrace 4 I_i (I_i +1) - |Y_i|^2\rbrace v_i^2.\nn
\eea
\endgroup\par
In terms of $v_i,~I_i$ and $Y_i$, the $\rho$ parameter can be written as,
\begingroup
\allowdisplaybreaks \beq
\rho = \frac{\sum_i \lbrace 4 I_i (I_i +1) - |Y_i|^2 \rbrace v_i^2}{2~\sum_i |Y_i|^2 v_i^2}.\nn
\eeq
\endgroup\par
In the singlet scalar extension of the SM, both isospin ($I$) and hypercharge ($Y$) of the extra scalar are zero. So there is no modification of the masses of gauge bosons ($W^\pm,~Z$) and the same is true for $\rho$ at the tree-level. Although, in the doublet extension, with hypercharge 1, the masses of the vector bosons get modified as $M_Z^2=\frac{1}{4} (g_1^2 + g_2^2)(v_1^2+ v_2^2)$  and $M_W^2=\frac{1}{4} g_2^2(v_1^2+ v_2^2)$. Then the $\rho$ parameter for this model does not deviate from the SM at the tree-level. However, for the triplet extension of the SM, the $\rho$ parameter can be written as,
\begingroup
\allowdisplaybreaks \bea
\rho({ Y=0}) &=& \frac{v_1^2 + 2 v_2^2}{v_1^2},\nn\\
\rho({ Y=2}) &=& \frac{v_1^2 + 2 v_2^2}{v_1^2 + 4 v_2^2}.\nn
\eea
\begingroup\par
The recent experimental fit of $\rho$ parameter is $1.0004\pm 0.00024$~\cite{Agashe:2014kda}. This constraints the vacuum expectation value $v_2$ of the extra triplet scalar to be less than 4 GeV~\cite{NajimWWscat}.

Although in an extended scalar sector, the $\rho$ parameter may or may not get additional contribution in comparison to the SM, 
at loop level it is not immune to get corrected. The new physics effects can alter the vacuum polarization of the gauge bosons. These effects are encoded in the so-called Peskin–Takeuchi~\cite{Peskin:1991sw} parameters and are denoted by $S,~T$, and $U$. Alternatively these can also be expressed in terms of Altarelli-Barbieri~\cite{Altarelli:1990zd} parameters $\epsilon_1,~\epsilon_2$ and $\epsilon_3$. The correction to the $\rho$ parameter is related with $T$ as, $\alpha T = \Delta \rho$, where $\alpha$ is the fine structure constant. The generic expressions of these observables for the extended scalar sectors are given in Ref.~\cite{Lavoura:1993nq}. These electroweak precision observables~\cite{Baak:2014ora} are used in this thesis to put constraints on parameter space of new physics models.
\subsubsection{Stability of the electroweak vacuum}\label{stabofscalar}
The extended scalar sectors may have more than one scalar field in the potential. Due to the complicated structures of such scalar potentials, it is difficult to adjudge the stability of the minimum. A scalar potential is said to be bounded from below, if and only if the potential does not become negative infinity along any direction of the field space. 

To find the boundedness of the potential one has to use the copositiviy criteria of a symmetric matrix. A symmetric matrix ${\cal M}$ is copositive if the quadratic form $\chi^T {\cal M} ~\chi > 0$ for all vector $\chi > 0$, The notation $\chi > 0$ means that all the components of the vector $\chi$ are greater than zero. These criteria guarantee that the potential never becomes negative infinity and the vacuum is absolutely stable. 

Let us consider a scalar potential,
\begingroup
\allowdisplaybreaks \beq
V(\phi_i)=  \sum_{i,j}^N m_{ij}^2 \phi_i \phi_j +\sum_{i,j,k,l}^N \lambda_{ijkl} \phi_i \phi_j \phi_k \phi_l.\nn
\eeq
\endgroup\par
For high field values, the first term can be neglected. The coefficient of the terms of the potential which contains $odd$ number of fields, are taken to be zero. So, one can write the potential in biquadratic in fields  and can apply the copositivity criteria to find the conditions for the boundedness of the scalar potential. The scalar potential is given by, 
\begingroup
\allowdisplaybreaks \beq
V(\phi_i)=  \sum_{i,j}^N \lambda_{ij} |\phi_i |^2 |\phi_j |^2. \nn
\eeq
\endgroup\par
This potential can be written as in the form of a symmetric matrix $\chi^T {\cal M} ~\chi$, where $\chi$ consists of $|\phi_i |^2$ and ${\cal M}$ contains $\lambda_{ij}$, here $\lambda_{ij}$ is symmetric under exchange of the indices. The scalar potential is bounded from below if~\cite{Chakrabortty:2013mha},
\begingroup
\allowdisplaybreaks \beq
\lambda_{ii}>0, ~~~ \lambda_{ij} +\sqrt{\lambda_{ii} \lambda_{jj}} >0 ~~~~~~~{\rm for}~~ i,j=1,2.\nn
\eeq
\endgroup\par
However, the copositivity criteria are no longer valid when the scalar potential become negative and has extra new minima along any direction of the field space. In this case a minimum located at low field values is said to be metastable if the transition time from this minimum to any other minima of the potential is greater than the lifetime of the Universe, otherwise, it is unstable. The modification of copositivity criteria with metastable electroweak vacuum for the extended scalar sector will be discussed in Chapter~\ref{chap:MetaSMextended}.
\subsubsection{Unitarity of the scattering matrix}
\label{uniscalarg}
Unitarity bound on the extended scalar sectors can be calculated from the scattering-matrix of different processes. The technique was developed by Benjamin W. Lee, C. Quigg, and H. B. Thacker for the SM and it can also be applied to various kinds of extended scalar sectors. The scattering-matrix for the extended scalar sector consists of different scalar-scalar, gauge boson-gauge boson, gauge boson-scalar scattering amplitudes. Using the Born approximation for partial waves, the scattering cross-section for any process can be written as,
\begingroup
\allowdisplaybreaks \beq
\sigma = \frac{16 \pi}{s} \sum_l^\infty ~ (2 l + 1) |a_l(s)|^2,\nn
\eeq
\endgroup\par
where $s=4 E_{CM}^2$ is the Mandelstam variable, where $E_{CM}$ is the center of mass energy of the incoming particles. $a_l$ is the partial wave coefficients corresponding to specific angular momentum values $l$. This leads to the following unitarity constraint, $Re(a_l) < \frac{1}{2}$. At high energy the dominant contribution to the $a_l$, i.e., to the amplitude of the two-body scattering processes $a,b \ra c,d$ comes from the diagram involving the quartic couplings. Far away from the resonance, the other contributions to the amplitude from the scalar mediated $s$-, $t$-, and $u$-channel processes are negligibly small. Also in the high energy limit, the amplitude of scattering process involving longitudinal gauge bosons can be approximated by the scalar amplitude in which gauge bosons are replaced by their corresponding Goldstone bosons. This is known as equivalence theorem~\cite{equiv}. So to test unitarity of the models with extended scalar sectors, one can construct the scattering-matrix which consists of only the quartic couplings at very high energies. Unitary constraints demand that the eigenvalues of the scattering-matrix should be less than $8\pi$. The detailed calculations will be shown for Higgs triplet model with a hyperchargeless scalar triplet in Chapter~\ref{chap:EWSBextended}.

\vspace{0.1cm}
From the above discussions in Sections~\ref{stabofscalar} and \ref{uniscalarg}, it is clear that the requirement of stability of the scalar potential gives a lower bound on the quartic coupling, whereas unitarity of scattering-matrix gives the upper bound on the quartic coupling of the extended scalar sector.

These are some of the generic constraints on the extended scalar sector beyond the SM. Additional bounds on such new physics models can be put from other phenomenological studies of dark matter, neutrino mass etc. The parameter spaces of these new models are constrained from non-observation of these scalar signals at the direct search experiments at high-energetic colliders like LEP, LHC, etc. The direct search bound on the 2HDM, HTM($Y=0$), and HTM($Y=2$) will be discussed in the Chapter~\ref{chap:EWSBextended}.
\section{Organization of the Thesis}
 The organization of the thesis is as follows:

This chapter starts with a brief introduction to the standard model of particle physics. After that, the deficiencies of the SM have been discussed from the theoretical as well as experimental points of view. Generic bounds on the extended scalar sectors from the stability of the electroweak vacuum, unitarity of the scattering matrix and the electroweak precision experiments have been reviewed. Compositions of the physical and unphysical scalar fields in the extended scalar sectors have also been discussed.

An extra scalar sector protected by a discrete symmetry $Z_2$ can solve the puzzle of nature of dark matter in the Universe. In the second Chapter, a brief summary of the indications of presence of the dark matter from the astrophysical observations have been discussed. The calculation of relic density of the dark matter has been reviewed. The direct and indirect detection of the dark matter in experiments has also been mentioned in brief.
 
In the third Chapter, various kinds of extended scalar sector such as singlet extension of the SM, type-II two Higgs doublet model and Higgs triplet model with two different hypercharges $Y=0~{\rm and}~2$, have been reviewed. After electroweak symmetry breaking, the masses of all physical particles have been calculated in these new models. In the presence of a new extended scalar sector, the modified couplings of the Higgs-like scalar to other SM particles have been shown in terms of the mixing angles of the new extended scalar fields to the fields of SM doublet. The previously measured generic bounds on the extended scalar sectors have been applied on these models. 

The extended scalar sector can be probed through longitudinal vector gauge boson scattering and it has been discussed in the fourth Chapter. The generic expressions of the longitudinal vector gauge boson scattering amplitude of the processes like ($a$)~ $W_L^\pm  W_L^\mp \ra W_L^\pm  W_L^\mp$, ($b$) $W_L^\pm  W_L^\pm \ra W_L^\pm  W_L^\pm$, ($c$) $W_L^\pm  W_L^\mp \ra Z_L  Z_L$, ($d$) $W_L^\pm  Z_L \ra W_L^\pm  Z_L$, and ($e$) $Z_L  Z_L \ra Z_L  Z_L$ have been presented in the context of extended scalar sectors such as type-II two Higgs doublet model, and Higgs triplet models with $Y=0$ and $Y=2$ scalar triplets~\cite{NajimWWscat}. These scattering processes can be useful to distinguish these extended scalar sectors from the SM as well as between one another.

In the fifth Chapter, the stability of the electroweak vacuum of the Higgs scalar potential in the SM has been reviewed. It has been shown how the Higgs scalar potential evolves with the running energy from the electroweak scale to the Planck scale $M_{Pl}$. In this study, the Higgs scalar potential~\cite{Degrassi:2012ry} up to two-loop quantum corrections is used and it has been improved by three-loop renormalization group running of the coupling parameters. As a part of this thesis work a computer code has been developed to analyze a scalar potential at higher energies. In the SM, the formation of the second, deeper minimum near the Planck scale has been discussed. The detailed calculation of tunneling probability of the electroweak minimum to the deeper minimum around the Planck scale is also provided. It has been shown that the present measured values of the SM parameters imply that the stability of electroweak minimum is excluded at $\sim3\sigma$.
 
As the SM is extended by additional scalar multiplets, the stability of the electroweak minimum of the Higgs potential gets improved. The extended scalar sectors with a discrete symmetry $Z_2$ can also provide a viable dark matter candidate. In the sixth Chapter, the effective Higgs potential has been calculated to explore the stability of the electroweak vacuum in these extended scalar sectors. The contributions to the effective Higgs potential from these new physics models have been taken at one-loop level only. In this scenario, the stability of the electroweak minimum of the new effective Higgs potential in the different extensions of the scalar sector of the SM, namely with a real singlet~\cite{Khan:2014kba}, inert doublet~\cite{Khan:2015ipa} or an inert triplet with hypercharge $Y = 0$~\cite{NajimTrip}, has been explored. In these extended sectors, parameter spaces have been identified that correspond to the stable and metastable electroweak vacuum and also satisfy the relic density constraints on dark matter from Planck experiment. The modified stability conditions have also been shown when electroweak vacuum is metastable.
 
The last Chapter provides the summary and the conclusions of this thesis work.
\chapter{Dark matter: Relic Density, Direct and Indirect detections}
\label{chap:RelicDark}
\linespread{0.1}
\graphicspath{{Chapter2/}}
\pagestyle{headings}
\noindent\rule{15cm}{1.5pt}
\section{Introduction}
The last few years have seen a revolution in cosmology and astrophysics. It is confirmed that the Universe is filled with not only dark matter (DM) but also the even more enigmatic dark energy. 

The analysis of observational data of the satellite-based experiments, Wilkinson Microwave Anisotropy Probe (WMAP)~\cite{Bennett:2012zja} and Planck~\cite{Ade:2013zuv} that look for such very tiny anisotropies in Cosmic Microwave Background Radiation (CMBR) suggest that the total mass-energy of the Universe contains $69\%$ dark energy, $27\%$ dark matter and $4\%$ ordinary matter. The nature of these two dominant components of the Universe is currently one of the biggest mysteries in the  modern particle and astrophysics. Dark matter is a hypothetical type of matter, first proposed by Fritz Zwicky in 1933. Dark matter is mysterious and invisible in nature as it can neither be seen by our eyes nor observed by any telescope. Dark matter is electrically neutral $\textendash$ it neither emits nor absorbs any electromagnetic radiation. If the dark matter would have electromagnetic and strong interactions with normal matter, it would have formed isotopes~\cite{nbynH} of estimated abundance ${n} > 10^{-10} ~{n_H}$, which contradicts the present upper limit of hydrogen isotopes abundance $n_H$. The dark matter has mass so it can interact with the ordinary matter through the gravitational interaction. 

There are many convincing evidences about the existence of the dark matter in the Universe. A few of these will be discussed in the next section.

\section{Evidence of Dark Matter}
There are several observations to support the argument that the Universe contains a large amount of matter. In the year 1933, Fritz Zwicky was the first astronomer to propose the existence of dark matter. He was studying a very large cluster of galaxies nearest to the Earth: the Coma cluster. He used the Virial theorem, an equation which relates the average kinetic energy of a system to its total potential energy, to determine the gravitational mass of the cluster. He measured the total mass of the luminous object (stars and gas) in the galaxies to find that the mass of the luminous matter was not enough to keep the cluster bound, and was several times smaller than the inferred gravitational mass. Zwicky concluded that there must be non-luminous matter present in the galaxies.
\subsection{Galactic Rotation Curves}
Similar discrepancy was observed in galactic rotation curves, which is the strongest evidence of the presence of the dark matter in the galaxies. In the 1970s, Kent Ford and Vera Rubin discovered that rotation curves of galaxies are flat (see Fig.~\ref{grc}).
For the analysis of a spiral galaxy, one measures the rotational velocity $v(r)$ of stars, gas etc. in the galaxy as a function of their distance $r$ from the galactic center.
The spiral galaxy has a dense central region and the density of the visible mass is reduced as one goes away from the central region. From standard Newtonian dynamics, one would expect a Keplerian decline of the rotation curve as one goes away from the dense central region of the galaxy. The observations show that instead of Keplerian decline the velocities are rather constant after a certain distance $r$ from the galactic center as in Fig.~\ref{grc}.
The flat rotation curves have now been observed for almost all galaxies, including our galaxy, the Milky Way. If the galaxy contains far more unknown mass than the luminous object, then this flat rotation curve can be explained.
It was found that more than 95$\%$ of the mass of galaxies consists of unknown dark matter.
\begin{figure}[ht]
\begin{center}
{
\includegraphics[width=2.7in,height=2.5in, angle=0]{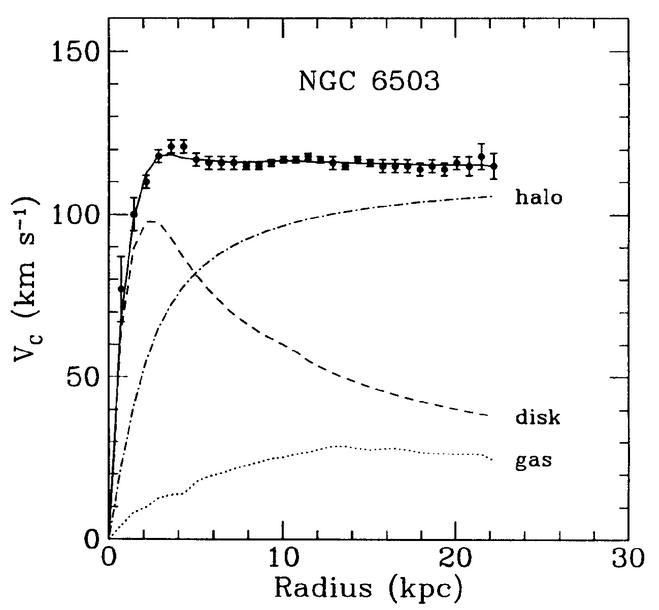}}
\caption{\textit{Galactic rotation curve for NGC 6503 dwarf spiral galaxy. Image credit: Katherine Freese~\cite{Freese:2008cz}.}}
\label{grc}
\end{center}
\end{figure}
\subsection{Gravitational Lensing and Bullet Cluster}
 \begin{figure}[ht]
 \begin{center}
 \subfigure[]{
 \includegraphics[width=2.7in,height=2.7in, angle=0]{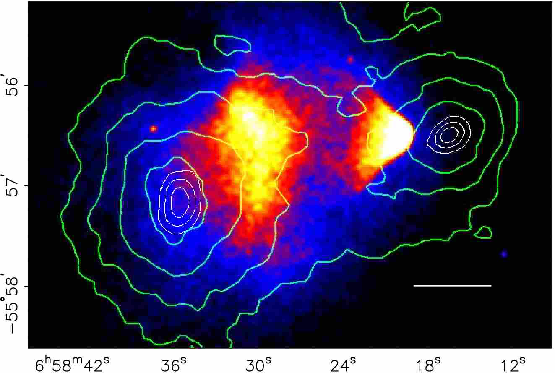}}
 \subfigure[]{
 \includegraphics[width=2.7in,height=2.7in, angle=0]{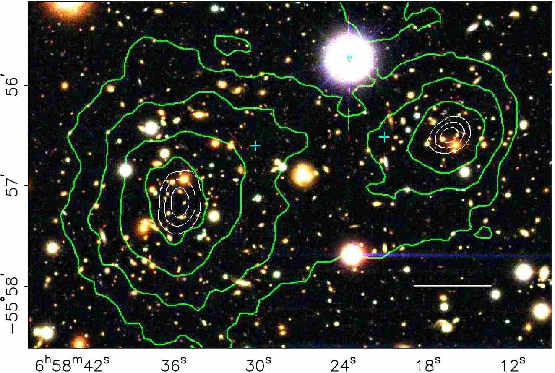}}
 \caption{\label{figBullet} \textit{$(a)$ the left panel is a X-ray satellite Chandra image of the cluster of the visible matter and $(b)$ The right panel shows a direct image of the cluster obtained with the 6.5-m Magellan telescope in the Las Campanas Observatory. The contour lines are drown for the images produces due to the gravitational lensing by the dark matter in the galaxies. Image credit: D. Clowe et al.~\cite{Clowe:2006eq}. } }
 \end{center}
 \end{figure}
The bending of light while it passes through the vicinity of a gravitating mass gives rise to the lensing effect. This phenomenon is known as gravitational lensing.
If a luminous object is present in the background of the gravitating mass at a suitable distance, then the lensing effect will create a distorted or multiple images.
The observance of such lensing effects in the galaxies by unseen matter indicates the presence of dark matter.
The huge amount of unseen dark matter present in the galaxies will produce ring images (Einstein's rings) through strong gravitational lensing effect. Weak lensing by smaller astronomical objects such as planets, stars, will produce distorted images. From the lensing effect, the unseen dark matter mass present in the galaxies can be estimated using the lens equation of Einstein's general theory of relativity. Weak and strong gravitational lensing phenomena have been used for discovering one of the most prolific evidence of dark matter in the ``Bullet Cluster'' in the cluster 1E0657-558. The bullet cluster was created due the collisions of two giant galaxies in the Universe. When the smaller galaxy passed through the core of the larger galaxy, the baryonic mass distribution of the smaller one suffered distortion in shape due to the enormity of the collision and it took the shape of a bullet. The X-ray analyses reveal the baryonic mass distribution in the two colliding clusters. The X-ray image of baryonic distributions are shown in Fig.~\ref{figBullet}(a).
The colors indicate the X-ray temperature of the plasma: blue is coolest and white is hottest. Direct image of the cluster obtained with the 6.5-m Magellan telescope at the Las Campanas Observatory is shown in Fig.~\ref{figBullet}(b). The green contour line in Fig.~\ref{figBullet}(a) and (b) form due to the gravitational lensing of unknown matter (dark matter) presence in the galaxies. One can see from these figures that the dominant population of the baryonic mass is in X-ray gas which is well separated from the respective dark matter halo of the cluster.
This analysis indicates the dark matter halos in the galaxies, when passed through each other, remain unperturbed and undistorted. The phenomenon of the ``Bullet Cluster'' provides an observational evidence of the existence of dark matter in the galaxies and indicates that the dark matter is almost collisionless.
\subsection{The Cosmic Microwave Background}
\begin{figure}[ht]
\begin{center}
{
\includegraphics[width=2.7in,height=2.5in, angle=0]{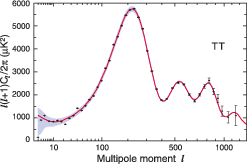}}
\caption{\textit{ The figure shows temperature-temperature (TT) angular power spectrum of CMBR from the nine years WMAP data. Image credit: C. L. Bennett et al. [WMAP Collaboration]~\cite{Bennett:2012zja}.}}
\label{CMBR}
\end{center}
\end{figure}
The Cosmic Microwave Background Radiation (CMBR) provides a snapshot of the oldest light after the `last scattering' in our Universe.
The last scattering took place when the Universe was just 380,000 years old. Hence, the CMBR is the earliest photograph of our Universe.
The fluctuations in CMBR provide the information of the structure formations of the Universe.
The anisotropies in the CMBR are capable of telling us the geometry (or curvature), the baryon content, the dark matter and dark energy content of the Universe, the value of the Hubble constant, whether inflation occurred in the early Universe, etc. In Fig.~\ref{CMBR}, the correlation function $C_l$ of temperature anisotropy of CMBR, plotted with respect to multipoles ($l$) shows multiple peak at different $l$, where $\theta=\frac{2\pi}{l}$ is the angle between two directions.
This temperature anisotropy is useful to extract the physical properties of our Universe. The angular scale corresponding to the first peak in Fig.~\ref{CMBR} provides the precise measurement of the curvature of the Universe.
The third peak contains the information about the dark matter density in the Universe. From the ratio of the odd peaks to the even peaks, one can get information about the baryon density in the Universe.
\section{Components of Dark Matter}
So far, astrophysical and cosmological data can only tell us how much dark matter is there in the Universe, i.e., the total mass density of the dark matter and the fact that it does not interact electromagnetically and strongly. Although we are convinced that dark matter really exists, there is still no consensus on what it is composed of. The possibilities include the dense baryonic matter and the non-baryonic matter.

There are baryonic components of dark matter like Massive Compact Halo Objects (MACHOs): condensed objects such as black holes, white dwarfs, very faint stars, or non-luminous objects like planets. Using the microlensing of the light, the MACHO~\cite{Alcock:1998fx} and the EROS~\cite{Tisserand:2006zx} collaborations, concluded that the MACHOs could add a few percent to the known mass discrepancy in the Galaxy halo, observed in galactic rotation curves. So if there is a baryonic component to dark matter, then it has to be quite small.

The non-baryonic dark matter particles can be grouped into three categories on the basis of their velocities, namely hot dark matter (HDM), warm dark matter (WDM), and cold dark matter (CDM). The hot dark matter mass is very small and moves almost with the speed of the light.
The neutrinos can be considered as the best candidate for the hot dark matter. There are several reasons that the neutrino cannot be considered as a viable dark matter candidate which alone fulfill the relic abundance of the Universe.
There is an upper limit on neutrino mass from tritium beta decay experiments $m_\nu<2$ eV~\cite{Doe:2013jfe}.
Combined with the data from the neutrino oscillation experiments, one can find the contributions of neutrino to the matter density is $\Omega_{\nu}h^2<0.064$~\cite{Murayama:2007ek} in the Universe.
This is not enough to satisfy the total relic abundance of the Universe. For the neutrinos to dominate the halo of dwarf galaxies, one need to pack them so much that it would violate Pauli exclusion principle. The neutrino mass should be $m_\nu>500$ eV~\cite{Lin:1983vq}, to avoid this problem.
The neutrinos are so light that they are moving at the speed of light, so it is very difficult to form gravitationally bound structures from fast moving objects. The velocity of the objects will always be far above the local escape velocities. If all dark matter were neutrinos, then it would take an enormous time to form gravitationally bound structures.
So neutrino cannot be a prominent dark matter candidate. The speed of warm dark matter, e,g., sterile neutrino is rather small compared with the speed of the light. The current data require a mass range of warm dark matter to be in between $0.3$ keV and $3$ keV.
The warm dark matter is also a dark matter candidate, however, alone it is unable to explain the formation of large-scale structure. The cold dark matter moves very slowly.
Currently, all viable models of structure formation indicate that the Universe is dominated by cold dark matter. It is very difficult to determine the constituents of CDM. The candidates fall roughly into two categories. ($a$) Axions were proposed in the late 1970s to solve the strong-$CP$ problem in quantum chromodynamics (QCD). The Axions are favored as dark matter candidates as they interact very weakly. The best example of a nonthermal dark matter candidate is the Axion.
If it was produced thermally, then it cannot fulfill the relic abundance of the Universe. Although the Axions are of very light mass $\sim{\cal O}(10^{-5})$ eV, however, they behave like a cold dark matter. This was shown by authors of Ref.~\cite{Hwang:1996xd} using the proper time averaging of the perturbed oscillating scalar (Axion) field. ($b$) Weakly Interacting Massive Particle (WIMP) is another viable dark matter candidate. It is considered that these particles were created thermally in the early Universe.

It was observed that the Universe on large scales is not filled uniformly with matter, instead, we have large empty spaces (voids) separated by narrow filaments (gravitational interactions) and clumps of matter. This clumpy structure requires that the particles were moving slowly at the time the structure formed. This hints towards a fairly massive dark matter particle. The WIMP is the best viable candidate for the dark matter. If the WIMPs exist, mathematical modeling shows there must be about five times more of these than normal matter, which coincides with the abundance of dark matter that we observe in the Universe.
The annihilation cross-section of the WIMP dark matter provides the information that they interact with one another or the other standard particles via a force similar in strength to the weak nuclear force.
This is known as the ``WIMP miracle''.
Also it can explain the gamma-ray excess observed at Fermi LAT~\cite{Abdo:2009zk}, positron excess observed at AMS-02~\cite{Aguilar:2013qda} and PAMELA~\cite{Adriani:2012paa}. These are the motivations to adopt the WIMP as the leading candidate for the cosmological dark matter.

There is also another kind of cold dark matter candidate which may fulfill the relic abundance of the Universe, for example, the strongly interacting massive particle (SIMP). These particles  interact more strongly with each other. The model only works if SIMPs interact very weakly with ordinary matter.

Only WIMP dark matter has been considered in this thesis. Hence,
in the following sections, the relic density calculation of the WIMPs will be addressed and the direct and indirect detection of dark matter will be briefly reviewed.
\section{Relic Density of Dark Matter}

In the early Universe, WIMPs were created and they were in thermal equilibrium with photons and other particles at a very high temperature.
As the expansion continued, the temperature of the Universe kept falling down, so that the lighter particles did not have sufficient kinetic energy to produce the heavier particles. At some point, the massive particle density would drop low enough, such that, the probability of one WIMP finding another became small. These kinds of situation of particles are called ``freeze-out". The density of a specific
particle at the time of freeze-out is known as the ``relic-density" of this particle since its abundance remains same till today. One can calculate the exact freeze-out temperature of the particles by equating the reaction rate $\Gamma=n \sigma v$ and the rate of expansion of the Universe, i.e., the Hubble rate.

The reaction rate of particles per unit volume is given by,
\begp
\allowdisplaybreaks \beq
\Gamma=n \left\langle \sigma v\right\rangle.
\label{eq:decyaRate}
\eeq
\eegp\par
This is one of the most important quantities which is used to calculate the relic abundance of dark matter, where $\left\langle \sigma v\right\rangle$ is the thermally averaged annihilation cross-section and $n$ is the number density of the dark matter. In radiation dominated Universe, the expansion rate is given by,
\begp
\allowdisplaybreaks \beq
H=\frac{\dot{a}}{a}= g_{*}^{1/2}\frac{T^2}{\mpl} \left(\frac{\pi^2}{90} \right)^{1/2},
\label{eq:Hubble}
\eeq
\eegp\par
where $g_{*}$ is the effective degrees of freedom. At the ``freeze-out" temperature $T_f$ the annihilation stopped, i.e., the expansion rate become exactly equal to the annihilation rate. Comparing the eqns.~\ref{eq:decyaRate} and~\ref{eq:Hubble}, we get,
\begp
\allowdisplaybreaks \beq
n(T_f)\simeq g_{*}^{1/2}\frac{T_f^2}{ \left\langle \sigma v\right\rangle \mpl}.\nn
\eeq
\eegp\par
Using the Boltzmann equation, one can calculate the number density $n$ and freeze-out temperature of WIMP dark matter in the present Universe. In this case, one has to take into account (1) the rate of expansion of the Universe, (2) annihilation and co-annihilation of the dark matter particles, and (3) decay of the dark matter particles.
\begp
\allowdisplaybreaks \beq
\frac{dn}{dt}=-3 H n + \left\langle \sigma v\right\rangle (n^2 -(n^{eq})^2).
\label{eq:boltz1}
\eeq
\eegp\par
One can find the equilibrium number density of the dark matter using the classical Maxwell-Boltzmann distributions. For non-relativistic case, i.e, when temperature of the Universe is less than the mass of the dark matter, the equilibrium number density can be written as,
\begp
\allowdisplaybreaks
\bea
n^{eq}&=& \int \frac{d^3p}{(2\pi)^3} e^{-E/T}, \qquad {\rm with}\qquad E=M_{DM} + \frac{{\rm \bf P}_{DM}^2}{2 M_{DM}}.\nn\\
&=&e^{-x}\frac{M_{DM}^3}{(2\pi x)^{3/2}}\nn
\eea
\eegp
where  $x=\frac{M_{DM}}{T}$ and ${\rm \bf P}_{DM}$ is the momentum of dark matter. 

In the radiation dominated case, $H=\frac{1}{2 t}$, so using eqn.~\ref{eq:Hubble} we can get $dt$ as,
\begp
\allowdisplaybreaks \beq
dt=\frac{xdx}{H}.\nn
\eeq
\eegp\par
One can write the eqn.~\ref{eq:boltz1} as,
\begp
\allowdisplaybreaks \bea
\frac{dY}{dx}&=&-\frac{(1.32 g_{*}^{1/2} \mpl M_{DM})}{x^2}\left\langle \sigma v\right\rangle (Y^2 -Y_{eq}^2),
\label{eq:boltz2}
\eea
\eegp
where $Y$ is the yield of WIMPs and it is defined as the ratio of number density with the entropy density at the temperature $T$: $Y=\frac{n}{s}$. For dark matter mass $M_{DM}$, the entropy density can be written as $s=g_{*} (\frac{M_{DM}}{x})^3 \left(\frac{2 \pi^2}{45}\right)^{1/2}$, and the yield at time of thermal equilibrium is $Y_{eq}=\frac{n^{eq}}{s}=0.145 ~e^{-x} x^{3/2}$.

One can further simplify the eqn.~\ref{eq:boltz2} as,
\begp
\allowdisplaybreaks \bea
\frac{dz}{dx}&=&-\frac{1}{x^2}(z^2 -z_{eq}^2),
\label{eq:boltz3}
\eea
\eegp
where $z=(1.32 g_{*}^{1/2} \mpl M_{DM})\left\langle \sigma v\right\rangle Y$ and $z_{eq}= 1.914 ~g_{*}^{1/2} \mpl M_{DM} \left\langle \sigma v\right\rangle e^{-x} x^{3/2}$. $Y_0 \equiv Y(T=T_0)$ or $z_0 \equiv z(T=T_0)$ can be found by integrating the above equation from $x=0$ (i.e., $T=\infty$) to $x=x_0 \left( \equiv \frac{M_{DM}}{T_0}\right) $, where $T=T_0$ is temperature of the Universe today.

The relic abundance of dark matter in the present Universe can be written as,
\begp
\allowdisplaybreaks \beq
\Omega h^2=\frac{ M_{DM} Y_{0} }{\rho_c/s_0},
\nn
\eeq
\eegp
where $s_0\sim{\rm 2890 ~cm^{-3}}$ is the current entropy density and $\rho_c\sim {\rm 1.05\times10^{-5}} h^2 {\rm ~GeV cm^{-3}}$ is critical density of the Universe and $h=0.72$ is the Hubble parameter.

One can calculate the freeze-out temperature of the WIMP dark matter from the Boltzmann eqn.~\ref{eq:boltz3} by analytic approximations. Approximate solutions can be found under the assumption that below the freeze-out $T_f$ temperature $z \gg z_{eq}$ and $z \simeq z_{eq}$ above it.

For $x > x_f$, limit, the eqn.~\ref{eq:boltz3} can be written  as, 
\begp
\allowdisplaybreaks \beq
\int_{x_f}^{x} \frac{dz}{z^2}=-\int_{x_f}^{x} \frac{dx}{x^2}.\nn
\eeq
\eegp\par
In the limit $x\ra\infty$, $z(x_f) \gg z(x)$, one can get $z(\infty)=x_f$.

Similarly in the other limit $x < x_f$ one can get, $z_{eq}(x_f) \simeq x_f$, which implies,
\begp
\allowdisplaybreaks \beq
x_f \simeq 1.914 ~g_{*}^{1/2} \mpl M_{DM} \left\langle \sigma v\right\rangle e^{-x_f} x_f^{3/2}.\nn
\eeq
\eegp\par
Solving the above equation one can approximately calculate~\cite{Murayama:2007ek} $x_f$ as,
\begp
\allowdisplaybreaks \beq
x_f \approx 24 + log \left( \frac{M_{DM}}{100 ~{\rm GeV}} \right)+ log \left( \frac{\left\langle \sigma v\right\rangle}{10^{-9} ~{\rm GeV}^{-2}} \right) - \frac{1}{2} log \left( \frac{g_{*}}{100} \right)\nn
\eeq
\eegp\par
From this one can get the freeze-out temperature $T_f = \frac{M_{DM}}{x_f}$ of the dark matter.
\section{Direct Detections of Dark Matter}
One can also detect dark matter directly or indirectly from experiments. With the fact that the WIMP dark matter interacts with the matter weakly, there are many experiments~\cite{Savage:2008er, Aprile:2012nq,Aprile:2011hi,dmtools,Akerib:2013tjd} which are trying to detect the dark matter directly. If WIMPs scatter from atomic nucleus then it deposits energy in the detector given by,
\begp
\allowdisplaybreaks \beq
E_{deposit} = \frac{1}{2} M_{DM} v^2.\nn
\eeq
\eegp\par
The energy deposition can also be written as,
\begp
\allowdisplaybreaks \beq
E_{deposit} = \frac{\mu^2 v^2}{m_N}(1-\cos\theta).\nn
\eeq
\eegp\par
In the Earth frame, the mean velocity $v$ of the WIMPs relative to the target nucleus is about 220 km/s, $\mu$ is the reduced mass of the WIMP of mass $M_{DM}$ and the nucleus of mass $m_N$, and $\theta$ is the scattering angle.
As the dark matter is weakly interacting, it may rarely bump into the nucleus of a detector atom and deposit energy which may create a signature at the detector. The amount of energy of a WIMP with mass $M_{DM}=100$ GeV would deposit in the detector is $E_{deposit}\simeq 27$ keV.
It is very difficult to pick out the DM signature against the background from natural radioactivity. Natural radioactivity generally emits energies around MeV range. 
We need a radioactively clean and cosmic radiation free environment for the detection of the dark matter in experiments. The experimental set up must be placed in the deep underground to get shielded from the cosmic-ray or any other backgrounds.
\begin{figure}[ht]
\begin{center}
{
\includegraphics[width=2.7in,height=2.7in, angle=0]{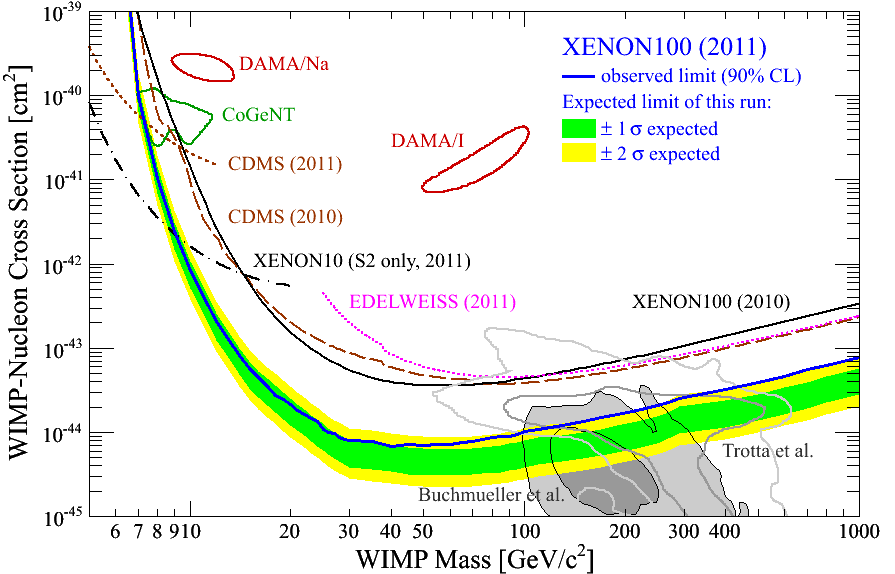}}
\caption{\textit{Limit on WIMP cross-sections (normalized to a single nucleon) for spin-independent coupling versus mass from different experiments. Most stringent limit comes from the LUX-2013 experiments. Image credit: Aprile E. {\it et al.}~\cite{Aprile:2012nq}}}
\label{direcdet}
\end{center}
\end{figure}

Dark matter detections are of two types: spin-dependent and spin-independent. In the spin-independent case, the scattering cross-section is proportional to the square of the atomic mass A,
whereas the cross-sections for spin-dependent scattering are proportional to $J(J+1)$, where $J$ is the spin of the target nucleus. The experimental sensitivity of spin-dependent cross-section is far below than that of spin-independent cross-sections as far as the dark matter is concerned~\cite{Agrawal:2010fh}.
As in this thesis in various models with extended scalar sectors, a lightest neutral scalar particle is considered as a viable WIMP dark matter, only the spin-independent cross-sections are considered.

The spin-independent elastic scattering cross-section of a WIMP with the nucleus is given by,
\begp
\allowdisplaybreaks \beq
\sigma = \frac{4 M_{DM}^2 m_N^2}{\pi(M_{DM}+m_N)^2} [Z f_p + (A-Z) f_n]^2,\nn
\eeq
\eegp
where $A$ and $Z$ are the atomic mass and atomic number of the target nucleus. $f_p,~f_n \approx 0.3$ are the form factors of the proton and the neutron.
 
Presently non-observation of dark matter in direct detections from experiments XENON~\cite{Aprile:2012nq, Aprile:2011hi}, LUX~\cite{Akerib:2013tjd} set a limit on WIMP-nucleon scattering cross-section for a given dark matter mass (see Fig.~\ref{direcdet}). Currently, most stringent bound is set by the LUX experiment. These experiments also ruled out the previous claim of finding signature of dark matter around 10 GeV by experiments DAMA/LIBRA~\cite{Bernabei:2010mq}, CoGeNT~\cite{  Aalseth:2012if}, CDMS~\cite{Agnese:2013rvf} etc. 

Various kinds of theoretical models consider dark matter particles $\sim {\cal O} (100)$ GeV which can easily be produced at the LHC. If the dark matter were created at the LHC, they would escape through the detectors without creating any signature.
However, the dark matter would carry away energy and momentum, so one could infer their existence from the amount of energy and momentum ``missing” after a collision.
\section{Indirect Detections of Dark Matter}
Indirect detection techniques are quite different for detection of the dark matter. If the dark matter and its antiparticle are the same,
then dark matter can annihilate to form known standard model particles such as photons (gamma-rays), electrons $e^-$, positrons $e^+$ etc. The dark matter can then be detected indirectly through products of such annihilations.
Various kinds of the detector placed in the orbits around the Earth, for example, Fermi Gamma-ray Space Telescope (FGST), Alpha Magnetic Spectrometer (AMS), PAMELA etc., observed the excess of gamma-ray and positron excess.
These observations cannot be explained from the known sources. The dark matter with different mass ranges can be an answer to this puzzle.
It is guessed that this excess is formed due to the annihilation of dark matter at the highly populated dark matter regions in the Universe like galactic center.
From particle physics point of view, the processes like $DM,~DM \rightarrow \gamma\gamma,~ e^+ e^-$ etc. have been used to explain such excess.
These processes are model dependent. The WIMP dark matter with different mass and coupling of a pair of dark matter to the SM particle can be considered to explain these high energetic gamma-rays excess from the galactic center and positron excess in the cosmic ray.

In the following subsections, the gamma-ray excess and the positron excess will be discussed.
\subsection{Gamma-ray Excess from Galactic Center}
\begin{figure}[ht]
\begin{center}
{
\includegraphics[width=3.7in,height=2.7in, angle=0]{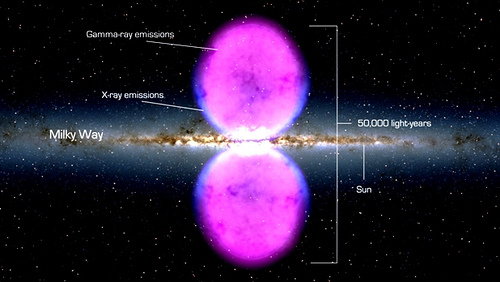}}
\caption{\textit{ The Fermi bubble around the galactic center. Image credit: NASA/Fermi
LAT.}}
\label{FermiLat}
\end{center}
\end{figure}
Gamma-ray excess at the galactic center has been observed by Fermi Gamma-ray Space Telescope (FGST)~\cite{Abdo:2009zk}. The FGST is the space-based gamma-ray observatory, consisting of a Large Area Telescope (LAT) and a Gamma-ray Burst Monitor (GBM). The Fermi LAT contains  the gamma-ray detector that can detect photons within the energy range of 20 MeV to 300 GeV.
This detector first observed that the galactic center has the bubble shaped gamma-ray lobes, extending 25000 light-years above and below the galactic plane (see Fig.~\ref{FermiLat}). The data from $1.25^\circ$ to $10^\circ$ around the galactic center can be explained with known sources of gamma-ray.
But the gamma-ray spectrum within $1.25^\circ$ shows an excess.
The dark matter self-annihilations into the gamma-rays at the galactic center give an explanation of this excess~\cite{Hooper:2010mq}. 
\subsection{Positron Excess in the Cosmic Ray}
\begin{figure}[ht]
\begin{center}
{
\includegraphics[width=3.7in,height=2.7in, angle=0]{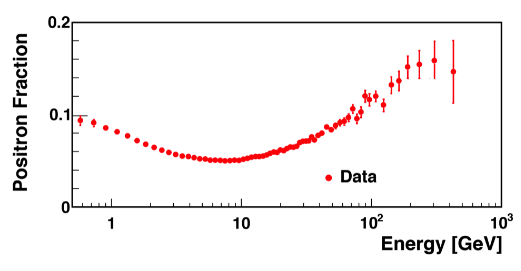}}
\caption{\textit{ The variations of fraction of positron flux with energy. Image credit: AMS-02.~\cite{Aguilar:2013qda}}}
\label{AMS02}
\end{center}
\end{figure}
Alpha Magnetic Spectrometer-02~\cite{Aguilar:2013qda} is placed at the International Space Station (ISS). It has the capability to detect photon, electron, positron, antiproton etc. in the cosmic rays. Also, it  provides the information of the energy of the detected particle. Presently it is detecting the particles in cosmic ray in the energy range of 0.5 GeV to 500 GeV. AMS has measured the positron fraction $\frac{\Phi_{e^+}}{\Phi_{e^+}+ \Phi_{e^-}}$, where $\Phi$ are the flux of the particles in the cosmic ray. The most striking result is the observation of a rise in the positron fraction starting at $\sim$10 GeV and extending at least to 350 GeV (see Fig.~\ref{AMS02} ). A possible explanation for this observation can be the dark matter annihilations~\cite{Geng:2014dea}.

However, the cosmic particles backgrounds are still poorly understood and it is not possible at the moment to make a definite statement about the origin of this excess. In this work, the detailed study of the indirect detection of the dark matter is kept aside.
\section{Summary}
Current observations support a remarkably simple model of the Universe consisting of baryonic matter, dark matter, and dark energy.
Several observations such as galactic rotation curves, gravitational lensing in bullet cluster, and the study of cosmic microwave background radiations support that the Universe contains unknown massive dark matter.
The properties of dark matter are as much of a mystery now as they were in 1930's.
In this chapter, the WIMPs and several other possible dark matter candidates which may contribute to the abundance of the dark matter in the Universe have been briefly discussed.
The detailed relic density calculations of the WIMP dark matter have been reviewed.
The direct and indirect detections of the dark matter have been shown.
The density of the dark matter in the Universe, formation of large-scale structure, observed $\gamma$-, positron-excess etc. cannot be explained without extending the SM of particle physics.
The simplest extensions involve dark matter, composed of  the new particle(s) that may explain these observations.  

In this thesis, the SM is extended with a $SU(2)_L$ singlet, doublet or triplet scalar.
If a discrete symmetry $Z_2$ is imposed on the extra scalar sector such that the standard model particles do not couple with an $odd$ number of these extra scalars, then the lightest neutral stable scalar can serve as a viable dark matter candidate, which may fulfill the relic abundance of the Universe.
The mass of the dark matter and the coupling of a pair of them to the standard model particle(s) correspond to a WIMP dark matter.
The detailed study of the scalar dark matter candidate as WIMPs will be discussed in Chapter~\ref{chap:MetaSMextended}.
\chapter{Electroweak Symmetry Breaking in Extended Higgs sector}
\label{chap:EWSBextended}
\linespread{0.1}
\graphicspath{{Chapter3/}}
\pagestyle{headings}
\noindent\rule{15cm}{1.5pt} 
\section{Introduction}
In 2012, the ATLAS and CMS collaborations at the Large Hadron Collider (LHC) at the CERN announced the discovery of a new scalar resonance with a mass $\sim125$ GeV. Although the properties of the scalar detected at LHC are consistent with the Higgs boson predicted by the
Standard Model (SM), the experimental data still allow the extension of the SM Higgs sector with one or more $SU(2)_L$ scalar multiplets. The main goal of this thesis is to explore various models with extended scalar sectors using weak vector boson scattering processes and from the stability of the scalar potential. Imposing a discrete symmetry $Z_2$ on these models, the lightest neutral $Z_2$-odd scalar can serve as a viable dark matter candidate which may fulfill the relic abundance of the dark matter in the Universe.
In this chapter, the structure of the scalar potential in these new models and, how the new scalar fields interact with the SM particles and between one another will be discussed. Various theoretical and experimental bounds on the extended scalar sectors will be reviewed as well.

In the next section, the SM extended with a real singlet scalar will be discussed. With and without imposing an extra discrete symmetry $Z_2$ to the extended sectors, how the particle gets masses through the electroweak symmetry breaking will be shown. In Sections~\ref{DoubletEXT},~\ref{TriY0Ext} and~\ref{TriY2Ext}, similar studies of models extended with a doublet or a triplet scalar with hypercharges $Y=0$ and 2 will be elaborated.
\section{The Standard Model with a real singlet scalar}\label{SigletEXT}
The simplest way to extend the Higgs sector of SM is to consider the addition of an extra real scalar $S$. The scalar $S$ is a singlet under the SM gauge group. The Lagrangian, invariant under a $Z_2$ symmetry $S\ra -S$, can be written as~\cite{McDonald:1993ex},
\begp
\allowdisplaybreaks
\begin{equation}
{\cal L}=\left(D_\mu\Phi\right)^\dagger\left(D^\mu\Phi\right)
+\frac{1}{2}\partial_\mu S \partial^\mu S -V(\Phi, S),\nn
\end{equation}
\eegp
with the scalar potential,
\begingroup
\allowdisplaybreaks \beq
V(\Phi, S)= -  {m}^2 |\Phi|^2 + \lambda ~|\Phi|^4 + \frac{1}{2} \overline{m}_S^2 S^2 +\frac{\kappa}{2} |\Phi|^2 S^2 +
 \frac{\lambda_S}{4!} S^4 \,.
\label{scalarSMpot}
\eeq
\begingroup
$\Phi$ is the SM complex doublet with vacuum expectation value (VEV) $v_d$,
\begingroup
\allowdisplaybreaks \begin{eqnarray}
\Phi = \left(\begin{array}{cc}
\phi^+ \\ \phi^0\end{array}\right)\, , \quad {\rm with} \quad \phi^0=\frac{1}{\sqrt{2}}(h^0+v_d+ i G^0)\nn
\end{eqnarray}
\endgroup
If the $Z_2$ symmetry break spontaneously then the singlet scalar also acquires VEV $ v_S $ such that $S=s+v_S$, i.e., the potential can have minimum at ($v_d,~v_S$) in $h^0 s$-plane. The required conditions are,
\begp
\allowdisplaybreaks
\bea
m^2 &=& {\lambda} {v_d}^2+\frac{\kappa {v_S}^2}{2},\nn\\
\overline{m}_S^2 &=& -\frac{\kappa {v_d}^2}{2 }-\frac{{\lambda_S} {v_S}^2}{6}.\nn
\eea
\eegp\par
The coefficient $\kappa$ and the VEVs, $v_d$ and $v_S$, govern the degree of mixing between $s$ and $h^0$ of the SM doublet. The mass of the scalar fields are determined by the parameters, $m$, $\overline{m}_S$, $\lambda$, $\kappa$ and the VEVs $v_d,~v_S$. Using the minimization conditions we can write the mass matrix of $h^0$ and $s$,
\begp
\allowdisplaybreaks
\bea
\left(\begin{array}{cc} h^0 & s \end{array}\right) \left(\begin{array}{cc}
A_{\phi\phi} & C_{\phi S}\\
C_{\phi S} & B_{S S}
\end{array}\right) \left(\begin{array}{c}
h^0 \\ s \end{array}\right)\label{Sigdiag}
\eea
\eegp
with,
\begp
\allowdisplaybreaks
\begin{eqnarray}
A_{\phi\phi} &=& 2 \lambda v_d^2,\nn\\
B_{S S} &=& \frac{1}{3} \lambda_S v_S^2 ,\nn\\
C_{\phi S} &=&  \kappa v_d v_S.\nn
\end{eqnarray}
\eegp\par
The mass eigenstates are obtained by diagonalizing the mass matrix in eqn.~\ref{Sigdiag} with a rotation of the ($h^0~s$) basis,
\begp
\allowdisplaybreaks
\bea
\left(\begin{array}{c} 
h \\H \end{array}\right) &=& \left(\begin{array}{cc}
\cos\alpha^{\prime} & \sin\alpha^{\prime}\\
-\sin\alpha^{\prime} & \cos\alpha^{\prime}
\end{array}\right) \left(\begin{array}{c}
h^0 \\ s \end{array}\right).\nn
\eea
\eegp\par
The mixing angle ($\alpha^\prime$) between the scalars can be written as,
\begp
\allowdisplaybreaks
\beq
\tan 2\alpha^{\prime} = \frac{2 C_{\phi S}}{B_{SS}-A_{\phi\phi}}.\nn
\eeq
\eegp\par
The masses of the scalars are,
\begp
\allowdisplaybreaks
\bea
M^2_{h} &=& \frac{1}{2}\left[(B_{SS}+A_{\phi\phi})-\sqrt{(B_{SS}-A_{\phi\phi})^{2}+4 C_{\phi S}^{2}}\right],\nn\\
M^2_{H} &=& \frac{1}{2}\left[(B_{SS}+A_{\phi\phi})+\sqrt{(B_{SS}-A_{\phi\phi})^{2}+4 C_{\phi S}^{2}}\right].\nn
\eea
\eegp\par
Here $h$ is considered to be the observed Higgs-like scalar particle at $\sim$ 125 GeV and $H$ is a heavy scalar particle yet to be observed. The couplings of the Higgs-like scalar to the known fermions and gauge bosons are also modified. The new modified couplings are $\cos\alpha^{\prime}$ times the couplings as in the SM. The couplings of the heavy Higgs to SM gauge bosons and fermions multiplied with $\sin\alpha^{\prime}$: For example, $ (\cos \alpha^\prime g_2 M_W)~h W^+ W^-$, $ (\sin \alpha^\prime g_2 M_W)~H W^+ W^-$ etc.

\subsection{Constraints on the SM$+S$ model}
The parameter space of this model is constrained by various kinds of theoretical considerations like absolute vacuum stability and unitarity of the scattering matrix. Also, the LHC puts severe constraints on this model. In the following, the constraints on the model will be discussed.

\subsubsection{Constraints from stability of the scalar potential}
The stability of the scalar potential requires that the potential should not become
\newpage
unbounded from below, i.e, it should not approach negative infinity along any direction of the field space at large field values. At very large field the quadratic terms of the scalar potential in eqn.~\ref{scalarSMpot} are negligibly small compared to the quartic terms, so the scalar potential can be written as,
\begp
\allowdisplaybreaks
\bea
V(h^0,~s) &=& \frac{1}{4} {\lambda} ({h^0})^4+\frac{\kappa}{4} ({h^0})^2 (s)^2+\frac{1}{24} {\lambda_S} (s)^4.\nn
\eea
\eegp
We can further simplify the above equation as,
\begp
\allowdisplaybreaks
\beq
V(h^0,~s) = \frac{1}{4}\left\lbrace \sqrt{\lambda} (h^0)^2 + \frac{\sqrt{\lambda_S}}{\sqrt{6}} (s)^2 \right\rbrace^2 + \frac{1}{4}\left\lbrace\kappa +   \sqrt{\frac{2 \lambda \lambda_S}{3}}\right\rbrace ({h^0})^2 (s)^2.
\label{scalpotstability}
\eeq
\eegp
The scalar potential in eqn.~\ref{scalpotstability} is bounded from below if,
\begp
\allowdisplaybreaks
\beq
\lambda(\Lambda) > 0, \quad \lambda_S(\Lambda) > 0 \quad {\rm and} \quad \kappa(\Lambda) + \sqrt{\frac{2 \lambda(\Lambda) \lambda_S(\Lambda)}{3}} > 0.
\nn
\eeq
\eegp
where the coupling constants are evaluated at a scale $\Lambda$ using RG equations. If the scalar potential has a metastable electroweak vacuum, then the above conditions will be modified and are shown in Chapter~\ref{chap:MetaSMextended}.
\subsubsection{Perturbativity constraints}
To ensure that the radiatively improved scalar potential $V(\Phi, S)$ of the SM+$S$ model remains perturbative at any given energy scale, one must impose the following upper bounds on the couplings $\lambda,\kappa$ and $\lambda_{S}$ of scalar potential $V(\Phi, S)$ as,
\begp
\allowdisplaybreaks
\beq
\mid \lambda(\Lambda), \kappa(\Lambda), \lambda_{S}(\Lambda)\mid \leq 4 \pi.\nn
\eeq
\eegp
\subsubsection{Constraints from unitarity of the scattering matrix}
The parameters of the scalar potential of this model are severely constrained by the unitarity of the scattering matrix (S-matrix) which consists of the quartic couplings of the scalar potential. At very high field values, one can obtain the scattering matrix by using various scalar-scalar, gauge boson-gauge boson, and scalar-gauge boson scatterings. The unitarity of the S-matrix demands that the eigenvalues of the scattering matrix should be less than $8\pi$. In this model, the unitary bounds are obtained from the scattering matrices as~\cite{Cynolter:2004cq}, 
\begp
\allowdisplaybreaks
\bea
\lambda \leq 8 \pi  \quad {\rm and} \quad \Big| 12 {\lambda}+{\lambda_S} \pm \sqrt{16 \kappa^2+(-12 {\lambda}+{\lambda_S})^2}\Big| \leq 32 \pi.\nn
\eea
\eegp
\subsubsection{Constraints from the LHC}
The signal strength measurement of the SM Higgs-like scalar with mass $125$ GeV implies $\sin \alpha^\prime$ less than $0.25$~\cite{Pelliccioni:2015hva}. At the Large Hadron Collider, the direct searches of the heavy Higgs $H$ boson in $gg\ra H \ra W^{*+} W^{*-}$ and  $gg\ra H \ra Z^* Z^*$ ( $W^{*\pm},Z^*$ decay into fermions) channels put a stringent bound~\cite{Pelliccioni:2015hva} on the mass of the heavy scalar. The mass range of $145-1000$ GeV is thus excluded at 95$\%$ C.L.
\subsection{Dark matter in SM$+S$}
If the imposed $Z_2$ symmetry is exact then it prevents the extra scalar to acquire VEV. The potential can have minimum only along the Higgs field direction, i.e., the electroweak symmetry breaking driven only by the SM Higgs doublet. This extra scalar field does not mix with the SM Higgs and $odd$ number of scalars do not couple with the SM particles. As a result, this scalar is stable and is considered to be a viable dark matter candidate. The dark matter studies in singlet extended SM will be taken up in Chapter~\ref{chap:MetaSMextended}. 
\section{Two Higgs doublet model}\label{DoubletEXT}
In this section, two Higgs doublet model will be briefly reviewed. In this model, an extra scalar doublet $\Phi^{\prime}$ is added with the SM doublet $\Phi$. Both of these fields possess hypercharge $+1$. Kinetic part of the Lagrangian of these two scalar fields, invariant under ${ SU(2)_L\times U(1)}_Y$ gauge group, can be written as,
\begp
\allowdisplaybreaks
\begin{equation}
{\cal L^{\rm 2HDM}_{\rm Kinetic}}=\left(D_\mu\Phi\right)^\dagger\left(D^\mu\Phi\right)+\left(D_\mu{\Phi'}\right)^\dagger\left(D^\mu{\Phi'}\right)\, ,\nn
\end{equation}
\eegp
where $D_\mu$ is the covariant derivative, given in eqn.~\ref{coVderi}.

The scalars are conventionally written as,
\begp
\allowdisplaybreaks
\begin{eqnarray}
\Phi = \begin{pmatrix}
\phi^+ \\  \phi^0
\end{pmatrix},
~~~{\rm and}~~~
{\Phi'}=\begin{pmatrix}
{\phi'}^{+} \\  {\phi'}^{0}
\end{pmatrix}.\nn
\end{eqnarray}
\eegp\par
The most general potential with these two scalar fields is given by~\cite{Branco:2011iw},
\begp
\allowdisplaybreaks
\begin{eqnarray}
V(\Phi,\Phi^{\prime}) &=&
m_{11}^2\, \left(\Phi^\dagger \Phi\right)
+ m^2_{22}\, \left(\Phi^{\prime\dagger} \Phi^{\prime}\right) -
 m^2_{12}\, \left(\Phi^\dagger \Phi^{\prime} + \Phi^{\prime\dagger} \Phi\right)
+ \lambda_1 \left( \Phi^\dagger \Phi \right)^2
\nonumber \\
&&+ \lambda_2 \left( \Phi^{\prime\dagger} \Phi^{\prime} \right)^2
+ \lambda_3\, \left(\Phi^\dagger \Phi\,\right) \left(\Phi^{\prime\dagger} \Phi^{\prime}\right)
+ \lambda_4\, \left(\Phi^\dagger \Phi^{\prime}\,\right)\left(\Phi^{\prime\dagger} \Phi\right)
\nonumber \\
&&+ \frac{\lambda_5}{2} \left[
\left( \Phi^\dagger\Phi^{\prime} \right)^2
+ \left( \Phi^{\prime\dagger}\Phi \right)^2 \right]
+\lambda_6\, \left(\Phi^\dagger \Phi\right)\, \left(\Phi^\dagger\Phi^{\prime} + \Phi^{\prime\dagger}\Phi\right) \nonumber \\
&&+ \lambda_7\, \left(\Phi^{\prime\dagger} \Phi^{\prime}\right)\, \left(\Phi^\dagger\Phi^{\prime} + \Phi^{\prime\dagger}\Phi\right).
\label{pot}
\end{eqnarray}
\eegp\par
The parameters $m_{11}, m_{22}$ and $\lambda_i$ ($i=1,\ldots 4$) are real, whereas $m_{12}$, $\lambda_5$, $\lambda_{6}$ and $\lambda_7$ could be complex in general.
If these parameters are complex, then imaginary parts of these parameters give rise to explicit $CP$-violation in the Higgs sector, as not all the imaginary parts can be removed by re-phasing transformations.
In this particular incarnation of 2HDM, a fermion can couple to both $\Phi$ and $\Phi^{\prime}$. However, this can lead to unacceptably large tree level flavor changing neutral currents (FCNC) \cite{PhysRevD.15.1958,PhysRevD.15.1966,Arhrib:2005nx,Dery:2013aba}. One can avoid such FCNCs by imposing a $Z_2$ symmetry, namely $\Phi \to -\Phi$ and $\Phi^{\prime} \to \Phi^{\prime}$.
The $Z_2$ symmetry is \emph{exact} when $m_{12}$, $\lambda_{6}$ and $\lambda_7$ vanish. Here only $\lambda_5$ can be complex.
But it becomes real after re-phasing one of the scalar doublets so that the scalar potential becomes $CP$-conserving. Furthermore, this symmetry is said to be broken \emph{softly} if $m_{12}\ne0$, i.e., it is violated in the quadratic terms only, but it is conserved in quartic terms i.e., $\lambda_{6}=0$ and $\lambda_7=0$.
At this point, we would like to mention that we are interested in a specific scheme of coupling of fermions to the doublets. This scheme is known as the \emph{Type-II} 2HDM, where the $down$-$type$ quarks and the charged leptons couple to $\Phi$ and the $up$-$type$ quarks, to $\Phi^{\prime}$\cite{Pich:2009sp}. This can be achieved by demanding $\Phi\to -\Phi$ and $\psi_{R}^{i}\to -\psi_{R}^{i}$ under the same $Z_2$ symmetry, where $\psi$ stands for charged leptons or $down$-$type$ quarks and $i$ represents the generation index.

Expanding the neutral fields $\phi^0$ and $\phi^{\prime 0}$ around their VEV,
\begp
\allowdisplaybreaks
\begin{eqnarray}
\phi^0 &=&  \frac{1}{\sqrt 2}(v_d+h^0+i\zeta_d),
\label{phio}
\\
\phi^{\prime 0} &=& 
\frac{1}{\sqrt{2}}(v^{\prime}_d+h^{\prime 0}+i\zeta^{\prime}_d).
\label{phiop}
\end{eqnarray}
\eegp\par
After electroweak symmetry breaking (EWSB), we have the following minimizing conditions~\cite{Eriksson:2009ws}, 
\begp
\allowdisplaybreaks
\bea
m_{11}^2 &=& m_{12}^2\tan\beta-\frac{1}{2}v^2\Bigl(2 \lambda_1\cos^2\beta+\left(\lambda_3+\lambda_4+\lambda_5\right)\sin^2\beta\Bigr),
\label{eq:m11}\nn\\
m_{22}^2 &=& m_{12}^2\cot\beta-\frac{1}{2}v^2\Bigl(2 \lambda_2\sin^2\beta+\left(\lambda_3+\lambda_4+\lambda_5\right)\cos^2\beta\Bigr),\label{eq:m22}\nn
\eea
\eegp
with $\tan\beta=\frac{v^{\prime}_d}{v_d}$. The two VEVs $v_d$ and $v^{\prime}_d$ of the doublets contribute to the weakly interacting gauge boson masses at the tree-level and the masses of weak gauge bosons of 2HDM are given as:
\begp
\allowdisplaybreaks
\begin{eqnarray}
M_W^2 = \frac{g^2_2}{4}v^2,~~M_Z^2 = \frac{g^2_2}{4\cos^2\theta_W}v^2,~~{\rm where}~~v^2=(v_d^2+v^{\prime 2}_d)=(246~{\rm GeV})^2\nn, 
\end{eqnarray}
\eegp
where $\theta_W$ is the Weinberg angle.
\subsection{Scalar Masses and Mixing for 2HDM}
After electroweak symmetry breaking, the squared mass matrix can be expressed as $8\times 8$ for the scalars. This matrix composed of four $2\times 2$ submatrices with bases, ($\phi^+, {\phi^\prime}^{+}$), (${\phi}^{-}, {\phi^\prime}^{-}$), ($h^0, {h^\prime}^{0}$), ($\zeta_d, \zeta_{d}^\prime$). After rotating these fields into the mass basis, we get five physical mass eigenstates ($H^\pm,h,H,A$) and remaining three states are massless Goldstone bosons ($G^\pm,G^0$) eaten up to give mass to the SM gauge bosons $W^\pm,Z$. The mass eigenvalues for the physical scalar for 2HDM are given by,
\begp
\allowdisplaybreaks
\begin{eqnarray}
\label{e:masq2hdm}
M^2_{A} &=& \frac{m_{12}^2}{\sin{\beta} \cos{\beta}}-v^2\lambda_5, \\
\label{e:mHpmsq2hdm}\nn
M_{H^{\pm}}^2 &=& m^2_{A}+\frac{1}{2}v^2\left(\lambda_5-\lambda_4\right),\\
\label{e:mhsq2hdm}\nn
M^2_{h} &=& \frac{1}{2}\left[(B+A)-\sqrt{(B-A)^{2}+4 C^{2}}\right],\\
\label{e:mHsq2hdm}\nn
M^2_{H} &=& \frac{1}{2}\left[(B+A)+\sqrt{(B-A)^{2}+4 C^{2}}\right],\\
\label{e:tan2al2hdm}\nn
{\rm with}~~~
\tan 2\alpha &=& \frac{2 C}{B-A},\label{e:scalarmasses2hdm}\nn
\end{eqnarray}
\eegp
where we have defined,
\begp
\allowdisplaybreaks
\begin{eqnarray}
A &=& M^2_{A} \sin^2{\beta}+v^{2} (2\lambda_1 \cos^2{\beta}+\lambda_5 \sin^2{\beta}),\nn\\
B &=& M^2_{A} \cos^2{\beta}+v^{2} (2\lambda_2 \sin^2{\beta}+\lambda_5 \cos^2{\beta}),\nn\\
C &=& -M_{A}^2 \sin{\beta} \cos{\beta} +v^{2} (\lambda_3+\lambda_4) \sin{\beta} \cos{\beta}.\nn
\end{eqnarray}
\eegp\par
The mixing between the two doublets in the charged, $CP$-even and $CP$-odd scalar sectors for 2HDM are respectively given by, 
\begp
\allowdisplaybreaks
\begin{eqnarray}
\begin{pmatrix}
G^\pm \\H^\pm \end{pmatrix} &=& \begin{pmatrix}
\cos\beta & \sin\beta\\
-\sin\beta & \cos\beta
\end{pmatrix}
 \begin{pmatrix}
\phi^\pm \\ \phi^{\prime\pm}  \end{pmatrix},\nn\\
\begin{pmatrix} 
h \\H \end{pmatrix} &=& \begin{pmatrix}
\cos\alpha & \sin\alpha\\
-\sin\alpha & \cos\alpha
\end{pmatrix}
\begin{pmatrix}
h^0 \\ h^{\prime 0} \end{pmatrix},\nn\\
\begin{pmatrix}
G^0 \\ A \end{pmatrix} &=& \begin{pmatrix}
\cos\beta & \sin\beta\\
-\sin\beta & \cos\beta
\end{pmatrix}
\begin{pmatrix}
\zeta_d \\ \zeta^{\prime}_d \end{pmatrix}.
\end{eqnarray}\nn
\eegp
\subsection{Constraints on 2HDM}
Varieties of considerations, stemming both from theoretical consistency conditions and from phenomenological bounds, constrain the 2HDM. In the following sections, various constraints such as vacuum stability, unitarity of scattering matrix and electroweak precision measurements on this model will be summarized.
\subsubsection{Constraints from stability of the scalar potential}
The tree-level scalar potential $V(\Phi_1,\Phi_2)$ is stable and bounded from below if~\cite{Deshpande:1977rw} 
\begp
\allowdisplaybreaks
\beq
\lambda_{1,2}(\Lambda) \geq 0, \quad  \lambda_{3}(\Lambda) \geq -2 \sqrt{ \lambda_{1}(\Lambda)\lambda_{2}(\Lambda)}, \quad  \lambda_{L,S}(\Lambda) \geq - \sqrt{ \lambda_{1}(\Lambda)\lambda_{2}(\Lambda)}  \label{stabilitybound}
\eeq
\eegp
where the coupling constants are evaluated at a scale $\Lambda$ using RG equations. However, these conditions become nonfunctional if $\lambda_1$ becomes negative at some energy scale to render the EW vacuum metastable. Under such circumstances we need to handle metastability constraints on the potential differently, which we pursue in Chapter~\ref{chap:MetaSMextended}. 
\subsubsection{Perturbativity constraints}
The radiatively improved scalar potential remain perturbative by requiring that all quartic couplings of $V(\Phi_1,\Phi_2)$ satisfy the following:
\begp
\allowdisplaybreaks
\beq
\mid \lambda_{1,2,3,4,5}(\Lambda)\mid \leq 4 \pi. 
\label{pertIDM}
\eeq
\begp

These conditions put an upper bound on the couplings of the scalar potential at an energy scale $\Lambda$.
\subsubsection{Constraints from unitarity of the scattering matrix}
Unitarity bounds on  $\lambda_i$ are obtained considering scalar-scalar, gauge boson-gauge boson, and scalar-gauge boson scatterings~\cite{Lee:1977eg}. The constraints come from the eigenvalues of the corresponding S-matrix~\cite{Arhrib:2012ia}:
\begp
\allowdisplaybreaks
\bea
| \lambda_3 \pm \lambda_4 | \leq 8\pi , ~~~~~ | \lambda_3 \pm \lambda_5 |  \leq 8\pi\nn\\
 |  \lambda_3+ 2 \lambda_4 \pm 3\lambda_5 | \leq 8 \pi\nn\\
 \Big | -\lambda_1 - \lambda_2 \pm \sqrt{(\lambda_1 - \lambda_2)^2 + \lambda_4^2} \Big | \leq 8 \pi
\\
 \Big | -3\lambda_1 - 3\lambda_2 \pm \sqrt{9(\lambda_1 - \lambda_2)^2 + (2\lambda_3 + \lambda_4)^2} \Big | \leq 8 \pi
\nn\\ 
 \Big |  -\lambda_1 - \lambda_2 \pm \sqrt{(\lambda_1 - \lambda_2)^2 + \lambda_5^2} \Big | \leq 8 \pi.\nn
\label{unitary}
\eea
\eegp
\subsubsection{Constraints from electroweak precision experiments}
In the 2HDM, the contributions to the $S, ~T$ and $U$ parameters can be written as in Ref.~\cite{He:2001tp},
\begp
\allowdisplaybreaks
\bea
S_{2HDM} &=& \frac{1}{\pi M_{Z}^2} \bigg[\sin^2(\beta-\alpha) \mathcal{B}_{22}(M_{Z}^2, M_{H}^2,M_{A}^2) - \mathcal{B}_{22}(M_{Z}^2, M_{H^{\pm}}^2,M_{H^{\pm}}^2) \nonumber \\
 && +\; \cos^2(\beta-\alpha)\bigg\{\mathcal{B}_{22}(M_{Z}^2, M_{h}^2,M_{A}^2)+\mathcal{B}_{22}(M_{Z}^2, M_{Z}^2,M_{H}^2)-\mathcal{B}_{22}(M_{Z}^2, M_{Z}^2,M_{h}^2)\nonumber \\
 && -\; M_{Z}^2\mathcal{B}_0(M_{Z}^2, M_{Z}^2,M_{H}^2)+M_{Z}^2\mathcal{B}_0(M_{Z}^2, M_{Z}^2,M_{h}^2)\bigg\} \bigg]\,,\nn\\
T_{2HDM} &=& \frac{1}{16\pi M_{W}^2 \sin^2\theta_W}\bigg[F(M_{H^{\pm}}^2,M_{A}^2)
+\sin^2(\beta-\alpha)\bigg\{F(M_{H^{\pm}}^2,M_{H}^2)-F(M_{A}^2,M_{H}^2)\bigg\} \nonumber \\
 &&\hspace{-0.05cm} +\; \cos^2(\beta-\alpha)\bigg\{F(M_{H^{\pm}}^2,M_{h}^2)-F(M_{A}^2,M_{h}^2)+F(M_{W}^2,M_{H}^2)-F(M_{W}^2,M_{h}^2) \nn\\
 && - F(M_{Z}^2,M_{H}^2) + F(M_{Z}^2,M_{h}^2) + 4 M_{Z}^2\overline{B}_0(M_{Z}^2, M_{H}^2,M_{h}^2) \nonumber \\
&&- 4 M_{W}^2 \overline{B}_0(M_{W}^2, M_{H}^2,M_{h}^2)\bigg\}\bigg]\,, \\\label{STU2HDM}
U_{2HDM} &=& -S_{2HDM} + \frac{1}{\pi M_{Z}^2} \bigg[ \mathcal{B}_{22}(M_{W}^2, M_{A}^2,M_{H^{\pm}}^2)-2\mathcal{B}_{22}(M_{W}^2, M_{H^{\pm}}^2,M_{H^{\pm}}^2) \nonumber \\
 && +\; \sin^2(\beta-\alpha)\mathcal{B}_{22}(M_{W}^2, M_{H}^2,M_{H^{\pm}}^2) \nonumber \\
 && +\; \cos^2(\beta-\alpha)\bigg\{\mathcal{B}_{22}(M_{W}^2, M_{h}^2,M_{H^{\pm}}^2)+\mathcal{B}_{22}(M_{W}^2, M_{W}^2,M_{H}^2)-\mathcal{B}_{22}(M_{W}^2, M_{W}^2,M_{h}^2)\nonumber \\
 && -\;  M_{W}^2 \mathcal{B}_0(M_{W}^2, M_{W}^2,M_{H}^2)+M_{W}^2\mathcal{B}_0(M_{W}^2, M_{W}^2,M_{h}^2)\bigg\} \bigg]\nn\,,
\eea
\eegp
where the functions $\mathcal{B}_{22},~\mathcal{B}_0,~\overline{B}_0$ and $F$ are defined as follows,
\begp
\allowdisplaybreaks
\bea 
  \mathcal{B}_{22}(q^2,m_1^2,m_2^2)&=& \frac{q^2}{24}[2 \ln q^2 + \ln(x_1x_2)-6F(x_1,x_2)\nn\\&&+\{(x_1-x_2)^3-3(x_1^2-x_2^2)+
  3(x_1-x_2)\} \ln \left( \frac{x_1}{ x_2}\right)\nn\\&&-\{2(x_1-x_2)^2-8(x_1+x_2)+\frac{10}{3}\} \\&& -  \{(x_1-x_2)^2-2(x_1+x_2)+1\}f(x_1,x_2)],\nn\\
\mathcal{B}_0(q^2,m_1^2,m_2^2) &=& 1 +\frac{1}{2} \bigg\{\frac{(x_1 + x_2)}{(x_1 - x_2)} - (x_1 - x_2)\bigg\} \ln \left( \frac{x_1}{x_2} \right)+ \frac{1}{2} f(x_1, x_2)\nn,
 \eea
\eegp 
\begp
\allowdisplaybreaks
\bea 
f(x_1,x_2)&=&\left\{
\begin{array}{ll}
-2\sqrt{\Delta}\big\{ \arctan\frac{x_1-x_2+1}{\sqrt{\Delta}}
-\arctan\frac{x_1-x_2-1}{\sqrt{\Delta}}\big\}\,,
&(\Delta>0)\,  \\ 
0\,,
&(\Delta=0)\,\\\label{fxy}
\sqrt{-\Delta}\ln\frac{x_1+x_2-1+\sqrt{-\Delta}}{x_1+x_2-1-\sqrt{-\Delta}}\,,
&(\Delta<0)\,,
\end{array}
\right.
 \eea
\eegp 
with $\Delta=2(x_1+x_2)-(x_1-x_2)^2-1$, $x_i\equiv m_i^2/q^2$ and
\begp
\allowdisplaybreaks
\bea
\overline{B}_0(m_1^2,m_2^2,m_3^2) &=& \frac{m_1^2\ln m_1^2 - m_3^2\ln
    m_3^2}{m_1^2-m_3^2} - \frac{m_1^2\ln m_1^2 - m_2^2 \ln
    m_2^2}{m_1^2-m_2^2},\nn\\
F(m_1^2,m_2^2) &=& F(m_2^2,m_1^2) 
= \frac{m_1^2+m_2^2}2 -\frac{m_1^2 m_2^2}{m_1^2-m_2^2}\ln \left(\frac{m_1^2}{m_2^2}\right). \label{Fm1m2}   
 \eea
\eegp\par
The experimental values of $S,~T$ and $U$ parameters from the precision electroweak measurements for the fixed Higgs mass and top quark mass to be at $125$ GeV and $173$ GeV respectively is given in the Ref.\;\cite{Baak:2014ora}.
\begp
\allowdisplaybreaks
\begin{equation}
\Delta S = 0.05 \pm 0.11,~~~\Delta T = 0.09 \pm 0.13,~~~ \Delta U = 0.01 \pm 0.11,
\label{STU1}
\end{equation}
\eegp
with correlation coefficients of $+0.90$ between $\Delta S$ and $\Delta T$, $-0.59$ between $\Delta S$ and $\Delta U$, $-0.83$ between $\Delta T$ and $\Delta U$. This would constrain the parameter space of 2HDM. If an exact $Z_2$ symmetry is imposed on the extra doublet then the eqn.~\ref{STU2HDM} will be modified as eqn.~\ref{STparam} of Chapter~\ref{chap:MetaSMextended}. 
\subsubsection{Constraints from the LHC}
Measured values of the Higgs signal strengths into different decay channels are consistent with the corresponding standard model expectations. Using the experimental data one gets $\sin(\beta-\alpha)\approx \pm1$ with $\tan\beta=\frac{v_2}{v_1}\approx	4$.
This limit is known as $alignment~limit$ of the 2HDM, in which the couplings of $h$ to vector bosons are SM-like.
In this limit, the nonstandard scalar masses are relatively unconstrained. $B$ meson decay ($B\ra X_s \gamma$, $X_s$ denotes a strange hadronic final state) which receives contributions from the charged Higgs at one-loop, imposes a stringent constraint on the Higgs boson mass $M_{H^+} > 300$ GeV~\cite{Mahmoudi:2009zx}.

\subsection{Dark Matter in 2HDM}
In the above context of 2HDM, we have chosen $Z_2$ to be softly broken such that the quartic couplings, $\lambda_{6,7}$ are absent in the potential. The scalar fields of both the doublets can mix and also couple with the fermions. If the $Z_2$ symmetry is unbroken then the term $m_{12}$ vanish and also this symmetry prevents the extra doublet to acquire a VEV, hence these scalar fields of the extra doublet do not mix with the fields of the SM scalar doublet.
No SM particle can couple with an $odd$ number of scalar fields of the extra doublet. The extra doublet $(\phi^{\prime+}, ~\phi^{\prime0})^{T}$ can be taken as the physical basis $(H^{+}, ~(H+iA)/\sqrt{2})^{\rm T}$.
The lightest neutral particle of the new scalar sector can be considered as a viable dark matter candidate. The detailed analysis related to the dark matter issues and metastability scenario will be presented in Chapter~\ref{chap:MetaSMextended}. 
\section{Higgs triplet model with hypercharge $Y=0$}\label{TriY0Ext}
In this model a real isospin $I=1$ and hypercharge $Y=0$ triplet $\widetilde{\Phi}$ is added with the SM Higgs doublet $\Phi$. This kind of model is known as Higgs triplet model (HTM). The kinetic part of the Lagrangian with these scalar fields is given by \cite{Chen:2008jg},
\begp
\allowdisplaybreaks
\begin{equation}
{\cal L^{\rm Y=0}_{\rm Kinetic}}=\left(D_\mu\Phi\right)^\dagger\left(D^\mu\Phi\right)
+\frac{1}{2}\left(D_\mu\widetilde{\Phi}\right)^\dagger\left(D^\mu\widetilde{\Phi}\right)\, ,\nn
\end{equation}
\eegp
where
\begp
\allowdisplaybreaks
\begin{equation}
D_\mu\widetilde{\Phi}=\biggl(\partial_\mu +i g_2 t_aW^a_\mu\biggr)\widetilde{\Phi}~~~{\rm with}~~~\widetilde{\Phi}=\begin{pmatrix}
\eta^+\\
\eta^0\\
-\eta^- \end{pmatrix}~,\ \label{TripPhiY0}
\end{equation}
\eegp
and
\begp
\allowdisplaybreaks
\begin{equation}
t_1=\frac{1}{\sqrt{2}}
\begin{pmatrix}
0&1&0\\
1&0&1\\
0&1&0
\end{pmatrix},\quad
t_2=\frac{1}{\sqrt{2}}
\begin{pmatrix}
0&-i&0\\
i&0&-i\\
0&i&0
\end{pmatrix},\quad
t_3=\begin{pmatrix}
1&0&0\\
0&0&0\\
0&0&-1
\end{pmatrix}\, \nn.
\end{equation}
\eegp\par
The most general ${ SU(2)_L\times U(1)}_Y$ scalar potential with SM Higgs doublet and
a real scalar triplet is given by,
\begp
\allowdisplaybreaks
\begin{eqnarray}
V(\Phi,\widetilde{\Phi})&=&\mu_1^2\left(\Phi^\dagger \Phi\right)
+\frac{\mu_2^2}{2} \left(\widetilde{\Phi}^\dagger \widetilde{\Phi}\right)
+{\widetilde{\lambda}_1} \left(\Phi^\dagger \Phi\right)^2 
+\frac{\widetilde{\lambda}_2}{ 4} \left(\widetilde{\Phi}^\dagger \widetilde{\Phi}\right)^2 
\nonumber \\ &&
+\frac{\widetilde{\lambda}_3}{ 2}\left(\Phi^\dagger \Phi\right)\left(\widetilde{\Phi}^\dagger \widetilde{\Phi}\right)
+\widetilde{\lambda}_4\Phi^\dagger \sigma^a \Phi\widetilde{\Phi}_a\, ,
\label{ScalarpotHTM}
\end{eqnarray}
\eegp
where $\widetilde{\Phi}_a=\left(\frac{1}{\sqrt{2}}(\eta^+ +\eta^-),~ \frac{1}{\sqrt{2}}(\eta^+ -\eta^-),~ \eta^0\right)$.

Now expanding neutral component of the triplet field around the VEV $v_t^\prime$ and using eqn.~\ref{phio} and after EWSB we have the following minimization conditions:
\begp
\allowdisplaybreaks
\bea
\mu_1^2 &=&\frac{1}{2} \left( 2 \widetilde{\lambda}_4 v^{\prime}_t -  (2\widetilde{ \lambda}_1 v_d^2 +\widetilde{\lambda}_3 v^{\prime 2}_t)\right),\nn\\
\mu_2^2 &=& \frac{1}{2 v^{\prime}_t}\left( \widetilde{\lambda}_4 v_d^2 - \widetilde{\lambda}_3 v_d^2 v^{\prime}_t - 2 \widetilde{\lambda}_2 v^{\prime 3}_t\right)\nn.
\eea
\eegp\par
In this scenario the masses for $W$ and $Z$ bosons are given by,
\begp
\allowdisplaybreaks
\begin{eqnarray}\label{Y0gaugemass}
M_W^2 = \frac{g^2_2}{ 4}v^2,
~~~~
M_Z^2 = \frac{g^2_2}{ 4 \cos^2\theta_W}v_d^2\, ,
\end{eqnarray}
\eegp
where VEVs are related to the SM VEV by $v^2=v_d^2+4 v^{\prime 2}_t=(246~{\rm GeV})^2$.

It is quite evident from eqn.~\ref{Y0gaugemass} that this model violates custodial symmetry at tree level, as,
\begp
\allowdisplaybreaks
\begin{eqnarray}
\rho &=& \frac{M_W^2}{ M_Z^2 \cos^2\theta_W}
=
1+4 \frac{v^{\prime 2}_t}{ v_d^2}
\, \nn.
\end{eqnarray}
\eegp\par
The experimental value of $\rho$ parameter is $1.0004\pm0.00024$~\cite{Agashe:2014kda}. Hence, $\delta \rho\approx0.0004\pm0.00024$. At tree-level, this puts a stringent constraint on the triplet scalar VEV, $v^{\prime}_t <4$ GeV at 3$\sigma$.
\subsection{Scalar Masses and Mixing for HTM ($Y=0$)}
After spontaneous symmetry breaking, there are two  physical neutral $CP$-even Higgs $h$, $H$, a pair of charged Higgs $H^\pm$ and three Goldstone bosons $G^\pm$, $G^0~(\equiv\zeta_d)$. The masses of the physical particles are given by,
\begp
\allowdisplaybreaks
\bea
M_{H^\pm}^2 &=&\widetilde{\lambda}_4 \frac{(v_d^2 + 4 v^{\prime 2}_t)}{2 v^{\prime}_t},\nn\\
M^2_{h} &=& \frac{1}{2}\left[(\widetilde{B}+\widetilde{A})-\sqrt{(\widetilde{B}-\widetilde{A})^{2}+4 \widetilde{C}^{2}}\right],\label{htmy0h0}\\
M^2_{H} &=& \frac{1}{2}\left[(\widetilde{B}+\widetilde{A})+\sqrt{(\widetilde{B}-\widetilde{A})^{2}+4 \widetilde{C}^{2}}\right],\nn
\eea
\eegp
where,
\begp
\allowdisplaybreaks
\bea
\widetilde{A} &=&{2 \lambda}_1 v_d^2,\nn\\
\widetilde{B} &=&\frac{{\lambda}_4 v_d^2 + 4 {\lambda}_2 v^{\prime 3}_t}{2 v^{\prime}_t},\label{ctilde}\\
\widetilde{C} &=&-{\lambda}_4 v_d + {\lambda}_3 v_d v^{\prime}_t.\nn
\eea
\eegp\par
The mixing between the SM doublet and the real triplet in the charged and $CP$-even scalar sectors for
this scenario are respectively given by,
\begp
\allowdisplaybreaks
\begin{eqnarray}
\begin{pmatrix}
G^\pm \\H^\pm \end{pmatrix} &=& \begin{pmatrix}
\cos\widetilde{\beta} & \sin\widetilde{\beta}\\
-\sin\widetilde{\beta} & \cos\widetilde{\beta}
\end{pmatrix}
\begin{pmatrix}
\phi^\pm \\ \eta^\pm \end{pmatrix},\nn\\
\begin{pmatrix}
h \\H \end{pmatrix} &=& \begin{pmatrix}
\cos\gamma & \sin\gamma\\
-\sin\gamma & \cos\gamma
\end{pmatrix}
\begin{pmatrix}
h^0 \\ \eta^0 \end{pmatrix}\nn,
\end{eqnarray}
\eegp 
with $\sin\gamma=\sqrt{\frac{\sqrt{(\widetilde{B}-\widetilde{A})^2 + 4\widetilde{C}^2}-(\widetilde{B}-\widetilde{A})}{2 \sqrt{(\widetilde{B}-\widetilde{A})^2 + 4\widetilde{C}^2}}}$~~~and~~~
$\tan\widetilde{\beta} =\frac{2 v^{\prime}_t}{v_d}$.

In large $\mu_2^2$ and small $v^{\prime}_t$ limit $\sin\gamma$ and $\sin\widetilde{\beta}$ can be written as,
\begp
\allowdisplaybreaks
\bea
\sin\gamma ~&=& \sqrt{\frac{1}{2}-\frac{1}{2\sqrt{1+ 16\frac{v^{\prime 2}_t}{v_d^2}}}}~\approx~0,\nn\\
\sin\widetilde{\beta} ~&=& \frac{2 v^{\prime}_t}{\sqrt{v_d^2 + 4 v^{\prime 2}_t}}~\approx~0 . \nn
\eea
\eegp\par
In this limit the $\widetilde{\lambda}$'s look like,
\begp
\allowdisplaybreaks
\bea
\widetilde{\lambda}_1&=&\frac{M_h^2}{2 v_d^2},\nn\\
\widetilde{\lambda}_2&=&\frac{2 (M_H^2 -M^2_{H^\pm})}{  v_d^2 \sin^2\widetilde{\beta}},\nn\\
\widetilde{\lambda}_3&=&\frac{2(M^2_{H^\pm}- (\sin\gamma/\sin\widetilde{\beta}) M^2_H)}{ v_d^2},\nn\\
\widetilde{\lambda}_4&=&\frac{\sin\widetilde{\beta} M^2_{H^\pm}}{v_d^2}.\nn
\eea
\eegp\par
Also, in the same limit, if $M_{H^\pm}$ and $M_H$ are very heavy compared to $M_h$, then both the charged and heavy neutral particle masses should be degenerate (see eqns.~\ref{htmy0h0} and \ref{ctilde}). For large differences in $M_{H^\pm}$ and $M_H$, the $\widetilde{\lambda}_{2,3}$ become non-perturbative (see Section~\ref{sec:perturbativity}), and in addition it may violate the unitarity (Section~\ref{sec:unitarity}) conditions.
\subsection{Constraints on the HTM ($Y=0$)}
\label{sec:constraintsHTM}
The parameter space of this model is constrained by theoretical considerations like  absolute vacuum stability, perturbativity, and unitarity of the scattering matrix. In the following, these theoretical bounds and the bounds from the electroweak precision measurements on the HTM ($Y=0$) will be discussed.
\subsubsection{Constraints from stability of the scalar potential}
A necessary condition for the stability of the vacuum comes from requiring that the scalar potential is bounded from below when the scalar fields become large in any direction of the field space. At the tree-level scalar potential potential $V(\Phi,\widetilde{\Phi})$ is bounded from below if
\begp
\allowdisplaybreaks
\beq
\widetilde{\lambda}_1(\Lambda) \geq 0, \quad  \widetilde{\lambda}_2(\Lambda) \geq 0, \quad  \widetilde{\lambda}_3(\Lambda) \geq - 2\sqrt{ \widetilde{\lambda}_1(\Lambda)\widetilde{\lambda}_2(\Lambda)}~,  
\label{stabilityboundHTM}\nn
\eeq
\eegp
where $\Lambda$ is an arbitrary energy scale. If the quantum corrections are included to the scalar potential, then the above conditions will be more complicated. The modification of the stability conditions of the scalar potential will be shown in Chapter~\ref{chap:MetaSMextended}.
\subsubsection{Perturbativity bounds}
\label{sec:perturbativity}
The radiatively improved scalar potential remain perturbative by requiring that all quartic couplings of $V(\Phi,\widetilde{\Phi})$ satisfy the following relations, 
\begp
\allowdisplaybreaks
\beq
\mid \widetilde{\lambda}_{1,2,3} (\Lambda)\mid \leq 4 \pi ~~~{\rm and}~~~\Big | \frac{\widetilde{\lambda}_4 (\Lambda)}{\Lambda} \Big | \leq 4 \pi~ . \nn
\eeq
\eegp

On applying such conditions, one implies upper bounds on the values of the couplings $\widetilde{\lambda}'s$ at low as well as high scales.
\subsubsection{Constraints from unitarity of the scattering matrix}
\label{sec:unitarity}
In this section, the derivation of unitary bounds on the quartic couplings of the scalar potential from the scattering matrix will be discussed. To the best of our knowledge, the full expressions of unitarity bounds on the scalar quartic couplings in this model have not yet been presented in the literature. It has been shown in Section~\ref{uniscalarg} of Chapter~\ref{chap:Intro} that the scattering matrix consists of only the quartic couplings of scalars, gauge bosons, and scalar-gauge bosons~\cite{Lee:1977eg}. 

At very high energies, the equivalence theorem implies that the longitudinal modes of the gauge bosons are equivalent to the corresponding Goldstone bosons, e.g., the amplitude of $W^+_L W^-_L \ra W^+_L W^-_L$ scattering is approximated by $G^+ G^- \ra G^+ G^-$. In this energy limit, pure scalar scattering processes are considered to obtain the unitarity bound on the quartic couplings of the scalar potential. The scalar quartic couplings in the physical bases, $G^\pm,~G^0~,H^\pm,$ $h,~H$, are complicated functions of $\widetilde{\lambda}$'s,$~\gamma,~ \widetilde{\beta}$. For example, the $hhhh$ vertex is 6($ \lambda_1 \cos^4\gamma + \lambda_3 \cos^2\gamma \sin^2\gamma + \lambda_2 \sin^4\gamma$). It is difficult to calculate the unitary bounds in the physical bases. So to simplify the quartic coupling, one can consider the non-physical scalar fields before electroweak symmetry breaking, $\phi^\pm,~\eta^\pm,~ \zeta_d,~h^0,~\eta^0$. Here the crucial point is that the S-matrix expressed in terms of the physical fields can be transformed into an S-matrix for the non-physical fields by making a unitary transformation~\cite{Kanemura:1993hm}.

Different quartic couplings in non-physical bases obtained by expanding the scalar potential of eqn.~\ref{ScalarpotHTM}, are given by,

\begin{minipage}{8cm}
\bea
\{\zeta_d~\zeta_d~\zeta_d~\zeta_d\}&=&6 \widetilde{\lambda}_{1},\nn\\
\left\{\phi^{+}~\phi^{+}~\phi^{-}~\phi^{-}\right\}&=&4 \widetilde{\lambda}_{1},\nn\\
\left\{\phi^{+}~\phi^{-}~h^0~h^0\right\}&=&2\widetilde{\lambda}_{1},\nn\\
\{\zeta_d~\zeta_d~\eta^0~\eta^0\}&=&\widetilde{\lambda}_{3},\nn\\
\{h^0~h^0~\eta^0~\eta^0\}&=&\widetilde{\lambda}_{3},\nn\\
\left\{\zeta_d~\zeta_d~\eta^{+}~\eta^{-}\right\}&=&\widetilde{\lambda}_{3},\nn\\
\left\{h^0~h^0~\eta^{+}~\eta^{-}\right\}&=&\widetilde{\lambda}_{3},\nn
\eea
\end{minipage}
\hspace{-4cm}
\begin{minipage}{11cm}
\bea
\left\{\zeta_d~\zeta_d~\phi^{+}~\phi^{-}\right\}&=&2 \widetilde{\lambda}_{1},\nn\\
\{\zeta_d~\zeta_d~h^0~h^0\}&=&2 \widetilde{\lambda}_{1},\nn\\
\{h^0~h^0~h^0~h^0\}&=&6 \widetilde{\lambda}_{1}, \nn\\
\left\{\phi^{+}~\phi^{-}~\eta^0~\eta^0\right\}&=&
\widetilde{\lambda}_{3},~\label{UntiHTM}\\
\{\eta^0~\eta^0~\eta^0~\eta^0\}&=& 6\widetilde{\lambda}_{2},\nn\\
\left\{\phi^{+}~\phi^{-}~\eta^{+}~\eta^{-}\right\}&=&
\widetilde{\lambda}_{3},\nn\\
\left\{\eta^0~\eta^0~\eta^{+}~\eta^{-}\right\}&=&2 \widetilde{\lambda}_{2},\nn
\eea
\end{minipage}
\begp
\allowdisplaybreaks
\bea
\left\{\eta^{+}~\eta^{+}~\eta^{-}~\eta^{-}\right\}&=&
4\widetilde{\lambda}_{2}.\nn ~~~~~~~~~~~~~~~~~~~~~~~~~~~~~~~~~~~~~~~~~~~~~~~~~~~
\eea
\eegp\par
The full set of these non-physical scalar scattering processes can be expressed as a $16\times16$ S-matrix. This matrix is composed of three submatrices of dimensions $6\times6$, $5\times5$, and $5\times5$ which have different initial and final states.

The first $6\times6$ sub-matrix ${\cal M}_1$ corresponds to scattering processes whose initial and final states are one of these: $(h^0~\phi^+,~\zeta_d~\phi^+$ $,~\eta^0~\phi^+,~h^0~\phi^+,~\zeta_d~\eta^+,~\eta^0~\eta^+)$. Using the Feynman rules in eqn.~\ref{UntiHTM}, one can obtain ${\cal M}_1$ as,
${\cal M}_1$=diag$(2\widetilde{\lambda}_1,~2\widetilde{\lambda}_1,
~2\widetilde{\lambda}_1$,$~\widetilde{\lambda}_3$,\\~$\widetilde{\lambda}_3$, $~\widetilde{\lambda}_3$).

The sub-matrix ${\cal M}_2$ corresponds to scattering process with one of the following initial and final states: $(h^0~\zeta_d,~\phi^+~\eta^-,~\eta^+~\phi^-,
~\eta^0~\zeta_d,~h^0~\eta^0)$. Similarly, one can calculate ${\cal M}_2$, it take the following form: ${\cal M}_2$=diag$(2\widetilde{\lambda}_1,~\widetilde{\lambda}_3,
~\widetilde{\lambda}_3,~\widetilde{\lambda}_3,
~\widetilde{\lambda}_3)$.

The third sub-matrix corresponds to scattering with one of the following initial and final states
$(\phi^+~\phi^-,~\eta^+~\eta^-,$ $~\frac{\zeta_d~\zeta_d}{\sqrt{2}},~\frac{h^0~h^0}{\sqrt{2}},~\frac{\eta^0~\eta^0}{\sqrt{2}})$. The factor $\frac{1}{\sqrt{2}}$ appeared due to statistics of identical particles. So one can find ${\cal M}_3$ with the help of the Feynman rules in eqn.~\ref{UntiHTM} and is given by,
\begp
\allowdisplaybreaks
\beq
{\cal M}_3=
\begin{pmatrix}
 4 \widetilde{\lambda}_1 & \widetilde{\lambda}_3 &  \sqrt{2} \widetilde{\lambda}_1 & \sqrt{2}\widetilde{\lambda}_1 & \frac{\widetilde{\lambda}_3}{ \sqrt{2}} \\
 \widetilde{\lambda}_3 & 4 \widetilde{\lambda}_2 & \frac{\widetilde{\lambda}_3}{ \sqrt{2}} & \frac{\widetilde{\lambda}_3}{ \sqrt{2}} &  \sqrt{2} \widetilde{\lambda}_2 \\
 \sqrt{2} {\widetilde{\lambda}_1} & \frac{\widetilde{\lambda}_3}{ \sqrt{2}} & 3 \widetilde{\lambda}_1 & \widetilde{\lambda}_1 & \frac{\widetilde{\lambda}_3}{2} \\
 \sqrt{2} \widetilde{\lambda}_1 & \frac{\widetilde{\lambda}_3}{ \sqrt{2}} & \widetilde{\lambda}_1 & 3 \widetilde{\lambda}_1 & \frac{\widetilde{\lambda}_3}{2} \\
 \frac{\widetilde{\lambda}_3}{ \sqrt{2}} & \sqrt{2} \widetilde{\lambda}_2 & \frac{\widetilde{\lambda}_3}{2} & \frac{\widetilde{\lambda}_3}{2} & 3 \widetilde{\lambda}_2
\end{pmatrix}\nn
\eeq
\eegp\par
The eigenvalues of ${\cal M}_3$ are : $\left\{ 2 \widetilde{\lambda}_1, 2 \widetilde{\lambda}_1,~2 \widetilde{\lambda}_2,~\frac{1}{2} \left(6 \widetilde{\lambda}_1+5 \widetilde{\lambda}_2\pm\sqrt{(6 \widetilde{\lambda}_1-5 \widetilde{\lambda}_2)^2+12 \widetilde{\lambda}_3^2}\right)\right\}$.

Unitary constraints of the scattering processes demand that the eigenvalues ${e_i}$'s ($i$=1,..,16) of the scattering-matrix should be less than $8\pi$.
\subsubsection{Constraints from electroweak precision experiments}
Electroweak precision data has imposed bounds on new physics models via Peskin-Takeuchi~\cite{Peskin:1991sw} $S, ~T, ~U$ parameters. The additional contributions from this model are given by~\cite{Forshaw:2003kh, Forshaw:2001xq},
\begp
\allowdisplaybreaks
\bea
S^{ Y=0}_{\rm HTM}\!\!\!\!& \simeq &\!\!\!\!  0, \nn\\
T^{ Y=0}_{\rm HTM} \!\!\!\!&=&\!\!\!\! \frac{1}{8\pi} \, \frac{1}{\sin^2\theta_W \cos^2\theta_W} \left[ \frac{M^2_H + M^{2}_{H^\pm}}{M^{2}_{Z}} \;
- \; \frac{2  M^{2}_{H^\pm} M^2_H}{M^{2}_{Z}(M^2_H -  M^{2}_{H^\pm})} \log\left(\frac{M^2_H}{M^{2}_{H^\pm}}\right)
\right] \nonumber \\
 \!\!\!\!&\simeq&\!\!\!\!\frac{1}{6\pi} \, \frac{1}{\sin^2\theta_W \cos^2\theta_W} \; \frac{(\Delta M)^{2}}{M^{2}_{Z}}, \nn\\
U^{ Y=0}_{\rm HTM} \!\!\!\!&=&\!\!\!\! -\frac{1}{3 \pi} \left( M^4_H\log \left( \frac{M^2_H}{M^{2}_{H^\pm}} \right) \frac{ (3 M^{2}_{H^\pm}-M^2_H)}{(M^2_H-M^{2}_{H^\pm})^3} + \frac{5(M^4_H+M^{4}_{H^\pm})-22 M^{2}_{H^\pm} M^2_H}{6(M^2_H-M^{2}_{H^\pm})^2} \right) \nonumber \\ \!\!\!\!&\simeq&\!\!\!\! \frac{\Delta M}{3 \pi M_{H^\pm}}.\nn
\eea
\eegp\par
$S$ is proportional to $\sin\widetilde{\beta}$. In the limit $\widetilde{\beta}\ra0$, the contribution to the $S$ parameter from the triplet scalar fields is negligible. For $\widetilde{\beta}\ra0$ and $M_{H^\pm,H}\gg M_h$ the charged particle ${H^\pm}$ and heavier $CP$-even Higgs $H$ are almost degenerate in mass, i.e., $\Delta M$ is very small. Also the contributions to the $T~{\rm and}~U$ parameters from this model are negligible. 
\subsubsection{Constraints from the LHC}
There is no direct search bound on the masses of the Higgs triplet model with hypercharge zero. One can get an indirect bound on the mass of the charged Higgs scalar from the diphoton excess. For $M_{H^\pm}<200$ GeV, $H^\pm$ will contribute significantly to the $\sigma(pp\ra\gamma\gamma)$ and $\sigma(pp\ra Z\gamma)$ cross-sections measured at the LHC.
\subsection{Dark Matter in HTM ($Y=0$)}
Like in the case of an additional singlet or doublet scalars, if one imposes a discrete symmetry on the extra triplet scalar then the neutral component of the scalar triplet can be considered as a viable dark matter candidate. In Chapter~\ref{chap:MetaSMextended}, the detailed study of $Z_2$-odd triplet scalar from metastability of electroweak vacuum as well as dark matter point of view will be discussed.
\section{Higgs triplet model with hypercharge $Y=2$ }\label{TriY2Ext}
The detailed discussion of the Higgs triplet model with hypercharge $Y=2$ will be provided in this section. The additional triplet is denoted by $\Delta = (\Delta_1,\Delta_2,\Delta_3)$, which can be written in the bi-doublet from as,
\begp
\allowdisplaybreaks
\begin{eqnarray}
\Delta = \frac{\sigma^i}{\sqrt 2}\Delta_i = \begin{pmatrix}
\delta^+/\sqrt 2 & \delta^{++}\\
\delta^0 & -\delta^+/\sqrt 2
\end{pmatrix},\nn
\end{eqnarray}
\eegp
where $\Delta_1=(\delta^{++}+\delta^0)/\sqrt 2,~\Delta_2=i(\delta^{++}-\delta^0)/\sqrt 2,~\Delta_3=\delta^+$. The Lagrangian of this model is given in the following, 
\begp
\allowdisplaybreaks
\begin{eqnarray}
{\cal L}^{Y=2} = {\cal L}^{Y=2}_{\rm Yukawa}+{\cal L}^{Y=2}_{\rm Kinetic}-V(\Phi,\Delta).
\label{lag}
\end{eqnarray}
\eegp
with the relevant kinetic and Yukawa interaction terms are respectively
\begp
\allowdisplaybreaks
\begin{eqnarray}
\label{y2kinetic}
{\cal L}^{Y=2}_{\rm kinetic} &=&\left(D_\mu\Phi\right)^\dag \left(D^\mu\Phi\right) +{\rm Tr}\left[\left(D_\mu \Delta\right)^\dag \left(D^\mu\Delta\right)\right], \\
\label{y2Yukawa}
{\cal L}^{Y=2}_{\rm Yukawa} &=& {\cal L}_{\rm Yukawa}^{\rm SM}-\frac{1}{\sqrt 2}\left(Y_\nu\right)_{ij} L_i^{\sf T}Ci\sigma_2\Delta L_j+{\rm h.c.}\,.
\end{eqnarray}
\eegp
$Y_\nu$ represents neutrino Yukawa coupling, and $C$ is the charge conjugation operator. Covariant derivative of the scalar triplet field is given by,
\begp
\allowdisplaybreaks
\begin{equation} 
D_\mu \Delta = \partial_\mu \Delta + i\frac{g_2}{2}[\sigma^a W_\mu^a,\Delta]+ig_1B_\mu \Delta \qquad (a=1,2,3).\nn
\end{equation}
\eegp\par
The scalar potential in eqn.~\ref{lag} is given by \cite{Arhrib:2011uy},
\begp
\allowdisplaybreaks
\begin{eqnarray} 
V(\Phi,\Delta) &=& -m_\Phi^2(\Phi^\dag \Phi)+\frac{\lambda^{\prime}_1}{4}(\Phi^\dag \Phi)^2+M^2_\Delta {\rm Tr}(\Delta ^\dag \Delta)+ \left(\mu \Phi^{\sf T}i\sigma_2\Delta^\dag\Phi+{\rm h.c.}\right)\,\nonumber\\
&& +\lambda^{\prime}_2\left[{\rm Tr}(\Delta^\dag \Delta)\right]^2+\lambda^{\prime}_3{\rm Tr}[(\Delta ^\dag \Delta)^2]+\lambda^{\prime}_4\Phi^\dag\Delta\Delta^\dag\Phi \nonumber\\
&&+\lambda^{\prime}_5(\Phi^\dag\Phi){\rm Tr}(\Delta ^\dag \Delta).
\label{eq:Vpd}
\end{eqnarray}
\eegp
$\lambda^{\prime}_i$ ($i=1,\ldots 5$) are real dimensionless coupling constants, while $m_\Phi, M_\Delta~{\rm and}~\mu$ are real mass parameters of the above potential. Here $\mu$ term is the only term which can generate a $CP$-phase, as the other terms of the potential are self-conjugate. However, that phase can be absorbed by redefining the fields $\Phi$ and $\Delta$. Furthermore, we assume that $m_\Phi^2 > 0$ for spontaneous symmetry breaking of above mentioned gauge group. After EWSB, the scalar potential (see eqn.~\ref{eq:Vpd}) is expanded around the VEVs of neutral scalar $CP$-even scalar fields: around $v_d$ as in eqn.~\ref{phio} and around $v_t$ as $\frac{1}{\sqrt{2}}(v_t+\xi^0+i\zeta_t)$ respectively. At the minimum of the scalar potential, the following conditions\cite{Arhrib:2011uy} are satisfied.
\begp
\allowdisplaybreaks
\begin{eqnarray}
m_\Phi^2 &=& \frac{\lambda^{\prime}_1 v^2_d}{4}-\sqrt{2}\mu v_t + \frac{(\lambda^{\prime}_5+\lambda^{\prime}_4)}{2}v_t^2,\label{eq:ewsb1}\nn\\ 
M_\Delta^2 &=& \frac{2\mu v^2_d-\sqrt{2}(\lambda^{\prime}_5+\lambda^{\prime}_4)v^2_d v_t-2\sqrt{2}(\lambda^{\prime}_2+\lambda^{\prime}_3)v^3_t}{2\sqrt{2}v_t}.
\label{eq:ewsb2}\nn
\end{eqnarray}
\eegp\par
Here the triplet VEV ($v_t$) contributes to the weakly interacting gauge boson masses at the tree-level as in the previous models. In this scenario the masses of weak gauge bosons are given as:
\begp
\allowdisplaybreaks
\begin{eqnarray}\label{y2gaugemass}
M_W^2 = \frac{g^2_2}{4}(v_d^2+2v^2_t),~~M_Z^2 = \frac{g^2_2}{4\cos^2\theta_W}(v_d^2+4v_t^2).
\end{eqnarray}
\eegp\par
In the following sections, the particle spectrum, the theoretical and experimental constraints on this model will be discussed.
\subsection{Scalar Masses and Mixing for HTM ($Y=2$)} 
After minimization of the scalar potential $V(\Phi,~\Delta)$ in eqn.~\ref{eq:Vpd} with respect to the VEVs, a $10\times 10$ squared mass matrix appeared for the scalars. This matrix is composed of four $2\times2$ and two $1\times1$ matrices. Among the ten eigenstates, seven are physical ($H^{\pm\pm},~H^\pm,~h,~H,~A$) and they are massive. Remaining three states are massless Goldstone bosons $G^\pm,~G^0$, eaten up to give mass to the SM gauge bosons $W^\pm,~Z$. The mass eigenvalues for the physical scalars are given by,
\begp
\allowdisplaybreaks
\begin{eqnarray}
M^2_{H^{\pm\pm}} &=& \frac{\sqrt{2}\mu v^2_d-\lambda^{\prime}_4v^2_d v_t-2\lambda^{\prime}_3v^3_t}{2v_t},\nn\\
M^2_{H^\pm} &=& \frac{(v_d^2+2v^2_t)(2\sqrt{2}\mu-\lambda^{\prime}_4 v_t)}{4v_t},\label{sing-mass}\nn\\
M^2_{A} &=& \frac{\mu(v_d^2+4v^2_t)}{\sqrt{2}v_t},\\
M^2_{h} &=& \frac{1}{2}\left(B'+A'-\sqrt{(B'-A')^2+4C'^2}\right),\nn\\
M^2_{H} &=& \frac{1}{2}\left(B'+A'+\sqrt{(B'-A')^2+4C'^2}\right),\label{eq:scamass}\nn
\end{eqnarray}
\eegp
${\rm with}~~
A' = \frac{\lambda^{\prime}_1}{2} v_d^2,~~
B' = \frac{\sqrt{2}\mu v^2_d+4(\lambda^{\prime}_2+\lambda^{\prime}_3)v^3_t}{2v_t},~~
C' = v_d[-\sqrt{2}\mu+(\lambda^{\prime}_5+\lambda^{\prime}_4)v_t]$.

The mixing between the doublet and triplet scalar fields in the charged, $CP$-even and $CP$-odd scalar sectors are respectively denoted by,
\begp
\allowdisplaybreaks
\begin{eqnarray}
\begin{pmatrix}
G^\pm \\H^\pm \end{pmatrix} &=& \begin{pmatrix}
\cos\beta' & \sin\beta'\\
-\sin\beta' & \cos\beta'
\end{pmatrix}
\begin{pmatrix}
\phi^\pm \\ \delta^\pm \end{pmatrix},\nn\\
\begin{pmatrix} 
h \\H \end{pmatrix} &=& \begin{pmatrix}
\cos\gamma' & \sin\gamma'\\
-\sin\gamma' & \cos\gamma'
\end{pmatrix}
\begin{pmatrix}
\eta^0 \\ \xi^0 \end{pmatrix},\nn\\
\begin{pmatrix}
G^0 \\ A \end{pmatrix} &=& \begin{pmatrix}
\cos\delta' & \sin\delta'\\
-\sin\delta' & \cos\delta'
\end{pmatrix}
\begin{pmatrix}
\zeta_d \\ \zeta_t \end{pmatrix},\nn
\end{eqnarray}
\begp
where the mixing angles are given by,
\begp
\allowdisplaybreaks
\begin{eqnarray}
\tan\beta' &=& \frac{\sqrt 2 v_t}{v_d},\label{mix1}\nn\\
\tan\delta' &=& \frac{2v_t}{v_d},
\label{mix2}\nn\\
\tan{2\gamma'} &=& \frac{2C'}{B'-A'}.\nn
\end{eqnarray}
\eegp
\subsection{Constraints on the HTM ($Y=2$)}\label{stab}
Let us  summarize the constraints on the scalar potential of the  $Y=2$ triplet model given by eqn.~\ref{eq:Vpd} in order to sustain the stability of the electroweak vacuum, the conservation of the tree-level unitarity in various scattering processes, etc.
\subsubsection{Constraints from stability of scalar potential}
Stability of the electroweak vacuum of the scalar potential in eqn.~\ref{eq:Vpd} requires that it be bounded from below i.e., there is no direction in field space along which the potential tends to minus infinity. The necessary and sufficient conditions are~\cite{Arhrib:2011uy}:
\begp
\allowdisplaybreaks
\bea
&&\lambda^{\prime}_1 (\Lambda) \geq 0,~~\lambda^{\prime}_2(\Lambda)+\lambda^{\prime}_3(\Lambda)\geq 0, ~~\lambda^{\prime}_2(\Lambda)+\frac{\lambda^{\prime}_3(\Lambda)}{2} \geq 0,\nn \nn\\
&&\lambda^{\prime}_5(\Lambda)+ \sqrt{\lambda^{\prime}_1(\Lambda)(\lambda^{\prime}_2(\Lambda)+\lambda^{\prime}_3(\Lambda))}\geq 0,\nn\\
&&\lambda^{\prime}_5(\Lambda)+ \sqrt{\lambda^{\prime}_1(\Lambda)\left(\lambda^{\prime}_2(\Lambda)+\frac{\lambda^{\prime}_3(\Lambda)}{2}\right)} \geq 0,\nn\\
&&\lambda^{\prime}_5(\Lambda)+\lambda^{\prime}_4(\Lambda)+\sqrt{\lambda^{\prime}_1(\Lambda)(\lambda^{\prime}_2(\Lambda)+\lambda^{\prime}_3(\Lambda))}\geq 0,\nn\\ &&{\rm and}~~\lambda^{\prime}_5(\Lambda)+\lambda^{\prime}_4(\Lambda)+ \sqrt{\lambda^{\prime}_1(\Lambda)\left(\lambda^{\prime}_2(\Lambda)+\frac{\lambda^{\prime}_3(\Lambda)}{2}\right)} \geq 0.\nn
\label{eq:lstab}
\eea
\eegp
where the coupling constants are evaluated at an arbitrary running scale $\Lambda$.
\subsubsection{Perturbativity constraints}
For the HTM ($Y=2$) to behave as a perturbative quantum field theory at any given scale, one must impose the conditions on the radiatively improved scalar potential $V(\Phi,\Delta)$ as, 
\begp
\allowdisplaybreaks
\beq
\mid \lambda^{\prime}_{1,2,3,4,5}(\Lambda)\mid \leq 4 \pi, ~~~~~~~\Big | \frac{\mu(\Lambda)}{\Lambda}\Big | \leq 4 \pi.
\label{pertHT2}\nn
\eeq
\eegp

These conditions imply an upper bound on the couplings $\lambda's$ and $\mu$ at an energy scale $\Lambda$. 
\subsubsection{Constraints from unitarity of the scattering matrix}
The tree-level unitarity of the S-matrix for elastic scattering imposes the following constraints~\cite{Arhrib:2011uy}:
\begp
\allowdisplaybreaks
\begin{eqnarray}
&&|\lambda^{\prime}_5+\lambda^{\prime}_4| \leq 8\pi,~~
|\lambda^{\prime}_5| \leq 8\pi\,,~~
|2\lambda^{\prime}_5+3\lambda^{\prime}_4|\leq 16\pi,\nn\\
&&|\lambda^{\prime}_1| \leq 16\pi\,\nonumber,~~
|\lambda^{\prime}_2| \leq 4\pi\,~~
|\lambda^{\prime}_2+\lambda^{\prime}_3| \leq 4\pi,\\
&& \Big |\lambda^{\prime}_1+4\lambda^{\prime}_2+8\lambda^{\prime}_3\pm\sqrt{(\lambda^{\prime}_1-4\lambda^{\prime}_2-8\lambda^{\prime}_3)^2+16\lambda^{\prime 2}_4}\Big |\leq 32\pi, \nn\\
&&\Big |3\lambda^{\prime}_1+16\lambda^{\prime}_2+12\lambda^{\prime}_3
\pm\sqrt{(3\lambda^\prime_1-16\lambda^{\prime}_2-
12\lambda^{\prime}_3)^2+24(2\lambda^{\prime}_5
+\lambda^{\prime}_4)^2} \Big |\leq 32\pi,\nonumber \\
&&|2\lambda^{\prime}_5-\lambda^{\prime}_4|\leq 16\pi\,~~{\rm and}~~|2\lambda^{\prime}_2-\lambda^{\prime}_3|\leq 8\pi.\nonumber
\label{eq:luni}
\end{eqnarray}
\eegp
\subsubsection{Constraints from the neutrino mass and the electroweak precision experiments}\label{prey2}
Here we would like to discuss the constraints on the parameters coming from electroweak precision tests and from the neutrino mass. In this model, using eqn.~\ref{y2Yukawa}, neutrino mass can be expressed as $(M_\nu)_{ij} = v_t ~(Y_\nu)_{ij}$. Consequently for realistic neutrino masses its calls for a tiny value of $v_t$ or $Y_\nu$ or both. 

If $Y_\nu \sim {\cal O}(1)$, then to satisfy sub-eV neutrino mass, $v_t$ should be ${\cal O}(10^{-9})$ GeV. This can be achieved via the type-II seesaw mechanism. Using eqn.~\ref{eq:ewsb2}, for $v_t << v_d$, the triplet VEV can be expressed as $v_t \equiv \frac{\mu v_d^2}{M_{\Delta}^2}$. So one can choose a large $M_{\Delta}$ to get a neutrino mass of ${\cal O}(0.1)$ eV. This mechanism provides neutrino mass naturally. At this limit, the couplings of the new scalar particles with the SM particles are vanishingly small. One can take another extreme limit on $v_t$ using $\rho$-parameter at the tree-level. In this model, the $\rho$-parameter can be calculated using the eqn.~\ref{y2gaugemass} and is given by,
\begp
\allowdisplaybreaks
\begin{eqnarray}\label{y2rho}
\rho = \frac{M_W^2}{M_Z^2\cos^2\theta_W} = \frac{1+\frac{2v_t^2}{v_d^2}}{1+\frac{4v_t^2}{v_d^2}},\nn
\end{eqnarray}
\eegp
and thereby changing the $\rho$-parameter at the tree-level from its SM value $\rho=1$. The electroweak precision measurements require the $\rho$-parameter to be very close to unity $1.0004\pm0.00024$~\cite{Agashe:2014kda}. This puts a severe restriction on the VEV of neutral $CP$-even component of the scalar triplet as, $v_t < 4$ GeV. In order to satisfy the neutrino mass, $Y_\nu$ should be ${\cal O}(10^{-9})$ for $v_t \sim {\cal O}$(1) GeV. So in this limit, the explanation of neutrino masses is quite unnatural. In this case, the coupling of the new extra scalar particles to SM particles are not as small as in the previous case with $v_t \approx10^{-9}$ GeV. The model benchmark points have been chosen with $v_t=3$ GeV in Chapter \ref{chap:VVscatt}.

At loop-level, the contributions of the scalar triplet with hypercharge $Y=2$ to the $S$, $T$ and $U$ parameters are given by~\cite{Lavoura:1993nq, Chun:2013vca}, 
\begp
\allowdisplaybreaks
\begin{eqnarray}
S^{Y=2}_{HTM} &=& -\frac{1}{3\pi} \ln\frac{m_{+1}^2}{ m_{-1}^2}
-\frac{2}{\pi} \sum_{T_3=-1}^{+1} (T_3 - Q s_W^2)^2 \,
\xi\left(\frac{m_{T_3}^2}{ m_Z^2}, \frac{m_{T_3}^2}{ m_Z^2}\right), \nn\\
T^{Y=2}_{HTM} &=& \frac{1}{ 16\pi c_W^2 s_W^2} \sum_{T_3=-1}^{+1} \left(2-T_3(T_3-1)\right)\,
F\left(\frac{m_{T_3}^2}{ m_Z^2}, \frac{m_{T_3-1}^2}{ m_Z^2}\right), \nonumber\\
U^{Y=2}_{HTM} &=& \frac{1}{6\pi} \ln \frac{m_{0}^4 }{ m_{+1}^2 m_{-1}^2}
+\frac{1}{\pi} \sum_{T_3=-1}^{+1} \left[ 2(T_3 - Q s_W^2)^2\,
\xi\left(\frac{m_{T_3}^2}{ m_Z^2}, \frac{m_{T_3}^2}{ m_Z^2}\right) \right.\nonumber\\
&& \left. ~~~~~~~~~~~~~~~~~~~~~~~~~~~~
-(2-T_3(T_3-1))\, \xi\left(\frac{m_{T_3}^2}{ m_W^2}, \frac{m_{T_3}^2}{ m_W^2}\right) \right], \nonumber
\end{eqnarray}
\eegp
where $m_{+1,0,-1} \equiv M_{H^{++},H^+,H^0}$ and the function $\xi(x,y)$ is defined as~\cite{Lavoura:1993nq},
\begp
\allowdisplaybreaks
\begin{eqnarray}
\xi (x_1, x_2) & = &
\frac{4}{9} - \frac{5}{12} (x_1 + x_2) + \frac{1}{6} (x_1-x_2)^2
\nonumber\\
            &  &
+ \frac{1}{4} \left[ x_1^2 - x_2^2 - \frac{1}{3} (x_1-x_2)^3 -
\frac{x_1^2 + x_2^2}{x_1 - x_2} \right] \ln \frac{x_1}{x_2}
\nonumber\\
            &  &
- \frac{1}{12} \Delta(x_1, x_2) f(x_1, x_2)\, .
\label{eq:csi}\nn
\end{eqnarray}
\eegp
The definitions of $\Delta$, $f$ and $F$ can be found in eqns.~\ref{fxy} and \ref{Fm1m2}.

These parameters can be used for constraining the parameter space of this model from the electroweak precision data~\cite{Baak:2014ora}.
\subsubsection{Constraints from the LHC}

For $v_t \sim 1$ GeV, the direct search of $H^{\pm\pm}$ via $pp \ra H^{++} H^{--}, ~H^{\pm\pm}\ra W^{*\pm} W^{*\pm}\ra$ \newpage$ \mu^{\pm} \nu_\mu \mu^{\pm}\nu_\mu$ process at the LHC, puts a lower bound on $M_{H^{\pm\pm}}>84$ GeV~\cite{Kanemura:2014ipa}.
If constraints like the stability, unitarity, $T$-parameter and $\mu_{\gamma\gamma}$ at LHC are considered, 
then one can obtain the following lower bounds on the nonstandard scalar masses: $M_{H^+}>130$ GeV, $M_{A,H}>150$ GeV~\cite{Das:2016bir}.

\subsection{Dark matter in HTM ($Y=2$)}
If a discrete $Z_2$ symmetry is imposed on the triplet scalar such that the couplings of an $odd$ number of scalar fields of the triplet with the SM particles are prohibited, then lightest of $H$ and $A$ can serve as a viable DM candidate which may saturate the measured DM relic density of the Universe.
In this model, the dark matter candidate can annihilate to the SM particles via the exchange of a Higgs or a $Z$ boson through $s$-channel diagrams and $H$, $A$ and $H^\pm$ mediated $t$- and $u$-channel diagrams. As the dark matter particle can interact with the nucleons through Higgs and $Z$ mediated $t$-channel exchanges, the dark matter direct detection cross-sections rather large in this model. It was shown in Ref.~\cite{Araki:2011hm}, a significant portions of the parameter space is excluded~\cite{Araki:2011hm} from direct detection experimental data from XENON and LUX.
\section{Summary}
In this chapter, several extensions of the scalar sector namely, SM$+S$, 2HDM, HTM with hypercharge $Y=0,~2$ have been explored. At a time, it has been considered that one such extended sector as a new physics option beyond the SM.
It is assumed that the electroweak symmetry breaking driven by the $CP$-even component the SM Higgs doublet and $CP$-even component of the extra scalar sector. In this case, both the $CP$-even scalar gets VEV and 
scalar field(s) of the extended sector can be mixed with the scalar fields of the SM Higgs doublet.
The Goldstone boson, which is a combination of the scalar fields of the SM doublet and extra scalar sector, is the longitudinal component of corresponding vector boson.
Also, it has been shown that the extra scalar fields can have direct couplings with the SM particles, or these may get generated after the electroweak symmetry breaking.
The other combinations become new physical scalar fields. It has been seen that depending on the isospin $I$ and hypercharge $Y$, the models with extended scalar sectors have different numbers of neutral and charged scalar fields.
Particle spectrum for different scalar extensions of the standard model has been calculated.
For each extension of scalar sector of the standard model, various theoretical and phenomenological constraints have been shown.
Using the above constraints, one can get allowed parameter space of these models.
In this work, several benchmark points have been chosen, for which the vector boson scattering cross-section distributions with CM energy of these new models such as 2HDM, HTM with hypercharge $Y=0,2$ have been shown in Chapter~\ref{chap:VVscatt}. 
It has also been discussed in this chapter if one impose a discrete symmetry $Z_2$ on the extended scalar such that $odd$ number of scalar fields do not couple with the SM particles then the lightest neutral scalar particle becomes stable. This scalar field can be taken as a dark matter candidate which may fulfill the relic abundance of the Universe that will be discussed in Chapter~\ref{chap:MetaSMextended}.

\chapter{Vector boson scattering in extended Higgs sector}
\label{chap:VVscatt}
\linespread{0.1}
\graphicspath{{Chapter4/}}
\pagestyle{headings}
\noindent\rule{15cm}{1.5pt} 
\section{Introduction}
It has been discussed in the earlier Chapters that the measurement of properties of this scalar boson at Large Hadron Collider (LHC), is consistent with the minimal choice of the scalar sector, consisting of a single complex doublet. However, the data still allow an extended scalar sector, which, in turn, can accommodate a more elaborate mechanism for the EWSB. One immediate extension of this kind is the presence of either additional scalar doublet(s) or higher multiplet(s) of $SU(2)_L$.
Even a marginal role of such scalars can in principle be probed in the upcoming experiments, utilizing their interaction with the electroweak gauge bosons.\footnote {It should be noted that the new scalars may not always participate in the EWSB, \eg as in the inert scalar model \cite{Khan:2014kba,Khan:2015ipa}.}

If indeed there are additional scalars that couple with the $W$-and $Z$-bosons, longitudinal vector boson scattering~(VBS) including scalar exchanges should provide a complementary way to direct search methods to probe into the scalar sector. In the SM, the Higgs boson helps preserve the unitarity of the $S$-matrix for the longitudinal electroweak vector boson scattering $V_L V_L \ra V_L V_L$.
The Higgs boson mediated diagram precisely cancels the residual $s$-dependence (where $\sqrt{s}$ denotes the energy in the center-of-mass frame), thus taming the high energy behavior of the cross-section appropriately~\cite{Dawson:1998yi}.
With an extended scalar sector, the preservation of unitarity could be a more complex process. Several factors then modify the $\sqrt{s}$-dependence of the $V_{L}V_{L}$ scattering.
The first of these is the modification, albeit small, of the strength of the 125~GeV scalar to gauge boson pairs. Secondly, the extent of the influence of other scalars present in an extended scenario depends on their gauge quantum numbers and on the theoretical scenario in general.
Thirdly, the observed mass of the 125~GeV scalar makes it kinematically impossible for it to participate as an $s$-channel resonance in $V_{L}V_{L}$ scattering processes.  However, such resonant peaks may in general occur when heavier additional scalars enter into the arena.

One can thus expect that the $\sqrt{s}$-dependence of $V_{L}V_{L}$
scattering cross-sections will be modified with respect to
SM-expectations as a result of the above effects. Such modifications
have been formulated in terms of certain general parameters in some
recent studies~\cite{Bhattacharyya:2012tj, Choudhury:2012tk, Chang:2013aya, Cheung:2008zh}. An apparent non-unitarity of the scattering
matrix may be noticed here when the SM-like scalar with modified
interaction strength is participating as the only scalar.
However, unitarity is restored once the complete particle spectrum is taken into
consideration.
We emphasize here that the three effects mentioned
above leave the signature of the specific non-standard EWSB sector in
the modified energy-dependence, as long as the new scalars have their masses
within or about the TeV-range.

To further elaborate our point, the purpose of writing this chapter is twofold. Using resonances at various $V_L V_L$ (where $V \equiv W^{\pm}~{\rm or}~Z$) scattering processes, we illustrate that it may be possible to distinguish between different extensions of the scalar sector, once the high-energy run of the LHC continues long enough. The shapes of the energy-dependence curves, especially the presence of resonant peaks, can shed light on the relevant scalar spectra of these models.  We use for illustration some popular extensions like the Type-II two Higgs doublet model (2HDM) and real as well as complex Higgs triplet models (HTM). It is shown in the ensuing study how the $\sqrt{s}$-dependence of the cross-sections reflect the characteristics of each of these scenarios so long as the additional scalars lie within about 2 TeV. This supplements rather faithfully other LHC-based phenomenology, and thus spurs the improvement of techniques to extract the dependence of $V_L V_L$ scattering cross-sections on $\sqrt{s}$. The other goal is
to present analytical expressions for $V_L V_L$ scattering amplitudes in these otherwise well-motivated models. To the best of our knowledge, these full expressions have not yet been presented in the literature.
  
\section{Polarization of Vector Bosons} 
To handle the electroweak sector, one of the main focus of this work is the polarization of the massive vector bosons $W^\pm$ and $Z$. In this section, the longitudinal and transverse modes of the vector bosons will be defined. 

The Lagrangian of a vector boson fields with mass $m$ is given by,
\begp
\allowdisplaybreaks
\beq
\mathcal{L} = - \frac{1}{4} F^{\mu\nu} F_{\mu\nu} - \frac{m^2}{2} A^\mu A_\mu.
\label{DegLag}
\eeq
\eegp
This Lagrangian is not gauge invariant because the mass term violates the $U(1)$ symmetry. The field strength tensor is given by, $F^{\mu\nu} = \partial^\mu A^\nu - \partial^\nu A^\mu $. Using the Euler-Lagrange equation $\partial^\mu \left( \frac{\partial \mathcal{L}}{\partial^\mu A^\nu} \right) - \frac{\partial \mathcal{L}}{\partial A^\nu} = 0$, the equation of motion of the gauge fields can be obtained for Lagrangian of eqn.~\ref{DegLag} as,		
\begp
\allowdisplaybreaks
 \beq
-\partial^\mu F_{\mu\nu} - m^2 A_\nu = 0.
\label{LagEOM}
\eeq
\eegp
In the momentum space, these equations can be written as,
\begp
\allowdisplaybreaks
\beq
- p^2 A_\nu + p_\nu p^\mu A_\mu  - m^2 A_\nu = 0.\nn
\eeq
\eegp
Now multiply this equation with $p^\nu$ to get,
\begp
\allowdisplaybreaks
\bea
- p^2 p^\nu A_\nu + p^2 p^\mu A_\mu  - m^2 p^\nu A_\nu &=& 0\nn\\
 \Rightarrow \hspace{2cm}p^\nu A_\nu &=& 0.\nn
\eea
\eegp
This reduces the numbers of independent components of the polarization vector of a massive vector boson to three.

For the massless field, one write eqn.~\ref{LagEOM} as,
\begp
\allowdisplaybreaks
\beq
\partial^\mu F_{\mu\nu} =0 \Rightarrow p^2 A_\nu - p_\nu p^\mu A_\mu =0.
\eeq
\eegp
These are the Maxwell equations which also give, $p^\mu A_\mu=0$, i.e., three degrees of freedom remains for a massless vector field. Masslessness of the photon demands the corresponding Lagrangian is invariant under a gauge transformation,
\begp
\allowdisplaybreaks
\beq
A_\mu \ra A_\mu + \partial_\mu f,
\eeq
\eegp
where $f$ is an function that satisfies $\square^2 f=0$. Now, one can choose $f$ such that $A_0(p)=0$, which is not possible for massive vector fields. So for the massless photon, we have an extra constraint on the vector fields due to the gauge invariance. Thus, the photon has only two polarization vectors which are transverse to the three-momentum of the photon.
 
Now the polarization vector will be calculated in terms of momentum and energy for a massive vector boson. The components of the polarization vector in $x$-, $y$-, and $z$-directions are chosen as,
\begp
\allowdisplaybreaks \bea
\epsilon_x^\mu &=& (0,1,0,0),\nn\\
\epsilon_y^\mu &=& (0,0,1,0),\nn\\
\epsilon_z^\mu &=& (x_1,x_2,x_3,x_4).\nn
\eea
\eegp
The polarization vector follows,
\begp
\allowdisplaybreaks
 \beq
\epsilon_i . \epsilon_j  = - \delta_{ij}, \quad {\rm and} \quad p.\epsilon(p) = 0,
\label{polarule}
\eeq
\eegp
where $p_\mu=(E,p_x,p_y,p_z)$ is the four momentum of the vector boson. Using eqn.~\ref{polarule}, we get, $x_2=x_3=0$ and,
\begp
\allowdisplaybreaks
 \bea
\epsilon_z . \epsilon_z = -1 \quad &\Rightarrow & \quad x_1^2 -x_4^2 = -1,\nn\\
p . \epsilon(p) = 0 \quad & \Rightarrow & \quad p_x=0,~p_y=0 ~{\rm and} ~x_4 = \frac{E}{p_z} x_1. \nn
\eea
\eegp
Thus, we get $x_1 = \frac{|{\bf p}|}{m}$, with $|{\bf p}| = \sqrt{p_x^2+p_y^2+p_z^2}$ and $x_4 = \frac{E}{m}$. The polarization vector along the $z$-axis can be written as,
\begp
\allowdisplaybreaks \beq
\epsilon_z = \left( \frac{|\bf p|}{m},\frac{E}{m} \hat{p} \right).
\eeq
\eegp
This is the longitudinal component and the combinations  $\epsilon^\pm = \frac{1}{\sqrt{2}} (\epsilon_x \pm i \epsilon_y)$ are the transverse components of the polarization vector corresponding to a massive vector.

From the above, it is evident that the longitudinal polarization of vector bosons grows with energy, whereas transverse components are independent of energy. Hence, at high energies $E \gg m$ one expects that longitudinal mode of the gauge bosons dominates over transverse modes. This is the difference between the longitudinal and transverse components of the gauge bosons.
\section{Connection with Electroweak Symmetry Breaking}
In the SM, after EWSB the charged scalar fields $G^\pm$ and $G^0$ of the scalar doublet are eaten by the $W^\pm$ and $Z$ gauge bosons respectively, which become massive and also gain a third, longitudinal polarization.
The longitudinal component of massive gauge boson is a Goldstone scalar field that belongs to the SM doublet. So through vector bosons scattering one can explore the Higgs sector.

In the extended scalar sectors such as doublet-, or triplet-extensions, one of the linear combination of charged scalar fields is eaten by the $W$ boson which becomes massive, other combinations of fields become massive charged scalar fields.
Similarly, a combination of pseudo scalars become the longitudinal part of massive $Z$ gauge boson and the other combinations survive as physical pseudoscalars.
One can use VBS processes to find the nature of the electroweak symmetry breaking in the presence of these extra scalars.
In future colliders, if we find any direct signature of the extended scalar sector then this study will supplement our knowledge about these scalars obtained from direct searches.

In the following section, the study of the scattering cross-section using the partial wave analysis will be reviewed.
\vspace{-0.2cm}
\section{Scattering cross-section: Partial Wave Analysis}
Let us consider a mono-energetic beam of particles being scattered by a target located at $(0, 0, 0)$. We will work here in the spherical coordinate system,
\begp
\allowdisplaybreaks \beq
(r \sin\theta \cos\phi ,r \sin\theta \sin\phi, r  \cos\theta)\nn
\eeq
\eegp
It is assumed that the detector covers a solid angle $d\Omega = \sin\theta d\theta d\phi$ in the direction ($\theta,\phi$) from the scattering center and the incoming beam travels along the $z$-axis. The wave vector is given by,
\begp
\allowdisplaybreaks \beq
\vec{k} = \frac{\sqrt{2 m E}}{\hbar} \hat{z},\nn
\eeq
\eegp
where $E$ is the energy of the incident beam. The number of particles per unit time entering the detector is $Nd\Omega$. The differential cross-section can be written as,
\begp
\allowdisplaybreaks \beq
\frac{d\sigma}{d\Omega} = \frac{dn}{F_{in}},
\label{difcross1}
\eeq
\eegp
which is defined as the number $dn$ of particles scattered into the direction ($\theta,\phi$) per unit time, per unit solid angle, normalized by the incident flux. Here $F_{in}$ is the incident flux of the beam, which is defined as the number of particles per unit time, crossing a unit area placed normal to the direction of incidence. In the next, we will calculate the differential cross-section.

The plane wave of the incident particles are defined as $\psi = A e^{i k z}$, with normalization $A$. The wave encounters a scattering potential producing an outgoing spherical wave. At large distances from the scattering center, one can decompose the wave function $\psi(r)$ into a part $e^{i k z}$ describing the incident beam and a part $\psi_{sc}$ for the scattered particles. As the collision is elastic, i.e., the energies of the incident and scattered particle are the same, one can write the wave function as,
\begp
\allowdisplaybreaks \beq
\psi(r)\Big |_{r\ra \infty} \approx e^{i k z} + \underbrace {f(\theta,\phi)\frac{e^{i k r}}{r}}_{\psi_{sc}}.
\label{scatterdwave}
\eeq
\eegp
The function $f(\theta,\phi)$ represents the amplitude of the scattering and depends on the interaction of the incident particles. For momentum dependent coupling between particles, the function $f$ is expressed as $f\equiv f(k,\theta,\phi)$ to include dependence on the incident energy and momentum of the particles. If the target is azimuthally symmetric then $f$ is independent of $\phi$, thus $f\equiv f(k,\theta)$.

The current density is associated with the wave function, $\psi$, is given by,
\begp
\allowdisplaybreaks \beq
\vec{J} = \frac{\hbar}{2 m} \left[ \psi^* \vec{\nabla}\psi - \psi \vec{\nabla}\psi^* \right] = \frac{1}{m}Re\left[\frac{\hbar}{i} \psi^* \vec{\nabla}\psi \right].
\label{Currentdens}
\eeq
\eegp
Using eqn.~\ref{Currentdens}, we can calculate both the incident flux and the number of scattered particles. The incident flux $F_{in}$ with probability constant $C$, is given by,
\begp
\allowdisplaybreaks \beq
F_{in} = C J_{in} = C \frac{\hbar k}{m}. \nn
\eeq
\eegp
The scattered particle current density can be written as,
\begp
\allowdisplaybreaks \beq
J_{sc} =\frac{1}{m} |f(\theta,\phi)|^2\frac{\hbar k}{r^2} .
\eeq
\eegp
The number of scattered particles crossing the area $\vec{ds}=ds\hat{r}$ is given by,
\begp
\allowdisplaybreaks \bea
dn &=& C \vec{J}_{sc}.\vec{ds}\nn\\
   &=& C \frac{\hbar k}{m}|f(\theta,\phi)|^2 d\Omega\nn\\
   &=& F_{in}|f(\theta,\phi)|^2 d\Omega ,\nn
\eea
\eegp
where $d\Omega=\frac{ds}{r^2}$ the solid angle subtended by the area $ds$. Comparing the above equation with eqn.~\ref{difcross1}, we get the differential cross-section as,
\begp
\allowdisplaybreaks \beq
\frac{d\sigma}{d\Omega} = \frac{dn}{F_{in}} = |f(\theta,\phi)|^2.\nn
\eeq
\eegp
The function, $f(\theta,\phi)$, actually gives us information about the probability amplitude for scattering processes in a direction $(\theta,\phi)$. In the following, the ${\rm Schr\ddot{o}dinger}$ equation of the scattering particles will be solved to find the scattering amplitude.

Let us consider a target of fixed central potential $V(r)$. The angular momentum $\vec{L}$ of the particles is constant under the central potential. There exists a stationary state which is a common eigenstate of the operator Hamiltonian $H$, angular momentum $\vec{L}$ and $z$-component of angular momentum($L_z$). We can write the wave function as $\psi_{nlm}$. This can be written as,
\begp
\allowdisplaybreaks \beq
\psi_{nlm} = R_{l}(r) Y^m_l(\theta,\phi).\nn
\eeq
\eegp
As we have considered the azimuthal symmetry, the above equation is independent of $\phi$, i.e., $m=0$. The spherical harmonics are then given by,
\begp
\allowdisplaybreaks \beq
Y^0_l(\theta,\phi) = \sqrt{\frac{2l+1}{4\pi}}P_l(\cos\theta).\nn
\eeq
\eegp
Let us now solve the radial part of the wave function. From the ${\rm Schr\ddot{o}dinger}$ equation of the wave function $\psi$,
\begp
\allowdisplaybreaks \beq
\left[- \frac{\hbar^2}{2 m} \nabla^2 + V(r) \right] \psi = E \psi\nn
\eeq
\eegp
one gets,
\begp
\allowdisplaybreaks \bea
\left[ \frac{d^2}{dr^2} + \frac{2}{r} \frac{d}{dr} -\Big\{ \frac{l(l+1)}{r^2} +\frac{2 m}{\hbar^2} V(r) - k^2 \Big\}  \right] R_l(r) &=& 0\nn\\
\Rightarrow \frac{d^2(r R_l(r))}{dr^2} -\left[ \frac{l(l+1)}{r^2} +\frac{2 m}{\hbar^2} V(r) - k^2 \right](r R_l(r))  &=& 0\nn.
\eea
\eegp
The solution of the radial part $ R_l(r)$ of the wave function can be written as,
\begp
\allowdisplaybreaks \beq
R_l(r) = J_l j_l(kr) + N_l \eta_l(kr)\nn
\eeq
\eegp
where $j_l(kr)$ and $\eta_l(kr)$ are the Bessel and Neumann functions. $ J_l$ and $N_l$ are the normalization constants, independent of $r$. The scattering potential is assumed to be short-ranged, i.e., $V(r)\ra 0$ at large distance $r$ and the particles are free from the scattering potential. In the free region, $r \ra \infty$, the Bessel and Neumann functions can be written as,
\begp
\allowdisplaybreaks \beq
\lim_{r\ra \infty} j_l(kr) \ra \frac{\sin\left( kr - \frac{l\pi}{2} \right)}{kr} \quad {\rm and} \quad \lim_{r\ra \infty} \eta_l(kr) \ra \frac{\cos\left( kr - \frac{l\pi}{2} \right)}{kr}.
\label{BesNeu}
\eeq
\eegp
Using eqns.~\ref{BesNeu}, one gets the radial part of the wave function $\psi$,
\begp
\allowdisplaybreaks \beq
\lim_{r\ra \infty}R_l(r) = C_l \frac{\sin\left( kr - \frac{l\pi}{2} + \delta_l \right)}{kr}.\nn
\eeq
\eegp
The normalization constant $C_l$ and phase $\delta_l$ are related with the $J_l$ and $N_l$ as,
\begp
\allowdisplaybreaks \beq
C_l = \sqrt{J_l^2 + N_l^2} \quad {\rm and} \quad \delta_l = - \frac{N_l}{J_l}.\nn
\eeq
\eegp
$R_l(r)$ can be written in the form of incoming and outgoing waves as,
\begp
\allowdisplaybreaks \bea
R_l(r) &\ra & \frac{1}{kr}C_l  \frac{e^{\left( kr - \frac{l\pi}{2} + \delta_l \right)}-e^{-\left( kr - \frac{l\pi}{2} + \delta_l \right)}}{2i}\nn\\
&= & \frac{1}{2ikr}C_l e^{-i\delta_l}e^{-\frac{l\pi}{2}}\left( e^{kr} e^{2 i \delta_l}-i^l e^{-kr}\right).\nn
\eea
\eegp
The solution of the wave function of the free particle is given by
\begp
\allowdisplaybreaks \bea
\psi_{nlm}& = & \frac{1}{2ikr} \sum_{l=0}^\infty C_l e^{-i\delta_l}e^{-\frac{l\pi}{2}}\left( e^{kr} e^{2 i \delta_l}-i^l e^{-kr}\right) P_l(\cos\theta).
\label{psisol1}
\eea
\eegp
The wave function $\psi=e^{ikz}$ is independent of azimuthal angle $\phi$ so that only the $Y_{lm}$ with $m = 0$, which are proportional to the Legendre polynomials $P_l(\cos\theta)$, can contribute to the expansion as,
\begp
\allowdisplaybreaks \bea
e^{ikz} &=& \sum_{l=0}^\infty i^l (2l+1) j_l(kr) P_l(\cos\theta)\nn\\
        &=& \sum_{l=0}^\infty i^l (2l+1) \frac{\sin \left( kr - \frac{l\pi}{2}\right)}{kr} P_l(\cos\theta), \quad {\rm as}~~r\ra \infty\nn\\
        &=& \frac{1}{2ikr}\sum_{l=0}^\infty  (2l+1) e^{-\frac{l\pi}{2}}\left( e^{kr}-i^l e^{-kr}\right) P_l(\cos\theta).
        \label{scazwave}
\eea
\eegp
Using eqn.~\ref{scazwave} in eqn.~\ref{scatterdwave}, we get the wave function of the scattered particles as,
\begp
\allowdisplaybreaks 
\bea
\hspace{-0.6cm}  \psi(r)\Big |_{r\ra \infty} &\approx & e^{i k z} + {f(\theta,\phi)\frac{e^{i k r}}{r}}\nn\\
& = & \frac{1}{2ikr}\sum_{l=0}^\infty  (2l+1) e^{-\frac{l\pi}{2}}\left( e^{kr}-i^l e^{-kr}\right) P_l(\cos\theta) +  f(\theta,\phi)\frac{e^{i k r}}{r}.
\label{psisol2}
\eea
\eegp
Comparing eqn.~\ref{psisol1} and eqn.~\ref{psisol2}, one can get, $C_l = i^l (2l+1) e^{i\delta_l}$ and, 
\begp
\allowdisplaybreaks \bea
f(\theta,\phi) &=& \frac{1}{2ik} \sum_{l=0}^\infty  (2l+1) (e^{2 i \delta_l}-1) P_l(\cos\theta)\nn\\
                &=& \frac{1}{k} \sum_{l=0}^\infty  (2l+1) a_l P_l(\cos\theta).\label{amliF}
\eea
\eegp
One can find $a_l$ from the above equations as, 
\begp
\allowdisplaybreaks \beq
a_l = e^{2 i \delta_l}-1 = e^{i \delta_l} \sin{\delta_l}.
\eeq
\eegp
The real and imaginary part of $a_l$ follows equation of the unitary circle (Fig.~\ref{fig:UNItary}) as,
\begp
\allowdisplaybreaks \beq
|a_l| \leq 1 \Rightarrow [Re(a_l)]^2 + \left[ Im(a_l)-\frac{1}{2} \right]^2 \leq \left(\frac{1}{2}\right)^2.
\eeq
\eegp

So the scattering amplitude obeys the unitarity conditions. In the previous discussions, the quantum mechanical scattering cross-section has been calculated using the method of partial wave analysis. How this quantum mechanical scattering amplitude $f(\theta,\phi)$ is related to the Feynman amplitude in Quantum Field Theory (QFT) will be shown now.
\begin{figure}[h!]
\begin{center}
{
\includegraphics[width=3in,height=2.3in, angle=0]{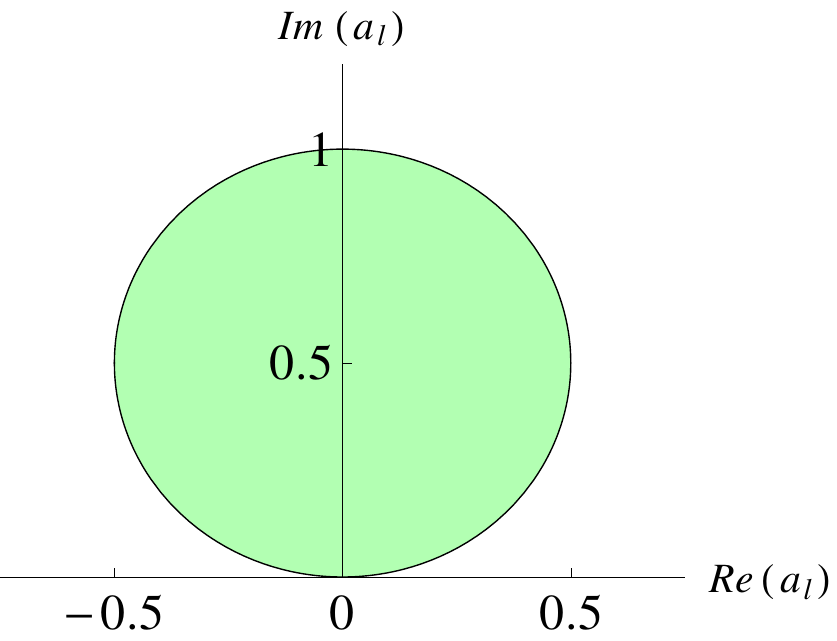}}
\caption{ \textit{Argand diagram of the function $a_l$ in $Re(a_l)-Im(a_l)$ plane is a unitary circle with radius $0.5$.}}
\label{fig:UNItary}
\end{center}
\end{figure}
\par
Let us consider a scattering process for which the Feynman amplitude ${\cal M}$ can be expressed as a function of CM energy $\sqrt{s}$ and scattering angle $\theta$ or equivalently, of the Lorentz-invariant Mandelstam variables $s$, $t$ and $u$. The differential cross-section is defined as,    
\begp
\allowdisplaybreaks \beq \label{dexse}
\frac{d\sigma}{d\Omega}=\frac{1}{64\pi^2s}|{\cal M}|^2=|f(\theta,\phi)|^2,
\eeq
\eegp
where ${\cal M}$ is the Feynman amplitude. Comparing with eqn.~\ref{amliF} one can write the following relation as,
\begp
\allowdisplaybreaks \beq
{\cal M} = 16 \pi \sum_{l=0}^\infty (2l+1) a_l P_l(\cos\theta).
\label{Ampli}
\eeq
\eegp
Using eqn.~\ref{Ampli} and putting into eqn.~\ref{dexse},
\begp
\allowdisplaybreaks \bea
\frac{d\sigma}{d\Omega} &=& \frac{1}{64\pi^2s} (16 \pi)^2 \sum_{l=0}^\infty (2l+1)^2 |a_l|^2 \left[ P_l(\cos\theta) \right] ^2\nn\\
&=& \frac{4}{s} \sum_{l=0}^\infty (2l+1)^2 |a_l|^2 \left[ P_l(\cos\theta)\right]^2 .\nn
\eea
\eegp
Now we get the total cross-section after integrating the differential cross-section over the solid angle ($d\Omega = -2\pi d(\cos\theta)$),
\begp
\allowdisplaybreaks \bea
\sigma &=& \frac{8 \pi}{s} \sum_{l=0}^\infty (2l+1)^2 |a_l|^2  \times \int_{-1}^{1} \left[ P_l(\cos\theta) \right]^2 d(\cos\theta)\nn\\
&=& \frac{8 \pi}{s} \sum_{l=0}^\infty (2l+1)^2 |a_l|^2 \times \frac{2}{(2l+1)} \nn\\
&=& \frac{16\pi}{s}\sum_{l=0}^\infty(2l+1)|a_l|^2.
\label{optcl}
\eea
\eegp
Multiplying both sides of eqn.~\ref{Ampli} with $P_l(\cos\theta)$ and integrating with respect to $\cos\theta$ from $-1$ to $+1$, one gets,
\begp
\allowdisplaybreaks \beq\label{unicond}
a_l = \frac{1}{32} \int_{-1}^{+1} {\cal M}(\theta)P_l(\cos\theta) d(\cos\theta).
\eeq
\eegp
where $a_l$ are the partial wave coefficients corresponding to specific angular momentum values $l$. If the amplitude at tree level increases with energy then the unitary bound is violated after certain energy, then the theory becomes sick and can indicate the incompleteness of theory.

In the SM, various vector bosons scattering processes such as $W_L^\pm W_L^\mp \ra W_L^\pm  W_L^\mp$, $W_L^\pm  W_L^\pm \ra W_L^\pm  W_L^\pm$, $W_L^\pm  W_L^\mp \ra Z_L  Z_L$, $W_L^\pm  Z_L \ra W_L^\pm Z_L$ and $Z_L  Z_L \ra Z_L Z_L$ have been reviewed. It has been checked that without a Higgs boson, the unitarity condition is not fulfilled at high energies. The inclusion of Higgs-mediated diagrams restores unitarity rather spectacularly. Any extended scalar sector is in general expected to satisfy the unitarity condition, unless one can come to terms with strongly coupled physics controlling electroweak interactions at high energy. Thus the $V_L V_L$ scattering cross-sections in a `well-behaved' new physics scenario should fall at high center-of-mass energies. However, if the scattering process involves the participation of an $s$-channel resonance at mass $M$, then one expects a peak at  $\sqrt{s} = M$, above which the cross-section should die down gradually. The energy-dependence of the cross-sections, along with the appearance (or otherwise) of such resonant peaks should thus be computed if one has to verify the imprints of new physics in VBS when the appropriate measurements are feasible.

In this work, the amplitudes have been calculated in different models, using the exact expressions for polarization vectors (see appendix~\ref{feynmanamp}), as we are dealing with the energy range ($\sim 200 $ GeV $\rightarrow 2000$ GeV). However, we have checked that at the high-energy limit, our results are consistent with calculations based on the equivalence theorem.
\section{Benchmark points in different models for $V_LV_L$ scattering}\label{BenchmarlModel}
In the Chapter~\ref{chap:EWSBextended} detailed theoretical and experimental constraints on extended scalar sectors namely type-II 2HDM, HTM ($Y=0$), and HTM ($Y=2$) had been discussed. Measurements of the couplings at LHC of SM-like Higgs with the vector bosons (see table~\ref{table1LHC} of Chapter~\ref{chap:Intro}) put indirect constraints on the models with an extended scalar sector. For example, a charged Higgs can contribute to $h\gamma\gamma$ at one loop. In our analysis, the heavier scalars are taken to be heavy so that $h\gamma\gamma$ constraints are not that important. $hWW$ and $hZZ$ coupling measurements at present agrees with SM values, thereby restricting couplings of the heavier scalars appreciably. As a result, 2HDM is pushed towards the alignment regime where couplings of heavier Higgs bosons with SM gauge bosons tend to vanish. We have taken care of all such constraints at 1$\sigma$ in our analysis. 

{\bf Benchmark points used for $V_LV_L$ scattering}:

\underline{type-II 2HDM}:   

$(i)$ $\cos(\beta-\alpha)=0.04$, $M_{H^\pm}\simeq M_H=500$ GeV, $\Gamma_{H^\pm}=3.6$ GeV, $\Gamma_{H}=7$ GeV 

$(ii)$ $\cos(\beta-\alpha)=0.08$, $M_{H^\pm}\simeq M_H=500$ GeV, $\Gamma_{H^\pm}=3.7$ GeV, $\Gamma_{H}=6.7$ GeV

$(iii)$ $\cos(\beta-\alpha)=0.02$, $M_{H^\pm}\simeq M_H=1500$ GeV, $\Gamma_{H^\pm}=25.4$ GeV, $\Gamma_{H}=26.8$ GeV

$(iv)$ $\cos(\beta-\alpha)=0.04$, $M_{H^\pm}\simeq M_H=1500$ GeV, $\Gamma_{H^\pm}=101.9$ GeV, $\Gamma_{H}=106.7$ GeV

\underline{HTM ($Y=0$)}: 

$(i)$ $\sin\gamma=0.023$, $v_t^\prime=3$ GeV, $M_{H^\pm}\simeq M_H=500$ GeV, $\Gamma_{H^\pm}=0.167$ GeV, $\Gamma_{H}=0.12$ GeV

$(ii)$ $\sin\gamma=0.024$, $v_t^\prime=3$ GeV, $M_{H^\pm}\simeq M_H=1500$ GeV, $\Gamma_{H^\pm}=5.8$ GeV, $\Gamma_{H}=4.8$ GeV

\underline{HTM ($Y=2$)}: 

$(i)$ $\sin\gamma^\prime=0.025$, $v_t=3$ GeV, $M_{H^{\pm\pm}}\simeq M_{H^\pm}\simeq M_H=500$ GeV, $\Gamma_{H^{\pm\pm}}=0.0001$ GeV, $\Gamma_{H^\pm}=0.006$ GeV, $\Gamma_{H}=0.023$ GeV

$(ii)$ $\sin\gamma^\prime=0.024$, $v_t=3$ GeV, $M_{H^{\pm\pm}}\simeq M_{H^\pm}\simeq M_H=1500$ GeV, $\Gamma_{H^{\pm\pm}}=0.0002$ GeV, $\Gamma_{H^\pm}=0.023$ GeV, $\Gamma_{H}=0.635$ GeV

~{\bf N.B.}:~The similar notations have been followed for the above models as in the Chapter~\ref{chap:EWSBextended}. 
\section{$V_LV_L$ scattering with extended scalar sectors}\label{NewModelsResults}
Next, we demonstrate how it is possible to distinguish among
2HDM, HTM~($Y=0$) and HTM~($Y=2$) using the five VBS processes: $W_L^+
W_L^- \ra W_L^+ W_L^-$, $W_L^+ W_L^- \ra Z_L Z_L$, $Z_L Z_L \ra Z_L
Z_L$, $W_L^+ W_L^+ \ra W_L^+ W_L^+$ and $W_L^+ Z_L\ra W_L^+ Z_L$. One
can immediately see that the mediating scalar can be a neutral
scalar, as also a singly or  doubly charged Higgs.
Thus the very constituents of 2HDM or triplet scenarios
are potential players in the game.

Two things turn out to be crucial here: (a) nature of the energy-dependence,
and (b) the center-of-mass energy at which the resonances occur.
The shape of the resonance depends on the
decay width, and hence, on the mass and the coupling of the resonating
scalar. Thus an identification of the resonance can guide one
to the theoretical scenario including the particle spectrum.

In any model with an extended scalar sector around a TeV, the very fact
that the $VVh$ interactions ($V \equiv W, Z$ and $h \equiv$ the 125 GeV scalar) are largely
SM-like makes the non-resonant additional contributions small.
In 2HDM, however, these constraints allow enough parameter space for
the heavier scalars
to have a large decay width so that the effect of resonances can be
felt for a wider range of $\sqrt{s}$. In HTM models, however, this is not the
case and the resonances are narrow.

 For a 2HDM scenario, the lighter $CP$-even scalar in the particle
 spectrum is usually interpreted as the SM-like Higgs. Going
 especially by the rate of decays into pairs of gauge bosons, the
 couplings of this state is expected to be `nearly SM-like', implying
 that a 2HDM can be feasible largely in the `alignment limit'. Recent
 LHC data are by and large consistent with this limit
 \cite{Das:2015mwa}. Hence we have performed our analysis almost in
 that limit. Among the five scattering modes, we have resonant peaks
 for only three channels, namely, $W_L^+ W_L^- \ra W_L^+ W_L^-$,
$W_L^+ W_L^- \ra Z_L
 Z_L$ and $Z_L Z_L \ra Z_L Z_L$ (see Fig.\;\ref{fig:2hdm})
 involving the heavier $CP$-even Higgs $H$. We have set its mass $M_H$ of
 at two benchmark values (500 GeV and 1500 GeV) which are given in Section~\ref{BenchmarlModel}.
The corresponding decay widths
($\Gamma_H$) can also be read off from the resonance peaks in Fig.\;\ref{fig:2hdm}. Using the high-energy scattering
 amplitudes given in Appendices \ref{feynmanamp} and
 \ref{highenergy}, we should be able to predict the shapes of
 plots which contain such resonance peaks. It is
 quite evident from the plots, that apart from the occurrence of the
 peaks, the cross-sections are almost SM-like, as expected in the
 alignment limit.

Let us decompose the aforementioned
  amplitude\footnote{This decomposition is not relevant
   for $Z_L Z_L \ra Z_L Z_L$,  since in this process there
   is no gauge boson mediated Feynman diagram.} as ${\cal M}= {\cal
   M}_{gauge,h}+{\cal M}_{H}$, where ${\cal M}_{H}$ is proportional to
 $1/(E_{CM}^2 - M_H^2)$. When $E_{CM} < M_H$, the $H$ mediated diagram
 interferes constructively with the remaining terms, due to which the
 interference contributions to cross-section increases with
 energy. On the other hand, for $E_{CM} > M_H$, ${\cal M}_{H}$ interferes
 destructively, and hence its contribution to the cross-section decreases.
In the high-energy
 limit ($E_{CM} \gg M_H$), the amplitudes can be expressed as a power
 series in the energy (see eqn.\;\ref{energyseries} of Appendix
 \ref{highenergy}). In this limit the terms proportional $E^4_{CM}$ as
 well as $E^2_{CM}$ of the amplitude become zero. The remaining terms
 are  either independent of energy or go in negative powers of energy, so that
 the cross-section decreases with rising energy, thus ensuring
 perturbative unitarity.
It has been seen that in expansion of $|{\cal M}|^2$, square of the term ${\cal M}_{H}$ which containing the factor $\frac{\Gamma_H^2M_H^2}{(E_{CM}^2 - M_H^2)^2+\Gamma_H^2 M_H^2}$, is always larger than the interference term that contains $\frac{ (E_{CM}^2 - M_H^2)}{(E_{CM}^2 - M_H^2)^2+\Gamma_H^2 M_H^2}$ near the pole. However, depending on the relative magnitudes of ${\cal M}_{gauge,h}$ and ${\cal M}_{H}$, the interference term may dominate over $|{\cal M}_{H}|^2$ away from the pole.  
 It should be also noted here that due to the absence of
 $W^\pm ZH^\pm$ couplings in 2HDM,
there are no Feynman diagrams
 mediated by $H^\pm$ ($s$-channel) for
 $W_L^+ Z_L\ra W_L^+ Z_L$. Moreover, 2HDM does not contain $H^{\pm\pm}$ which can mediated the process $W_L^+ W_L^+\ra W_L^+ W_L^+$.
 Therefore, for 2HDM we have no peaks for these two processes. The cross-sections for these processes are also similar to that of SM, due to the feeble coupling strength of $H$ with gauge bosons.

As has been mentioned already, we have also studied triplet models
with two different
values of the $U(1)$ hypercharge ($Y=0$ and 2). In these models we can have
interactions of charged scalars with pairs of gauge bosons.
\begin{figure}[h!]
 \begin{center}
 {
 \includegraphics[width=7cm,height=5.90cm, angle=0]{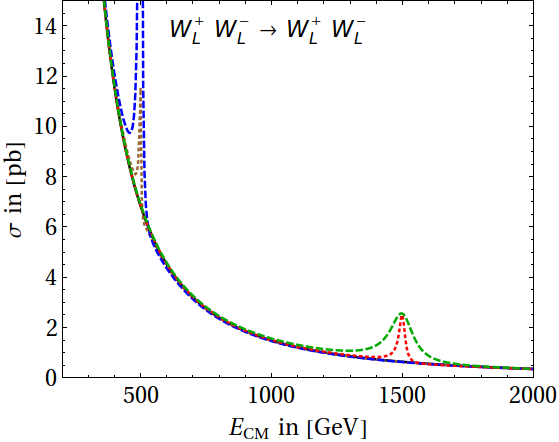}
 \includegraphics[width=7cm,height=6cm, angle=0]{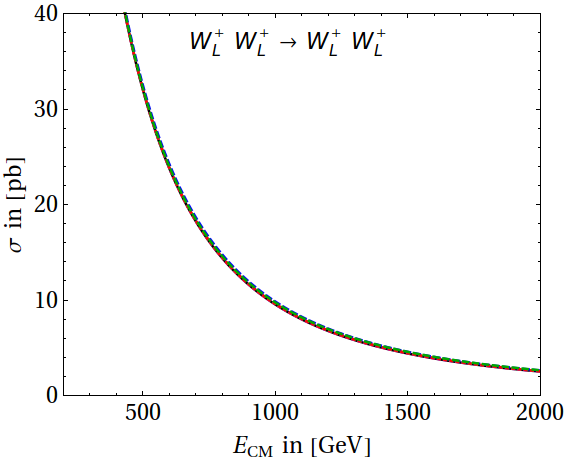}
 \includegraphics[width=7cm,height=6cm, angle=0]{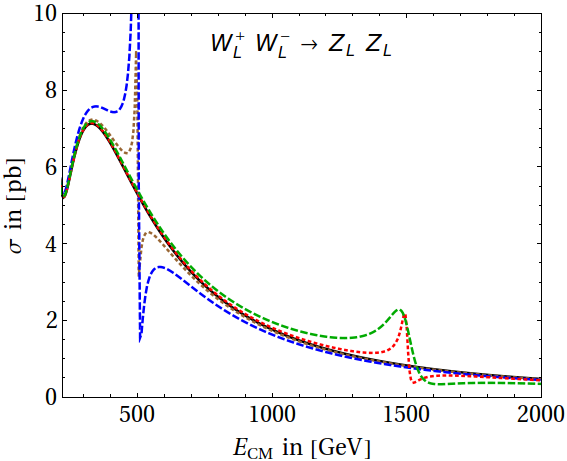}
 \includegraphics[width=7cm,height=6cm, angle=0]{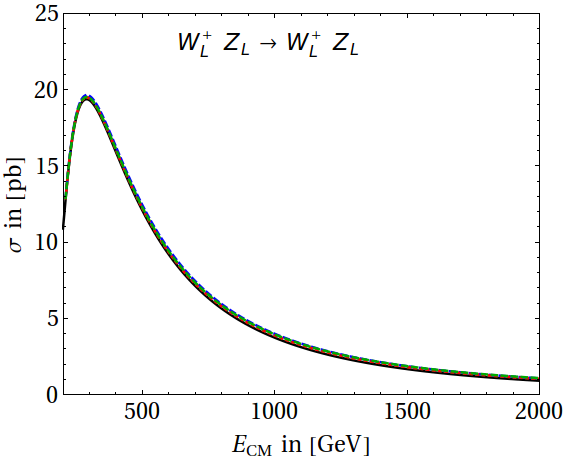} \\
\hspace{-1cm} \includegraphics[width=7cm,height=6cm, angle=0]{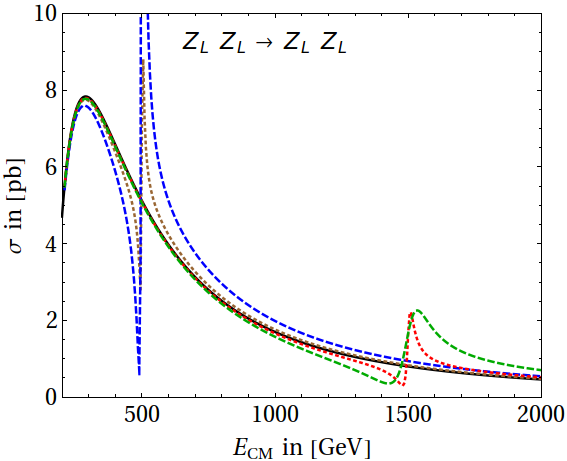}
\hskip 2pt \includegraphics[width=6cm,height=4.6cm, angle=0]{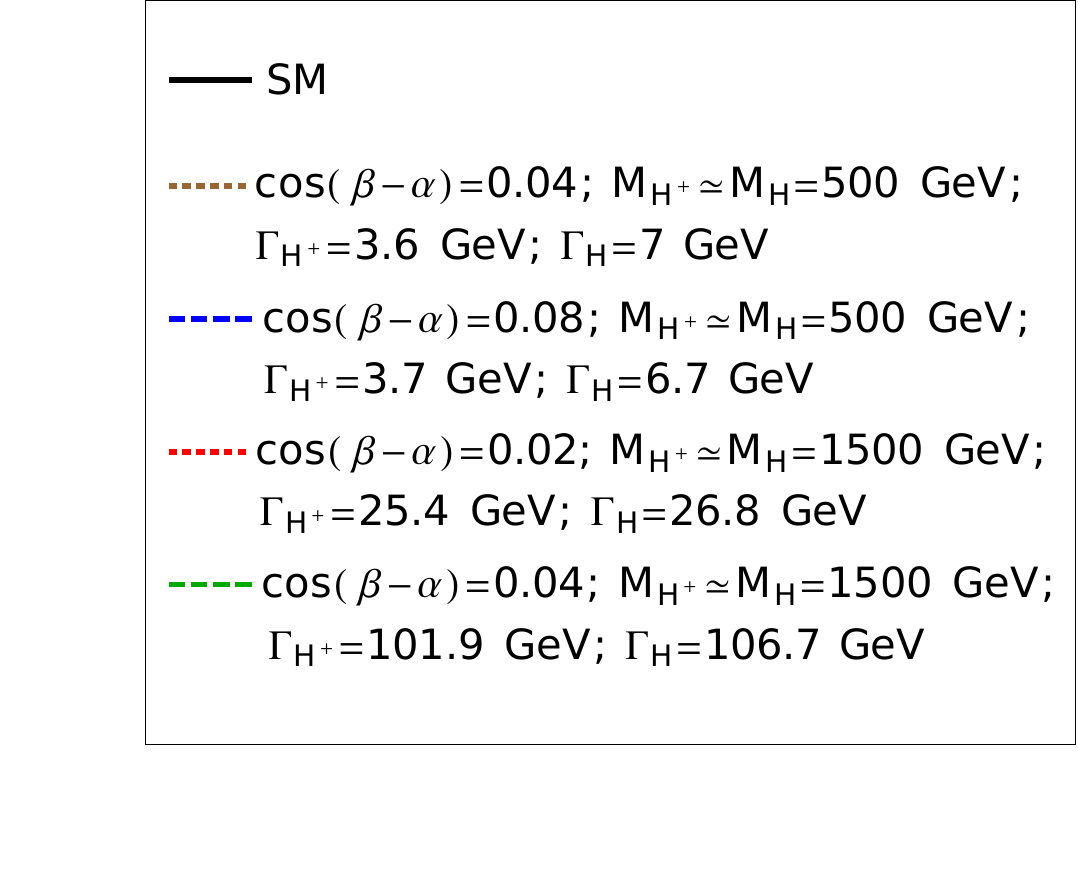}
 \caption{\label{fig:2hdm} \textit{Plots for~$VV$~scattering in 2HDM. } }
 }
 \end{center}
 \end{figure}
\begin{figure}[h!]
 \begin{center}
 {
 \includegraphics[width=7cm,height=5.9cm, angle=0]{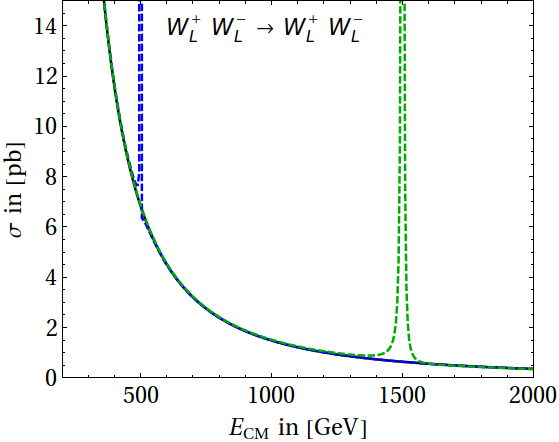}
 \includegraphics[width=7cm,height=6cm, angle=0]{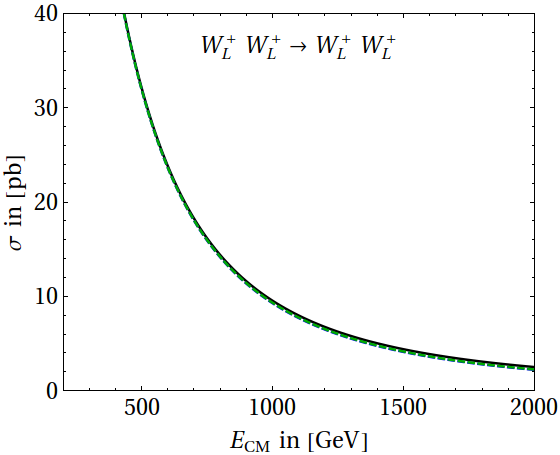}
 \includegraphics[width=7cm,height=6cm, angle=0]{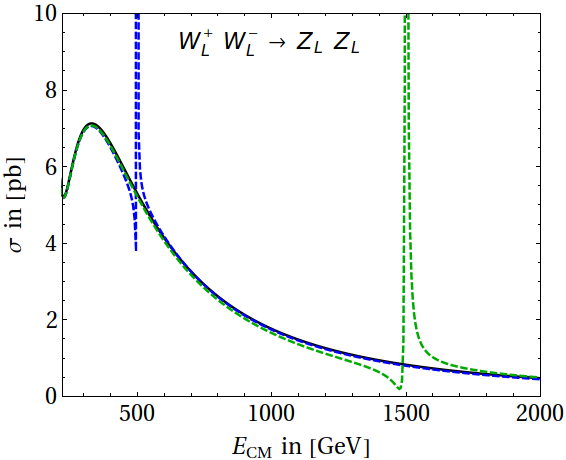}
 \includegraphics[width=7cm,height=6cm, angle=0]{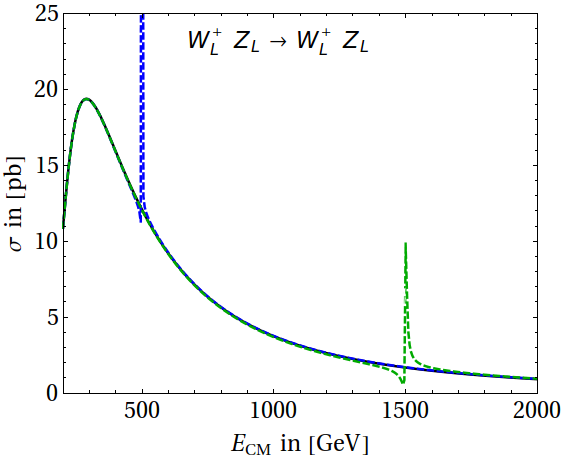}\\
\hspace{-1cm}\includegraphics[width=7cm,height=6cm, angle=0]{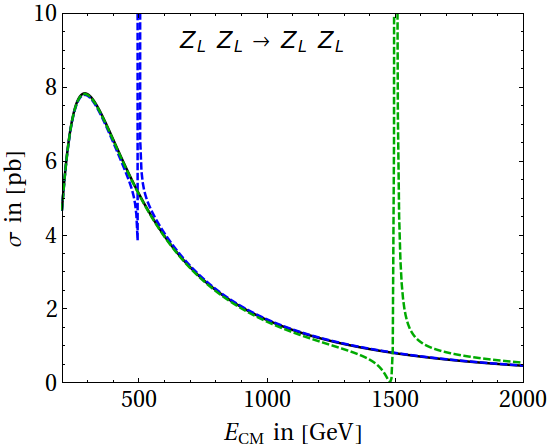}
\hskip 2pt
\includegraphics[width=6cm,height=4.6cm, angle=0]
{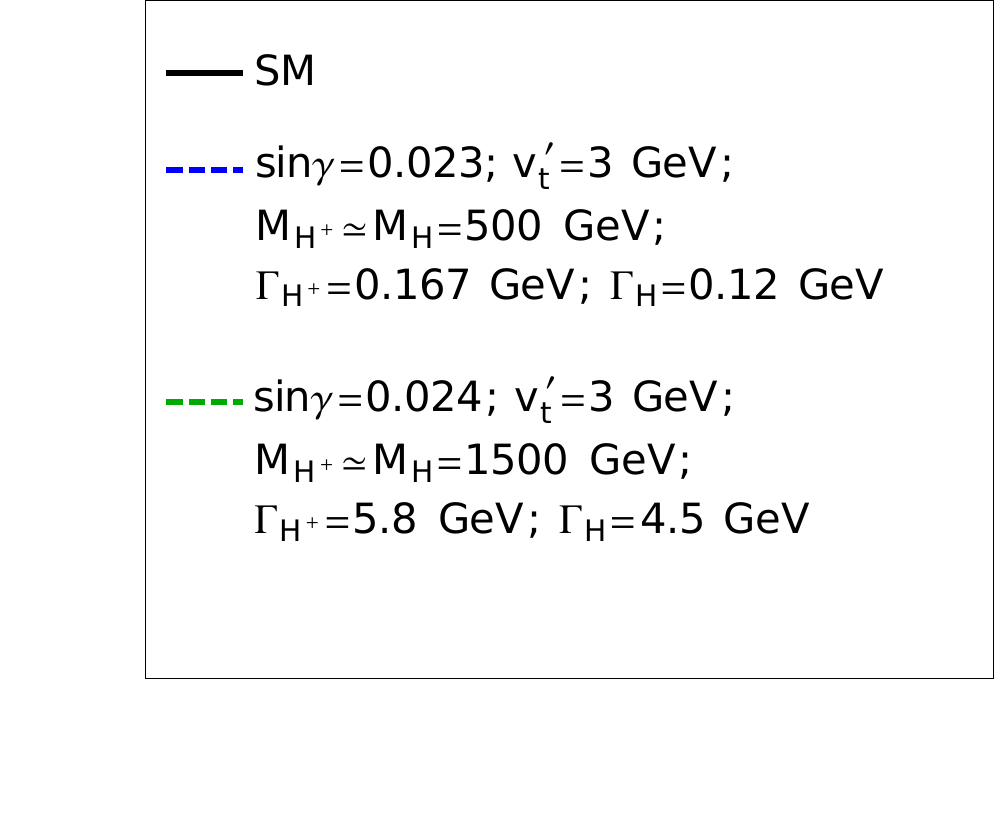} 
 \caption{\label{fig:htm0} \textit{Plots for~$VV$~scattering in Y=0, HTM. } }
}
 \end{center}
 \end{figure}
\begin{figure}[h!]
\begin{center}
{	
 \includegraphics[width=7cm,height=6cm, angle=0]{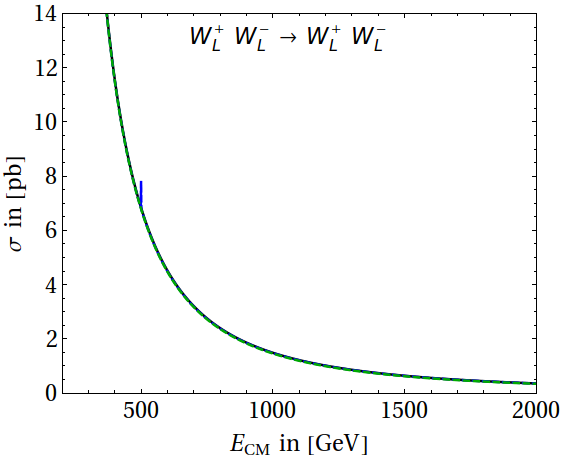}
 \includegraphics[width=7cm,height=6cm, angle=0]{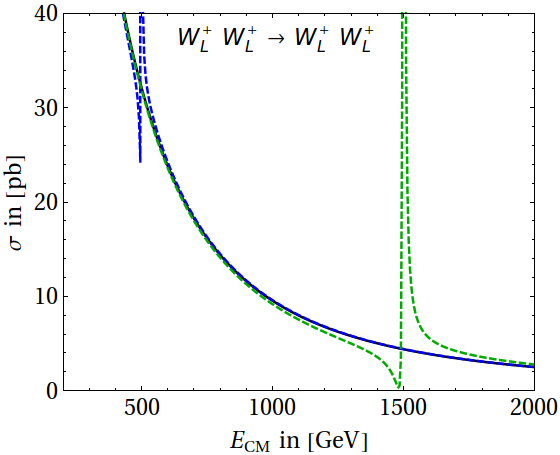}
 \includegraphics[width=7cm,height=6cm, angle=0]{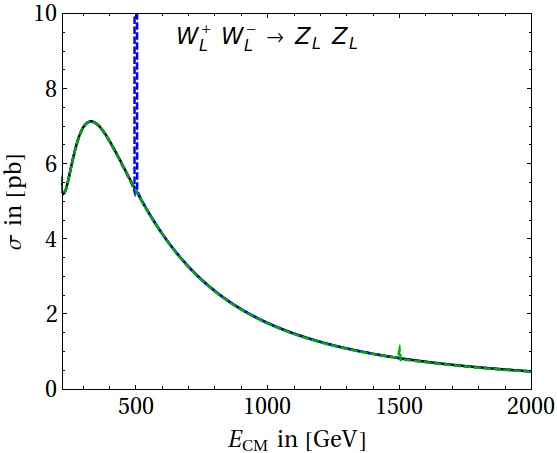}
 \includegraphics[width=7cm,height=6cm, angle=0]{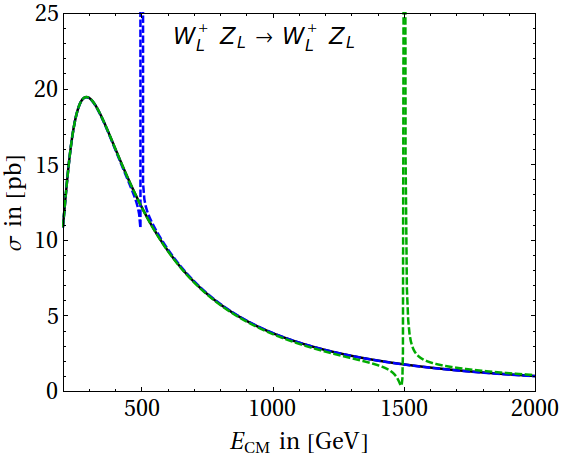}  \\
\hspace{-1cm} \includegraphics[width=7cm,height=6cm, angle=0]{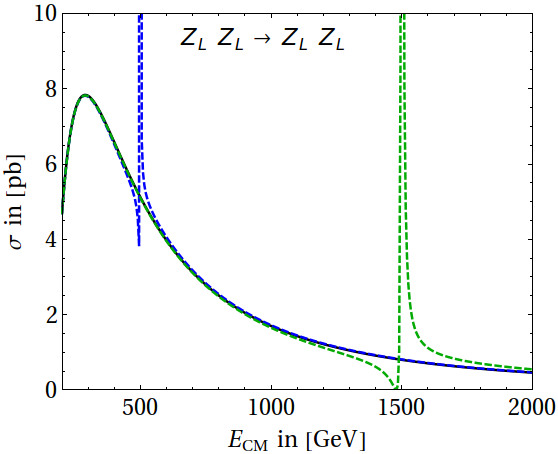}	
\hskip 2pt \includegraphics[width=6cm,height=4.6cm, angle=0]{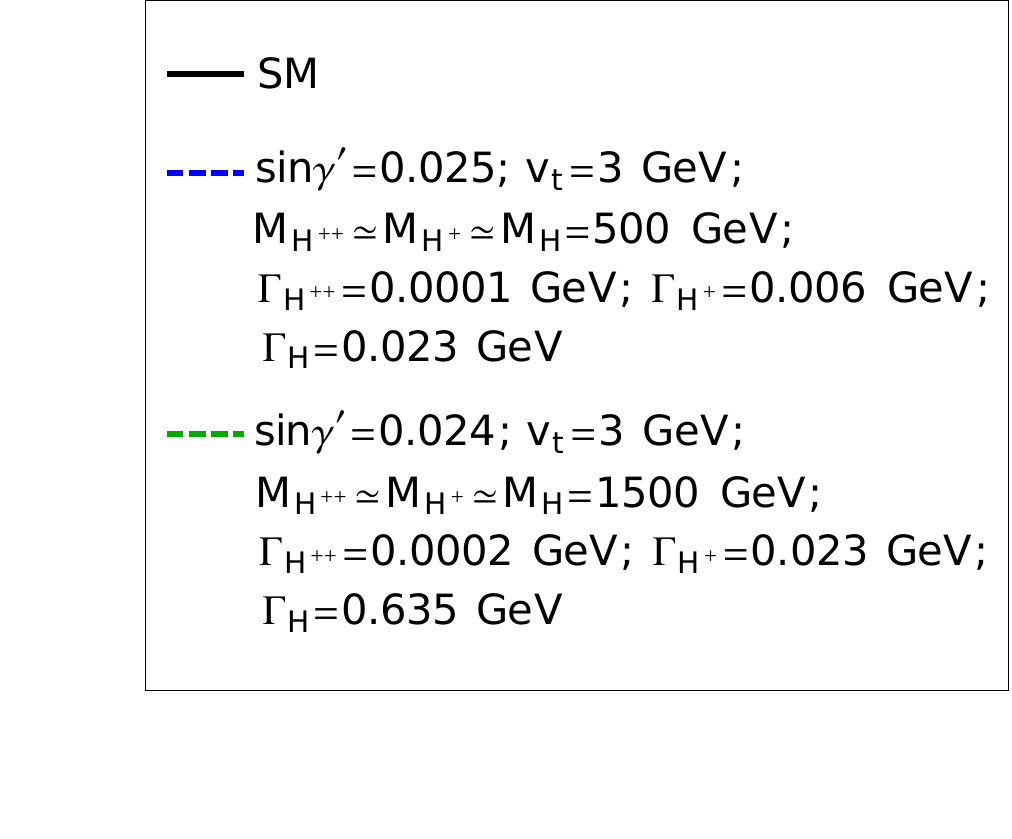}
\caption{\label{fig:htm2} \textit{Plots for~$VV$~scattering in Y=2, HTM. } }
 }
 \end{center}
 \end{figure}

Of these,
we primarily  focus on a $Y=2$ HTM. This scenario is relevant
in the context of the type-II seesaw mechanism of neutrino mass generation,
and it also arises in left-right symmetric gauge theories. Now the question is how to isolate such a scenario from a $Y=0$ HTM or even a type-II 2HDM models? our findings will be summarized in Table~\ref{table1} which clearly indicates that HTM ($Y=0$) and 2HDM can be distinguished via a $s$-channel  $H^+$ resonance in $W_L^+ Z_L\ra W_L^+ Z_L$ scattering process, as  $H^+$ couples to $W^+$ and $Z$ only in the triplet models.

The other aspect which might help in such a distinction is the width of the resonances: In 2HDM the allowed parameter space allows for a wider resonance than that in the HTM models. But width of the resonances does not help in identifying the hypercharge of the scalar triplet. 
\begin{table}[h!]
\begin{center}
    \begin{tabular}{ | c | c | c | c |}
         \hline
     Process &  ${\rm 2HDM}$ & ${\rm HTM}(Y=0)$ & ${\rm HTM}(Y=2)$ \\
    \hline
     $W_L^+ W_L^- \rightarrow W_L^+ W_L^-$ & $\cmark$, $(H)$& $\cmark$, $(H)$&  $\cmark$, $(H)$  \\
     \hline
     $W_L^+ W_L^+ \rightarrow W_L^+ W_L^+$ & $\xmark$ & $\xmark$ &  $\cmark$, $(H^{++})$  \\
     \hline
     $W_L^+ W_L^- \rightarrow Z_L Z_L$  & $\cmark$, $(H)$ & $\cmark$, $(H)$ &  $\cmark$, $(H)$  \\
     \hline
     $W_L^+ Z_L \rightarrow W_L^+ Z_L$  & $\xmark$ & $\cmark$, $(H^+)$ &  $\cmark$, $(H^+)$  \\
     \hline
     $Z_L Z_L \rightarrow Z_L Z_L$  & $\cmark$, $(H)$ & $\cmark$, $(H)$ & $\cmark$, $(H)$ \\
     \hline
     \end{tabular}
    \caption{ \textit{Different scattering processes and corresponding  mediator particles for resonance in various extended scalar sectors. $``\cmark"$ indicates presence of a resonance  where as  $``\xmark"$ corresponds to no resonance peak.} }
    \label{table1W}
\end{center}
\end{table}

As HTM ($Y=2$) contains a $H^{++}$ that can couple to a pair of $W^+$, in contrast to the 2HDM and HTM ($Y=0$) models, the distinguishing feature of this model would be a $s$-channel  $H^{++}$ resonance in $W_L^+ W_L^+\ra W_L^+ W_L^+$ scattering process. 

\section{Summary}\label{conclusionSUM}
In this chapter, the elastic scattering of different modes of longitudinally polarized gauge bosons in various extended scalar sectors has been considered. The exact expression of longitudinal polarization vectors has been used to determine the scattering cross-sections. Generally, the idea  of $V_LV_L$ scattering has been used for the proper understanding of the EWSB or in other words to perceive the importance of Higgs boson. In this analysis, a quantitative study of $V_L V_L$ scattering has been carried out to probe the various non-standard scalars in various BSM scenario at an intermediate energy-range.  

In this analysis, the parameter space of these models has been scanned via several theoretical as well as phenomenological constraints. From the allowed parameter space, several benchmark points have been picked up for which the CM energy dependence of scattering cross-sections of different VBS modes in several models have been shown. 

As far as model benchmark points are concerned, we have chosen only those regions of the parameter space which satisfy all the existing constraints.
The present LHC data have allowed only a small portion of the parameter space of these models, thus we are compelled to execute our analysis in the restricted parameter space.
As a result the cross-section is very much similar to the SM case away from the peaks. To differentiate between different extended scalar sectors from VBS one need to look for resonances at different VBS modes.

In this analysis to illustrate the effect of the non-standard scalar particles, two different masses have been chosen, $\sim 500$ GeV and other is at the relatively higher $\sim 1500$ GeV. Depending on the model parameters of a particular model and for a particular process the height and width of the peaks (if present) at the two different values of mass range of the heavier scalars are different from each other, as these are controlled by mass and corresponding decay width of those non-standard scalars which we want to probe via $V_L V_L$ scattering. If we consider a particular process then we can differentiate between the different models from the nature of these peaks. So this analysis provides an exclusive approach to discriminate the different extended scalar sectors.


\chapter{Metastability in Standard Model}
\label{chap:MetaSM}
\linespread{0.1}
\graphicspath{{Chapter5/}}
\pagestyle{headings}
\noindent\rule{15cm}{1.50pt} 
\section{Introduction}
If one assumes the standard model is valid up to the Planck $\mpl$ scale then the Higgs potential develops a second minimum near $\mpl$ that is much deeper than the electroweak (EW) vacuum in which we live. 
This implies absolute stability of the EW vacuum (minimum) is excluded at a confidence level of about 3$\sigma$. For the measured experimental values of the SM parameters the instability (the Higgs quartic coupling $\lambda$ becomes negative) occurs at scales larger than $10^{10}$ GeV. The instability problem does not necessarily lead to an inconsistency of our existence at the EW scale. The transition time from the EW vacuum to its deeper minimum is spectacularly greater than the lifetime of the Universe. New physics can change the stability of EW vacuum modifying the SM Higgs potential.

When extrapolating the known physics to higher energy scales, renormalization of the known parameters plays a crucial role. 
The parameters of the theory involved in a physical process, are dependent on the energy scales.

In this chapter, the metastability of the SM is revisited. The stability of the SM potential and mass bounds of the Higgs from the (meta)stability, instability, and perturbativity will be reviewed. Finally in the summary, the interpretation of the instability problem and possible hints about the new physics at or below the instability scale will be discussed.
\section{Effective Higgs potential in the Standard Model}
\label{sec:SMpot}
In this section, the structure of the Higgs potential from the EW scale to the Planck scale $\mpl$ will be analyzed. In this study, the Higgs scalar potential up to two-loop quantum corrections is used and it has been improved by three-loop renormalization group running of the coupling parameters. 

The SM tree-level Higgs potential is given by,
\begp
\allowdisplaybreaks
\beq
V_0^{\rm SM}(\phi)=-\frac{1}{2} m^2 \phi^2 + \frac{1}{4} \lambda \phi^4 .
\label{v0}\\
\eeq
\eegp
In Landau gauge using $\MS$ scheme, the SM Higgs potential up to two-loop can be found in Refs.~\cite{Ford:1992pn, Martin:2001vx, Degrassi:2012ry}.
The one-loop potential can be written as,
\begp
\allowdisplaybreaks \bea
V_1^{\rm SM}(\phi)&=&\sum_{i=1}^5 \frac{n_i}{64 \pi^2} M_i^4(\phi) \left[ \ln\frac{M_i^2(\phi)}{\mu^2(t)}-c_i\right],
\label{V1loop}
\eea
\eegp
where $n_i$ is the number of degrees of freedom. For scalars and gauge bosons, $n_i$ comes with a positive sign, whereas for fermions it is associated with a negative sign. Here $c_{H,G,f}=3/2$, $c_{W,Z}=5/6$ and 
$\mu(t)$ is related to the running parameter $t$ as, $\mu(t)=M_Z \exp(t)$. $M_i(\phi)$ is given by,
\begp
\allowdisplaybreaks \beq
M_i^2(\phi)= \kappa_i(t) \phi^2(t)-\kappa_i^{\prime}(t),
\nn 
\eeq
\eegp
$n_i$, $\kappa_i$ and $\kappa_i^{\prime}$ can be found in eqn.~(4) of Ref.~\cite{Casas:1994qy} (see also  Refs.~\cite{Altarelli:1994rb, Casas:1994us, Casas:1996aq, Quiros:1997vk}). It has been observed that the variation in the mass term $m^2$ of the Higgs potential from the EW scale to $\mpl$ is very small. One can neglect it for $\phi \gg 1 $ TeV. Also, the $\beta$-functions of eqns.~\ref{betag1}-\ref{betayt} are independent of $m^2$. In this work, the RGE of $m^2$ has not been considered. 
The running energy scale $\mu$ is replaced with the field $\phi$ in the potential \cite{Casas:1994qy} such that all the couplings of the SM remain within the perturbative domain.
For $\phi\gg v$, the quantum corrections to $V(\phi)$, are reabsorbed in the effective running coupling $\lambda_{\rm eff}$ (see eqn.~\ref{eq:effqurtic}) such that the effective potential becomes,
\begp
\allowdisplaybreaks
\beq
V_{\rm eff}(\phi) \simeq \frac{1}{4} \lambda_{\rm eff} \phi^4. \label{EFFPOT}
\eeq
\eegp
The matching conditions and RG evaluation of the SM couplings, which play an important role in the stability of the Higgs potential will be discussed in the next section.

\subsection{Matching conditions and RG evaluation of the SM couplings}
\label{MatchingSM}
Using matching conditions one can evaluate the coupling constants at the highest mass scale of the SM namely the top quark mass $M_t$ and then run them according to the  RGEs up to the Planck scale. To know their values at $M_t$, one should take into account various threshold corrections up to  $M_t$~\cite{Sirlin:1985ux, Melnikov:2000qh, Holthausen:2011aa},
\begp
\allowdisplaybreaks \bea
g_i(M_t)&=& g_i(M_Z) + \delta g_i(M_t).\label{Matchg}\\
\lambda(M_t)&=&\frac{M_h^2}{2 v^2} \left(1+\delta_H (M_t)\right).\label{MatchH}\\
y_t(M_t) &=& \frac{\sqrt{2} M_t}{v} \left( 1+\delta_t (M_t)\right).\label{MatchYt}
\eea
\eegp
All coupling constants are expressed in terms of pole masses (see table.~\ref{table1}).
To calculate $g_1 (M_t)$ and $g_2 (M_t)$, one-loop RGEs are enough. For $g_3 (M_t)$, first three-loop RGE running of $\alpha_s$ with five flavors excluding the top quark are used, and then the effect of top is included using prescriptions of an effective field theory. The leading term in four-loop RGE for $\alpha_S$ is also taken into account.  Amongst all Yukawa couplings, the running of $y_t$ is the most significant. $y_t(M_t)$ is related to the top pole mass $M_t$ by the matching condition eqn.~\ref{MatchYt}. In $\delta_t (M_t)$, three-loop QCD, one-loop EW, and two-loop ${\cal O}(\alpha \alpha_S)$ corrections are taken into account. Similarly, the relation between $\lambda(M_t)$ in $\MS$ scheme and Higgs pole mass $M_h$ is given in eqn.~\ref{MatchH}. In $\delta_H (M_t)$, one-loop EW, two-loop ${\cal O}(\alpha \alpha_S)$, two-loop ${\cal O}(y_t^4 g_3 ^2)$ and two-loop ${\cal O}(y_t^6)$  corrections are considered. 
The loop effects considered in these matching conditions are comparable to Refs.~\cite{Bezrukov:2012sa, Degrassi:2012ry}. After knowing the values of various coupling constants at $M_t$, full three-loop SM RGEs are used to run them up to $\mpl$.
\setlength\tablinesep{1pt}
\setlength\tabcolsep{2pt}
\begin{table}[h!]
\begin{center}{
    \begin{tabular}{ | c | c |  c | }
    \hline
     Physical Observable & Value & Reference \\
    \hline
     Pole mass of the $W^\pm$ boson $M_W$ & 80.384 $\pm$ 0.014 GeV & \cite{Wmass}  \\
     \hline
     Pole mass of the $Z$ boson $M_Z$& 91.1876 $\pm$ 0.0021 GeV & \cite{Zmass} \\
     \hline
     Pole mass of the Higgs $M_h$& 125.7 $\pm$ 0.3 GeV & \cite{HiggsMass}    \\
     \hline
     Pole mass of the top quark $M_t$& 173.1 $\pm$ 0.6 GeV & \cite{topmass}     \\
     \hline
     ${\rm VEV}$ & 246.21971 $\pm$ 0.00006 GeV & \cite{MuLan}  \\
     \hline
      $\alpha_S(M_Z)$ & 0.1184 $\pm$ 0.0007 &  \cite{Bethke:2012jm}\\
     \hline
             \end{tabular}}
    \caption{\textit{SM observables which can be taken as input to fix the SM fundamental parameters $ g_1, g_2,g_3, y_t$ and $\lambda$.}}
    \label{table1}
\end{center}
\end{table}

Using these matching conditions one can estimate all the coupling constants at the scale $M_t$ including the uncertainty in $M_t,~M_h$ and $\alpha_s(M_Z)$ (see table~\ref{table1}) as:
\begp
\allowdisplaybreaks \bea
 g_1(\mu=M_t)&=&0.358725+0.000007\left( \frac{M_t~{\rm [GeV]} -173.1}{0.6} \right) \label{eq:g1mt}
 \\
g_2(\mu=M_t)&=&0.64818-0.00002\left( \frac{M_t~{\rm [GeV]} -173.1}{0.6} \right)\\
\label{eq:g2mt}
 g_3(\mu=M_t)&=&1.16449 - 0.0003\left( \frac{M_t~{\rm [GeV]} -173.1}{0.6} \right) \nn\\
 &&\hspace{0cm}+ 0.0031\left(\frac{\alpha_S(M_Z)-0.1184}{0.0007}\right)\\
 \label{eq:g3mt}
 y_t(\mu=M_t)&=&0.935643+0.0033\left( \frac{M_t~{\rm [GeV]} -173.1}{0.6} \right) -0.0004\left(\frac{\alpha_S(M_Z)-0.1184}{0.0007}\right)\nn\\
 &&\hspace{0cm} - 0.00001 \left(\frac{M_h~{\rm [GeV]}-125.7}{0.3}\right)\qquad\\
 \label{eq:ytmt}
 \lambda(\mu=M_t)&=&0.127054-0.00003\left( \frac{M_t~{\rm [GeV]} -173.1}{0.6} \right) \nn\\
 &&\hspace{0cm} -  0.00001 \left(\frac{\alpha_S(M_Z)-0.1184}{0.0007}\right)+0.00061 \left(\frac{M_h{\rm [GeV]}-125.7}{0.3}\right)\qquad
 \label{eq:lammt}
\eea
\eegp
 \begin{figure}[h!]
 \begin{center}
 \subfigure[]{
 \includegraphics[width=2.7in,height=2.7in, angle=0]{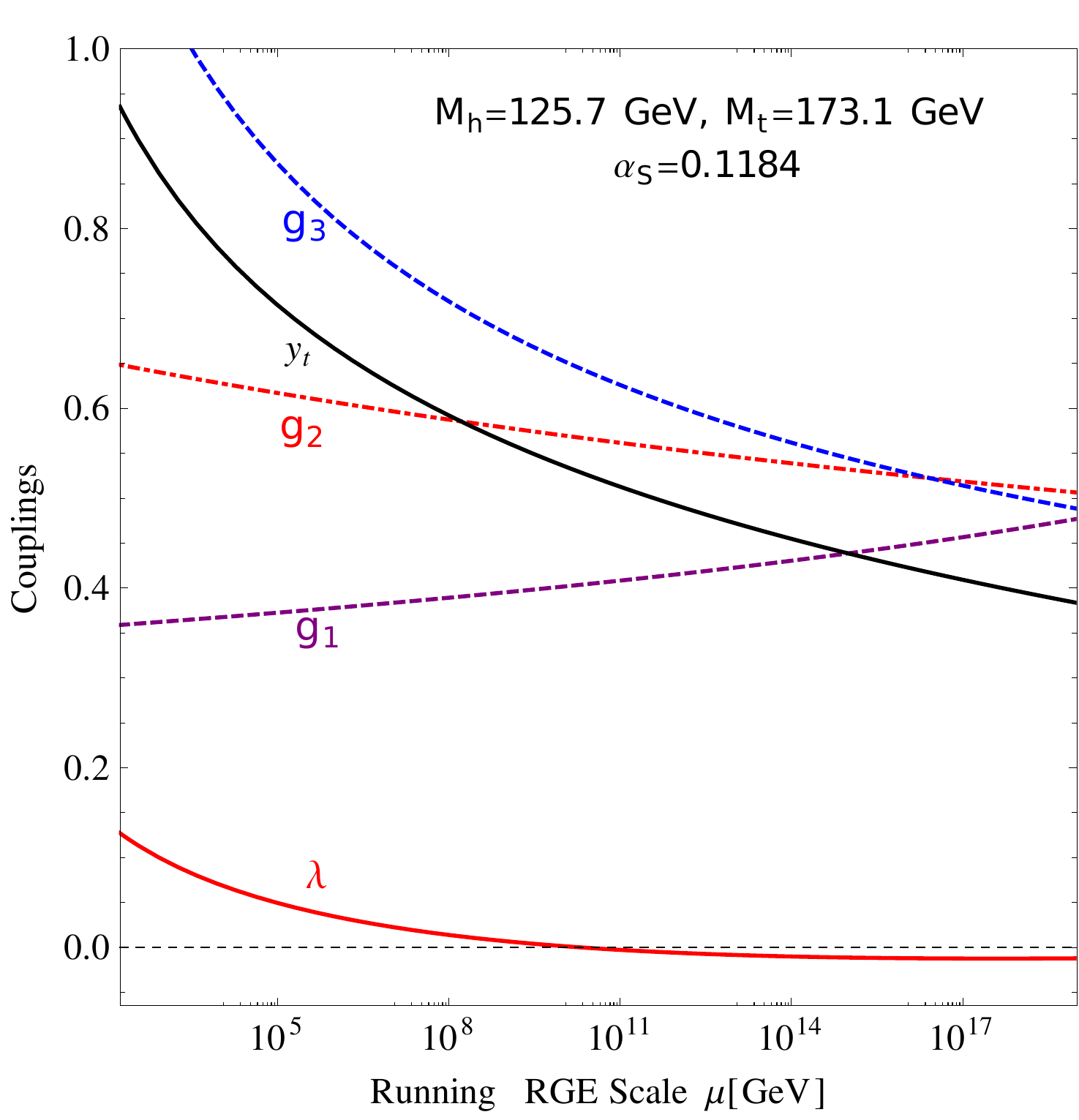}}
 \subfigure[]{
 \includegraphics[width=2.7in,height=2.7in, angle=0]{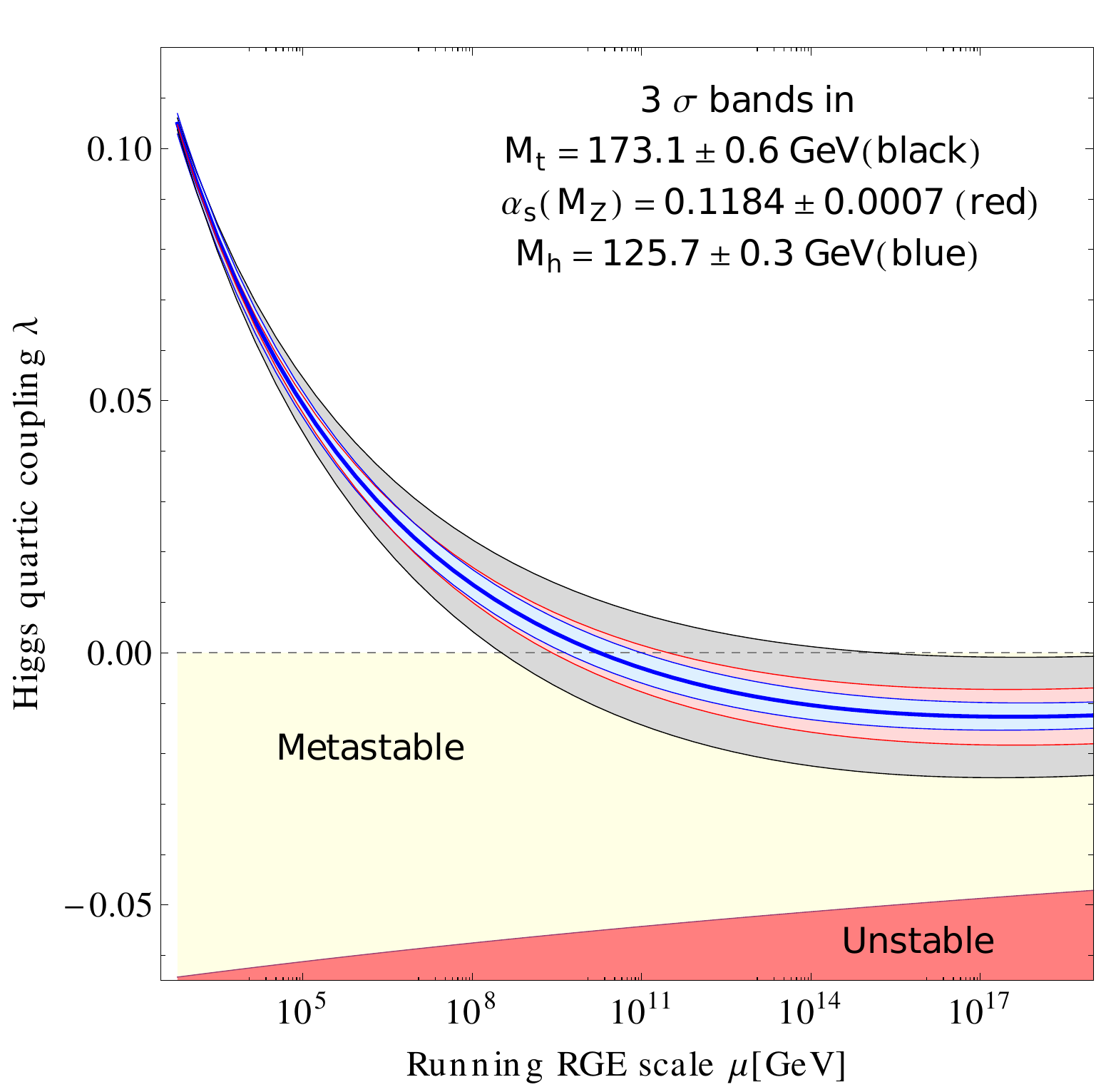}}
 \caption{\label{fig:couplings} \textit{ $({\bf a})$ RG evolution of the gauge couplings $g_1,g_2,g_3$, top Yukawa coupling $y_t$, Higgs self-quartic coupling $\lambda$ in $\MS$ scheme and $({\bf b})$ RG evolution of $\lambda$ in the {\rm SM} and in the panes $3\sigma$ bands for $M_t,~M_h$ and $\alpha_s(M_Z)$ are displayed.} }
 \end{center}
 \end{figure}
As running of all the SM couplings is being considered up to the Planck scale, the values at very high energies are extremely sensitive to the initial values at $M_t$.
These changes should be taken into account because it will also affect the stability of EW vacuum of Higgs potential. However, the stability of the Higgs potential does not alter due to the uncertainties in the other SM parameters at $M_t$. 
Setting all the couplings at $M_t$, using the RG-eqns.~\ref{betag1}-\ref{betalam}, one can obtain the SM couplings  at $\mpl$ as:
\begp
\allowdisplaybreaks \bea
 g_1(\mu=\mpl)&=&0.477685+0.00002\left( \frac{M_t~{\rm [GeV]} -173.1}{0.6} \right) \nn\\
 &&\hspace{0cm} +0.00001 \left(\frac{\alpha_S(M_Z)-0.1184}{0.0007}\right) \nn\\
 &&\hspace{0cm}+10^{-8} \left(\frac{M_h{\rm [GeV]}-125.7}{0.3}\right)\qquad\\
 \label{eq:g1mpl}
 g_2(\mu=\mpl)&=&0.505632-0.00001\left( \frac{M_t~{\rm [GeV]} -173.1}{0.6} \right) \nn\\
 &&\hspace{0cm} -0.00001 \left(\frac{\alpha_S(M_Z)-0.1184}{0.0007}\right) \nn\\
 &&\hspace{0cm}+ 10^{-9} \left(\frac{M_h{\rm [GeV]}-125.7}{0.3}\right)\qquad\\
 \label{eq:g2mpl}
 g_3(\mu=\mpl)&=&0.48714-0.00002\left( \frac{M_t~{\rm [GeV]} -173.1}{0.6} \right) \\
 &&\hspace{0cm} +0.0002 \left(\frac{\alpha_S(M_Z)-0.1184}{0.0007}\right) \nn\\
 &&\hspace{0cm}+ 10^{-9} \left(\frac{M_h{\rm [GeV]}-125.7}{0.3}\right)\qquad\\
 \label{eq:g3mpl}
y_t(\mu=\mpl)&=&0.38227+0.00304\left( \frac{M_t~{\rm [GeV]} -173.1}{0.6} \right)\nn\\
 &&\hspace{0cm}+0.0009 \left(\frac{\alpha_S(M_Z)-0.1184}{0.0007}\right) \nn\\
 &&\hspace{0cm}- 8\times 10^{-6} \left(\frac{M_h{\rm [GeV]}-125.7}{0.3}\right)\qquad\\
 \label{eq:ytmpl}
 \lambda(\mu=\mpl)&=&-0.0122748-0.003935\left( \frac{M_t~{\rm [GeV]} -173.1}{0.6} \right) \nn\\
 &&\hspace{0cm} -0.00209 \left(\frac{\alpha_S(M_Z)-0.1184}{0.0007}\right) \nn\\
 &&\hspace{0cm}+0.0008 \left(\frac{M_h{\rm [GeV]}-125.7}{0.3}\right)\qquad
 \label{eq:lammpl}
\eea
\eegp
The couplings $g_1,g_2,g_3,y_t$ and $\lambda$ have been plotted with the running energy in Fig.~\ref{fig:couplings}(a) and only the Higgs quartic coupling $\lambda$ has been shown in Fig.~\ref{fig:couplings}(b).
In Fig.~\ref{fig:couplings}(b), the blue solid line corresponds to the measured central values of $M_t,~M_h$ and $\alpha_S(M_Z)$, the blue band belongs to the 3$\sigma$ deviation in $M_h$, whereas the red and black correspond to the deviation in $\alpha_S(M_Z)$ and $M_t$ respectively.
As the running of the Yukawa couplings of the other quarks and leptons are very small, these are not included in this analysis.

As defined in eqn.~\ref{EFFPOT}, $\lambda_{\rm eff}$ differs from $\lambda$ as it takes care of quantum corrections.
It has been observed that for high energy scale $\phi\gg v$, the difference in $\lambda_{\rm eff}$ and $\lambda$ is not appreciable (see Fig.~\ref{fig:lambdaSM}). In the stability analysis, $\lambda_{\rm eff}$ has been used instead of $\lambda$.
\begin{figure}[h!]
\begin{center}
{
\includegraphics[width=2.7in,height=2.7in, angle=0]{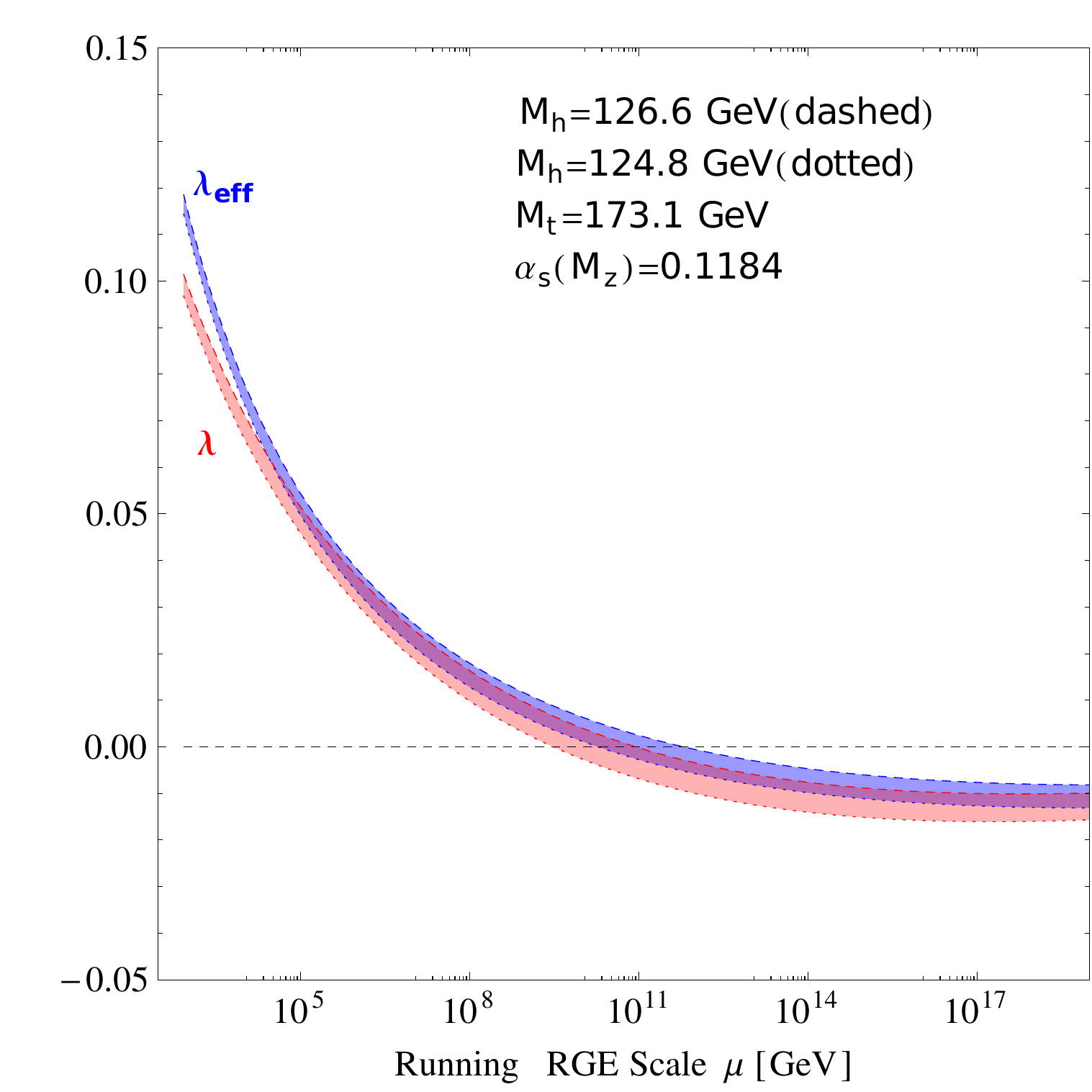}}
\caption{\label{fig:lambdaSM} \textit{ Higgs quartic coupling $\lambda$ and effective $\lambda_{\rm eff}$-function as a function of running RGE scale.} }
\end{center}
\end{figure}

The Higgs quartic coupling, $\lambda$ or $\lambda_{eff}$ becomes negative after $\mu \sim 10^{10}$ GeV. This corresponds to the metastability of the EW vacuum.
The detailed study of the metastability of the Higgs potential will be carried out in the next section.
\subsection{Metastability and Tunneling Probability}
\label{TunProb}
The Higgs quartic coupling determines the shape of the Higgs potential $V(\phi)$. How it changes with the running energy and why an extra deeper minimum in the Higgs potential is located near the Planck scale, will be discussed in the following.

One can realize from the $\beta$-function of the coupling constants of eqns.~\ref{betag1}-\ref{betayt} and Fig.~\ref{fig:couplings} that except the gauge coupling $g_1$, the other SM couplings are the decreasing functions of the running energy. The top Yukawa coupling $y_t$ decreases due to the dominant contributing term $- \frac{8 y_t  g_3^2}{16 \pi^2}$, which is present in $\beta_{y_t}$. Similarly in the $\beta$-function of the gauge couplings $g_2$ and $g_3$, the dominant contributions are $- \frac{7 g_3^2}{16 \pi^2}$ and $ - \frac{19 g_2^2}{96 \pi^2}$ respectively. But the running of the Higgs quartic coupling is of different nature than the other SM couplings.
\begin{figure}[h!]
\begin{center}
{
\includegraphics[width=2.7in,height=2.7in, angle=0]{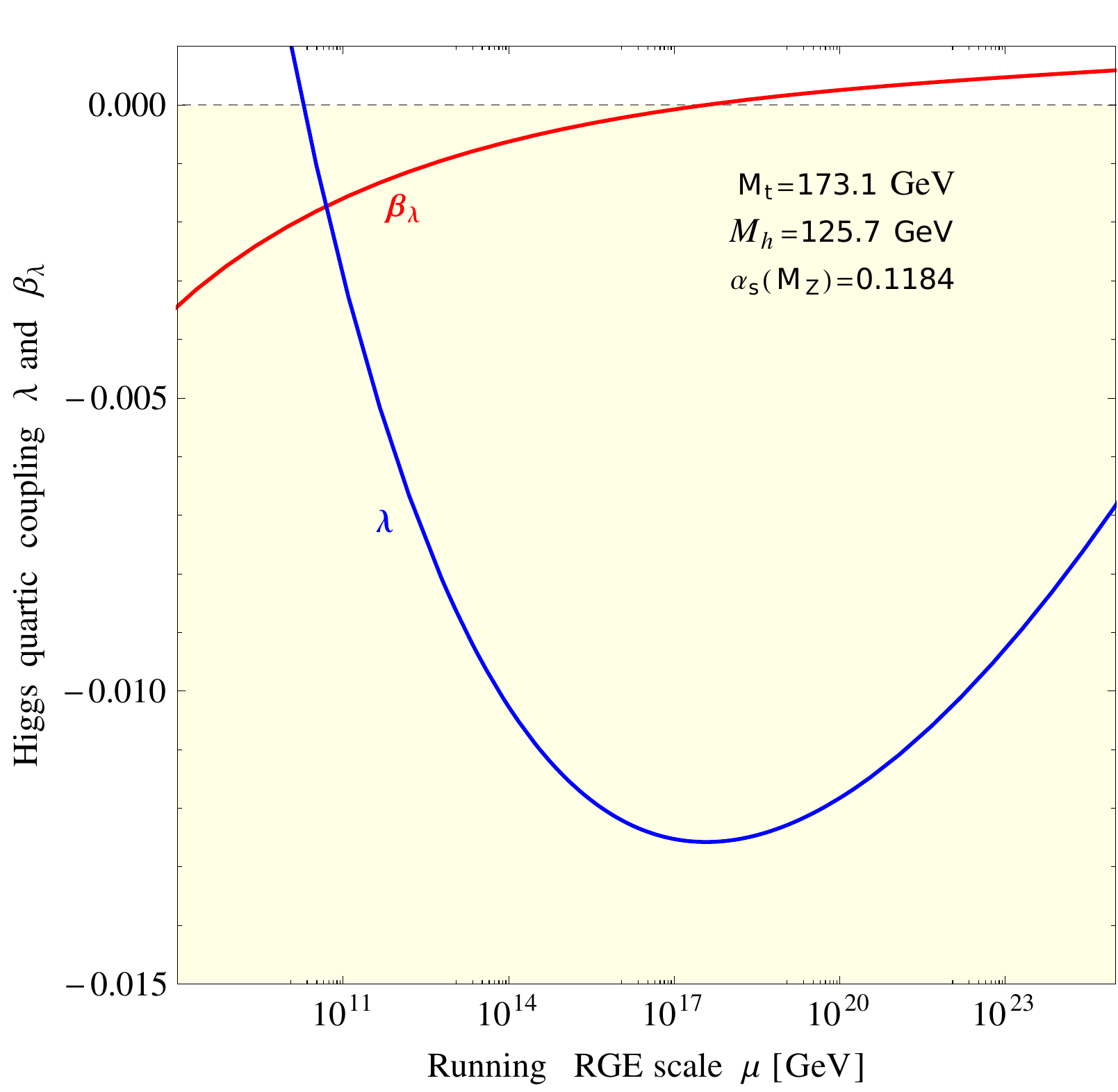}}
\caption{\label{fig:lambetamin} \textit{ Higgs quartic coupling $\lambda$ and corresponding $\beta$-function as a function of the running RGE scale.} }
\end{center}
\end{figure}

In Fig.~\ref{fig:lambetamin}, for the central values (including the other SM parameters) of $M_t=173.1$ GeV, $M_h=125.7$ GeV and $\alpha_S(M_Z)=0.1184$, the variation of Higgs quartic coupling $\lambda$ (blue) and corresponding $\beta$-function (red) with the running energy scale have been shown.
The $\lambda$ decreases up to $\sim 10^{17}$ GeV because $\beta_\lambda$ remains negative as it contains the dominant term, $- \frac{6 y_t^4}{16 \pi^2}$ (see eqn.~\ref{betalam}).
As $y_t$ also decreases with the running energy, $\beta_\lambda$ is an increasing function of the running energy.
It crosses zero around the energy $\mu_{min}\sim10^{17}$ GeV.
$\lambda$ increases after this energy scale. Therefore, the Higgs potential $V(\phi)$ has one extra minimum at that particular scale. The depth of the minimum is $\sim \frac{1}{4}|\lambda(\mu_{min})|\mu_{min}^4$, much deeper than the depth of the EW vacuum. So the false minimum, i.e., the EW vacuum could tunnel to the true (deeper) minimum. 

The vacuum tunneling probability of EW vacuum to the true minimum at the present epoch is given by~\cite{Coleman:1977py, Isidori:2001bm, Buttazzo:2013uya}
\begp
\allowdisplaybreaks \beq
{\cal P}_0=0.15 \frac{\Lambda_B^4}{H_0^4} e^{-S(\Lambda_B)},
\label{prob}\\
\eeq
\eegp
where $H_0=1.44\times10^{-42}$ GeV (natural units) is the Hubble parameter and $S(\Lambda_B)$ is the minimum action of the Higgs potential at energy $\Lambda_B$.
Now one can calculate the action $S$ of the Higgs potential using semi-classical approach. The Euclidean equations of motion of $\phi$ can be written as~\cite{Coleman:1977py, Isidori:2001bm, Buttazzo:2013uya},
\begp
\allowdisplaybreaks \begin{align}
-\partial_\mu \partial ^\mu \phi + \frac{\partial V_{\rm eff}(\phi)}{\partial \phi}=0\\
-\phi^{''}-\frac{3}{r}\phi^{'}+ |\lambda_{\rm eff}|~\phi^3=0,
\label{EOM}
\end{align}
\eegp
where $r=x_\mu x^\mu$ and the eqn.~\ref{EOM} satisfies the following boundary conditions, 
\begp
\allowdisplaybreaks \beq
\phi^{'}(0)=0~~~~~~{\rm and }~~~~\phi(\infty)=v\rightarrow 0~~.
\eeq
\eegp
Here the Euclidean solution of $\phi$ starting and end point is $\phi=v$ at Euclidean time $\tau(\equiv i t)=\mp \infty$. The choice of the origin for the Euclidean time $\tau$ can be made in such a way that the turning point $\phi=0$ is reached at $\tau=0$. Such a solution is called a ``bounce". With this boundary conditions, the Euclidean equation of motion can be solved analytically and the solution of eqn.~\ref{EOM} is given by,
\begp
\allowdisplaybreaks \beq
\phi(r)= \sqrt{\frac{2}{|\lambda_{\rm eff}|}} ~\frac{2 R}{r^2 + R^2},
\label{Solphi}
\eeq
\eegp
where $R$ is some arbitrary parameter, generally it is called the size of the bounce.

Action of the Higgs potential given by,
\begp
\allowdisplaybreaks \beq
S = \int d^4 x ~ \Big(\frac{1}{2}(\bigtriangledown\phi)^2-V_{\rm eff}(\phi)\Big).
\eeq
\eegp
Here $d^4 x=2 \pi^2 r^3 dr$ is the volume element of 4-dimensional sphere, using eqn.~\ref{Solphi}, one can get
\begp
\allowdisplaybreaks \bea
S &=&\frac{16 R^2}{|\lambda_{\rm eff}|}\int_0 ^\infty  2 \pi^2 r^3 dr~ \frac{ (r^2 - R^2)}{(r^2 + R^2)^4}\nonumber\\
&=&\frac{8 \pi^2}{3 ~|\lambda_{\rm eff}|}.
\label{Action}
\eea
\eegp
One can calculate tunneling time from EW vacuum to new deeper minimum using the eqn.~\ref{prob}. 
The lifetime $\tau_{\rm EW}$ is given by~\cite{Buttazzo:2013uya}, 
\begp
\allowdisplaybreaks \beq
\tau_{\rm EW} =\left( \frac{55}{3\pi}\right)^{1/4} \frac{e^{S(\Lambda_B)/4}} {\Lambda_B} \approx \frac{T_U} {{\cal P}_0^{1/4}},
\label{lifetew}
\eeq
\eegp

where, $T_U\approx\frac{0.96}{H_0}\approx13.7$ billion years is the lifetime of the Universe.

The minimum action of the Higgs potential is needed to calculate the EW vacuum decay time. For $\beta_{\lambda_{eff}}=0$, $\lambda_{eff}$ is minimum and negative, i.e., $S$ of eqn.~\ref{Action} become minimum. In this calculation the loop correction to the action is neglected, as in Ref.~\cite{Isidori:2001bm} it had been argued that setting the running scale to $R^{-1}$ significantly restricts the size of such corrections. Also the gravitational corrections~\cite{Coleman:1980aw,Isidori:2007vm} to the action is neglected as in Ref.~\cite{Khan:2014kba}. In Ref.~\cite{Isidori:2001bm} it was pointed out that thermal corrections are important at very high temperatures. Finite temperature effects to EW vacuum stability in the context of SM have been calculated in~\cite{Rose:2015lna}. It had been claimed in Ref.~\cite{Espinosa:1995se} that the parameter space corresponding to EW metastability gets further reduced. In this work zero temperature field theory has been used however.

If $\tau_{\rm EW}$ is greater than the lifetime of the universe $T_U$, then the EW vacuum is called metastable, i.e., $0 < {\cal P}_0<1$. This ensuing bounds on the effective Higgs quartic coupling~\cite{Isidori:2001bm, Khan:2014kba} as,
\begp
\allowdisplaybreaks \beq
\lambda_{\rm eff}(\Lambda_B)< 0 ~~~{\rm and}~~~\lambda_{\rm eff}(\Lambda_B) > \lambda_{\rm min}(\Lambda_B)=\frac{-0.06488}{1-0.00986 \ln\left( {v}/{\Lambda_B} \right)}\,.
\label{lamminSM}
\eeq
\eegp
On the other hand, if $\lambda_{\rm eff}(\Lambda_B)<\lambda_{\rm min}(\Lambda_B)$, then the vacuum is unstable, implies no existence of the Universe.
If $\lambda_{\rm eff}(\Lambda_B)>0$, i.e., ${\cal P}_0=0$, no transition will take place, i.e., the EW vacuum is stable. But the recent experimental data suggest that the EW vacuum remains in the metastable state, considering there is no new physics up to the Planck scale. Therefore, the EW vacuum could transit to the other minimum and such a transition can release an enormous amount of energy and which will destroy the present Universe.
\begin{figure}[h!]
\begin{center}
{
\includegraphics[width=2.7in,height=2.7in, angle=0]{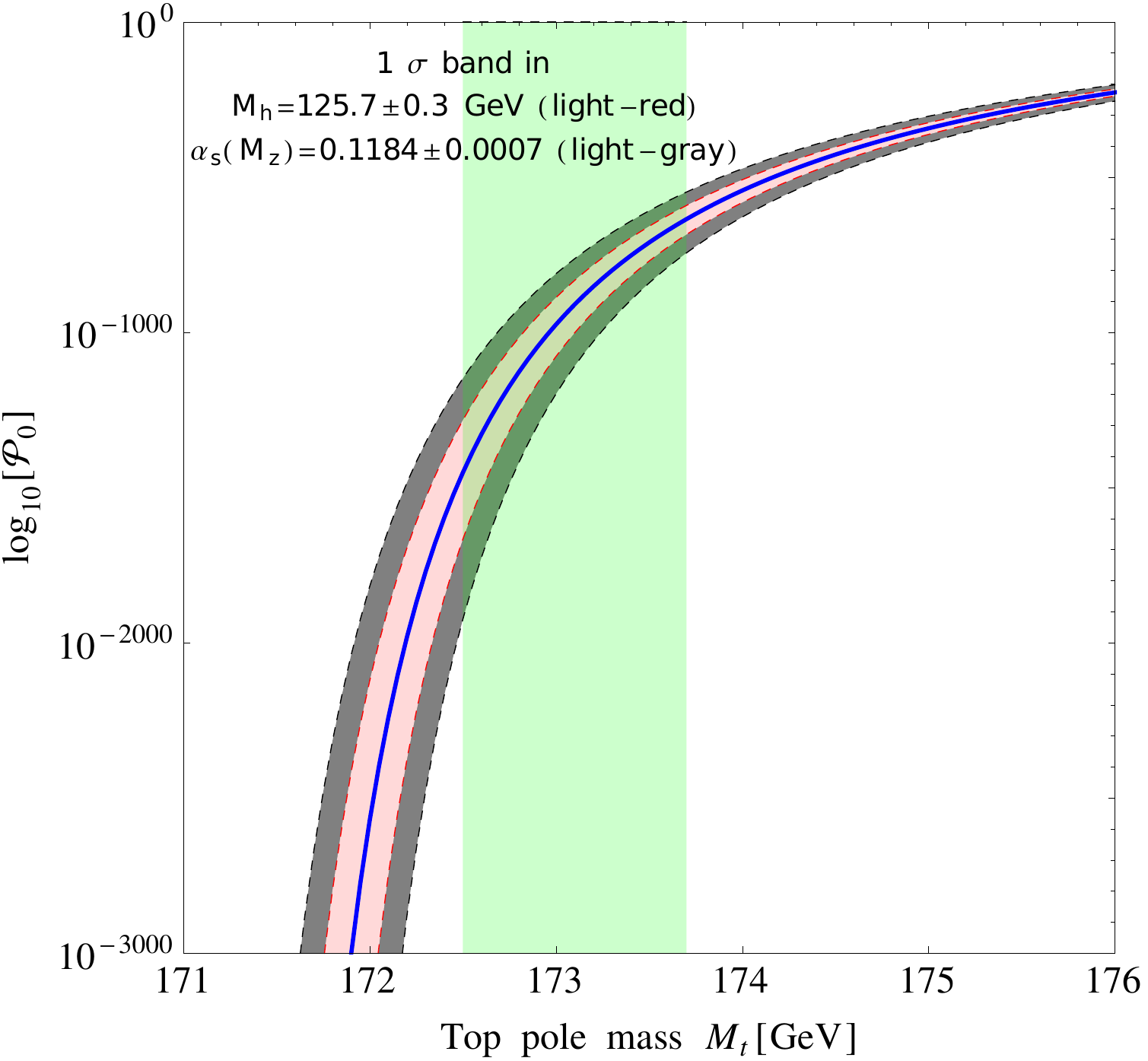}}
\caption{\label{fig:TunnelingSM} \textit{Tunneling probability ${\cal P}_0$ dependence on  $M_t$ in {\rm SM}. Light-green band stands for $M_t$ at $\pm 1\sigma$. The solid blue line corresponds to the central values of $M_h$ and $\alpha_S(M_Z)$, whereas the light-red band corresponds to the 1$\sigma$ deviation in Higgs mass and light-gray band for $\alpha_S(M_Z)$.}}
\end{center}
\end{figure}

The variations of the tunneling probability against the top mass $M_t$ has been shown in Fig.~\ref{fig:TunnelingSM}.
It is clear from the figure that the tunneling probability decreases with the decrease of $M_t$, i.e., the lifetime of EW vacuum increases.
For $M_t \lesssim 171$ the lifetime of EW vacuum become infinite as $\lambda > 0$ at all scales, i.e., the EW vacuum of the Higgs potential become absolutely stable.
One can obtain from Fig.~\ref{fig:TunnelingSM} that the experimental favored data imply that the lifetime of EW vacuum is $\sim 10^{100}-10^{500}$ years, much greater than the lifetime of our present Universe.
 \begin{figure}[h!]
 \begin{center}
 \subfigure[]{
 \includegraphics[width=2.7in,height=2.7in, angle=0]{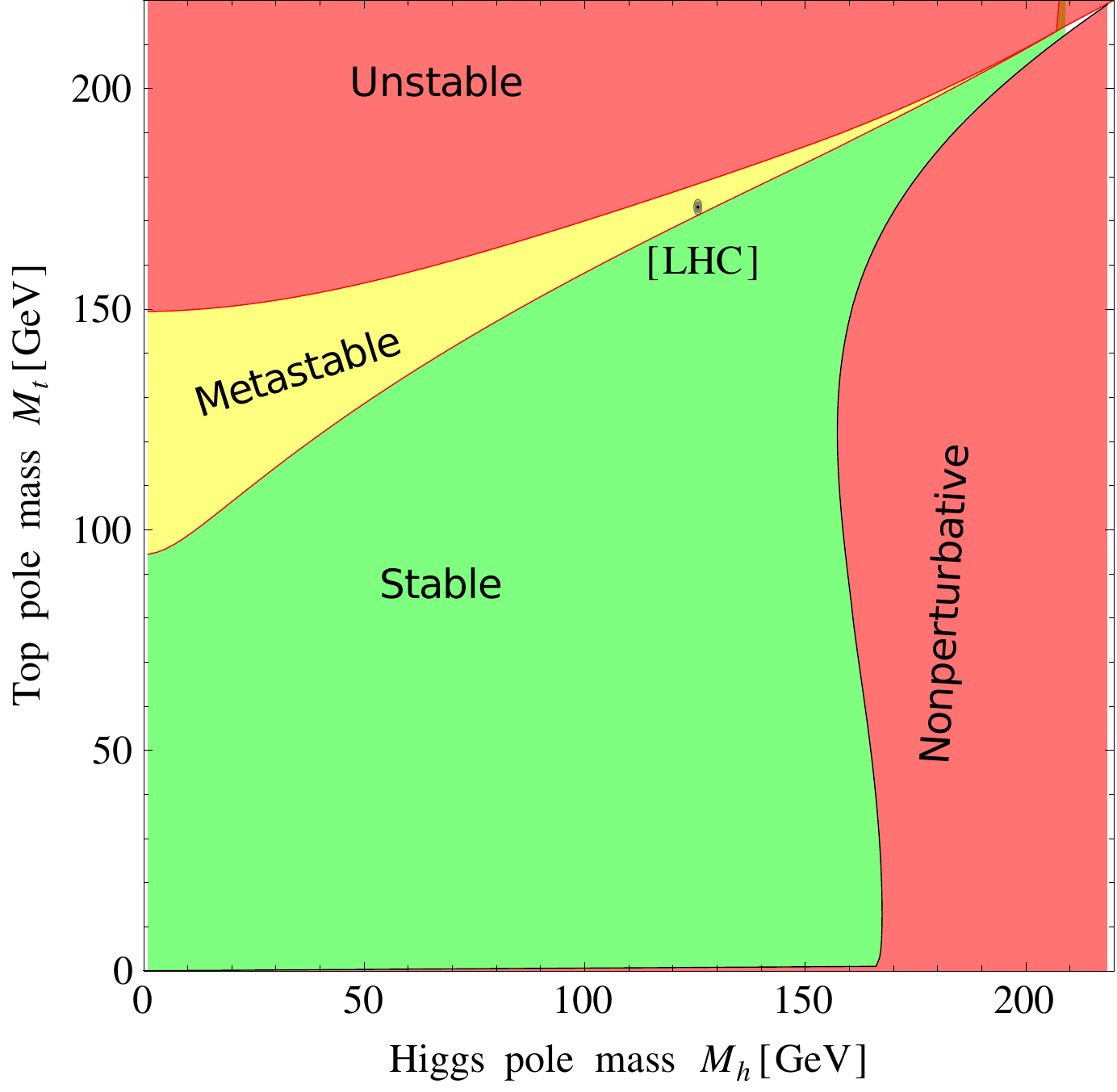}}
 \subfigure[]{
 \includegraphics[width=2.7in,height=2.7in, angle=0]{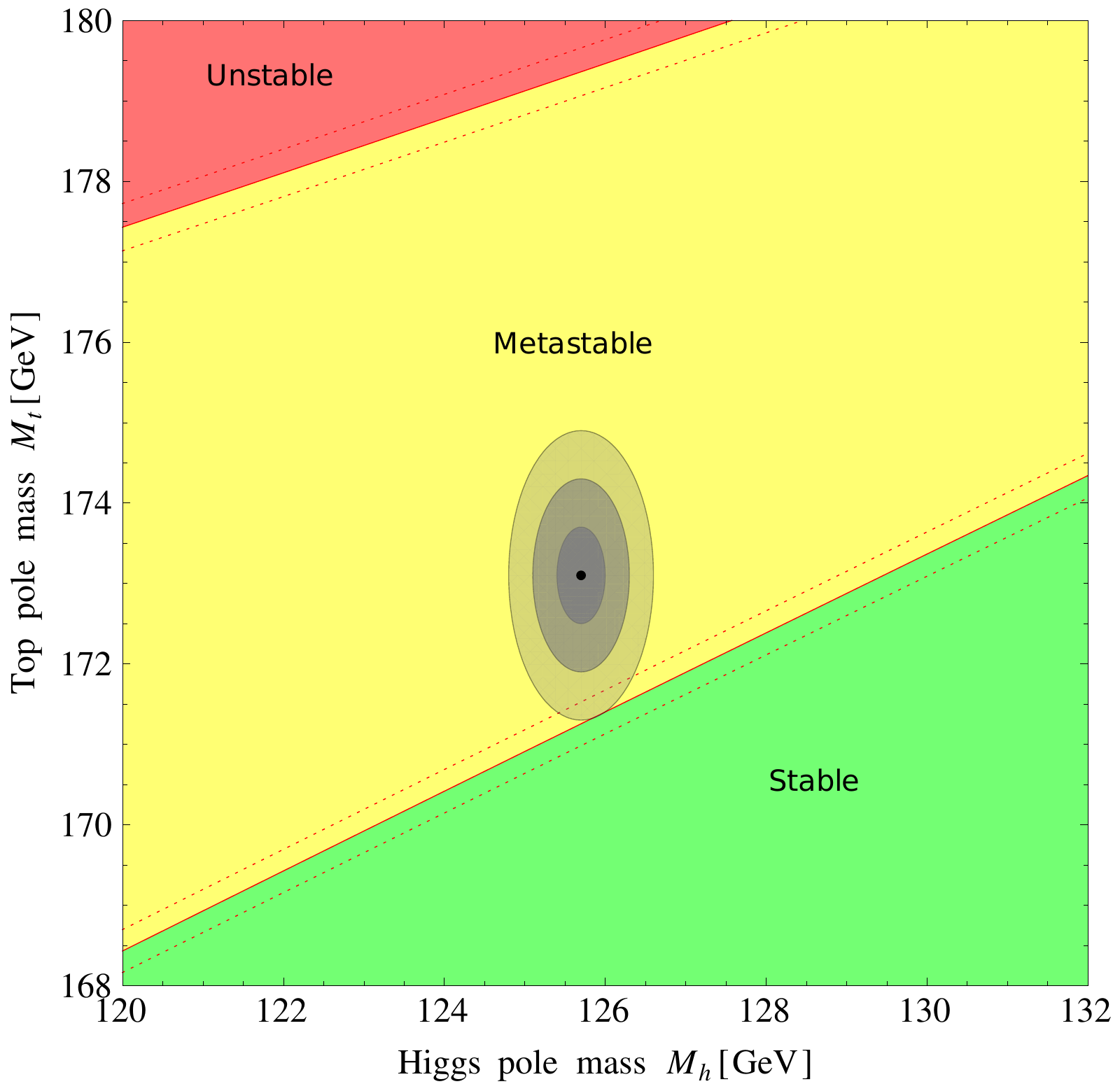}}
 \caption{\label{fig:Mt_MhSM} \textit{$({\bf a})$ In {\rm SM}, regions of absolute stability~(green), metastability~(yellow), instability~(red) of the EW vacuum in the $M_h - M_t$ plane phase diagram. $({\bf b})$ Zoomed in the region of the preferred experimental range of $M_h$ and $M_t$. The boundary lines $($red$)$ correspond to 1$\sigma$ variation in $\alpha_s(M_Z)$. The gray areas denote the experimentally favored zones for $M_h$ and $M_t$ at $1$, $2$ and  $3\sigma$. } }
 \end{center}
 \end{figure}

The stability of EW vacuum is best displayed with the aid of phase diagrams.
One can also identify the regions of EW stability and metastability in $M_t-M_h$ plane of Fig.~\ref{fig:Mt_MhSM} and in $\alpha_s(M_Z)-M_t$ plane of Fig.~\ref{fig:alpha-mt}.
In these figures, the unstable (red) and the metastable (yellow) region is separated by instability line which occurs when $\lambda(\mu)=\lambda_{\rm min}$.
The criticality ($\beta_{\lambda}(\mu)=\lambda(\mu)=0$) line separated the metastable and stable (green) regions.
The red region (right-side) of Fig.~\ref{fig:Mt_MhSM}(a) is excluded as the theory becomes non-perturbative at the Planck scale.
The Fig.~\ref{fig:Mt_MhSM}(a) illustrates the remarkable
coincidence for which the EW vacuum in the SM appears to live right in between the stable and unstable regions.
The Fig.~\ref{fig:Mt_MhSM}(b) is zoomed version of the Fig.~\ref{fig:Mt_MhSM}(a), here the gray ellipses are experimentally preferred range of $M_h$ and $M_t$ at $1$, $2$ and $3\sigma$.
From these figures one can conclude that for the Higgs mass $M_h<125.7$ GeV, the stability of the EW vacuum up to the Planck mass is excluded at 98$\%$ C.L. (one-sided).
\begin{figure}[h!]
\begin{center}
{\vspace{-0.17cm}
\includegraphics[width=2.7in,height=2.7in, angle=0]{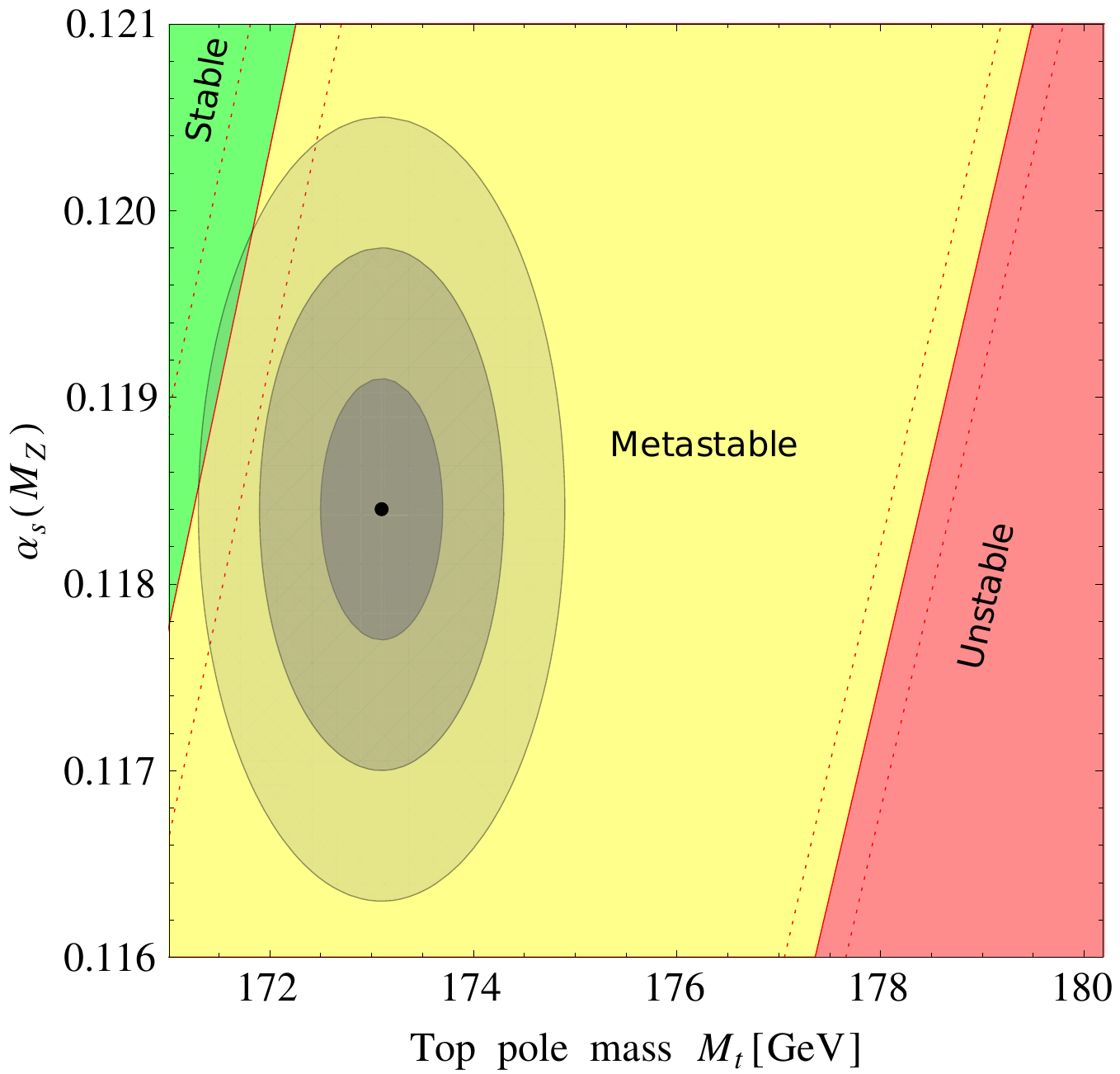}}
\caption{\label{fig:alpha-mt} \textit{ Phase diagram in $M_t-\alpha_S(M_Z)$ plane in {\rm SM}. Regions of absolute stability~(green), metastability~(yellow), instability~(red) of the EW vacuum are marked. The dotted lines correspond to $\pm 3\sigma$ variation in $M_h$ and the gray areas denote the experimental allowed region for $M_t$ and $\alpha_S(M_Z)$ at $1$, $2$ and  $3\sigma$.} }
\end{center}
\end{figure}
\subsection{Bounds on the Higgs mass from metastability and perturbativity}
As it has been seen from Fig.~\ref{fig:Mt_MhSM}, the measured values of $M_h$ and $M_t$ appear to be rather special, in the sense that they place the SM EW vacuum in a near-critical condition, i.e., at the border between stability and metastability.
The vacuum stability bound on the Higgs mass approximated as,
\begp
\allowdisplaybreaks \beq
 M_h~{\rm [GeV]}>129.46+1.12\left( \frac{M_t~{\rm [GeV]} -173.1}{0.6} \right) 
 - 0.56 \left(\frac{\alpha_S(M_Z)-0.1184}{0.0007}\right),
 \label{MassMh}
\eeq
\eegp

is obtained from the requirement $\lambda=\beta_{\lambda}=0$. One can see from eqn.~\ref{MassMh} that the main uncertainty comes from the top mass, $M_t$, so any improvement in the measurement of the top mass is of great importance for the question of EW vacuum stability.
\begin{figure}[h!]
\begin{center}
{
\includegraphics[width=2.7in,height=2.7in, angle=0]{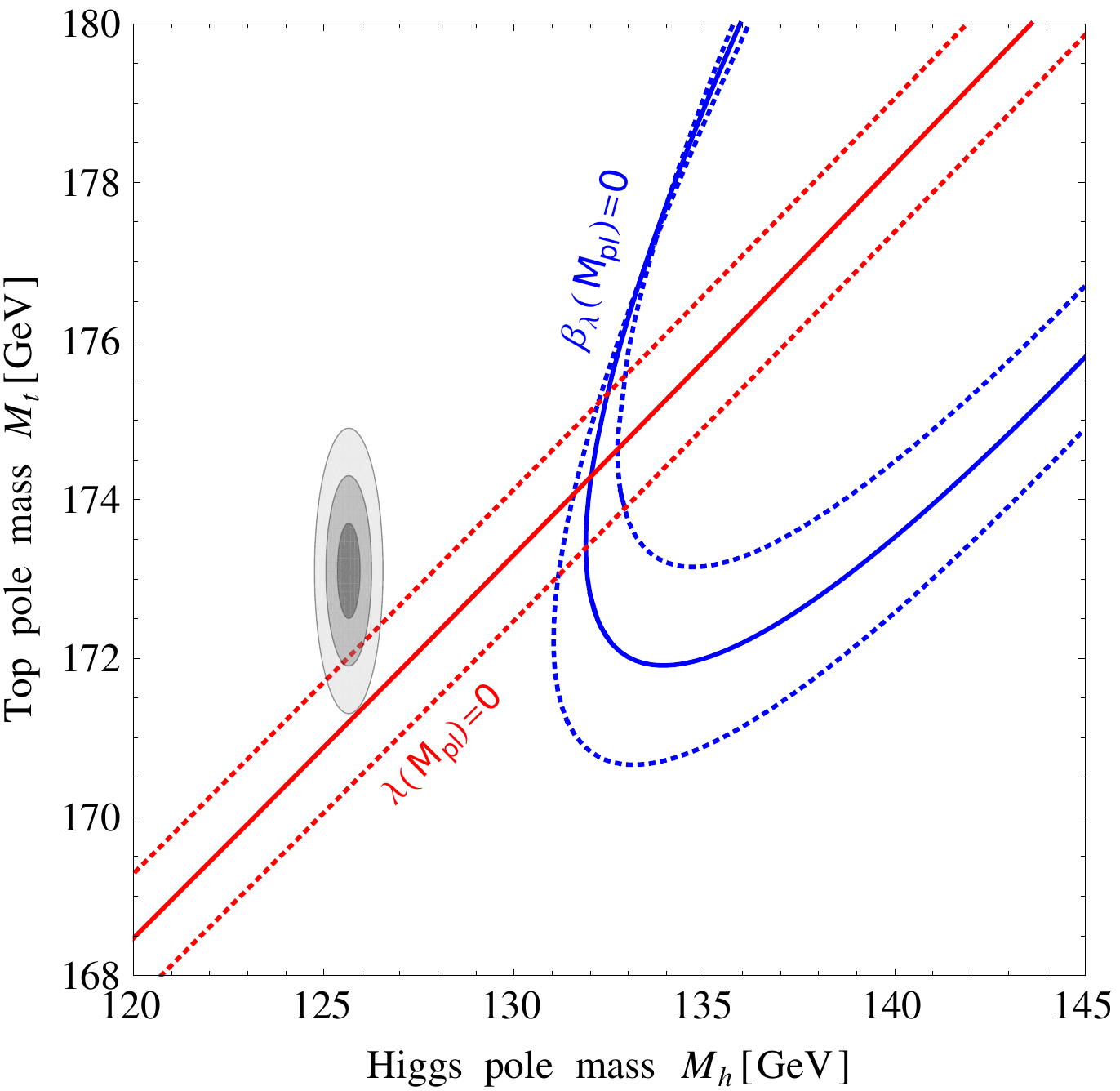}}
\caption{\label{fig:lambetampl} \textit{ Contour plot for $\lambda(\mpl) =0$ (red line) and  $\beta_{\lambda}(\mpl)=0$ (blue line). Dotted lines correspond to $\pm 3\sigma$ variation in $\alpha_s(M_Z)$. The gray areas denote the experimental allowed region for $M_h$ and $M_t$ at $1$, $2$ and  $3\sigma$.} }
\end{center}
\end{figure}
\par
The metastability bound is given by,
\begp
\allowdisplaybreaks \beq
 M_h~{\rm [GeV]}>109.73+1.84\left( \frac{M_t~{\rm [GeV]} -173.1}{0.6} \right) 
 - 0.88 \left(\frac{\alpha_S(M_Z)-0.1184}{0.0007}\right),
 \label{MassMhmeta}
  \eeq
  \eegp
 
comes from the requirement $\beta_\lambda=0$ and $\lambda=\lambda_{\rm min}$. To ensure perturbativity, one demand $\lambda(\mpl)< 4\pi $, which leads to
\begp
\allowdisplaybreaks \beq
 M_h~{\rm [GeV]}<172.23+0.36\left( \frac{M_t~{\rm [GeV]} -173.1}{0.6} \right) 
 - 0.12 \left(\frac{\alpha_S(M_Z)-0.1184}{0.0007}\right)\, .
 \label{MassMhpert}
\eeq
\eegp
 
Note that these bound are scheme independent, while the higher order RG equations, the potential calculation including radiative corrections or the threshold matching conditions are scheme and gauge dependent. The instability scale $\Lambda_I$ in Landau gauge with $\MS$ scheme has been found as,
\begp
\allowdisplaybreaks \bea
 \log_{10}\left(\frac{\Lambda_I}{{\rm GeV}}\right) &=& 10.277-0.82\left( \frac{M_t~{\rm [GeV]} -173.1}{0.6} \right) 
 + 0.33 \left(\frac{\alpha_S(M_Z)-0.1184}{0.0007}\right)\nn\\
 &&\hspace{4cm}+0.23 \left(\frac{M_h~{\rm [GeV]}-125.7}{0.3}\right)\, .
 \label{eq:instbilscaleSM}
\eea
\eegp
\subsection{Asymptotic safety in SM}
Shaposhnikov and Wetterich predicted~\cite{Shaposhnikov:2009pv} mass of the Higgs boson of $126$~GeV imposing the constraint $\lambda(\mpl)=\beta_{\lambda}(\mpl)=0$, in a scenario known as asymptotic safety of gravity. This corresponds to the fact that both the EW minimum and new minimum residing at the Planck scale, are degenerate. From Fig.~\ref{fig:lambetampl}, it clear that the present experimental data of the SM parameters does not allow this condition to be realized in the SM. This condition is satisfied at $M_h=132.033$ GeV and $M_t=174.294$ GeV for fixed central values of the other SM parameters.
\subsection{Veltman's conditions in SM}\label{VCsINSM}
The Higgs mass with quadratic divergence term in cutoff regularization scheme is given by,
\begp
\allowdisplaybreaks \bea
m_h^2 &=& m_{h,bare}^2 + \frac{\Lambda_{cut}^2}{16 \pi^2 v^2 }STr({\cal M}^2).\nn\\
STr({\cal M}^2) &=& 3(m_h^2 + m_Z^2 + 2 m_W^2 - 4 m_t^2).
                \label{VCcond}
\eea
\eegp
The quadratic divergences part gives rise to the hierarchy problem related to the Higgs mass. The condition for the absence of the quadratic divergences at one loop, $STr({\cal M}^2)=0$, is known as the Veltman's condition (VC). 
For high energy $\phi \gg v$ eqn.~\ref{VCcond} can be written as,
\begp
\allowdisplaybreaks \bea
\frac{STr({\cal M}^2)}{\phi^2} &=& 6 \lambda_1+\frac{9}{4} g_2^2+\frac{3}{4} g_1^2 -12 y_t^2=0.\nn
\eea
\eegp
Due to the large negative contribution from the term containing the top Yukawa coupling, it is not possible to satisfy VC at $\mpl$ given the experimental measurements of $M_t$ and $M_h$.
It has been seen that if one consider Higgs mass $M_h\approx 135\pm2.5$~GeV with top mass $M_t=173.1\pm 0.6$~GeV, the VCs in the SM at $\mpl$ can be satisfied, which is excluded at more than 5$\sigma$.
For $M_h\approx 135\pm2.5$ GeV the quadratic divergence part may cancel at one-loop but if two-loop corrections are included then the problem arises again.
Also, the VCs may be satisfied at a given scale but a nonzero quadratic divergence part in the Higgs mass may reappear for a different cut-off scale. This implies that the hierarchy problem is not solvable in the SM using the VC.
\section{Summary}
In this chapter, it has been shown that the ground state of  the Higgs potential depends on the SM parameters. With the present experimental values of the SM parameters, it has been observed that the Higgs quartic coupling $\lambda$, together with all other SM coupling constants, remains perturbative in the entire energy domain between the EW and the Planck scales.
It has also been seen that assuming the validity of the SM up to the Planck scale $\mpl$, the measured value
of $M_h \simeq 125$ GeV is near-critical, i.e., it places the EW vacuum right at the border between absolute stability and metastability. The absolute stability of the EW vacuum of the Higgs potential is excluded at about 98$\%$ confidence level.

In the SM, with the present measured values of the SM parameters, the asymptotic safety of gravity at the Planck scale $\mpl$ is not realized. Also the possible solution of hierarchy problem related to the Higgs mass, i.e., Veltman's condition cannot be achieved at any scale between EW and $\mpl$.
At $\mpl$ this condition is fulfilled for the Higgs mass $M_h>130$ GeV, which is excluded at more than 5$\sigma$ C.L.

New physics can modify the effective Higgs potential through the quantum corrections as well as the stability of the EW vacuum will also alter. In the next chapter, the detailed modifications of stability of the Higgs potential in the presence of new physics in the form of an additional scalar multiplet will be discussed.

\chapter{Metastability in Extended Scalar Sectors of the Standard Model}
\label{chap:MetaSMextended}
\linespread{0.1}
\graphicspath{{Chapter6/}}
\pagestyle{headings}
\noindent\rule{15cm}{1.5pt} 
\section{Introduction}
 
It has been seen in the previous chapter that life time of the Universe is finite in the SM. It is important to explore if the new extra scalar field(s) of an extended scalar sector has an answer to this puzzle in its reserve.
In this work a $SU(2)_L$ singlet~\cite{Khan:2015ipa} or a doublet~\cite{Khan:2014kba} or a hyperchargeless triplet~\cite{NajimTrip} scalar is added to the SM and a $Z_2$-symmetry is imposed on these new models such that the $odd$ number of new scalar fields do not couple with the SM particles.
As the lightest neutral particle of the additional scalar sector cannot decay, it becomes stable and serves as a viable dark matter candidate which may saturate the relic abundance of the dark matter in the Universe.
As these new models contain various kinds of new scalar fields, the structure of scalar potential and the absolute stability bounds of the EW vacuum of the scalar potential which had been shown in Chapter~\ref{chap:EWSBextended} are rather complicated than that in the SM due to the involvement of more parameters.

In this work, a detailed study of the metastability of the EW vacuum of Higgs potential in the extended scalar sector will be presented. Various kinds of phase diagrams will be drawn to illustrate the region of absolute stability, metastability, instability and the non-perturbativity on different parameter spaces of these extended scalar sectors. In the Section~\ref{SingletSMext}, the stability of the EW vacuum in the presence of a real singlet scalar will be discussed. In Sections~\ref{doubletexten} and~\ref{sec:ITM}, similar studies for the inert doublet model (IDM) and inert triplet model (ITM) will be elaborated.
\section{Singlet scalar extension of SM}\label{SingletSMext}
Recent cosmological and astrophysical evidences suggest presence of cold dark matter (DM). A simple choice is to add a gauge singlet real scalar $S$ to the SM~\cite{Burgess:2000yq}. An additional $Z_2$ symmetry ensures the stability of $S$. The scalar modifies the Higgs effective potential, and can ensure vacuum stability up to $\mpl$.
Such extensions of SM have been discussed in the literature~\cite{Haba:2013lga,Kadastik:2011aa, Gabrielli:2013hma,Gonderinger:2009jp,Clark:2009dc,Lerner:2009xg,Cheung:2012nb,Lebedev:2012zw, EliasMiro:2012ay,Antipin:2013bya, Khan:2012zw,Grzadkowski:2001vb, Branchina:2013jra, Lalak:2014qua, Branchina:2014usa,Eichhorn:2014qka} in the context of vacuum stability.

In the context of SM, detailed studies of metastability has shown in Chapter~\ref{chap:MetaSM}. 
Along the same line, in this section, the studies of the metastable vacuum has been extended to the SM+$S$ model~\cite{Khan:2014kba}.
\subsection{Effective potential and RGE running}
\label{EffpotScalar}
In this model, an extra real scalar singlet field $S$, odd under $Z_2$ symmetry, is added to the SM, providing a suitable candidate for dark matter. The corresponding Lagrangian density is given by,
\begp
\allowdisplaybreaks \beq
{\cal L}_{S}= \frac{1}{2} (\partial_{\mu} S) (\partial^{\mu} S) -V_0^S\nn
\eeq
\eegp
with,
\begp
\allowdisplaybreaks \beq
V_0^S= \frac{1}{2} \overline{m}_S^2 S^2 +\frac{\kappa}{2} |\Phi|^2 S^2 +
 \frac{\lambda_S}{4!} S^4 \, ,
 \label{Vscalar}
\eeq
\eegp
where, 
\begp
\allowdisplaybreaks \beq
\Phi=\begin{pmatrix}
G^+ \\ (h+v+iG^0)/\sqrt{2}
\end{pmatrix} \nn \, .
\eeq
\eegp
After spontaneous EW symmetry breaking, DM mass $M_S$ is expressed as $M_S^2=\overline{m}_S^2  +\kappa v^2/2$.

SM tree level Higgs potential
\begp
\allowdisplaybreaks \beq
V_0^{\rm SM}(\phi)=-\frac{1}{2} m^2 h^2 + \frac{1}{4} \lambda h^4\nn\\
\eeq
\eegp
is augmented by the 
SM+$S$ one-loop Higgs potential in Landau gauge using $\MS$ scheme, which is written as
\begp
\allowdisplaybreaks \beq
V_1^{{\rm SM}+S}(h)= V_1^{\rm SM}(h) + V_1^{S}(h). \nn
\eeq
\eegp
The expression of $V_1^{\rm SM}(h)$ is given\footnote{Here $h$ is used instead of $\phi$, representing the SM-like Higgs field.} in eqn.~\ref{V1loop} and the one loop contribution of the singlet scalar can be written as~\cite{Lerner:2009xg, Gonderinger:2012rd},
\begp
\allowdisplaybreaks \bea
V_1^{S}(h)&=&\frac{1}{64 \pi^2} M_S^4(h) \left[ \ln\left(\frac{M_S^2(h)}{\mu^2(t)} \right)-\frac{3}{2} \right], \nn
\eea
\eegp
where,
\begp
\allowdisplaybreaks \bea
M_S^2(h) &=& \overline{m}_S^2(t)  +\kappa(t) h^2(t)/2\, .
\nn
\eea
\eegp
SM contributions are taken at two-loop level~\cite{Ford:1992pn, Martin:2001vx, Degrassi:2012ry, Buttazzo:2013uya}, whereas the scalar contributions are considered at one-loop only. 

For $h\gg v$, the effective potential can be approximated as
\begp
\allowdisplaybreaks \beq
V_{\rm eff}^{{\rm SM}+S}(h) \simeq \lambda_{\rm eff}(h) \frac{h^4}{4}\, ,
\label{efflamS}\eeq
\eegp
with
\begp
\allowdisplaybreaks \beq
\lambda_{\rm eff}(h) = \lambda_{\rm eff}^{\rm SM}(h) +\lambda_{\rm eff}^{S}(h)\, ,
\label{lameffscalar}
\eeq
\eegp
where $\lambda_{\rm eff}^{\rm SM}(h)$ can be found in eqn.~\ref{eq:effqurtic} and,
\begp
\allowdisplaybreaks \bea
 \lambda_{\rm eff}^{S}(h)&=&e^{4\Gamma(h)} \left[\frac{\kappa^2}{64 \pi^2}  \left(\ln\left(\frac{\kappa}{2}\right)-\frac{3}{2}\right ) \right]\, . \label{efflamscalar}
\eea
\eegp
As quartic scalar interactions do not contribute to wave function renormalization at one-loop level, $S$ does not alter  $\gamma(\mu)$ of SM and anomalous dimension of $S$ is zero~\cite{Gonderinger:2009jp}.  
All running coupling constants are evaluated at $\mu=h$.

The beta functions for the new physics parameters $\kappa$ and $\lambda_S$ are given in eqn.~\ref{eq:betasinglet}. $\overline{m}_S$ also evolves with energy. But as the beta functions of other parameters do not involve $\overline{m}_S$, its beta function is not considered in this discussion.
Here, new physics effects are included in the RGEs at one-loop only.
\subsection{Singlet Scalar as a dark matter candidate}
\label{darksingletScalar}
As it has already been discussed that the new extended real singlet scalar is odd under $Z_2$ symmetry, i.e., under this symmetry, $S\rightarrow -S$, but Standard Model particles are invariant. Due to $Z_2$ symmetry $odd$ numbers of the scalar fields $S$ do not couple with the SM particles. The particle $S$ is stable and it can be considered as a viable dark matter candidate.
This model can provide dark matter with almost all possible mass ranges which allowed from the relic density constraints of WMAP~\cite{Bennett:2012zja} and Planck~\cite{Ade:2013zuv} data. 

After EW symmetry breaking one can write the potential (see eqn.~\ref{Vscalar}) explicitly as,
\begp
\allowdisplaybreaks \beq
V_0^S=\frac{1}{2}\left(\overline{m}_S^2+\frac{\kappa v^2}{2}\right) S^2 + \frac{k v}{2} h S^2+\frac{\kappa}{4} h^2 S^2+\frac{\lambda_S}{24} S^4.
\label{VscalarExpand}
\eeq
\eegp
$S$ is the scalar singlet under the SM gauge symmetry so there are no 4-point coupling like, $SSXX$ ($X$ is the SM vector boson or a fermion). The scalar field can annihilate to SM particles only $via$ Higgs exchange. $S$ is called a Higgs portal dark matter.
LHC has put a stringent bounds on the Higgs invisible decay width~\cite{Belanger:2013xza}. The decay width of $h$ to pair of $S$ is given by,
\begp
\allowdisplaybreaks \beq
\Gamma\left(h\rightarrow SS\right)= \frac{ v^2}{32\pi M_h }\kappa^2 \left(1-\frac{4 M_{\rm DM}^2}{M_{h}^2}\right)^{1/2}\,.
\label{decaywidthS}
\eeq
\eegp
Also the direct detection experiments, XENON\,100~\cite{Aprile:2011hi,Aprile:2012nq} and LUX~\cite{Akerib:2013tjd} puts a bound on the dark matter mass from the non-observation of dark matter-nucleon scattering. DM direct detection involve the $h$-mediated $t$-channel process, $SN\ra SN$. The scattering cross-section is given by,
\begp
\allowdisplaybreaks \beq
 \sigma_{S,N} = \frac{m_r^2}{4 \pi} f_N^2 m_N^2\left(\frac{\kappa}{M_{\rm DM} M_h^2}\right)^2 \label{directcsS}
\eeq
\eegp
where $f_N\approx0.3$ is the form factor of the nucleus. $m_r$ represents the reduced mass of the nucleus and the scattered dark matter particle.

With the invisible Higgs decay width at LHC and non-observation of dark matter in the direct detection experiment at the LUX and including indirect Fermi-LAT bounds, it has been shown in Refs.~\cite{Duerr:2015mva,Duerr:2015aka} that the dark matter mass below 50 GeV and $70-110$ GeV are excluded.
In the table~\ref{tableS1}, few benchmark points have been shown for low dark matter mass less than the half of the Higgs mass.
{\tt FeynRules}~\cite{Alloul:2013bka} along with {\tt micrOMEGAs}~\cite{Belanger:2010gh, Belanger:2013oya} have been used to compute relic density of scalar DM in SM+$S$ model. 
The main dominant contributions to the relic density in these region is $SS\ra b \bar{b}$.
\setlength\tablinesep{1pt}
\setlength\tabcolsep{10pt}
\begin{table}[h!]
\begin{center}
    \begin{tabular}{ | c | c |  c | c | c | c |}
    \hline
     $M_{DM}$ (GeV) & $\kappa$ & Relic density &$\sigma_{SI}$ ${\rm ({cm}^2)}$& Br($h\rightarrow S S$) in $\%$\\
    \hline
     $55$ & 0.007 & 0.1242 & $1.3\times10^{-46}$ & 1.83   \\
    \hline
     $56$ & 0.0045 & 0.1182 & $5.5\times10^{-47}$ & 0.72   \\
     \hline
     $58$ & 0.0018 & 0.1199 & $8.2\times10^{-48}$ & 0.121   \\
     \hline
     $60$ & 0.00075 & 0.1203 & $1.3\times10^{-48}$ & 0.013   \\
     \hline
             \end{tabular}
    \caption{\textit{ Benchmark points with dark matter mass $M_{DM}< M_h/2$, which is allowed from the relic density constraint on the DM of WMAP and Planck, $\Omega h^2=0.1198\pm 0.0026$ within 3$\sigma$ confidence level, direct detection LUX (2013) and Higgs invisible decay width from the LHC. }}
    \label{tableS1}
\end{center}
\end{table}

In Fig.~\ref{fig:relicScalar}, how the relic density changes with the dark matter mass in this model, has been shown.
The plot generated for three different Higgs portal coupling $\kappa(M_Z)=0.05$ (black), $\kappa(M_Z)=0.10$ (brown), and $\kappa(M_Z)=0.15$ (red). 
Here the blue band corresponds to the relic density constraints from WMAP and Planck data allowed at 3$\sigma$ confidence level. The light red band region is excluded from the Higgs invisible decay width. 
For the $\kappa(M_Z)=0.05$, one can see that there are four regions in dark matter mass which cross the blue band, i.e., satisfy the relic density in the right ballpark.

The dark matter mass near the 50 GeV, the dominant part in the annihilation of relic density is $SS \ra b\bar{b}$ (79$\%$), $SS\ra W^{\pm}W^{\mp*}$\footnote{The virtual $W^{\pm*}$ can decay to quarks and leptons.} (8$\%$), $SS \ra c\bar{c}$(7$\%$), however these regions are not allowed from Higgs invisible decay width as well as from the direct detection data. For 70 GeV, the process $SS \ra b\bar{b}$ is about 52$\%$ and $SS\ra W^{\pm}W^{\mp*}$ is 40$\%$ of total annihilation cross-section.
Whereas for the dark matter mass around 160 GeV, the dominant contributions in the annihilation are, $SS \ra W^{\pm}W^{\mp}$ (48$\%$), $SS \ra hh$ (30$\%$) and in $SS \ra ZZ$ (22$\%$) are allowed from direct searches. Similarly for other values of $\kappa(M_Z)$, the relic density in the right ballpark can be found near the dark matter mass 400 GeV and 500 GeV. In this case dominant contributions are $SS \ra hh,t\bar{t}$.
\begin{figure}[h!]
\begin{center}
{
\includegraphics[width=2.7in,height=2.7in, angle=0]{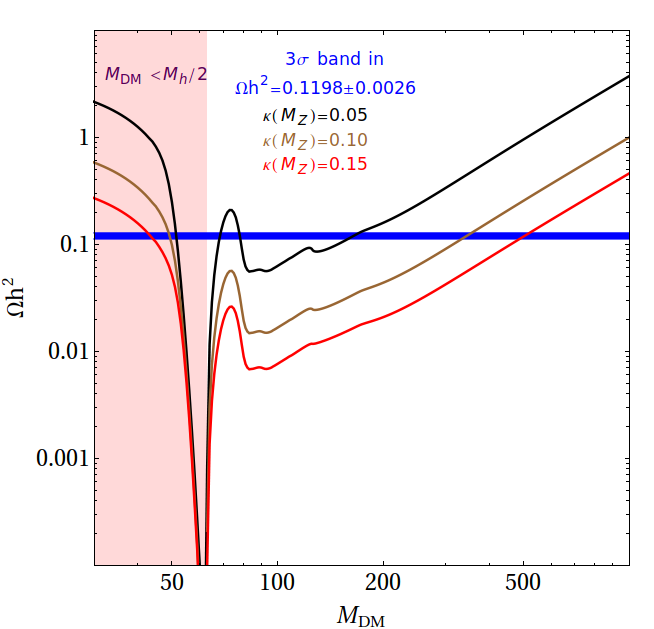}}
\caption{ \textit{ Dark matter relic density $\Omega h^2$ as a function of the dark matter mass $M_{DM}(\equiv M_S)$ for different values of the portal coupling: $\kappa(M_Z)=0.05$ (black), $\kappa(M_Z)=0.10$ (brown), and $\kappa(M_Z)=0.15$ (red). The thin blue band corresponds to the relic density of the dark matter, $\Omega h^2=0.1198 \pm 0.0026$ (3$\sigma$) of the present the Universe.}}
\label{fig:relicScalar}
\end{center}
\end{figure}

\begin{figure}[h!]
\begin{center}
{
\includegraphics[width=2.7in,height=2.7in, angle=0]{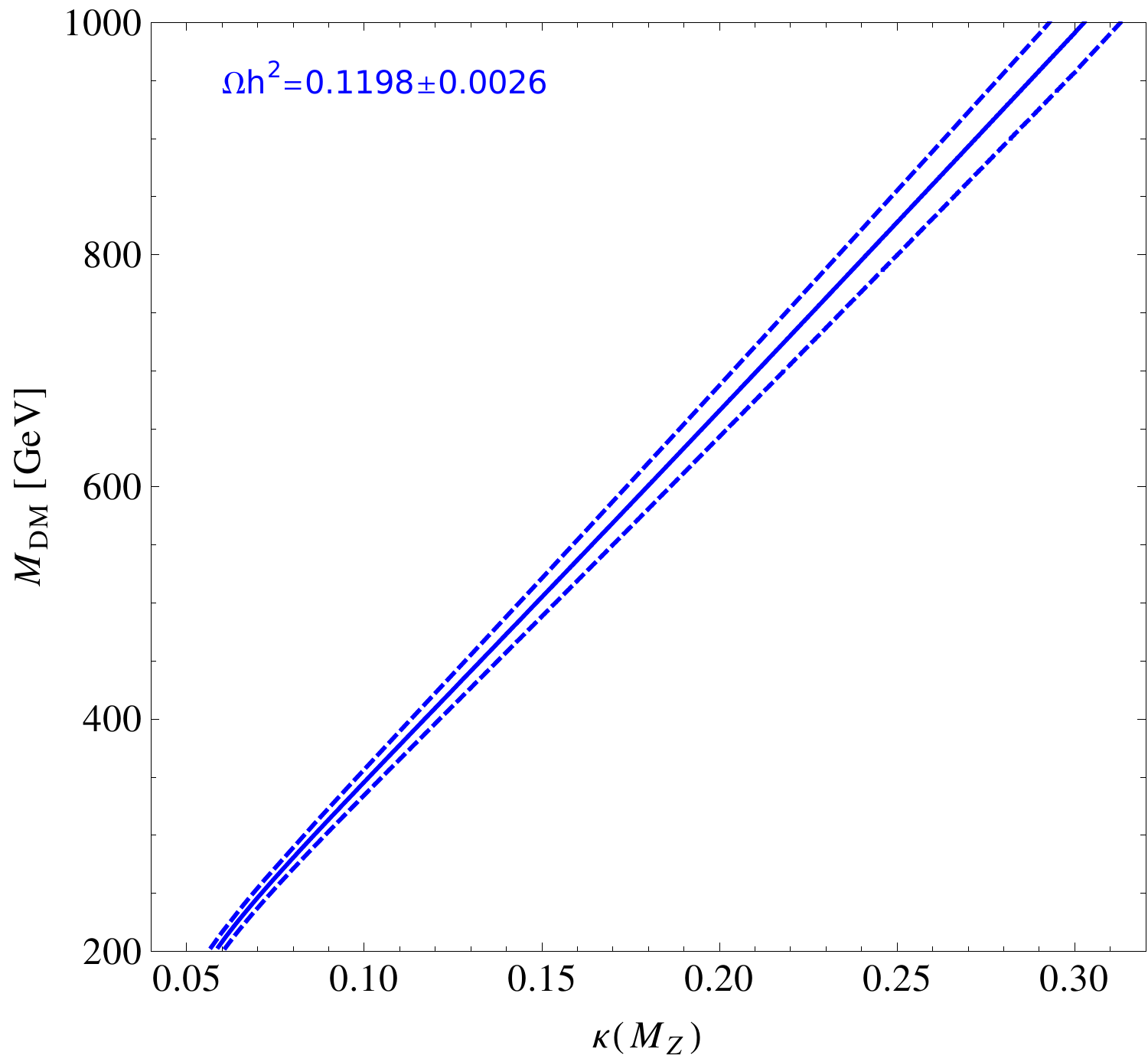}}
\caption{ \textit{ {\rm SM+}$S$ allowed parameter space in the $\kappa(M_Z)-M_{DM}$ plane in agreement with the relic density constraints, direct and indirect detection. Where the upper dotted, solid in the middle and the lower dotted blue line correspond to the relic density $\Omega h^2=0.112,0.1198$ and $0.1276$ respectively.}}
\label{fig:reliclmsvsmass}
\end{center}
\end{figure}

In Fig.~\ref{fig:reliclmsvsmass}, the relic density in $\kappa(M_Z)-M_{DM}$ plane for the high dark matter masses have been plotted.
In this plot the blue line (upper dotted) corresponds to the relic density $\Omega h^2=0.112$ (minimum relic density allowed from WMAP and Planck) and other two belong to $\Omega h^2=0.1198$ and $\Omega h^2=0.1276$ (maximum) respectively. The relic density band is almost like a straight line because for $M_{DM} > M_W, M_h$, the annihilation cross-section is proportional to $\frac{\kappa^2}{ M_{DM}^2}$.

\subsection{RGE running from $\mu=M_t$ to $\mpl$ in SM+$S$}
The similar technique of the matching conditions of Section~\ref{MatchingSM} has been used to calculate all the coupling at $\mu=M_t$. After knowing the values of various coupling constants at $M_t$, full three-loop SM RGEs and one-loop RGEs for the scalar $S$ have been used to run them up to $\mpl$.

\setlength\tablinesep{1pt}
\setlength\tabcolsep{5pt}
\begin{table}[h!]
\begin{center}
    \begin{tabular}{ | c | c | c | c |  c | c | c | c | c | c |}
    \hline
    $M_S$ (GeV)& $\kappa(M_Z)$ & $\lambda_S(M_Z)$ & $g_1$ & $g_2$ & $g_3$ & $y_t$ & $\lambda$ & $\kappa$ & $\lambda_S$\\
\hline
    620 & 0.185 & 1 & 0.478 & 0.506 & 0.487   & 0.382 & $-0.0029$ & 0.424 &  4.66 \\
    \hline
    795 & 0.239 & 0.389 & 0.478 & 0.506 & 0.487 & 0.382 & 0 & 0.412 & 1  \\         \hline
    \end{tabular}
    \caption{\textit{ Values of all {\rm SM+}$S$ coupling constants at $\mpl  = 1.2 \times 10^{19}$~GeV with $M_t=173.1$~GeV, $M_h=125.7$~GeV and $\alpha_S(M_Z)=0.1184$.}}
    \label{table2Scalar}
\end{center}
\end{table}

For SM+$S$, the running~\ref{eq:betasinglet} depends on the extra parameters $M_S$, $\kappa(M_Z)$ and $\lambda_S(M_Z)$. Assuming the values of SM parameters at $M_t$ as given in eqns.~\ref{eq:g1mt}$-$ \ref{eq:lammt}, for two different sets of $M_S$, $\kappa(M_Z)$ and $\lambda_S(M_Z)$, the values of all parameters at $\mpl$ in SM+$S$ model in Table~\ref{table2Scalar} has been presented.
The first set stands for our benchmark point, as described later. For this choice, $\lambda(\mpl)$ is negative. The second set is chosen such that $\lambda(\mpl)=0$.
Note that as the running of new physics parameters is not considered till $M_S$, in this case it is all the same to specify these parameters either at $M_Z$ or at $M_t$. 

In the SM, the gauge couplings $g_1,g_2,g_3$, top Yukawa coupling $y_t$ do not vanish at $\mpl$. The same is true in SM+$S$ model~\cite{Gabrielli:2013hma}, since running of these couplings are hardly affected by $S$, as displayed in Fig.~\ref{fig:SMScalar}.  Higgs portal coupling $\kappa$ and scalar self-quartic coupling $\lambda_S$ increase with energy. The rise of $\lambda_S$ is so rapid that it may render the theory nonperturbative at higher energies. For example, if $\lambda_S(M_Z)>1.3$, $\lambda_S$ becomes nonperturbative before $\mpl$.  Higgs self-quartic coupling $\lambda$ also gets affected by inclusion of $S$. But  the change is not visible in Fig.~\ref{fig:SMScalar}. 
 \begin{figure}[h!]
 \begin{center}
 \includegraphics[width=3.7in,height=3.1in, angle=0]{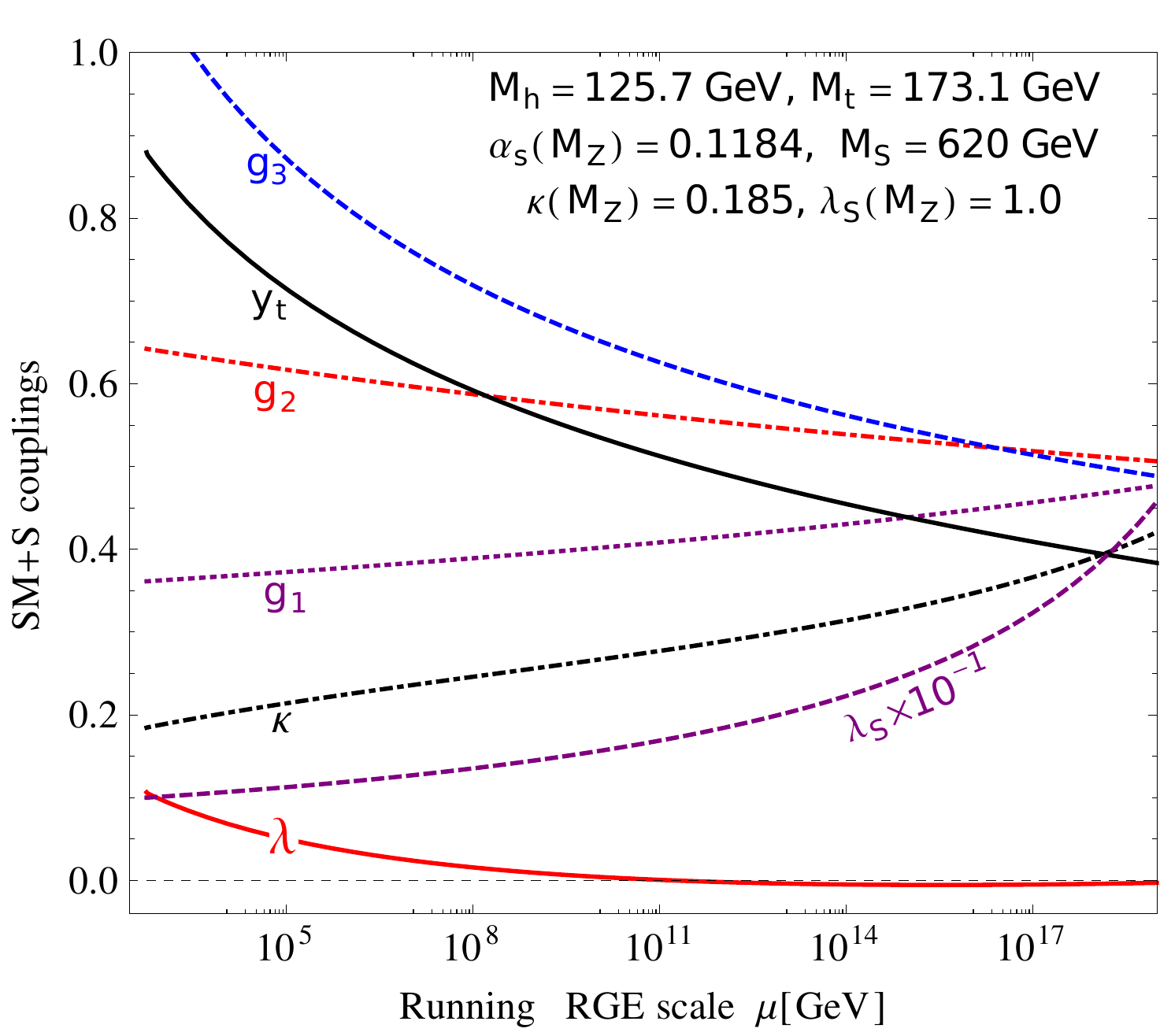}
 \caption{\label{fig:SMScalar} \textit{  {\rm SM+}$S$ RG evolution of the gauge couplings $g_1,g_2,g_3$, top Yukawa coupling $y_t$, Higgs self-quartic coupling $\lambda$, Higgs portal coupling $\kappa$ and scalar self-quartic coupling $\lambda_S$ in $\MS$ scheme. } }
 \end{center}
 \end{figure}
 \begin{figure}[h!]
 \begin{center}
 {\includegraphics[width=2.7in,height=2.7in, angle=0]{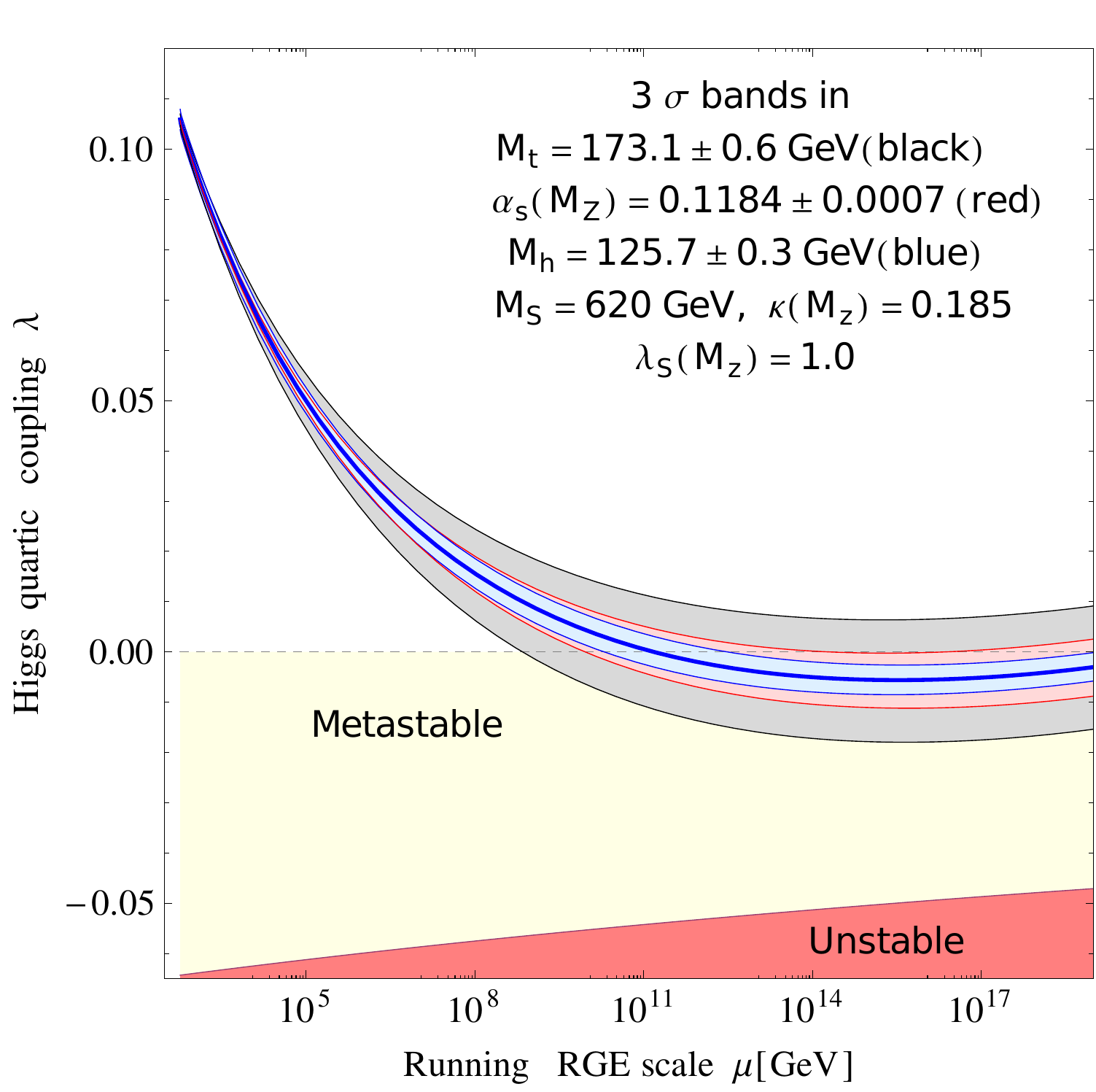}}
 \caption{\label{fig:lambda} \textit{ RG evolution of $\lambda$ in the {\rm SM+}$S$ for our benchmark point. $3\sigma$ bands for $M_t,M_h$ and $\alpha_s(M_Z)$ are displayed.} }
 \end{center}
 \end{figure}

As the issue of stability hinges on the value of $\lambda$ at higher energies, the running of $\lambda$ is focused for SM+$S$ model in Fig.~\ref{fig:lambda} as for the SM in Fig.~\ref{fig:couplings}(b), for a {\em benchmark point}  $M_S = 620$ GeV, $\kappa(M_Z)=0.185$ and $\lambda_S (M_Z)=1$.
It has been observed that the behavior of $\lambda$ running might change significantly, modifying instability scale $\Lambda_I$ to $1.68 \times 10^{11}$~GeV, whereas in SM $\Lambda_I\sim 1.9 \times 10^{10}$ GeV (see eqn.~\ref{eq:instbilscaleSM}).
It has the potential to push out the EW vacuum from metastability to a stable vacuum.
The benchmark point was chosen keeping in mind that the new physics effects are clearly visible, yet the vacuum is still in metastable state.
This point also satisfies the WMAP and Planck imposed DM relic density constraint $\Omega h^2=0.1198\pm 0.0026$~\cite{Ade:2013zuv}. 
 \begin{figure}[h!]
 \begin{center}
 \subfigure[]{
 \includegraphics[width=2.7in,height=2.7in, angle=0]{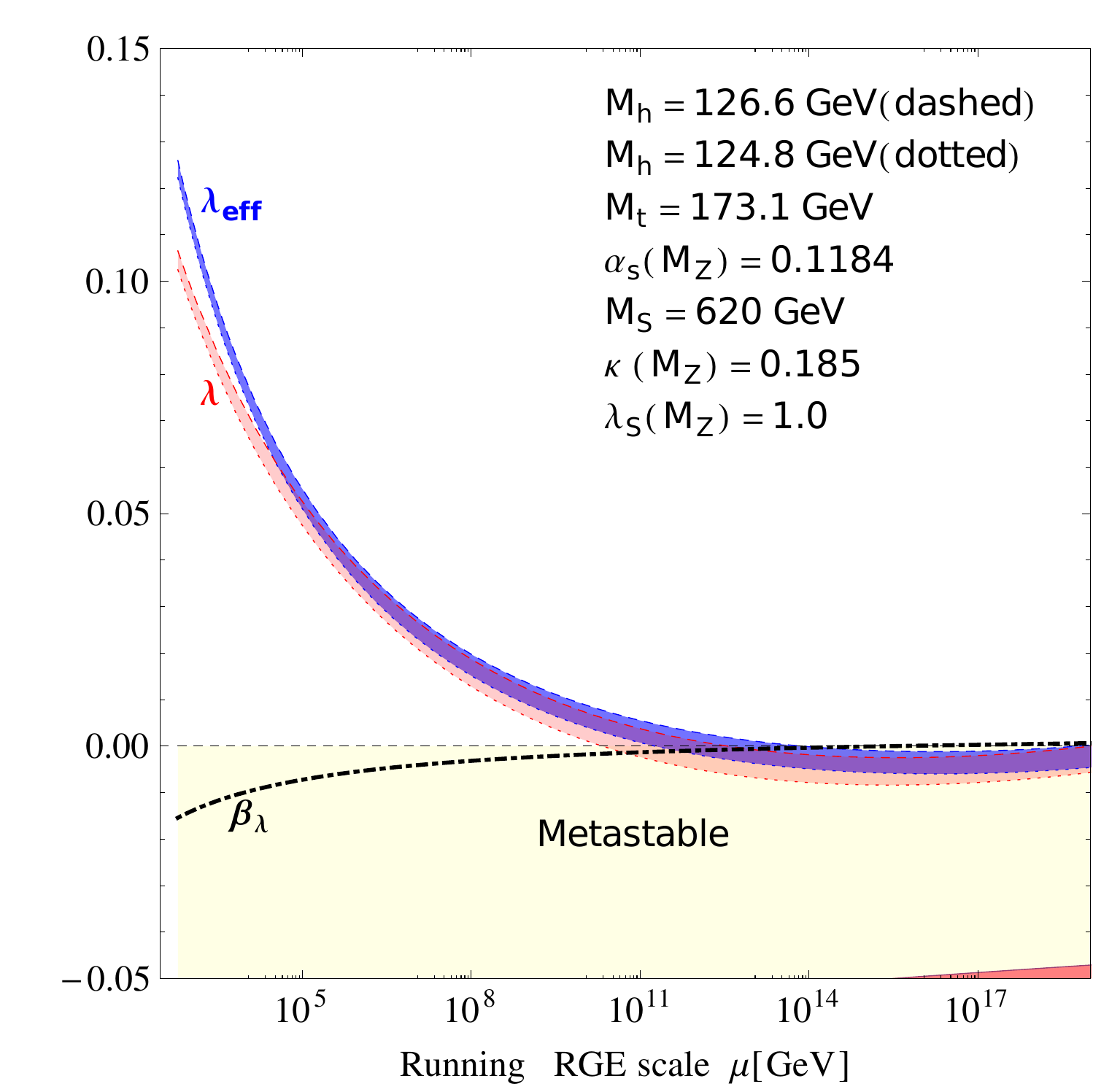}}
 \subfigure[]{
 \includegraphics[width=2.7in,height=2.7in, angle=0]{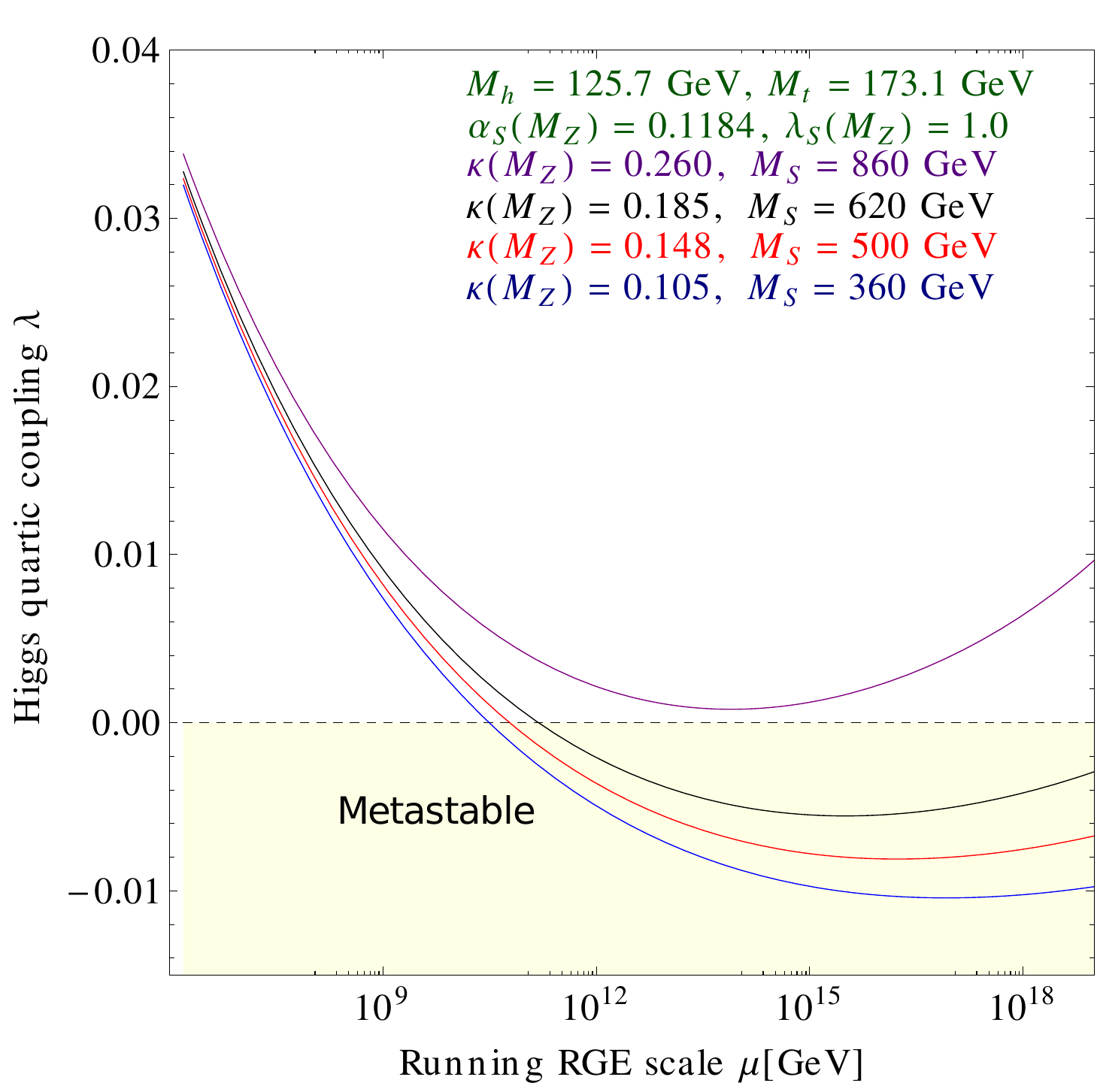}}
 \caption{\label{fig:lambdaef} \textit{$({\bf a})$ RG evolution of $\lambda$ (red band), $\lambda _{\rm eff}$ (blue band)  and  $\beta_\lambda$ (dot-dashed black) in {\rm SM+}$S$ for our benchmark point. $({\bf b})$~Evolution of  $\lambda$ for different $\kappa(M_Z)$. Each $\kappa(M_Z)$ corresponds to a specific $M_S$ to satisfy DM relic density $\Omega h^2 \approx$ 0.1198.} }
 \end{center}
 \end{figure}

As defined in eqn.~\ref{lameffscalar}, $\lambda_{\rm eff}$ differs from $\lambda$ as it takes care of loop corrections. In the SM, in Fig.~\ref{fig:lambdaSM}, one can see that the difference $\lambda_{\rm eff}-\lambda$ is always positive and negligible near $\mpl$. Similar features has also been seen in {\rm SM+}$S$ model (see Fig.~\ref{fig:lambdaef}(a)).
However, the instability scale $\Lambda_I$ changes significantly if $\lambda_{\rm eff}$ instead of $\lambda$ is chosen to work: In SM, the instability scale changes to $1.25 \times 10^{11}$~GeV and in {SM+}$S$, it becomes $1.7\times 10^{12}$~GeV.
Also $\beta_\lambda$ has been plotted to show that at high energies, $\lambda$, $\lambda_{\rm eff}$ and $\beta_\lambda$ all seem to vanish. 

In Fig.~\ref{fig:lambdaef}(b) the RGE running of $\lambda$ for various new physics parameters has been displayed to explicitly demonstrate that as $\kappa(M_Z)$ increases, for a given energy, $\lambda$ assumes a higher value~\cite{Kadastik:2011aa,Chen:2012faa}.
Finally, for some parameter space, $\lambda$ never turns negative, implying stability of the EW vacuum. It happens due to the $\kappa^2/2$ term in $\beta_\lambda$.  Due to this positive contribution, the presence of the scalar never drives EW vacuum towards instability.
Next, the tunneling probability will be calculated to demonstrate stability issues with EW vacuum in this model SM+$S$.
\subsection{Tunneling probability and Metastability in SM+$S$}
\label{chapt3}
The present data on $M_h$ and $M_t$ indicate that the Universe might be residing in a false vacuum in SM, waiting for a quantum tunneling to a true vacuum lying close to the Planck scale. 

The vacuum decay probability ${\cal P}_0$ of EW vacuum at the present epoch is given in eqn.~\ref{prob}. In $S(\Lambda_B)=\frac{8\pi^2}{3|\lambda(\Lambda_B)|}$, the effective Higgs quartic coupling as given in eqn.~\ref{lameffscalar} has been used.
 \begin{figure}[h!]
 \begin{center}
\subfigure[]{
 \includegraphics[width=2.7in,height=2.7in, angle=0]{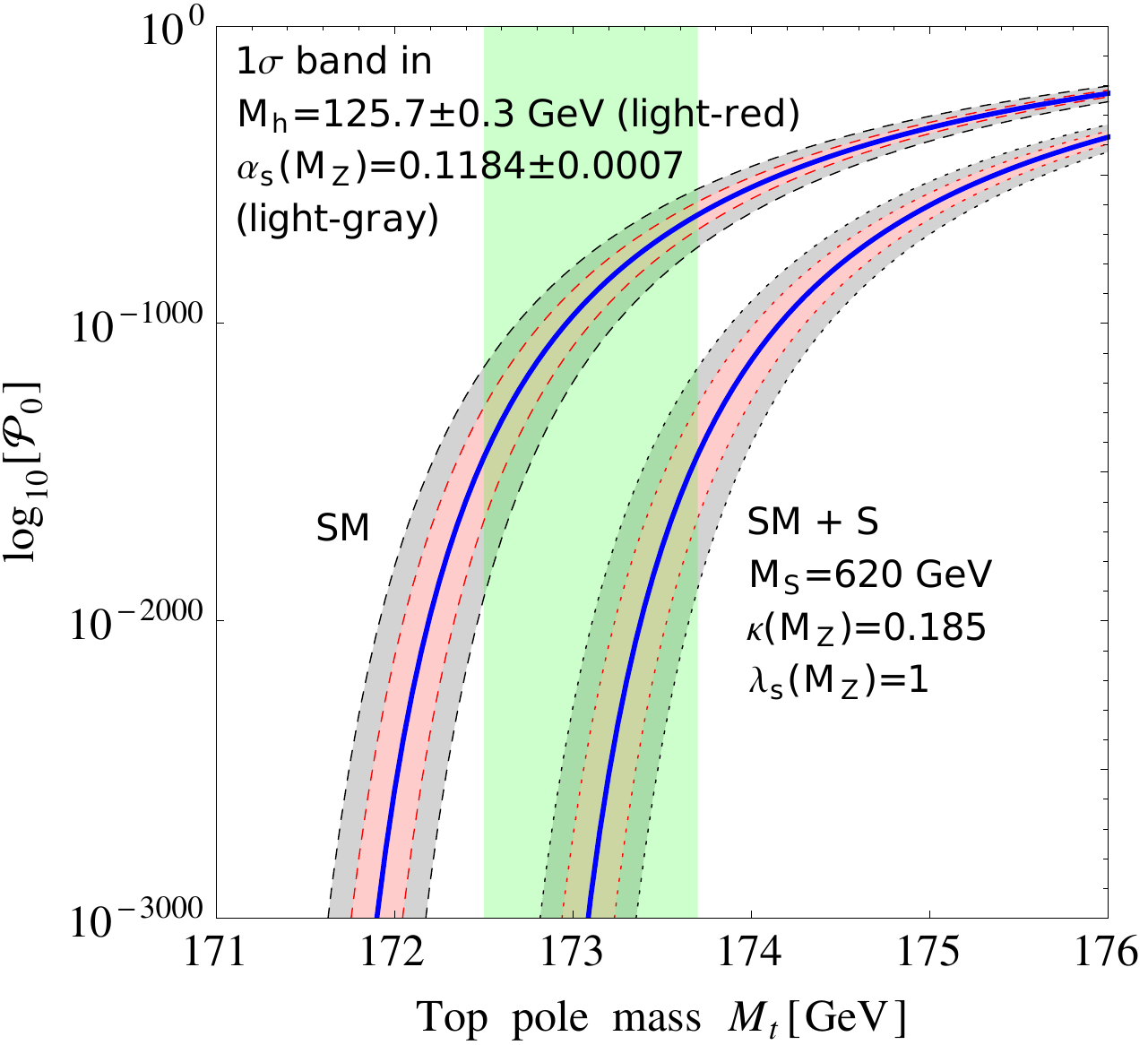}}
 \subfigure[]{
 \includegraphics[width=2.7in,height=2.7in, angle=0]{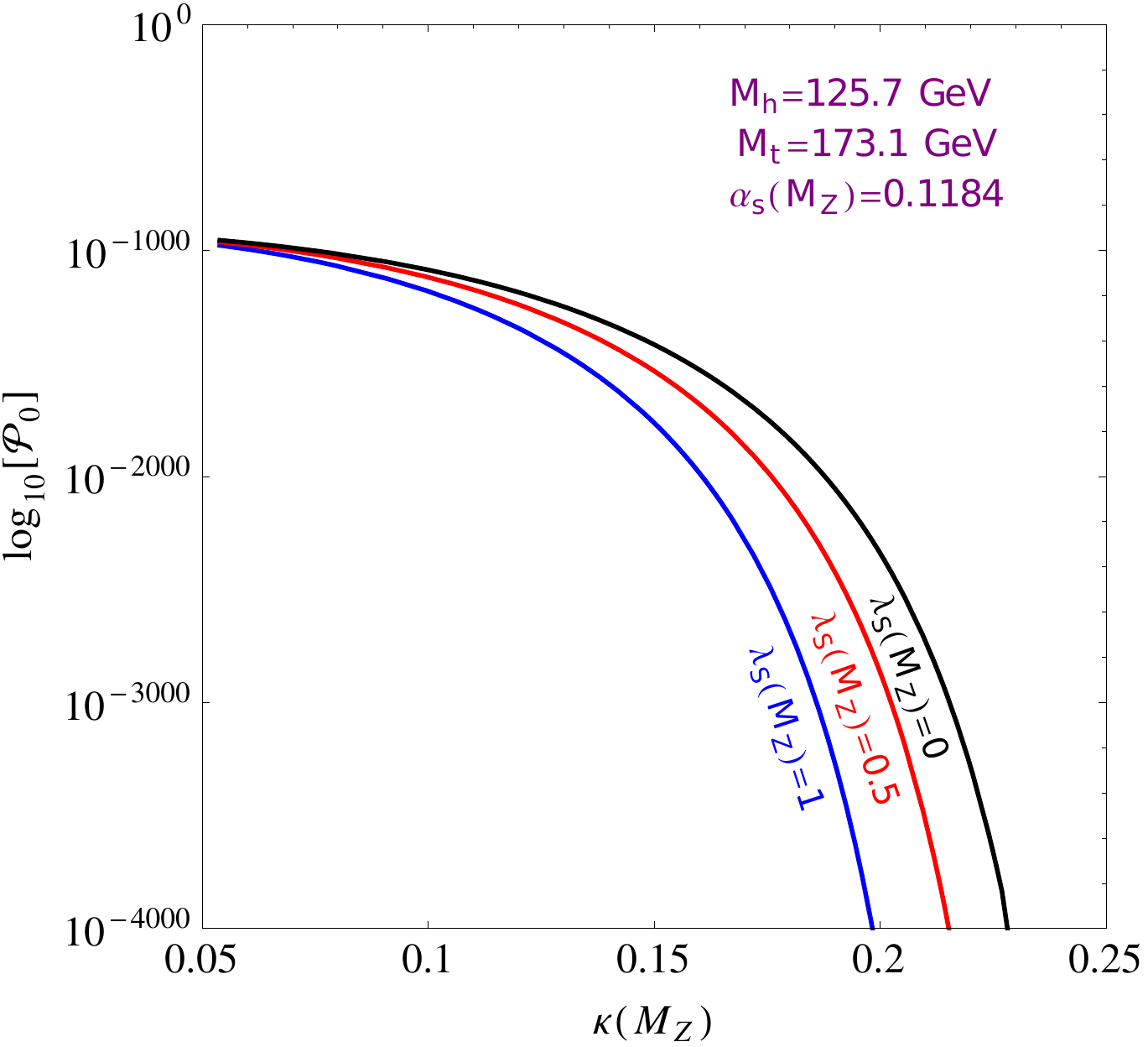}}
 \caption{\label{fig:TunS} \textit{ $({\bf a})$ Tunneling probability ${\cal P}_0$ as a function of  $M_t$. The left band corresponds to {\rm SM} and the right one to {\rm SM+}$S$ for our benchmark point. Light-green band stands for $M_t$ at $\pm 1\sigma$. $({\bf b})$ ${\cal P}_0$ as a function of $\kappa(M_Z)$ for various $\lambda_S(M_Z)$.} }
 \end{center}
 \end{figure}
The metastability and perturvativity constraints in this model are shown here.

\begin{itemize}
\item
If $\lambda(\Lambda_B)>\frac{4\pi}{3}$, $|\kappa|>8\pi$, $|\lambda_S| > 8\pi$, then the theory is nonperturbative. 
\item
If $\lambda(\Lambda_B)>0$, then the vacuum is stable. 
\item
If $0>\lambda(\Lambda_B)>\lambda_{\rm min}(\Lambda_B)$, then the vacuum is metastable. 
\item
If $\lambda(\Lambda_B)<\lambda_{\rm min}(\Lambda_B)$, then the vacuum is unstable. 
\item
If $\lambda_S<0$, the potential is unbounded from below along the $S$-direction. 
\item
If $\kappa<0$, the potential is unbounded from below along a direction in between $S$ and $H$.  
\end{itemize}
In Fig.~\ref{fig:TunS} tunneling probability ${\cal P}_0$ as a function of  $M_t$ has been plotted. To calculate ${\cal P}_0$, the minimum value of $\lambda_{\rm eff}$ (see eqn.~\ref{lameffscalar}) has been found and put the same in $S(\Lambda_B)$. The right band corresponds to the tunneling probability for our benchmark point. 
For comparison, the same plot for SM as the left band in Fig.~\ref{fig:TunS}(a) is shown. 1$\sigma$ error bands in $\alpha_S$ and in $M_h$ are also displayed. The error due to $\alpha_S$ is clearly more significant than that due to $M_h$.
It has been observed that for a given $M_t$, these new physics effects lower the tunneling probability. It bolsters our earlier observation that scalar $S$ helps the EW vacuum to come out of metastability.
It has been demonstrated in Fig.~\ref{fig:TunS}(b), where 
${\cal P}_0$ as a function of $\kappa(M_Z)$ for different choices of $\lambda_S(M_Z)$ has been plotted, assuming central values for $M_h$, $M_t$ and $\alpha_S$. It has been seen that for low values of $\kappa(M_Z)$, ${\cal P}_0$ tends to coincide with its SM value. For a given $\kappa(M_Z)$, for higher $\lambda_S(M_Z)$, ${\cal P}_0$ gets smaller, making the EW vacuum more stable.

\subsection{Phase diagrams in SM+$S$}
\label{chapt4}
The phase diagrams in different parameter planes for this model has been presented to demonstrate the stability of the EW vacuum.
 
 \begin{figure}[h!]
 \begin{center}
 \subfigure[]{
 \includegraphics[width=2.7in,height=2.7in, angle=0]{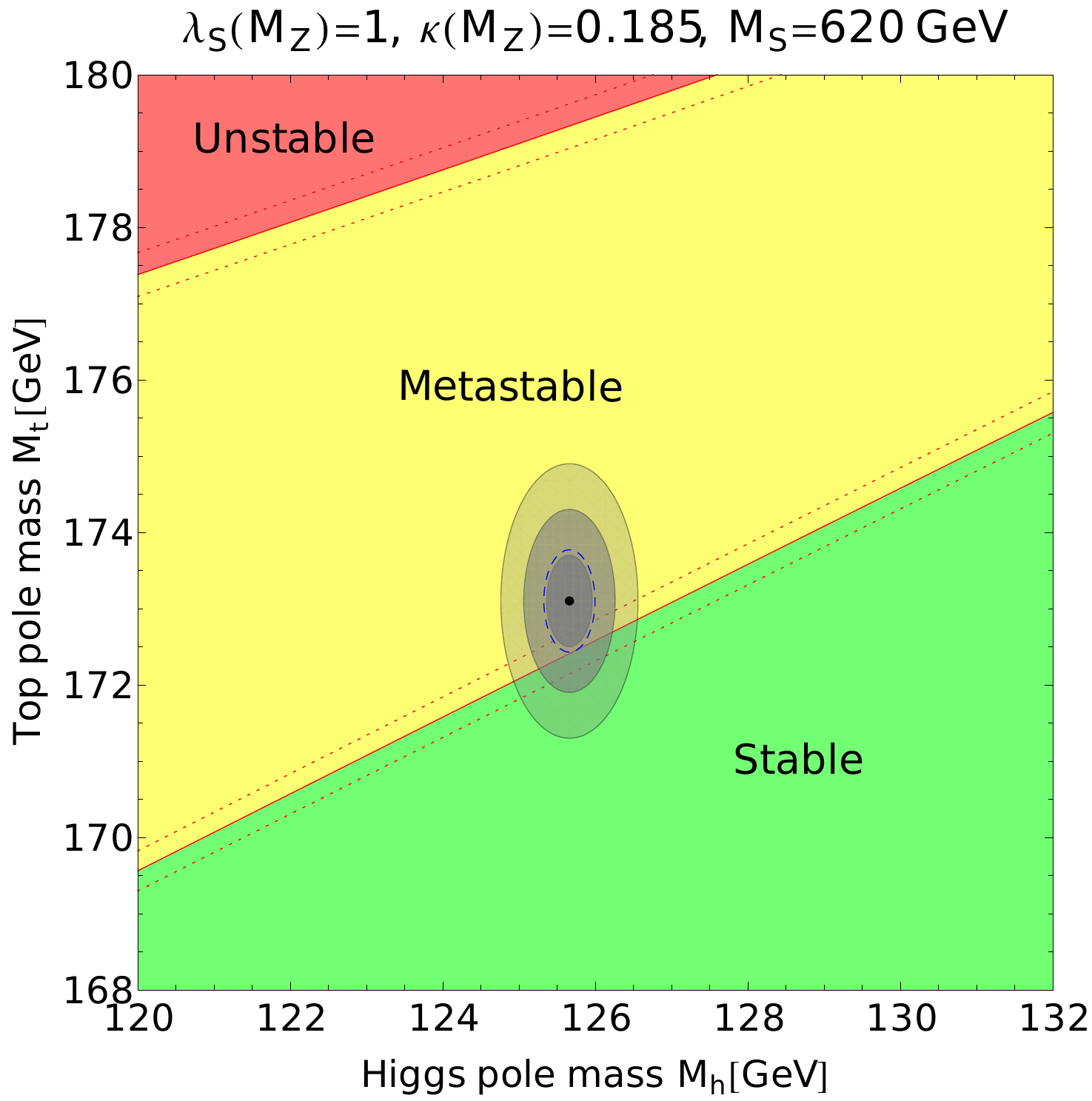}}
 \subfigure[]{
 \includegraphics[width=2.7in,height=2.7in, angle=0]{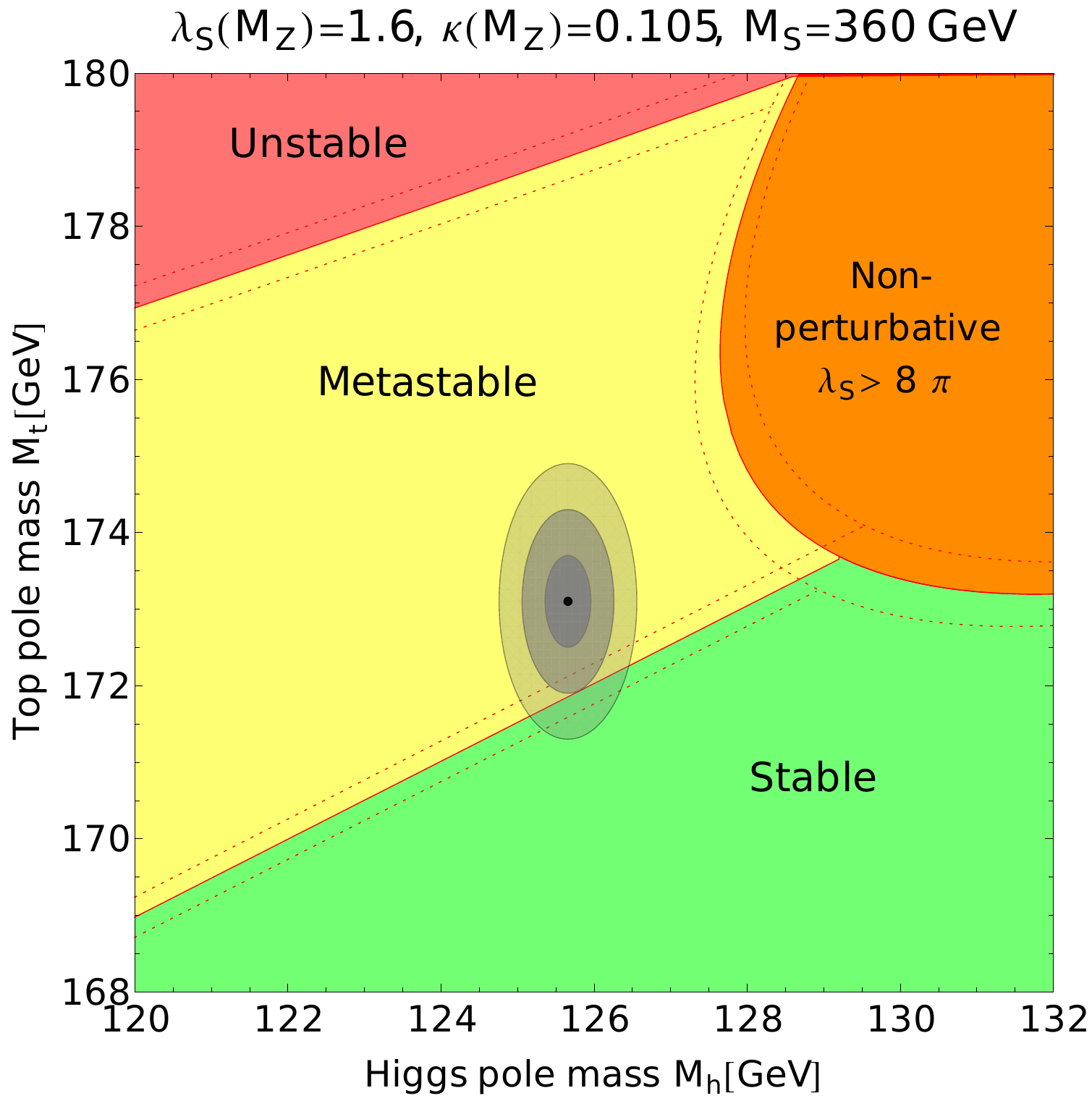}}
 \caption{\label{fig:Mt_MhScalar} \textit{$({\bf a})$ In {\rm SM+}$S$, regions of absolute stability~(green), metastability~(yellow), instability~(red) of the EW vacuum in the $M_h - M_t$ plane phase diagram is presented for the benchmark point $M_S = 620$~GeV, $\kappa(M_Z)=0.185$ and $\lambda_S (M_Z)=1$. 
 $({\bf b})$ Similar plot for $M_S = 360$~GeV, $\kappa(M_Z)= 0.105$ and $\lambda_S (M_Z) = 1.6$. The orange region  corresponds to nonperturbative zone for $\lambda_S$.
The three boundary lines (dotted, solid and dotted red) correspond to $\alpha_s(M_Z)=0.1184 \pm 0.0007$. The gray areas denote the experimentally favored zones for $M_h$ and $M_t$ at $1$, $2$ and  $3\sigma$. } }
 \end{center}
 \end{figure}

 \begin{figure}[h!]
 \begin{center}
 {\vspace{-0.2cm}
 \includegraphics[width=3in,height=2.7in, angle=0]{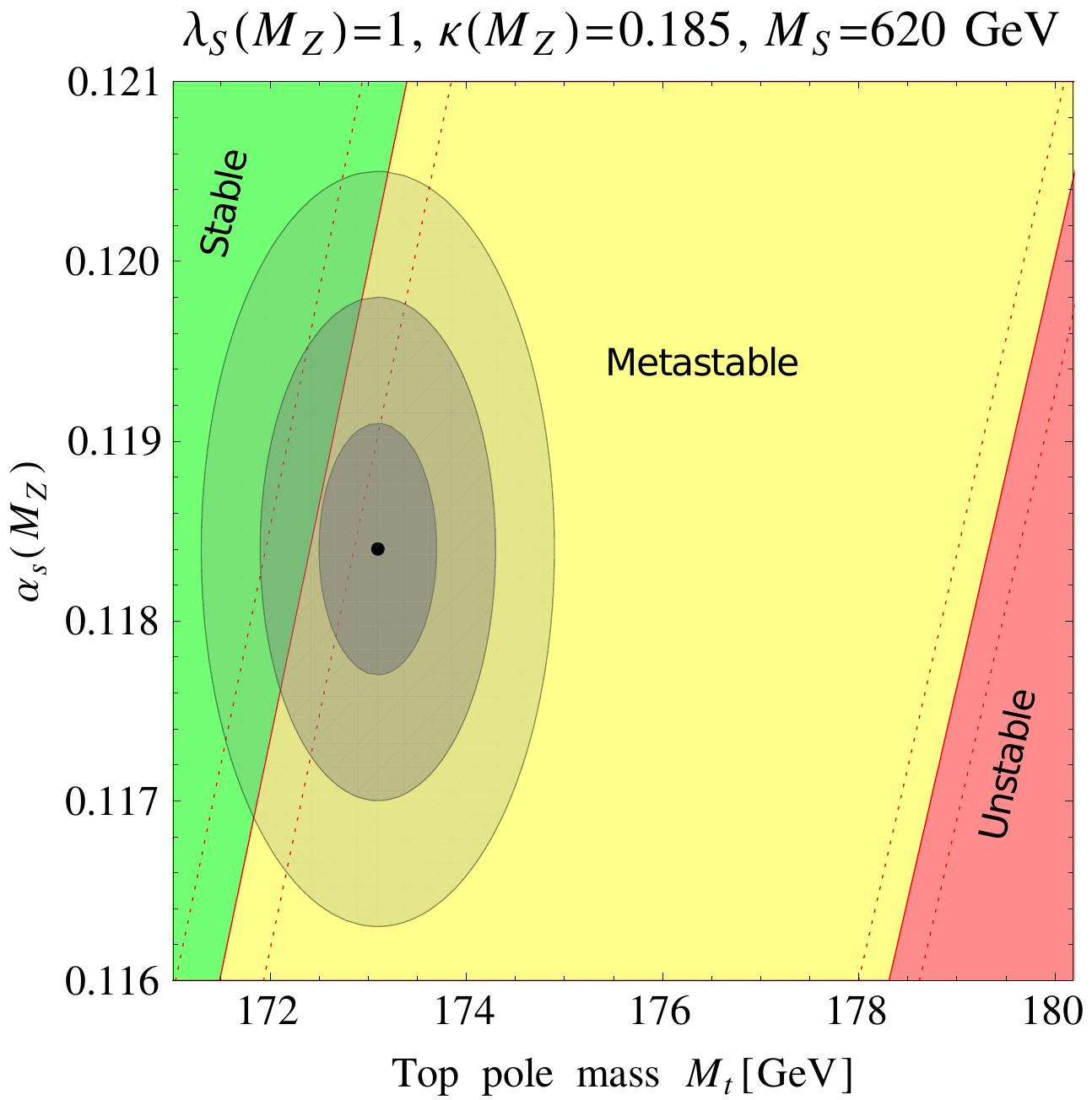}}
 \caption{\label{fig:Alpha_MtScalar} \textit{Phase diagram in $M_t-\alpha_S(M_Z)$ plane in {\rm SM+}$S$ model for our benchmark point. Regions of absolute stability~(green), metastability~(yellow), instability~(red) of the EW vacuum are marked. The dotted lines correspond to $\pm 3\sigma$ variation in $M_h$ and the gray areas denote the experimental allowed region for $M_t$ and $\alpha_S(M_Z)$ at $1$, $2$ and  $3\sigma$.} }
 \end{center}
 \end{figure}
In the SM, the phase diagram in $M_h-M_t$ plane is given in Fig.~\ref{fig:Mt_MhSM}. Given the measured errors on $M_t$ and $M_h$, the SM phase diagram indicates that the stability of EW vacuum is excluded\footnote{In a mass dependent renormalization scheme, such exclusion happens at $3.5\sigma$~\cite{Spencer-Smith:2014woa}.} at $\sim 3\sigma$. However, the extra scalar in this model modifies these findings, as illustrate by
the phase diagram in the $M_h - M_t$ plane for SM+$S$ in Fig.~\ref{fig:Mt_MhScalar}. For our benchmark point, it has been seen that the boundaries shift towards higher values of $M_t$, so that the EW vacuum stability is excluded only at 1.1$\sigma$, indicated by the blue-dashed ellipse.
For  $M_S = 360$~GeV, $\kappa(M_Z)= 0.105$ and $\lambda_S(M_Z) = 1.6$, the plot is redrawn to highlight the fact that $\lambda_S$ might turn out to be too large, so that the theory becomes nonperturbative (marked as the orange region in Fig.~\ref{fig:Mt_MhScalar}(b)).
Here EW vacuum stability is excluded at 2$\sigma$. All these boundaries separating various stability regions in the phase diagram depend on $\alpha_S$. 1$\sigma$ bands for the same is also displayed in these figures.

Given the sizable error on $\alpha_S(M_Z)$, it is instructive to draw the phase diagram in the $M_t-\alpha_S(M_Z)$ plane as well. This diagram for SM is available in Fig.~\ref{fig:alpha-mt}. The same is presented in Fig.~\ref{fig:Alpha_MtScalar} for this model using our benchmark point. With increase of $\kappa(M_Z)$ and/or $\lambda_S(M_Z)$, the boundaries between different stability regions shift towards right, allowing the EW vacuum to be more stable.

 \begin{figure}[h!]
 \begin{center}
 \includegraphics[width=3in,height=2.7in, angle=0]{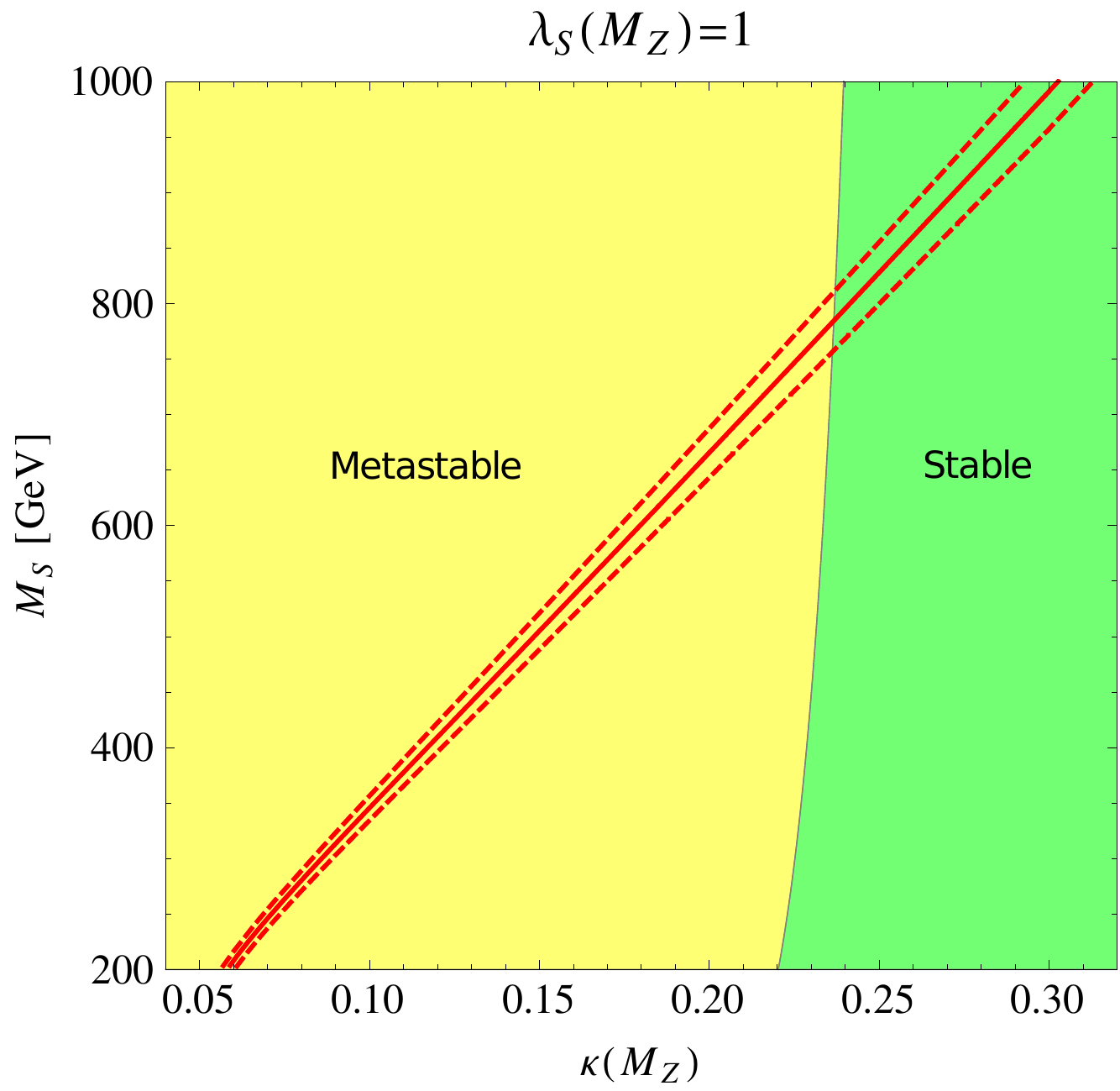}
 \caption{\label{fig:Mskappa} \textit{ Phase diagram in $\kappa(M_Z)-M_S$ plane in {\rm SM+}$S$. Regions of absolute stability (green), metastability (yellow) of the EW vacuum for $\lambda_S (M_Z) = 1$ are shown. The red solid line corresponds to $\Omega h^2 \approx$ 0.1198. The red dashed lines correspond to 3$\sigma$ error on $\Omega h^2$.} }
 \end{center}
 \end{figure}

The phase diagram for $\kappa(M_Z)-M_S$ plane is displayed in Fig.~\ref{fig:Mskappa}. As addition of the scalar does not drive the EW vacuum towards instability, there is no unstable region marked on the plot.
Between the dashed lines, the allowed region is marked ensuing from relic density constraints. 

In SM, the vacuum stability, metastability and perturbativity bound are given in eqns.~\ref{MassMh}$-$ \ref{MassMhpert}.
In SM+$S$ model, change in $M_h$ bounds with respect to $\kappa(M_Z)$ was considered in Refs.~\cite{Gonderinger:2009jp, Eichhorn:2014qka} for different cut-off scales, considering stability aspects only.
As shown in Fig.~\ref{fig:Mhkappa}, in presence of the scalar $S$, these bounds shift to lower values for larger $\kappa(M_Z)$.
For large values of $\kappa(M_Z)$, depending on the choice of $\lambda_S$ at $M_Z$, $\lambda_S(\mpl)$ may become so large that the theory becomes nonperturbative.
This imposes further constraints on the parameter space, shown as the curved line representing $\lambda_S(\mpl)=8\pi$. As before, for a given $\kappa(M_Z)$, $M_S$ is chosen in such a way that  $\Omega h^2\approx 0.1198$ for $M_h=125.7$~GeV. However, in the plot, as $M_h$ changes, $\Omega h^2$ also changes. But this variation is contained within 3$\sigma$. 
 \begin{figure}[h!]
 \begin{center}
 \includegraphics[width=3.3in,height=2.7in, angle=0]{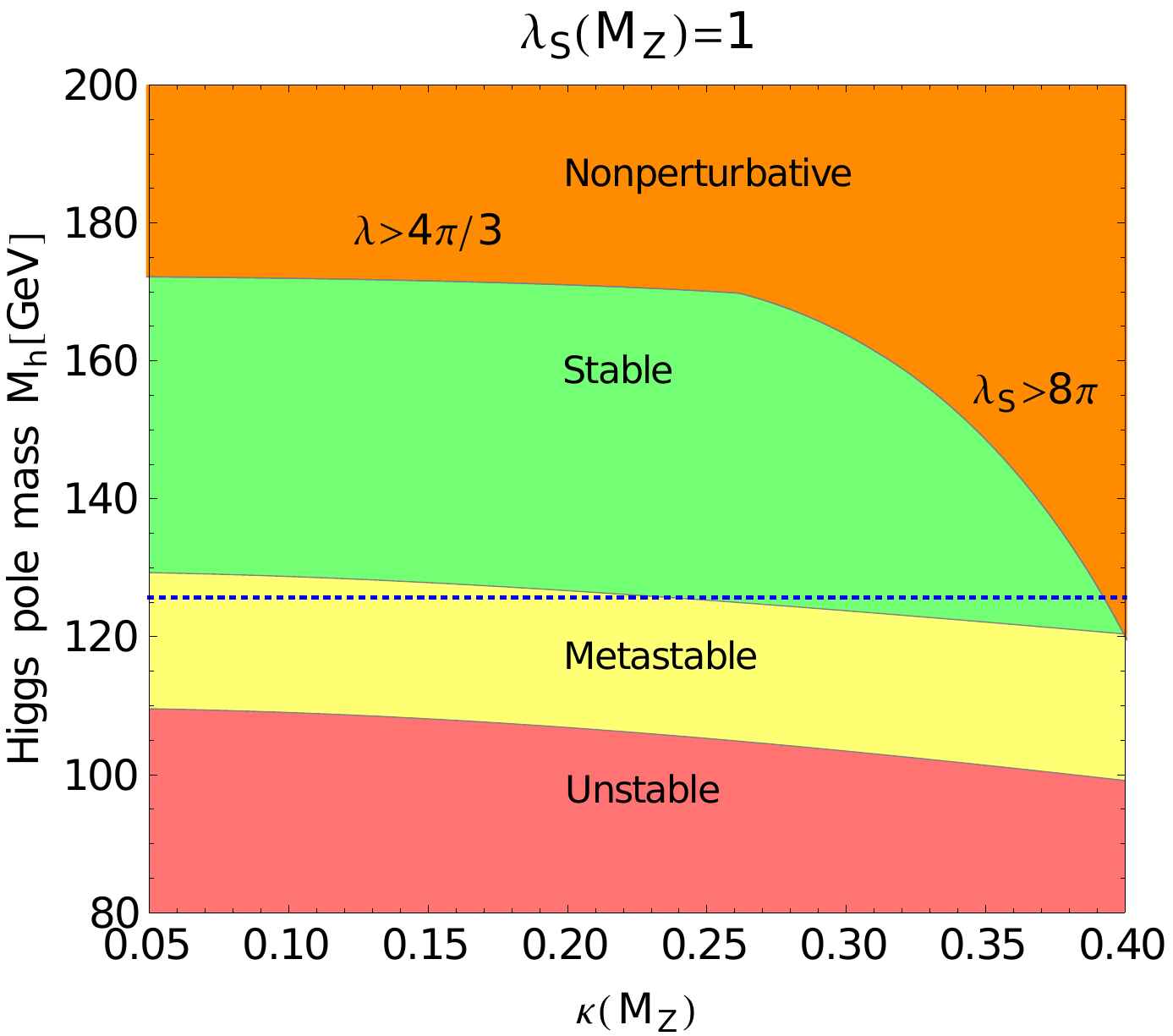}
 \caption{\label{fig:Mhkappa} \textit{ Phase diagram in $\kappa(M_Z)- M_h$ plane in {\rm SM+}$S$. Regions of absolute stability (green), metastability (yellow) and instability (red) of the EW vacuum for $\lambda_S (M_Z) = 1$ are displayed. $\lambda$ and/or $\lambda_S$ are/is nonperturbative in the orange region. The blue dashed line corresponds to $M_h=125.7$~GeV. } }
 \end{center}
 \end{figure}
\subsection{Confidence level of vacuum stability in SM+$S$}
As new physics effects do change the stability of EW vacuum, it is important to show the change in the confidence level at which stability is excluded or allowed.
In Fig.~\ref{fig:kappaconfidence}, the confidence level against $\kappa(M_Z)$ has been plotted for $M_t=173.1$~GeV, $M_h=125.7$~GeV and $\alpha_S(M_Z)=0.1184$. $M_S$ is dictated by $\kappa(M_Z)$ to satisfy $\Omega h^2\approx 0.1198$.
For $\lambda_S(M_Z)=1$, It has been seen that the EW vacuum becomes stable for  $\kappa(M_Z)=0.24$ onward. For a lower $\lambda_S(M_Z)$, this point shifts to a higher value. If $\lambda_S(M_Z)=0$, stability is assured for $\kappa(M_Z)\ge 0.27$. Note that as $\kappa$ dependence in RGE running of $\lambda$ creeps in through the term $\kappa^2/2$ in $\beta_\lambda$, the stability strongly depends on $\kappa(M_Z)$. However, as  $\beta_\lambda$ depends on $\lambda_S$ only {\it via} $\kappa$ running, although $\lambda_S$ running is relatively strong, the stability of EW vacuum does not change appreciably when $\lambda_S(M_Z)$ is varied from $0$ to $1$.

 \begin{figure}[h!]
 \begin{center}
 {
 \includegraphics[width=3in,height=2.7in, angle=0]{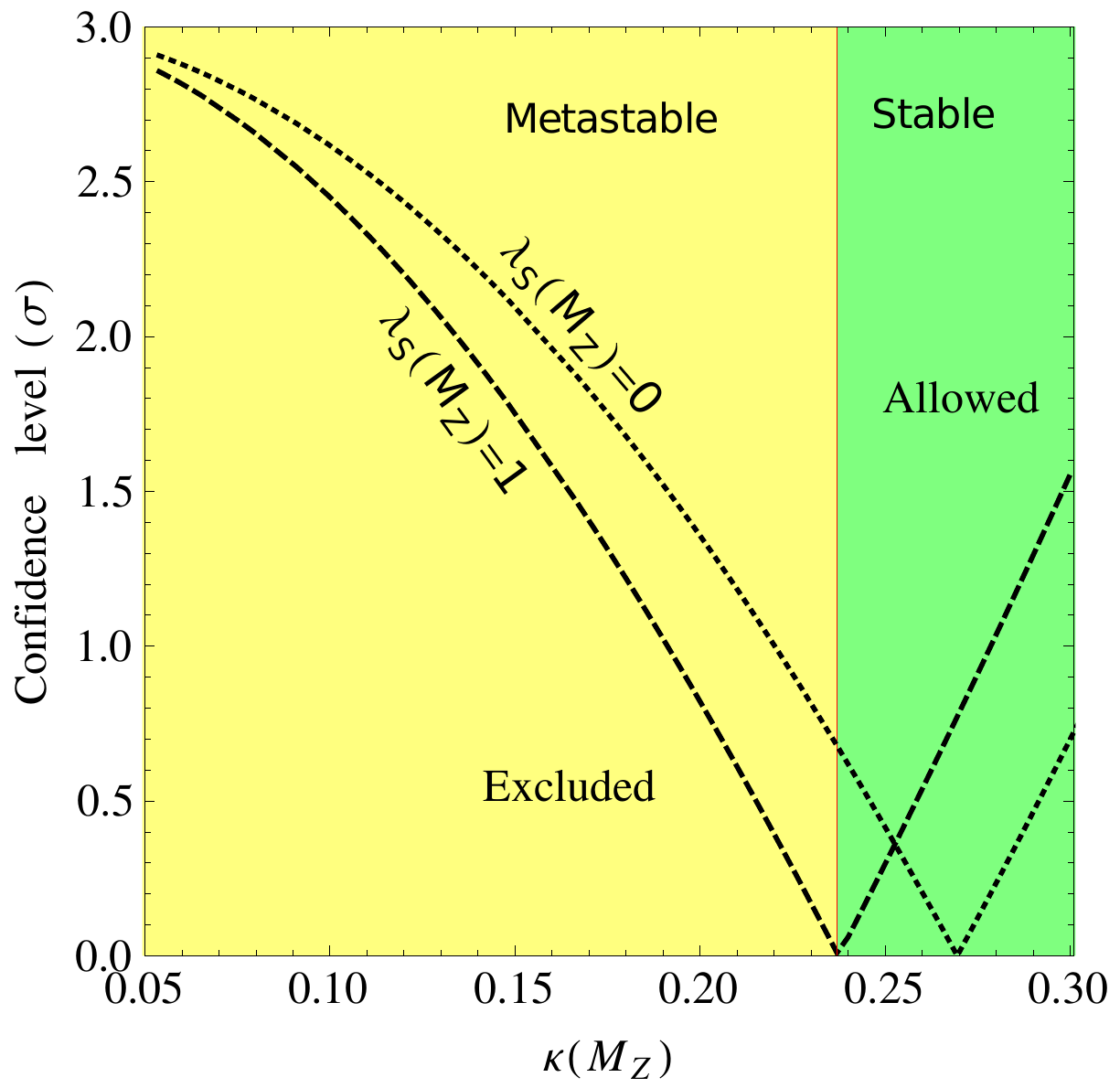}}
 \caption{\label{fig:kappaconfidence} \textit{ Dependence of confidence level (one-sided) at which EW vacuum stability is excluded/allowed on $\kappa(M_Z)$ in {\rm SM+}$S$. Regions of absolute stability (green) and metastability (yellow) of EW vacuum are shown.} }
 \end{center}
 \end{figure}
\subsection{Asymptotic safety in SM+$S$}
 \begin{figure}[h!]
 \begin{center}
 \subfigure[]{
 \includegraphics[width=2.7in,height=2.7in, angle=0]{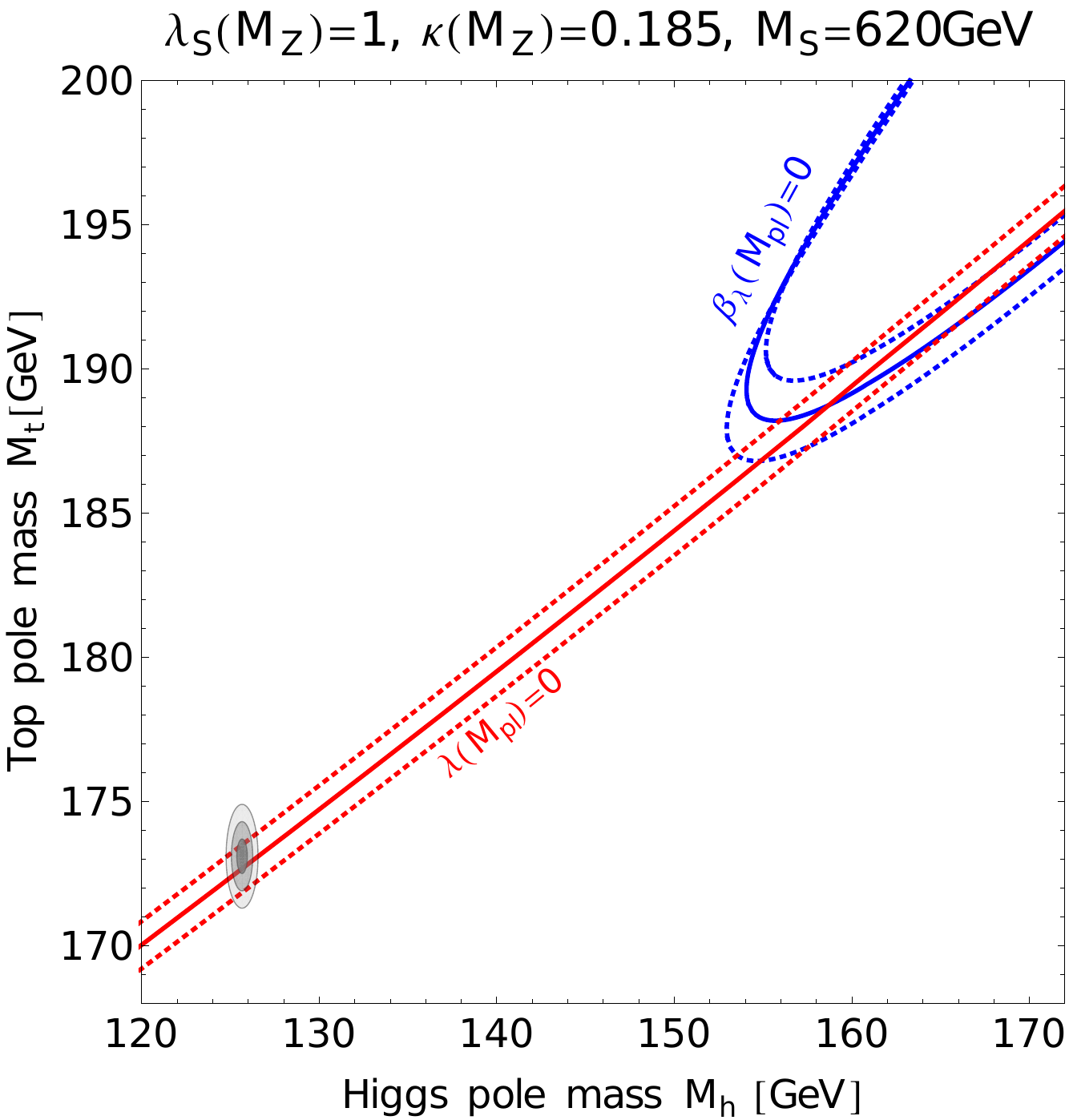}}
 \subfigure[]{
 \includegraphics[width=2.7in,height=2.7in, angle=0]{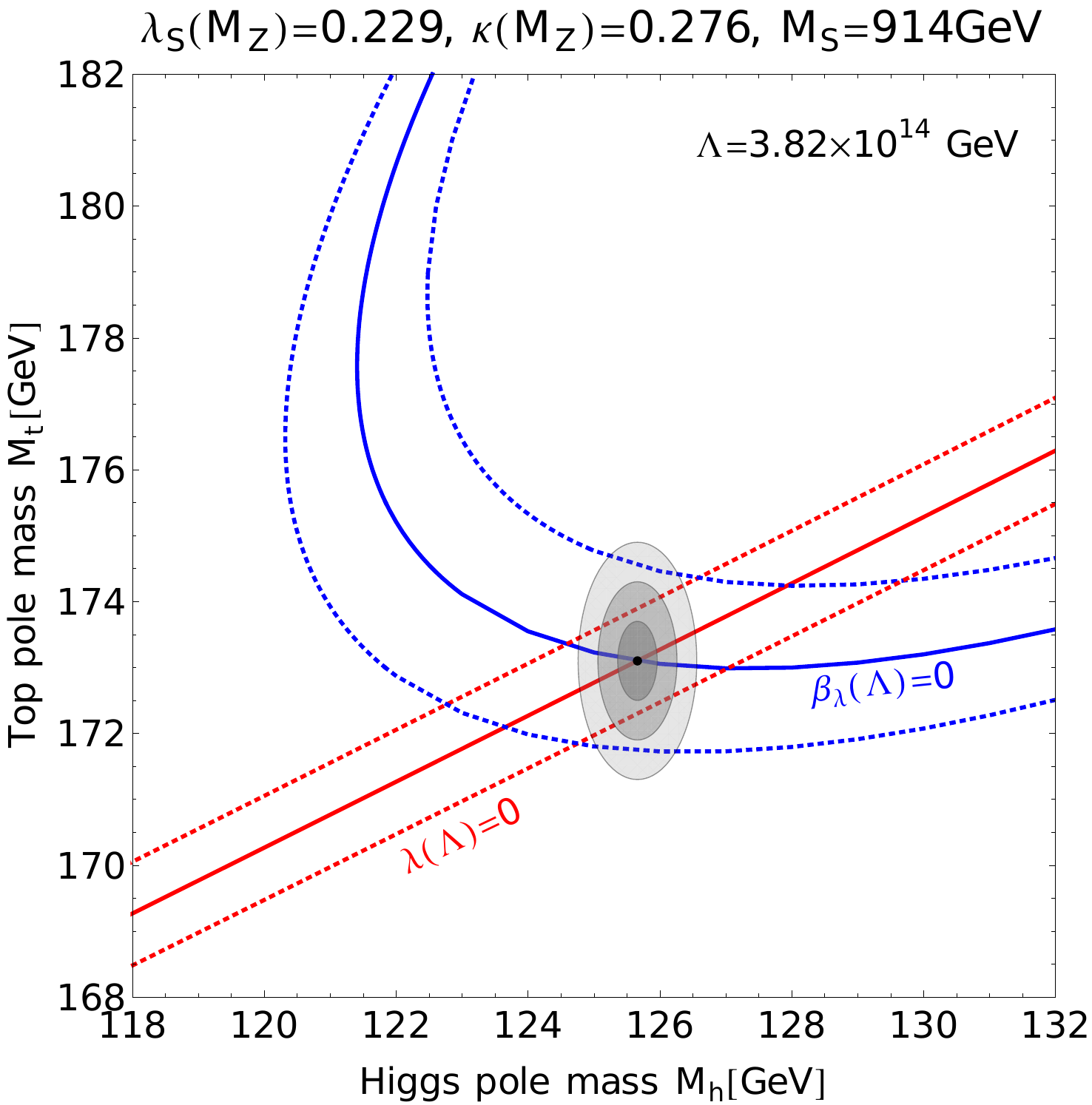}}
 \caption{\label{fig:asymp} In {\rm SM+}$S$ model, \textit{$({\bf a})$ Contour plot for $\lambda(\mpl) =0$ (red line) and  $\beta_{\lambda}(\mpl)=0$ (blue line) for our benchmark point. $({\bf b})$ Similar plot for $\lambda ({\Lambda}) =0$ and  $\beta_{\lambda}(\Lambda)=0$, where $\Lambda=3.8\times10^{14}$~GeV. Dotted lines correspond to $\pm 3\sigma$ variation in $\alpha_s(M_Z)$. The gray areas denote the experimental allowed region for $M_h$ and $M_t$ at $1$, $2$ and  $3\sigma$.} }
 \end{center}
 \end{figure}

Shaposhnikov and Wetterich predicted~\cite{Shaposhnikov:2009pv} mass of the Higgs boson of $126$~GeV imposing the constraint $\lambda(\mpl)=\beta_{\lambda}(\mpl)=0$, in a scenario known as asymptotic safety of gravity. As mentioned before, this corresponds to two degenerate vacua. In Fig.~\ref{fig:lambetampl}, it has been shown that the present error in $M_t$ and $M_h$ does not allow this condition to be realised in SM. In SM+$S$ model, the situation worsens~\cite{Haba:2013lga} and it has  demonstrated in Fig.~\ref{fig:asymp}(a) for our benchmark point. In presence of the scalar, the values of $M_t$ and $M_h$, required to satisfy this condition, are pushed far away from the experimentally favored numbers: For our benchmark point this condition is satisfied at  $M_h=140.8$ GeV and  $M_t=179.5$ GeV.

However, it is possible to meet this condition at a lower energy than at $\mpl$. In Fig.~\ref{fig:asymp}(b), it is demonstrated that at a different point in the parameter space: $M_S = 914$~GeV, $\kappa(M_Z)=0.276$ and $\lambda_S (M_Z)=0.229$, the condition $\lambda ({\Lambda}) =\beta_{\lambda}(\Lambda)=0$ is indeed satisfied at $\Lambda=3.8\times10^{14}$ GeV and is also consistent with experimentally allowed range for $M_t$ and $M_h$. The value of $\Lambda$ decreases with  $\lambda_S (M_Z)$ and $\kappa(M_Z)$. The corresponding value of $M_S$ is chosen to satisfy relic density of DM constraints. 
Also, it is difficult to simultaneously satisfy Veltman condition at $\mpl$~\cite{Haba:2013lga}. All these observations indicate that some new physics could be operational at very high energies  to take care of such issues.

\section{Doublet scalar extension of SM}\label{doubletexten}
The model is popularly known in the literature as the inert doublet~(ID) model, first proposed by Deshpande and Ma~\cite{Deshpande:1977rw}. 

In previous Section~\ref{SingletSMext}, stability analysis, including NNLO corrections, has done for extended singlet scalar DM model. The stability of the EW vacuum was shown to depend on new physics parameters. 
In this section such an analysis has been extended to the ID model. It has been assumed that ID DM is the only DM particle which saturates the entire DM relic density. In this context, the constraints on the parameters of the ID model have been reviewed.

A detailed study on the ID parameter space was performed in Refs.~\cite{Dolle:2009fn, Gustafsson:2012aj,Honorez:2010re,Goudelis:2013uca, Arhrib:2013ela,Chakrabarty:2015yia,Hambye:2007vf,Coleman:1973jx,Chakrabarty:2014aya, Das:2015mwa} indicating bounds from absolute EW vacuum stability, perturbativity, collider study, EW precision tests (EWPT), etc.
In this work~\cite{Khan:2015ipa}, the new parameter spaces have been found allowing the metastability of the Higgs potential.

\subsection{Inert Doublet Model}
\label{sec:IDM}
In this model, the standard model is extended by adding an extra ${ SU(2)_L}$ doublet scalar, odd under an additional discrete $Z_2$ symmetry. The $Z_2$ symmetry prohibits the inert doublet to acquire a vacuum expectation value.

The scalar potential of eqn.~\ref{pot} at the tree level with exact $Z_2$ symmetry, is given\footnote{Here the notation of Higgs quartic coupling $\lambda_1$ is used instead of $\lambda$.} by,
\begp
\allowdisplaybreaks \bea
V(\Phi_1,\Phi_2) &=& m_{11}^2 |\Phi_1|^2 + \lambda_1 |\Phi_1|^4+ m_{22}^2 |\Phi_2|^2 + \lambda_2 |\Phi_2|^4\nn\\
&&+ \lambda_3 |\Phi_1|^2 |\Phi_2|^2 
+  \lambda_4 |\Phi_1^\dagger \Phi_2|^2 + \frac{\lambda_5}{2} \left[ (\Phi_1^\dagger \Phi_2)^2 + { h.c.}\right]  \, ,
\label{Scalarpot}
\eea
\eegp
where the SM Higgs doublet $\Phi_1$ and the inert doublet $\Phi_2$ are given by
\begp
\allowdisplaybreaks \beq
	\Phi_1 ~=~ \left( \begin{array}{c} G^+ \\ \frac{1}{\sqrt{2}}\left(v+h+i G^0\right) \end{array} \right),
	\qquad
	\Phi_2 ~=~ \left( \begin{array}{c} H^+\\ \frac{1}{\sqrt{2}}\left(H+i A\right) \end{array} \right) \, \nn.
\eeq
\eegp
$\Phi_2$ contains a $CP$ even neutral scalar $H$, a $CP$ odd neutral scalar $A$, and a pair of charged scalar fields $H^\pm$. The $Z_2$ symmetry prohibits these particles to decay entirely to SM particles. The lightest of $H$ and $A$ can then serve as a DM candidate. 

After EW symmetry breaking, the scalar potential is given by
\begp
\allowdisplaybreaks \bea
V(h, H,A,H^\pm) &=&  \frac{1}{4} \left[ 2 m_{11}^2 (h+v)^2 + \lambda_1 (h+v)^4 +2 m_{22}^2 (A^2+H^2+2 H^+ H^-) \right. \nn \\
&& \left. + \lambda_2 (A^2 + H^2 + 2 H^+ H^-)^2  \right] \nn \\
&& + \frac{1}{2} (h+v)^2 \left[  \lambda_3 H^+ H^- 
+  \lambda_S  A^2  
+  \lambda_L  H^2 \right] \label{Scalarpot2}
\eea
\eegp
where,
\begp
\allowdisplaybreaks \bea
\lambda_{L,S}&=&\frac{1}{2}\left(\lambda_3+\lambda_4\pm\lambda_5\right) \, .
\eea
\eegp
Masses of these scalars are given by,
\begp
\allowdisplaybreaks \begin{align}
	M_{h}^2 &= m_{11}^2 + 3 \lambda_1 v^2,\nn \\
	M_{H}^2 &= m_{22}^2 +  \lambda_L v^2, \nn \\
	M_{A}^2 &= m_{22}^2 + \lambda_S v^2,\nn \\
	M_{H^\pm}^2 &= m_{22}^2 + \frac{1}{2} \lambda_3 v^2  \,\nn .
\end{align}
\eegp
For  $\lambda_4-\lambda_5<0$ and $\lambda_5>0$~($\lambda_4+\lambda_5<0$ and $\lambda_5<0$), 
$A$~($H$) is the lightest $Z_2$ odd particle (LOP).
In this work, $A$ has been taken as the LOP and hence, as a viable DM candidate.
Choice of $H$ as LOP will lead to similar results. 

For large DM mass, $M_A\gg M_Z$, appropriate relic density of DM is obtained if $M_A$, $M_H$, and $M_{H^\pm}$ are nearly degenerate that will be explained latter. Hence, in anticipation, one can define,
\begp
\allowdisplaybreaks \bea
	\Delta M_H&=& M_H-M_A, \nn\\
	\Delta M_{H^\pm}&=&M_{H^\pm}-M_A \nn\, .
\eea
\eegp
so that the new independent parameters for the ID model become $\{M_A, \Delta M_H, \Delta M_{H^\pm},\\ \lambda_2, \lambda_S\}$. Here $\lambda_S$ is chosen as $A$ is treated as the DM particle.

The one-loop effective potential for $h$ in the $\MS$ scheme and the Landau gauge is given by   
\begp
\allowdisplaybreaks \beq
V_1^{{\rm SM}+{\rm ID}}(h)= V_1^{\rm SM}(h) + V_1^{\rm ID}(h)
\eeq
\eegp
The SM one-loop effective Higgs potential $V_1^{\rm SM}(h)$ can be found in eqn.~\ref{V1loop} and the additional contribution to the one-loop effective potential due to the inert doublet is given by~\cite{Hambye:2007vf}
\begp
\allowdisplaybreaks \beq
V_1^{\rm ID}(h)= \sum_{j=H,A,H^+,H^-} \frac{1}{64 \pi^2} M_j^4(h) \left[ \ln\left(\frac{M_j^2(h)}{\mu^2(t)} \right)- \frac{3}{2} \right] 
\eeq
\eegp
where, 
\begp
\allowdisplaybreaks \beq
M_j^2(h)=\frac{1}{2} \,\lambda_{j}(t) \, h^2(t)+m_{22}^2(t) 
\eeq
\eegp
with $\lambda_{A}(t)=2 \lambda_{S}(t)$, $\lambda_{H}(t)=2 \lambda_{L}(t)$, and $\lambda_{H^\pm}(t)= \lambda_{3}(t)$.

In the present work, in the Higgs effective potential, SM contributions are taken at the two-loop level, whereas the ID scalar contributions are considered at one loop only. 

For $h \gg v$, the Higgs effective potential can be approximated as
\begp
\allowdisplaybreaks \beq
V_{\rm eff}^{{\rm SM}+{\rm ID}}(h) \simeq \lambda_{\rm 1,eff}(h) \frac{h^4}{4}\, ,
\label{efflamID}
\eeq\eegp
with
\begp
\allowdisplaybreaks \beq
\lambda_{\rm 1,eff}(h) = \lambda_{\rm 1,eff}^{\rm SM}(h) +\lambda_{\rm 1,eff}^{\rm ID}(h)\, ,
\label{lameffidm}
\eeq
\eegp
where $\lambda_{\rm 1,eff}^{\rm SM}$ can be found in eqn.~\ref{eq:effqurtic} and,
\begp
\allowdisplaybreaks \bea
 \lambda_{\rm 1,eff}^{\rm ID}(h)&=&\sum_{j=L,S,3} e^{4\Gamma(h)} \left[\frac{\delta_j \lambda_j^2}{64 \pi^2}  \left(\ln\left(\delta_j\lambda_j\right)-\frac{3}{2}\right ) \right]\, . 
 \label{efflamdoublet}
\eea
\eegp
Here $\delta_j=1$ when $j=L,S$;  $\delta_j=\frac{1}{2}$ for $j=3$; and anomalous dimension $\gamma(\mu)$ of the Higgs field takes care of  its wave function renormalization (see the Appendix).
As quartic scalar interactions do not contribute to wave function renormalization at the one-loop level, ID does not alter  $\gamma(\mu)$ of the SM. All running coupling constants are evaluated at $\mu=h$.

If DM mass $M_A$ is larger than $M_t$, then the ID starts to contribute after the energy scale $M_A$. For  $M_A<M_t$, the contributions of ID to the $\beta$-functions (see eqns.~\ref{betal_1}$-$ \ref{betal_5}) are rather negligible for the running from $M_A$ to $M_t$, as is evident from the expressions.

To compute the RG evolution of all the couplings, all the couplings with threshold corrections at $M_t$ has been calculated. In Table~\ref{table1IDM} a specific set of values of $\lambda_i$ at $M_t=173.1$~GeV and at $\mpl = 1.2 \times 10^{19}$~GeV for $M_h=125.7$~GeV and $\alpha_s\left(M_Z\right)=0.1184$ have been provided. In Fig.~\ref{fig:SMIDM} the running of the scalar couplings $(\lambda_i)$ has been shown for this set of parameters. It has been seen that for this specific choice of parameters, $\lambda_1$ assumes a small negative value leading to a metastable EW vacuum as discussed in the following sections. This set is chosen to reproduce the DM relic density in the right ballpark.
\setlength\tablinesep{2pt}
\begin{table}[h!]
\begin{center}
    \begin{tabular}{| c | c | c | c | c | c | c | c |}
    \hline
     &$\lambda_S$ & $\lambda_L$ & $\lambda_2$ & $\lambda_3$ & $\lambda_4$ & $\lambda_5$ & $\lambda_1$\\
\hline
   ~$M_{t}$ ~&~ 0.001 ~&~ 0.039 ~&~ 0.10 ~&~ 0.0399 ~&~ 0.00003   ~&~ 0.038 ~&~ 0.127~\\
            \hline
   $\mpl$ & 0.046 & 0.082 & 0.127 & 0.090 & 0.038   & 0.036 & $-0.009$\\
              \hline
    \end{tabular}
    \caption{A set of values of all ID model coupling constants at  $M_t$ and $\mpl$ for $M_{A}=573$ GeV, $\Delta M_{H^\pm}=1$ GeV, $\Delta M_{H}=2$ GeV, and $\lambda_S\left(M_Z\right)=0.001$.}
    \label{table1IDM}
\end{center}
\end{table}
\subsection{Constraints on ID model}
\label{sec:constraintsID}
ID model parameter space is constrained from theoretical considerations like  absolute vacuum stability, perturbativity and unitarity of the scattering matrix.  EW precision measurements and direct search limits at LEP, LHC put severe restrictions on the model. The recent measurements of Higgs decay width at the LHC put additional constraints. The requirement that the ID DM saturates the DM relic density all alone restricts the allowed parameter space considerably. The tree level vacuum stability bound of the ID model can be found in eqn.~\ref{stabilitybound}. The unitary (eqn.~\ref{unitary}) and perterbativity (eqn.~\ref{pertIDM}) bounds up to the  $\mpl$ have also been taken into account.
 \begin{figure}[h!]
 \begin{center}
 \includegraphics[width=3.1in,height=2.7in, angle=0]{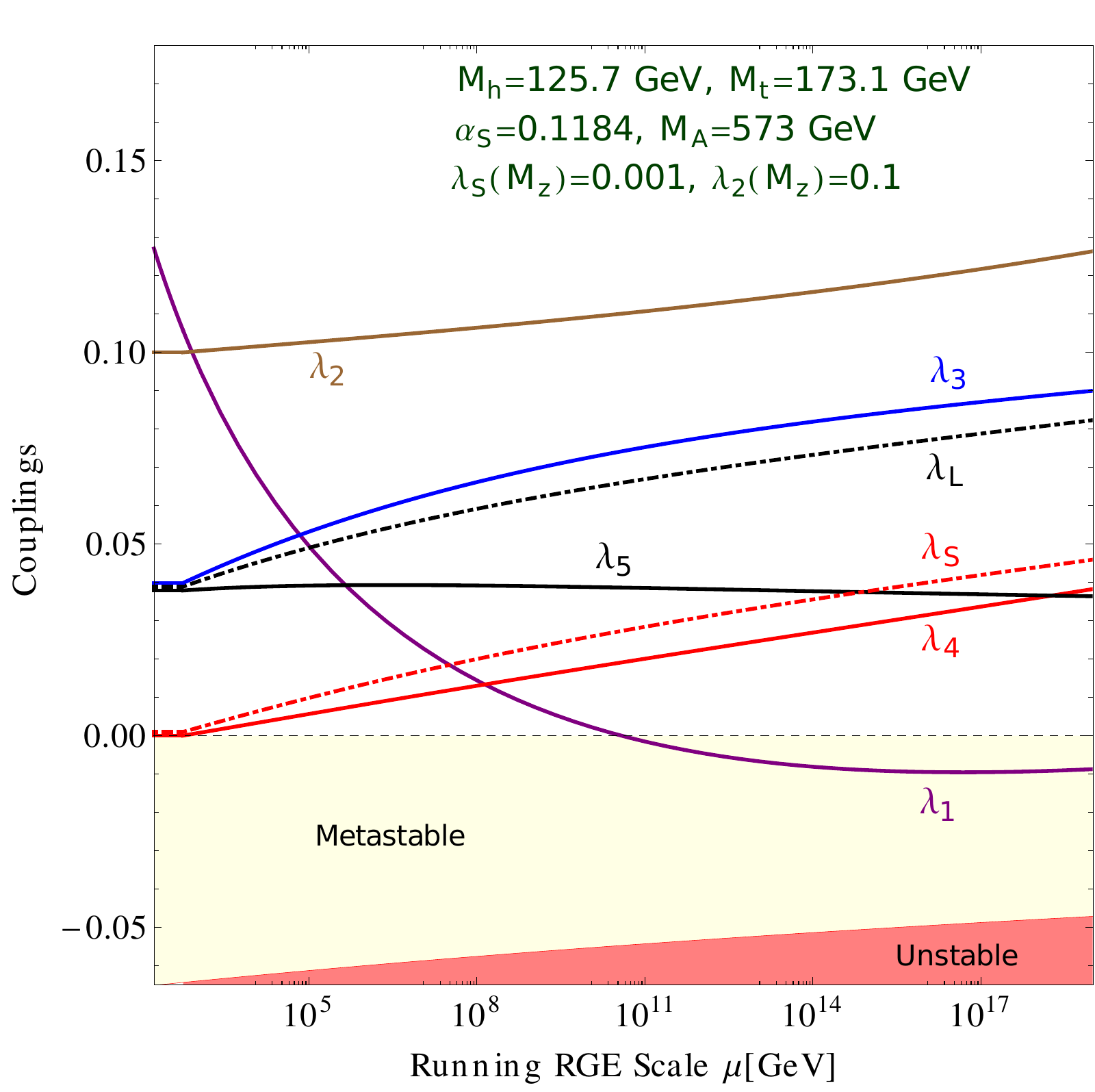}
 \caption{\label{fig:SMIDM} \textit{ {\rm IDM} RG evolution of the couplings $\lambda_i\left(i=1,..,5\right), \lambda_L, \lambda_S$ for the set of parameters in Table~\ref{table1IDM}. } }
 \end{center}
 \end{figure}
\subsubsection{Bounds from electroweak precision experiments}
Bounds ensuing from electroweak precision experiments are imposed on new physics models {\it via} Peskin-Takeuchi~\cite{Peskin:1991sw} $S,~ T,~ U$ parameters. The additional contributions from 2HDM can be found in eqn.~\ref{STU2HDM}, which are modified for ID model~\cite{Barbieri:2006dq,Arhrib:2012ia} as,
\begp
\allowdisplaybreaks \bea
 \Delta S &=& \frac{1}{2\pi}\Bigg[ \frac{1}{6}\ln\left(\frac{M^2_{H}}{M^2_{H^\pm}}\right) -
  \frac{5}{36} + \frac{M^2_{H} M^2_{A}}{3(M^2_{A}-M^2_{H})^2} + 
\frac{M^4_{A} (M^2_{A}-3M^2_{H})}{6(M^2_{A}-M^2_{H})^3} \ln \left(\frac{M^2_{A}}{M^2_{H}}\right)\Bigg], \nn\\
 \Delta T &=& \frac{1}{32\pi^2 \alpha v^2}\Bigg[ F\left(M^2_{H^\pm}, M^2_{A}\right)
+ F\left(M^2_{H^\pm}, M^2_{H}\right) - F\left(M^2_{A}, M^2_{H}\right)\Bigg]
\label{STparam}
\eea
\eegp
The NNLO global electroweak fit results of $\Delta S$, $\Delta T$ and $\Delta U$ are given eqn.~\ref{STU1} have been used, with a correlation coefficient of $+0.91$, fixing $\Delta U$ to zero. The contribution of the scalars in the ID model to $\Delta U$ is rather negligible. 
 \begin{figure}[h!]
 \begin{center}
 \includegraphics[width=2.7in,height=2.7in, angle=0]{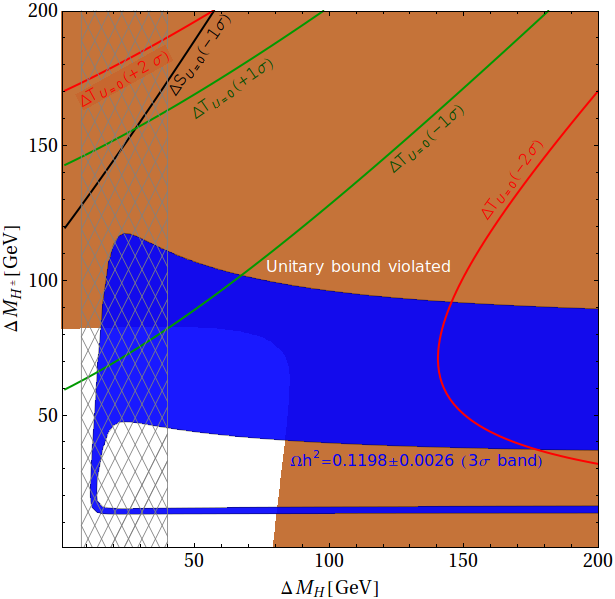}
  \caption{\label{fig:STUcheck} \textit{Allowed parameter space in $\Delta M_{H^\pm}-\Delta M_{H}$ plane for $M_A=70$~GeV and $\lambda_S=0.007$ in {\rm IDM}. Constraints from $S$ and $T$ parameters are shown by solid black, green and red lines. The blue band corresponds the 3$\sigma$ variation in $\Omega h^2=0.1198\pm0.0026$~\cite{Ade:2013zuv}. On the brown region the unitarity bound is violated on or before $\mpl$. The cross-hatched region is excluded from LEP\,II data. }} 
 \end{center}
 \end{figure} 
To assess the implications of $S$ and $T$ constraints on the ID model, in Fig.~\ref{fig:STUcheck}, in the $\Delta M_{H^\pm}-\Delta M_{H}$ plane, various constraints is displayed for $M_A=70$~GeV and $\lambda_S=0.007$.
This is the maximum value of $\lambda_S$ for the given DM mass, allowed by LUX~\cite{Akerib:2013tjd} direct detection data at 1$\sigma$.  The blue region allowed by relic density constraints shifts upwards for smaller $\lambda_S$.
Constraints on $\Delta S$ and $\Delta T$, as mentioned in eqn.~\ref{STU1}, are marked as black, green, and red  solid lines.
It has been seen that the $1\sigma$ bound on $\Delta T$ is the most stringent one. The black line corresponding to the lower limit on $\Delta S$ at $1\sigma$ can also be interesting. But the LEP\,II bounds, represented by the cross-hatched bar, take away a considerable part.
The line representing the $\Delta S$ upper limit is beyond the region considered and lies towards the bottom-right corner of the plot.
In this plot, increasing $M_H$ enhances $S$, but $T$ gets reduced.
In addition, if one demand unitarity constraints to be respected up to $\mpl$, assuming no other new physics shows up in between, the parameter space gets severely restricted.
In this plot, a small window is allowed by $\Delta T$ only at $2\sigma$, which satisfies DM relic density constraints. As mentioned earlier, this window gets further reduced with smaller $\lambda_S$ (the relic density band takes an ``L'' shape as shown in Fig.~\ref{fig:STU2}). 
\subsubsection{Direct search limits from LEP and LHC}
The decays $Z\ra AH$, $Z\ra H^+ H^-$, $W^\pm\ra A H^\pm$, and $W^\pm\ra H H^\pm$ are restricted from $Z$ and $W^\pm$ decay widths at LEP. It implies $M_{A} + M_{H} \geq M_{Z}$, $2M_{H^\pm}\geq M_{Z}$, and $M_{H^\pm} + M_{H,A} \geq M_{W}$. More constraints on the ID model can be extracted from chargino~\cite{Pierce:2007ut} and neutralino~\cite{Lundstrom:2008ai} search results at LEP\,II: The charged Higgs mass $M_{H^\pm}\geq 70$~GeV. The bound on $M_A$ is rather involved: If  $M_A<80$ GeV, then $\Delta M_H$ should be less than $\sim 8$ GeV, or else $M_H$ should be greater than $\sim 110$~GeV ({see Fig.~\ref{fig:STU1}}).
The Run 1 of LHC data provide significant constraints on the ID parameter space through the direct searches in final states with two leptons plus missing transverse energy~\cite{Belanger:2015kga}.
These analyses exclude inert scalar masses of up to about 55 GeV. For $M_{H^\pm}=85$ GeV and $M_{A}=55$ GeV, the $M_{H}\lesssim 145$ GeV are excluded at 4$\sigma$ and for $M_{A}>60$ GeV all masses of $H$ are allowed.
For $M_{H^\pm}>150$ GeV with $M_{A}=55$ GeV, $115\lesssim M_{H} \lesssim 160$ GeV region is excluded and $M_{A}>80$ GeV all masses of $H$ are allowed.
\subsubsection{Bounds from LHC diphoton signal strength}
In the ID model, Higgs to diphoton signal strength $\mu_{\gamma\gamma}$ is defined as 
\begp
\allowdisplaybreaks \beq
\mu_{\gamma\gamma} = \frac{\sigma(gg\ra h\ra\gamma\gamma)}{\sigma(gg\ra h\ra\gamma\gamma)_{\rm SM}}\approx \frac{Br(h \rightarrow {\gamma\gamma})_{\rm ID}}{Br(h \rightarrow {\gamma\gamma})_{\rm SM}}
\eeq
\eegp
using the narrow width approximation for the production cross-section of $\sigma(gg\ra h\ra \gamma\gamma)$ and the fact that $\sigma(gg\rightarrow h)$ in both the SM and ID are the same. 

Now if the ID particles have masses less than $M_h/2$, $h\rightarrow \rm ID,ID$ decays are allowed. In that case,
\begp
\allowdisplaybreaks \beq 
 \mu_{\gamma\gamma}= \frac{\Gamma(h\rightarrow \gamma\gamma)_{\rm ID}} {\Gamma(h\rightarrow \gamma\gamma)_{\rm SM}} ~ \frac{\Gamma_{\rm tot}(h\rightarrow \rm SM,SM)}{\Gamma_{\rm tot}(h\rightarrow {\rm SM,SM})+\Gamma_{{\rm tot}}(h \rightarrow {\rm ID,ID})}\, ,
 \label{mugagalow}
\eeq
\eegp
where~\cite{Cao:2007rm} 
\begp
\allowdisplaybreaks \beq
\Gamma\left(h\rightarrow \rm ID,ID\right)= \frac{ v^2}{16\pi M_h }\lambda_{\rm ID}^2 \left(1-\frac{4 M_{\rm ID}^2}{M_{h}^2}\right)^{1/2}\, ,
\label{decaywidth}
\eeq
\eegp
where for ${\rm ID}=A,H,H^\pm$, $\lambda_{\rm ID}=\lambda_S, \lambda_L, \sqrt{2}\,\lambda_3$.

In this case, the ID particles are heavier than $M_h/2$, 
\begp
\allowdisplaybreaks \beq 
 \mu_{\gamma\gamma}= \frac{\Gamma(h\rightarrow \gamma\gamma)_{\rm ID}}{\Gamma(h\rightarrow \gamma\gamma)_{\rm SM}}\, .
 \label{mugagahigh}
\eeq
\eegp
In the ID model, the $H^\pm$ gives additional contributions at one loop. 
The analytical expression is given by~\cite{Djouadi:2005gj, Swiezewska:2012eh,Krawczyk:2013jta}
\begp
\allowdisplaybreaks \beq
\Gamma(h\rightarrow \gamma\gamma)_{\rm ID}=
\frac{\alpha^2 m_h^3}{ 256\pi^3
v^2}\left|\sum_{f}N^c_fQ_f^2y_f
F_{1/2}(\tau_f)+ y_WF_1(\tau_W)
+Q_{H^{\pm}}^2 \frac{v\mu_{{hH^+H^-}}}{
2m_{H^{\pm}}^2} F_0(\tau_{H^{\pm}})\right|^2\
\label{hgaga}
\eeq
\eegp
where $\tau_i=m_h^2/4m_i^2$. $Q_{f}$, $Q_{H^{\pm}}$ denote electric charges of corresponding particles. $N_f^c$ is the color factor. $y_f$ and
$y_W$ denote Higgs couplings to $f\bar{f}$ and $WW$. $\mu_{hH^+H^-}= \lambda_3 v$ stands for the coupling constant of the $hH^+H^-$ vertex. The loop functions $F_{(0,\,1/2,\,1)}$  are defined as
\begp
\allowdisplaybreaks \begin{align}
F_{0}(\tau)&=-[\tau-f(\tau)]\tau^{-2}\, ,\nn\\
F_{1/2}(\tau)&=2[\tau+(\tau-1)f(\tau)]\tau^{-2}\, ,\nn\\
F_{1}(\tau)&=-[2\tau^2+3\tau+3(2\tau-1)f(\tau)]\tau^{-2}\, , \nn
\label{loopfn}
\end{align}
\eegp
where,
\begp
\allowdisplaybreaks \beq
f(\tau)= \bigg\{\begin{array}{ll}
(\sin^{-1}\sqrt{\tau})^2\,,\hspace{60pt}& \tau\leq 1\\
-\frac{1}{ 4}[\ln{1+\frac{\sqrt{1-\tau^{-1}}}{1}-\sqrt{1-\tau^{-1}}}-i\pi]^2\,,\quad
&\tau>1
\label{ftau}
\end{array}\;,\;
\eeq
\eegp
From the diphoton decay channel of the Higgs at the LHC, the measured values are  $\mu_{\gamma\gamma}$=$1.17\pm0.27$ from ATLAS~\cite{Aad:2014eha} and  $\mu_{\gamma\gamma}$=$1.14^{+0.26}_{-0.23}$ from CMS~\cite{Khachatryan:2014ira}. 

One can see that a positive $\lambda_3$ leads to a destructive interference between SM and ID contributions in eqn.~\ref{hgaga} and {\it vice versa}. Hence, for ID particles heavier than $M_h/2$, $\mu_{\gamma\gamma}<1$ ($\mu_{\gamma\gamma}>1$) when $\lambda_3$ is positive (negative). However, if these ID particles happen to be lighter than $M_h/2$, they might contribute to the invisible decay of the Higgs boson. Using the global fit result~\cite{Belanger:2013xza} that such an  invisible branching ratio is less than $\sim 20$\%, in eqn.~\ref{mugagalow}, the second ratio provides a suppression of $\sim 0.8$ -- $1$.

Now can negative $\lambda_3$ is allowed in the ID model? It will be discussed at the end of this section. 
The benchmark points with positive $\lambda_3$ is used here,
allowed at 1$\sigma$ by both CMS and ATLAS experiments. 

\subsubsection{Constraints from dark matter relic density and direct search limits}
As it has been seen that Higgs can decay to pair of inert particles (see eqns.~\ref{Scalarpot2},\ref{decaywidth}), so most of the parameter spaces with dark matter mass less than $M_h/2$, are strictly restricted from the Higgs invisible decay width of LHC and also from the direct detection experimental data. In this model the dark matter mass below 50 GeV are excluded from these constraints.
In Table~\ref{tableI1}, few benchmark points for IDM with the dark matter mass, $M_A(\equiv M_{DM})< M_h/2$ have been shown.

The heavy $CP$-even and charge Higgs mass have been fixed at 180 and 200 GeV respectively. The relic density, dark matter-nucleon cross-section, as well as the branching ratio of Higgs to pair of dark matter particles have been shown in the same table. For all these points, it has been seen that the dominant process in the dark matter annihilation is $AA\ra b \bar{b}$.
\setlength\tablinesep{1pt}
\setlength\tabcolsep{6pt}
\begin{table}[h!]
\begin{center}
    \begin{tabular}{ | c | c |  c | c | c | c |}
    \hline
     $M_{DM}(\equiv M_A)$ (GeV) & $\lambda_S$ & Relic density &$\sigma_{SI}$ ${\rm ({cm}^2)}$& Br($h\rightarrow S S$) in $\%$\\
    \hline
     $54.7$ & 0.004 & 0.1263 & $1.78\times10^{-46}$ & 3.67   \\
     \hline
     $56$ & 0.0022 & 0.1197 & $3.47\times10^{-48}$ & 1.05   \\
     \hline
     $58.8$ & 0.0006 & 0.1222 & $2.4\times10^{-48}$ & 0.06   \\
     \hline
     $60$ & 0.00035 & 0.1236 & $1.13\times10^{-48}$ & 0.017   \\
     \hline
             \end{tabular}
    \caption{\textit{ Benchmark points with dark matter mass $M_{DM}< M_h/2$, with $M_{H^\pm}=200$ ${\rm GeV}$ and $M_{H^\pm}=180$ ${\rm GeV}$, which is allowed from the relic density constraint of WMAP and Planck, $\Omega h^2=0.1198\pm 0.0026$ within 3$\sigma$ confidence level, direct detection LUX (2013) and Higgs invisible decay width from the LHC. }}
    \label{tableI1}
\end{center}
\end{table}

The ID dark matter candidate $A$ can self-annihilate into SM fermions. Once the DM mass is greater than the $W^\pm$-mass, so that the DM can annihilate into a pair of $W^\pm$ bosons, the cross-section increases significantly, thereby reducing DM relic density. Hence, it becomes difficult to saturate $\Omega h^2$ after $\sim 75$~GeV with positive $\lambda_S$, although for $M_A<75$~GeV, both signs of  $\lambda_S$ can be allowed to arrive at the right DM relic density $\Omega h^2$.

The role of the sign of $\lambda_S$ can be understood from the contributing diagrams to the  $AA\ra W^+ W^-$  annihilation processes. Four diagrams contribute: the $AAW^+ W^-$ vertex driven point interaction diagram (henceforth referred to as the $p$-channel diagram), $H^+$-mediated $t$- and $u$-channel diagrams, and the $h$-mediated $s$-channel diagram. For $AA\ra ZZ$  annihilation, the $t$- and $u$-channel diagrams are mediated by $H$. A negative $\lambda_S$ induces a destructive interference between the $s$-channel diagram with the rest, thereby suppressing $AA\ra W^+ W^-, ZZ$ processes. For DM masses of $75-100$~GeV, this can be used to get the appropriate $\Omega h^2$~\cite{Gustafsson:2012aj, LopezHonorez:2010tb}. To avoid large contributions from $t$- and $u$-channel diagrams and coannihilation diagrams, the values of $M_H$ and $M_{H^\pm}$ can be pushed to be rather large $\gtrsim 500$~GeV. However, to partially compensate the remaining $p$-channel diagram by the  $s$-channel one, $\lambda_S$ assumes a large negative value $\sim -0.1$, which is ruled out by the DM direct detection experiments. That is why in the ID model, DM can be realized below 75 GeV, a regime designated as the ``low'' DM mass region. 

 \begin{figure}[h!]
 \begin{center}
 \subfigure[]{
 \includegraphics[width=2.7in,height=2.7in, angle=0]{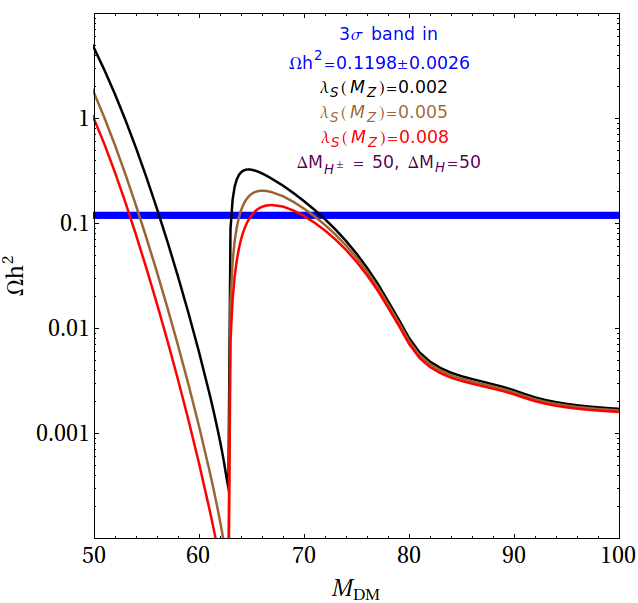}}
 \subfigure[]{
 \includegraphics[width=2.7in,height=2.7in, angle=0]{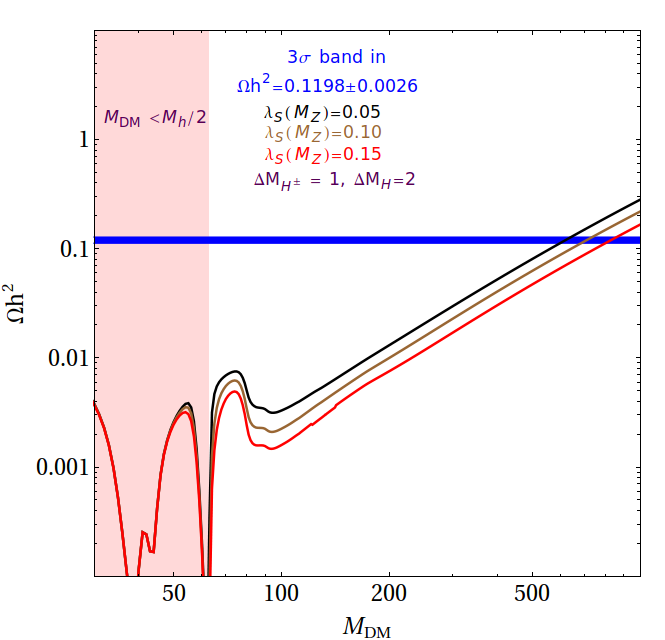}}
 \caption{\label{fig:relicvsmassIn} In {\rm IDM}, \textit{$({\bf a})$ Dark matter relic density $\Omega h^2$ as a function of the dark matter mass $M_{DM}(\equiv M_A)$ for different values of the Higgs portal coupling: $\lambda_S(M_Z)=0.002$ (black), $\lambda_S(M_Z)=0.005$ (brown), and $\lambda_S(M_Z)=0.008$ (red), with $\Delta M_{H^\pm}=\Delta M_{H}=50$ GeV. $({\bf b})$ For Higgs portal coupling: $\lambda_S(M_Z)=0.05$ (black), $\lambda_S(M_Z)=0.10$ (brown), and $\lambda_S(M_Z)=0.15$ (red), with $\Delta M_{H^\pm}=1$ GeV and $\Delta M_{H}=2$ GeV. The thin blue band corresponds to the relic density of the dark matter, $\Omega h^2=0.1198 \pm 0.0026$ (3$\sigma$) of the present the Universe. Light red band is disallowed from the Higgs invisible decay width.} }
 \end{center}
 \end{figure}
At ``high'' DM mass $M_A \gtrsim 500$~GeV, one can get the right $\Omega h^2$ due to a partial cancellation between different diagrams contributing to the $AA\ra W^+ W^-$ and $AA\ra ZZ$ annihilation processes. For example, in $AA\ra W^+ W^-$, the $p$-channel diagram tends to cancel with the $H^+$-mediated $t$- and $u$-channel diagrams~\cite{Hambye:2009pw, Goudelis:2013uca} in the limit $M_A\gg M_W$, and the sum of amplitudes of these diagrams in this limit is proportional to $M_{H^\pm}^2-M_A^2$. Hence, at high $M_A$, a partial cancellation between these diagrams is expected for nearly degenerate $M_{H^\pm}$ and $M_A$. Similarly, for $AA\ra ZZ$, a cancellation is possible when the masses $M_{H}$ and $M_A$ are close by. For $M_A \gtrsim 500$~GeV, keeping the mass differences of $M_{H^\pm}$ and $M_{H}$ with $M_A$ within $8$~GeV, such  cancellations help reproduce the correct $\Omega h^2$. It is nevertheless worth mentioning that such nearly degenerate masses will lead to coannihilation of  these $Z_2$ odd ID scalars~\cite{griest} to SM particles.
Despite such near-degeneracy, both $H$ and $H^+$, being charged under the same $Z_2$ as $A$, decay promptly to the LOP $A$, so that they do not become relics. {\tt FeynRules}~\cite{Alloul:2013bka} along with {\tt micrOMEGAs}~\cite{Belanger:2010gh, Belanger:2013oya} used to compute the relic density of $A$. 

In Fig.~\ref{fig:relicvsmassIn}, the relic density against the dark matter mass have been shown. For Fig.~\ref{fig:relicvsmassIn}(a), $\Delta M_{H^\pm}=\Delta M_{H}=50$ GeV have been chosen, so the relic density is mainly dominated by the self-annihilation of the dark matter candidate. For three different values of Higgs portal coupling, $\lambda_S=0.002,~0.005,~0.008$, the relic density turns out in the right ballpark for the dark matter mass $65-70$ GeV. The dominant annihilation channels are, $AA\ra b \bar{b},W^{\pm}W^{\mp*}$. Fig.~\ref{fig:relicvsmassIn}(b), the relic density is in the right ballpark with dark matter mass around 600 GeV. In this case both the co-annihilation and annihilation of the dark matter play important role. In this region inert particles annihilate mainly to $W^{\pm}W^{\mp},ZZ,hh$. The blue band corresponds to the constraints coming from the WMAP and Planck data. The region which is allowed from the relic density is also allowed from the direct search of the dark matter.

DM direct detection experiments involve the $h$-mediated $t$-channel process $AN\ra AN$ with a cross-section proportional to $\lambda_S^2/M_A^2$ in the limit $M_A\gg m_N$: 
\begp
\allowdisplaybreaks \beq
 \sigma_{A,N} = \frac{m_r^2}{\pi} f_N^2 m_N^2\left(\frac{\lambda_S}{M_{A} M_h^2}\right)^2 \label{directcs}
\eeq
\eegp
where $f_N\approx0.3$ is the form factor of the nucleus. $m_r$ represents the reduced mass of the nucleus and the scattered dark matter particle. 

Thus, $\lambda_S$ is constrained from non observation of DM signals at XENON\,100~\cite{Aprile:2011hi,Aprile:2012nq} and LUX~\cite{Akerib:2013tjd}. For $M_{A}= 70$~GeV, the ensuing bound from LUX~\cite{Akerib:2013tjd,dmtools} data at 1$\sigma$ is $|\lambda_S|< 0.007$.  

The constraint on $\lambda_S$ from DM direct detection experiments gets diluted with $M_A$ [see eqn.~\ref{directcs}]. Hence, for low DM mass, direct detection bounds are more effective. At high mass, the relic density constraints are likely to supersede these bounds. For example, for $M_{DM}= 573$~GeV, the upper limit on $|\lambda_S|$ is 0.138 from LUX. However, to satisfy the relic density constraints from the combined data of WMAP and Planck within 3$\sigma$, $\lambda_S$ can be as large as $0.07$ only.

Within the framework of the ID model, it is possible to explain the observations in various indirect DM detection experiments~\cite{Arhrib:2013ela, Modak:2015uda} for some regions of the parameter space.
In this work, such details have not included as such estimations involve proper understanding of the astrophysical backgrounds and an assumption of the DM halo profile which contain some arbitrariness. For a review of constraints on the ID model from astrophysical considerations see, for example, Ref.~\cite{Gustafsson:2010zz}.

\vskip 20pt
\noindent{\bf\underline {Sign of $\lambda_3$}}\\

Whether $\lambda_3$ can be taken as positive or negative depends on the following:
\begin{itemize}

\item
If the ID model is not the answer to the DM puzzle, so that both relic density and direct detection constraints can be evaded, no restriction exists on the possible sign of $\lambda_3(M_Z)$. Otherwise, the following two cases need be considered:
\begin{itemize}
\item
A negative $\lambda_3(M_Z)$ implies 
\beq
\lambda_S(M_Z) < -\frac{1}{v^2} (M_{H^\pm}^2-M_A^2) \nn\, .
\eeq
As $A$ is considered as the DM candidate, so that $M_A<M_{H^\pm}$, $\lambda_S(M_Z)$ is always negative when $\lambda_3(M_Z)<0$. 
For low DM mass, the splitting  $(M_{H^\pm}-M_A)\gtrsim 10$~GeV, as otherwise DM coannihilation processes cause an inappreciable  depletion in $\Omega h^2$. For $M_A=70$~GeV, this implies a lower bound $\lambda_S(M_Z)\lesssim -0.025$, which violates the DM direct detection bound $|\lambda_S|< 0.007$. Hence, for low DM mass, a negative $\lambda_3(M_Z)$  is not feasible.
\item
For high mass DM, the right relic density can be obtained when the splitting  $(M_{H^\pm}-M_A)\sim$ a few GeV or less. The above logic then implies that a negative $\lambda_3$ does not put any severe restriction on $\lambda_S$ to contradict DM direct detection bounds as earlier. Hence, for a high DM mass, $\lambda_3(M_Z)$ can assume both the signs. Moreover, due to propagator suppression for large $M_{H^\pm}$ in the $h\gamma\gamma$ vertex, the ID contribution to $\mu_{\gamma\gamma}$ is negligibly small and hence, the sign of $\lambda_3(M_Z)$ is not constrained by measurements on $\mu_{\gamma\gamma}$ as well. 
\end{itemize}
\item
If at any scale, $\lambda_3$ is negative while $\lambda_1>0$, then the bound~(\ref{stabilitybound}) must be respected. 
\item
If at some scale, $\lambda_1<0$, then a negative $\lambda_3$ makes the potential unbounded from below, as mentioned in the following section. This means one can start with a negative $\lambda_3(M_Z)$, but with RG evolution when $\lambda_1$ turns negative, $\lambda_3$ evaluated at that scale must be positive. Such parameter space does exist.
Here, a significant deviation has been found of this analysis from earlier analyses which did not allow a negative $\lambda_1$. For example, in Ref.~\cite{Goudelis:2013uca} a negative $\lambda_3(M_Z)$ was not allowed from stability of the Higgs potential if the theory has to be valid up to $10^{16}$~GeV together with relic density considerations. 
\end{itemize}
\subsection{Tunneling Probability and Metastability in IDM}
\label{sec:metastability}
 \begin{figure}[h!]
 \begin{center}
\subfigure[]{
 \includegraphics[width=2.7in,height=2.7in, angle=0]{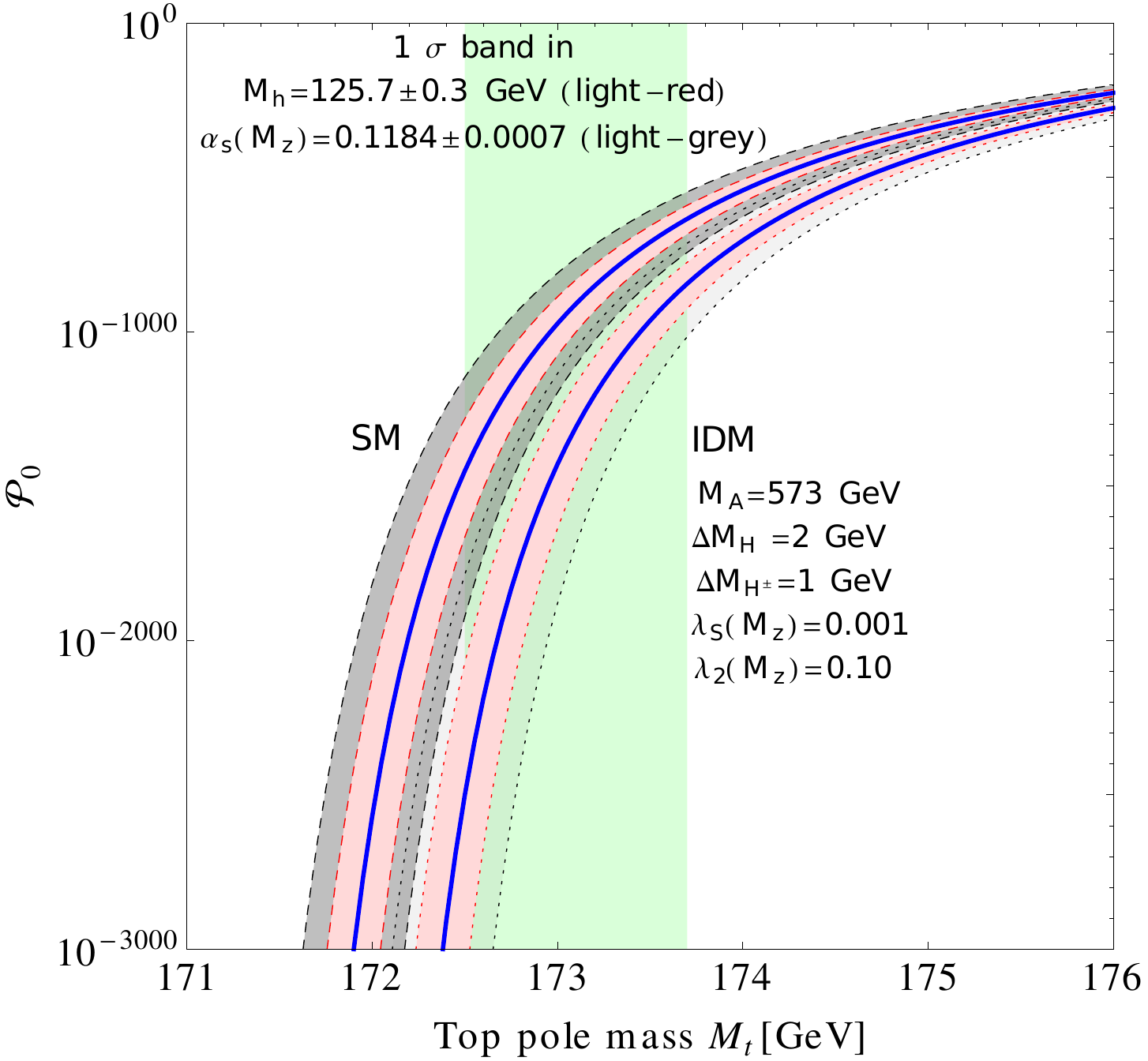}}
 \hskip 15pt
 \subfigure[]{
 \includegraphics[width=2.7in,height=2.7in, angle=0]{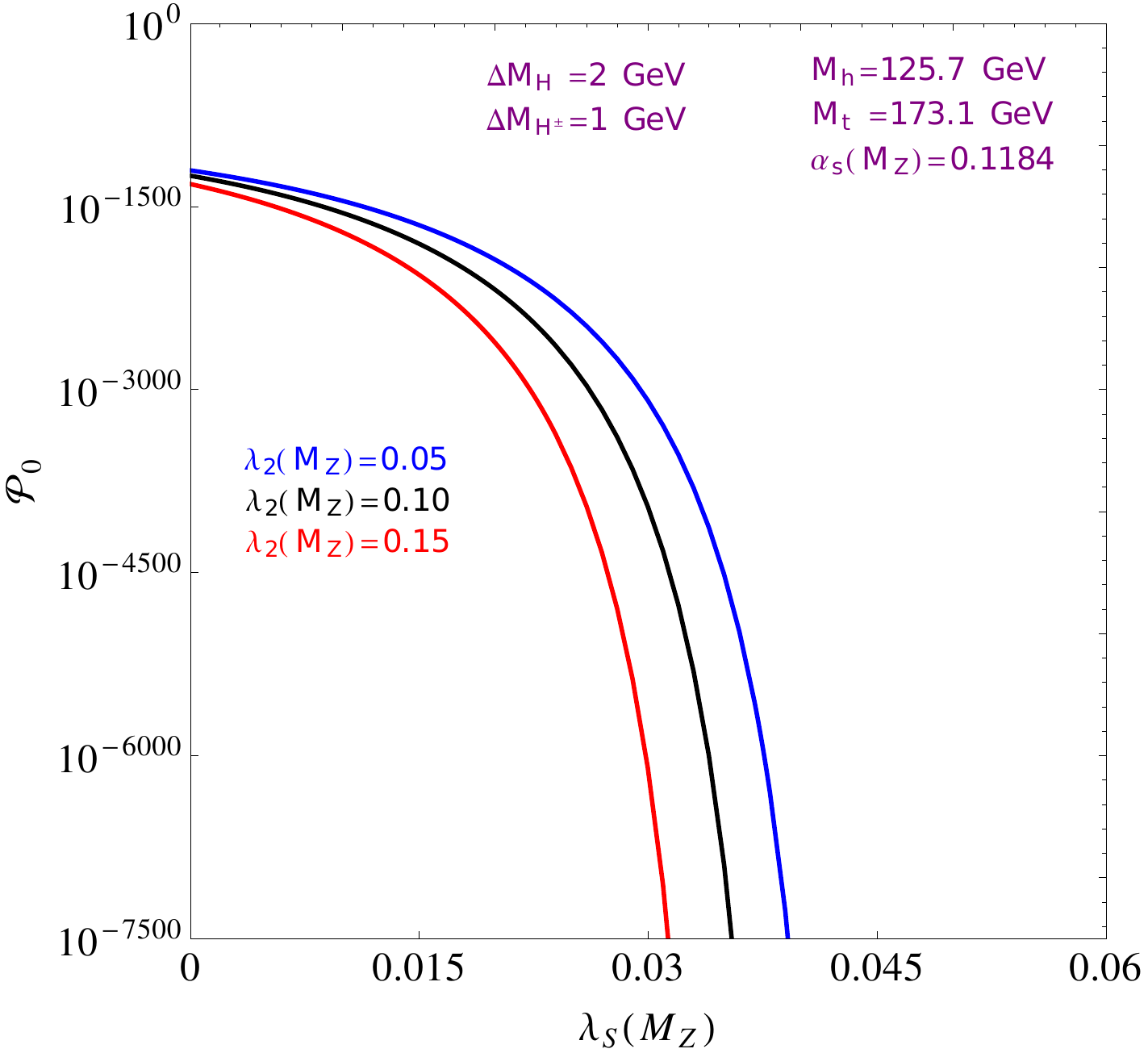}}
 \caption{\label{fig:TunIDM} \textit{ (a) Tunneling probability ${\cal P}_0$ dependence on  $M_t$. The left band (between dashed lines) corresponds to {\rm SM}. The right one (between dotted lines) is for $ID$ model for DM mass $M_A=573$~GeV.  Constraints from WMAP and Planck measured relic density, as well as XENON\,100 and LUX DM direct detection null results are respected for these specific choice of parameters. Light-green band stands for $M_t$ at $\pm 1\sigma$. (b) ${\cal P}_0$ is plotted against Higgs dark matter coupling $\lambda_S(M_Z)$ for different values of $\lambda_2(M_Z)$.}}
 \end{center}
 \end{figure}

The presence of the inert doublet induces additional contributions to $\beta_{\lambda_1}$ (see eqn.~\ref{betal_1}).  As a result, which is generic for all scalars, $\lambda_1$ receives a positive contribution compared to the SM, which pushes a metastable vacuum towards stability, implying a lower ${\cal P}_0$ (see eqn.~\ref{prob}).

Electroweak metastability in the ID model has been explored earlier in the literature, albeit in a different context~\cite{Barroso:2013awa,Barroso:2012mj,Barroso:2013kqa,Gil:2012ya}. If $H^+$ gets a VEV, there could exist another charge-violating minimum. If instead, $A$ receives a VEV, another $CP$-violating minimum could pop up. But these vacua always lie higher than the usual EW vacuum. If $Z_2$ is broken by introducing additional soft terms in the Lagrangian, then the new $Z_2$-violating minimum can be lower than the usual $Z_2$-preserving EW minimum. As in this work, $Z_2$ is an exact symmetry of the scalar potential, so such cases need not be considered. However, as mentioned earlier, if at some scale before $\mpl$, the sign of $\lambda_1$ becomes negative, there might exist a deeper minimum which is charge-, $CP$- and $Z_2$-preserving and lying in the SM Higgs $h$ direction. 

The EW vacuum is metastable or unstable, depends on the minimum value of $\lambda_1$ before $\mpl$, which can be understood as follows. For EW vacuum metastability, the decay lifetime should be greater than the lifetime of the Universe, implying ${\cal P}_0 < 1$ (eqn.~\ref{lamminSM}).
Hence the vacuum stability constraints on the ID model can now be reframed, when $\lambda_{\rm 1, eff}$ runs into negative values, implying metastability of the EW vacuum.
It is reminded to the reader that in the ID model, instability of the EW vacuum cannot be realized as addition of the scalars only improves the stability of the vacuum. 
\begin{itemize}
\item
If $0>\lambda_{\rm 1, eff}(\Lambda_B)>\lambda_{\rm 1,min}(\Lambda_B)$, then the vacuum is metastable. 
\item
If $\lambda_{\rm 1, eff}(\Lambda_B)<\lambda_{\rm 1,min}(\Lambda_B)$, then the vacuum is unstable. 
\item
If $\lambda_2<0$, then the potential is unbounded from below along the $H, A$ and $H^\pm$ direction. 
\item
If $\lambda_3(\Lambda_{\rm I})<0$, the potential is unbounded from below along a direction in between $H^\pm$ and $h$.
\item
If $\lambda_L(\Lambda_{\rm I})<0$, the potential is unbounded from below along a direction in between $H$ and $h$.
\item
If $\lambda_S(\Lambda_{\rm I})<0$, the potential is unbounded from below along a direction in between $A$ and $h$.
\end{itemize}
In the above, $\Lambda_{\rm I}$ represents any energy scale for which $\lambda_{\rm 1, eff}$ is negative and the conditions for unboundedness of the potential follow from eqn.~\ref{Scalarpot2}.
At this point note the significant deviations are being made in the allowed parameter space compared to the usual vacuum stability conditions:  According to eqn.~\ref{stabilitybound}, $\lambda_{3,L,S}$ can take slightly negative values.
But at a scale where $\lambda_{\rm 1,eff}$ is negative, with the new conditions $\lambda_{3,L,S}$ have to be positive.

The tunneling probability ${\cal P}_0$ is computed by putting the minimum value of $\lambda_{\rm 1, eff}$ in eqn.~\ref{Action} to minimize $S(\Lambda_B)$.
In Fig.~\ref{fig:TunIDM}$\rm{ \left(a\right)}$, ${\cal P}_0$ has been plotted as a function of $M_t$.
The right band corresponds to the tunneling probability for our benchmark point as in Table~\ref{table1IDM}. For comparison, ${\cal P}_0$ has been plotted for SM as the left band in Fig.~\ref{fig:TunIDM}$\rm{(a)}$. 1$\sigma$ error bands in $\alpha_s$ and $M_h$ are also shown. As expected, for a given $M_t$, the presence of ID lowers tunneling probability. This is also reflected in Fig.~\ref{fig:TunIDM}$\rm {\left(b\right)}$, where ${\cal P}_0$ has been plotted as a function of  $\lambda_S(M_Z)$ for different choices of $\lambda_2(M_Z)$, assuming  $M_h=125.7$~GeV, $M_t=173.1$~GeV, and $\alpha_s=0.1184$. Here DM mass $M_A$ is also varied with $\lambda_S$ to get  $\Omega h^2=0.1198$. For a given $\lambda_S(M_Z)$, the higher the value of $\lambda_2(M_Z)$, the smaller ${\cal P}_0$ gets, leading to a more stable EW vacuum.

\subsection{Phase diagrams in IDM}  
\label{sec:phasediag}
The stability of EW vacuum depends on the value of parameters at low scale, chosen to be $M_Z$. In order to show the explicit dependence of EW stability on various parameters, it is customary to present phase diagrams in various parameter spaces.
 \begin{figure}[h!]
 \begin{center}
 \includegraphics[width=2.7in,height=2.7in, angle=0]{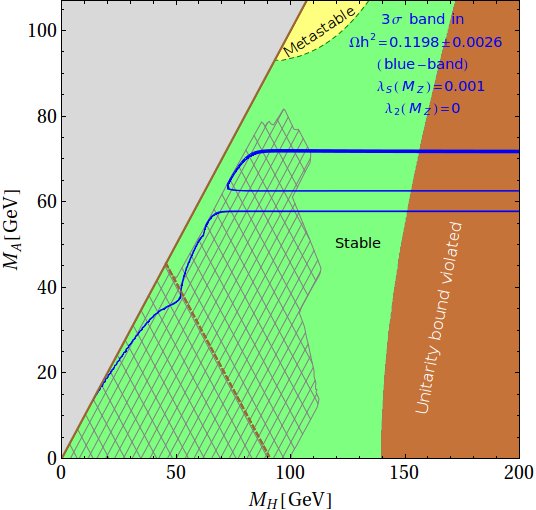}
 \caption{\label{fig:STU1} \textit{Constraints in $M_{H}-M_{A}$ plane in {\rm IDM}. The cross-hatched region  is excluded from LEP~{\rm \cite{Lundstrom:2008ai}}.  Choosing $M_{H^\pm}=120$~GeV and $\lambda_S(M_Z)=0.001$, relic density constraint is satisfied at 3$\sigma$ on the blue band. The green (yellow) region corresponds to EW vacuum stability (metastability). The solid brown line correspond to $M_H=M_A$. The gray area on the left to it is of no interest to us as $M_H>M_A$ has been chosen. The dashed brown line shows the LEP\,I limit. On the brown region, unitarity constraints are violated before $\mpl$. } }
 \end{center}
 \end{figure} 
In Fig.~\ref{fig:STU1} the LEP constraints has been shown in the $M_{H}- M_A$ plane as in Ref.~\cite{Lundstrom:2008ai}.
Identifying regions of EW stability and metastability, the plot has been updated here.
As a scenario is being considered where the ID model is valid till $\mpl$, there are further limits from unitarity.
The relic density constraint imposed by WMAP and Planck combined data is represented by the thin blue band. The choice of $\lambda_2(M_Z)$ does not have any impact on relic density calculations, but affects EW stability as expected. In this plot, for higher values of $\lambda_2(M_Z)$, the region corresponding to EW metastability will be smaller. The chosen parameters  satisfy the LUX direct detection bound. 

 \begin{figure}[h!]
 \begin{center}
 \includegraphics[width=2.7in,height=2.7in, angle=0]{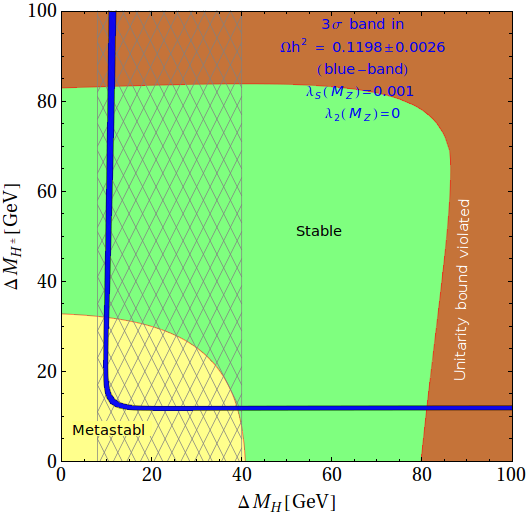}
 \caption{\label{fig:STU2} \textit{Phase diagram in $\Delta M_{H}-\Delta M_{H^\pm}$ plane for $M_A=70$~GeV in {\rm IDM}. The green and yellow regions correspond to EW vacuum stability and metastability respectively. The cross-hatched band is excluded from LEP. The brown region suffers from unitarity violation before $\mpl$. The blue band reflects relic density constraint at 3$\sigma$.
  } }
 \end{center}
 \end{figure} 

As small splitting among $M_A$, $M_H$, and $M_{H^\pm}$ leads to some cancellations among diagrams contributing to DM annihilation, $\Delta M_{H^\pm}$ and $\Delta M_{H}$ are often used as free parameters in the ID model. In Fig.~\ref{fig:STU2}, constraints on this parameter space for $M_A=70$~GeV has been presented. As before, the brown region corresponds to unitarity violation before $\mpl$. For small $\Delta M_{H^\pm}$ and $\Delta M_{H}$, $\lambda_{3,4,5}$ are required to be small, which leads to little deviation from SM metastability. The metastable region is shown by the yellow patch, which shrinks for larger $\lambda_2$. The blue band reflects the relic density constraint for $\lambda_S(M_Z)=0.001$. For such small $\lambda_S(M_Z)$, the $h$-mediated $s$-channel diagram in $AA\ra WW$ or $AA\ra ZZ$ contributes very little. $H^+$- or $H$-mediated $t$- and $u$-channel diagrams are also less important than the quartic vertex driven diagram due to propagator suppression. This explains the ``L'' shape of the blue band. For higher values of $\lambda_S(M_Z)$, the shape of the band changes and ultimately leads to a closed contour. It appears that due to LEP constraints, EW vacuum metastability is almost ruled out. Although the LEP constraint permits $\Delta M_H<8$~GeV, allowing a narrow strip towards the left, the relic density constraints cannot be satisfied on this strip as it leads to an increased rate of DM coannihilation processes, leading to a dip in $\Omega h^2$. But as it will be seen later, if $M_t$ and $\alpha_s$ are allowed to deviate from their respective central values, for some region in this parameter space, it is possible to realize a metastable EW vacuum.    

 \begin{figure}[h!]
 \begin{center}
 \subfigure[]{
 \includegraphics[width=2.7in,height=2.7in, angle=0]{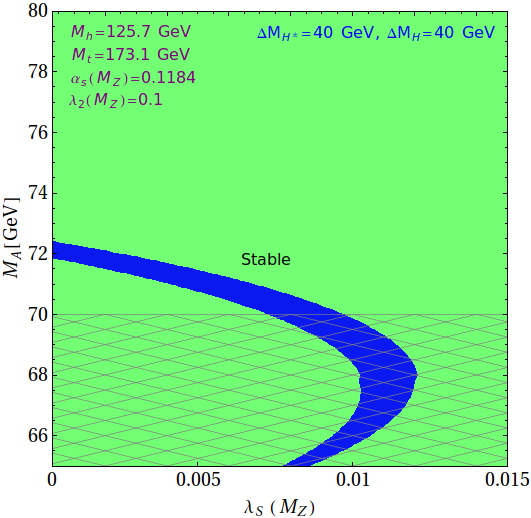}}
 \hskip 15pt
 \subfigure[]{
 \includegraphics[width=2.7in,height=2.7in, angle=0]{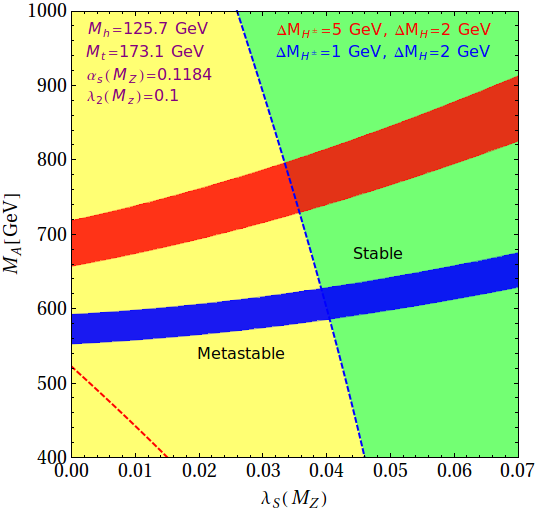}}
 \caption{\label{fig:MDM_LamS} \textit{Phase diagram in $\lambda_S(M_Z)-M_A$ plane for $\lambda_2(M_Z)=0.1$ in {\rm IDM}. Panel (a) stands for `low' DM mass. The blue band  corresponds the 3$\sigma$ variation in $\Omega h^2$ when $\Delta M_{H^\pm}=40$~GeV and $\Delta M_H=40$~GeV. LEP direct search constraints are represented by the cross-hatched band at the bottom. Entire green region imply EW vacuum stability. Panel (b) stands for `high' DM masses.  The relic density band (blue) now correspond to  $\Delta M_{H^\pm}=1$~GeV and $\Delta M_H=2$~GeV. The corresponding stable and metastable phases for EW vacuum are represented by green and yellow patches respectively.   The relic density band (red) corresponds to  $\Delta M_{H^\pm}=5$~GeV and $\Delta M_H=2$~GeV. For this, the boundary separating the EW phases is denoted by the red dashed line.} }
 \end{center}
 \end{figure}

To delineate the role of $M_H$ in EW vacuum stability, in Figs.~\ref{fig:STU1} and~\ref{fig:STU2}, $\lambda_2(M_Z)$ is chosen to be small. To demonstrate the effect of $\lambda_S$, phase diagrams will be presented in Fig.~\ref{fig:MDM_LamS} in the $\lambda_S(M_Z) - M_A$ plane. Panel (a) deals with low DM masses. For $\Delta M_{H^\pm}=40$~GeV and $\Delta M_H=40$~GeV, part of the allowed relic density band (blue) is allowed from LEP constraints (cross-hatched band). The entire parameter space corresponds to EW vacuum stability. Choosing small  $\Delta M_{H^\pm}$ and $\Delta M_H$, which imply small values of $\lambda_{3,4,5}$, can lead to metastability. But those regions are excluded by LEP. Again, metastability can creep in if $M_t$ and $\alpha_s$ are allowed to deviate from  their central values. 

In Fig.~\ref{fig:MDM_LamS}(b), the same parameter space has been studied for high DM masses. As mentioned before, to obtain the correct relic density, smaller mass splitting among various ID scalars needs to be chosen. For $\Delta M_{H^\pm}=1$~GeV and $\Delta M_H=2$~GeV, the 3$\sigma$ relic density constraint is shown as the blue band. The blue dashed line demarcates the boundary between stable (green) and metastable (yellow) phases of EW vacuum. The choice of small values of $\Delta M_{H^\pm}$ and $\Delta M_H$, in turn, leads to a large region pertaining to EW metastability. 

To illustrate the sensitivity to the mass splitting, in Fig.~\ref{fig:MDM_LamS}(b), another relic density band (red) has been presented for $\Delta M_{H^\pm}=5$~GeV and $\Delta M_H=2$~GeV. The corresponding boundary between the phases is denoted by the red dashed line. The region on the right implies EW stability (the green and yellow regions do not apply to this case). As for high DM masses, EW metastability can be attained for a sizable amount of the parameter space; $\lambda_2(M_Z)$ need not be chosen to be very small to maximize the metastable region for the sake of demonstration. 

The fact that for SM the EW vacuum stability is ruled out at $\sim 3\sigma$, is demonstrated by a phase space diagram in the $M_t-M_h$ plane~\cite{Degrassi:2012ry, Buttazzo:2013uya}. In Ref.~\cite{Khan:2014kba}, similar diagrams were presented for a singlet scalar extended SM. To demonstrate the impact of ID scalars to uplift the EW vacuum metastability, phase diagrams have been presented in the $M_t-M_h$ plane for two sets of benchmark points in Fig.~\ref{fig:Mt_MhIDM}. Panel (a) is drawn for $M_A = 70$~GeV, $\Delta M_{H^\pm}=11.8$~GeV, $\Delta M_H=45$~GeV, $\lambda_S(M_Z)=0.001$, and $\lambda_2 (M_Z) = 0.1$. For panel (b) the set of parameters in Table~\ref{table1IDM} is being used.
 \begin{figure}[h!]
 \begin{center}
 \subfigure[]{
 \includegraphics[width=2.7in,height=2.7in, angle=0]{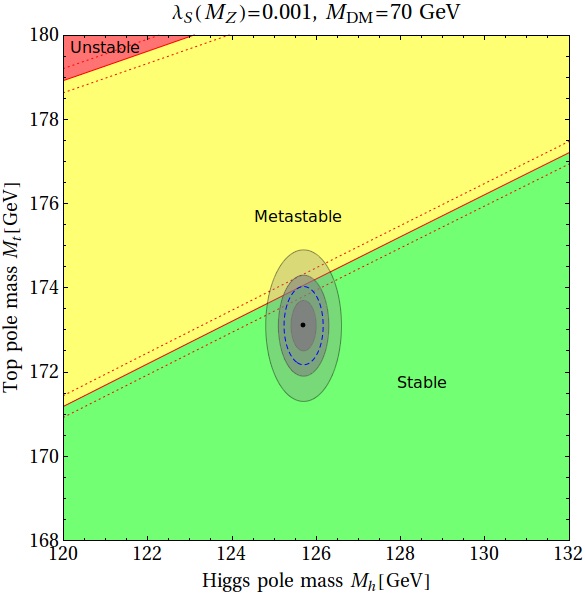}}
 \hskip 15pt
 \subfigure[]{
 \includegraphics[width=2.7in,height=2.7in, angle=0]{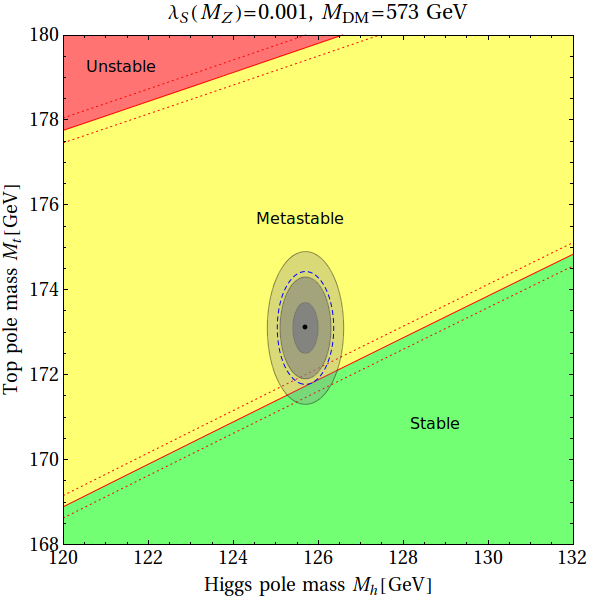}}
 \caption{\label{fig:Mt_MhIDM} \textit{Phase diagrams in $M_h - M_t$ plane in {\rm IDM}. Panels (a) and (b) stand for `low' and `high' DM masses respectively. Regions of 
absolute stability~(green), metastability~(yellow), instability~(red) of the EW vacuum are also marked. The gray zones represent error ellipses at  $1$, $2$ and  $3\sigma$. The three boundary lines (dotted, solid and dotted red) correspond to $\alpha_s(M_Z)=0.1184 \pm 0.0007$. Details of benchmark points are available in the text. } }
 \end{center}
 \end{figure}
Both sets of parameters are chosen so that they respect the WMAP and Planck combined results on DM relic density and the direct detection bounds from XENON\,100 and LUX. As in Ref.~\cite{Khan:2014kba}, the line demarcating the boundary between stable and metastable phases of EW vacuum is obtained by demanding that the two vacua be at the same depth, implying $\lambda_1(\Lambda_B)=\beta_{\lambda_1}(\Lambda_B)=0$. The line separating the metastable phase from the unstable one is drawn using the conditions $\beta_{\lambda_1}(\Lambda_B)=0$ and $\lambda_1(\Lambda_B)=\lambda_{1, \rm min}(\Lambda_B)$, as in eqn.~\ref{lamminSM}. The variations due to uncertainty in the measurement of $\alpha_s$ are marked as dotted red lines. In each panel, the dot representing central values for $M_h$ and $M_t$ is encircled by 1$\sigma$, 2$\sigma$, and 3$\sigma$ ellipses representing errors in their measurements.  According to Fig.~\ref{fig:Mt_MhIDM}(a), EW vacuum stability is allowed at 1.5$\sigma$, whereas in Fig.~\ref{fig:Mt_MhIDM}(b), it is excluded at 2.1$\sigma$, indicated by blue-dashed ellipses.

  \begin{figure}[h!]
 \begin{center}
 \subfigure[]{
 \includegraphics[width=2.7in,height=2.7in, angle=0]{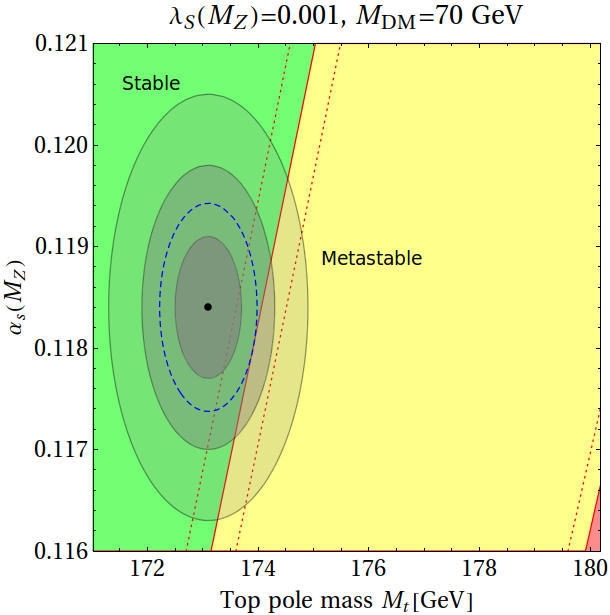}}
 \hskip 15pt
 \subfigure[]{
 \includegraphics[width=2.7in,height=2.7in, angle=0]{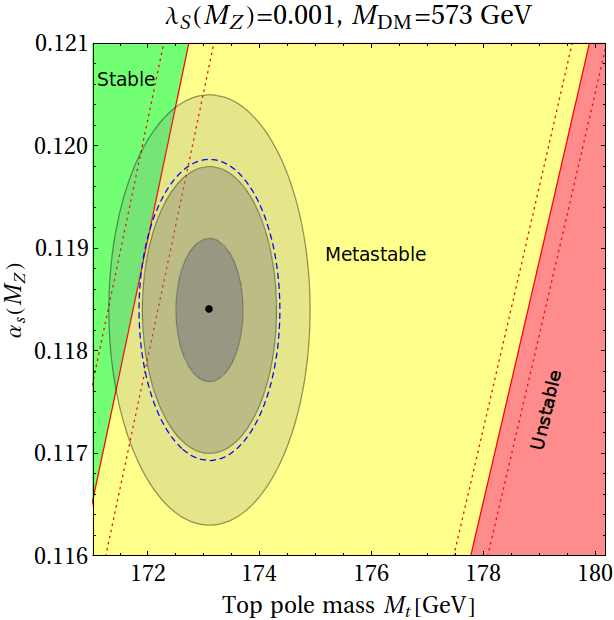}}
 \caption{\label{fig:Alpha_MtIDM} \textit{ Phase diagrams in $M_t-\alpha_s(M_Z)$ plane in {\rm IDM} for the same sets of benchmark points as in Fig.~\ref{fig:Mt_MhIDM}. Notations used are also the same as in Fig.~\ref{fig:Mt_MhIDM}. } }
 \end{center}
 \end{figure}
As in the literature SM EW phase diagrams are also presented in the $\alpha_s(M_Z)-M_t$ plane~\cite{Bezrukov:2012sa, EliasMiro:2011aa}, the same is done in the ID model as well. In Fig.~\ref{fig:Alpha_MtIDM}, the same sets of benchmark parameters has been used as in Fig.~\ref{fig:Mt_MhIDM}. As a consistency check, one can note that the EW vacuum is allowed or ruled out at the same confidence levels.
 \begin{figure}[h!]
 \begin{center}{
 \includegraphics[width=2.7in,height=2.7in, angle=0]{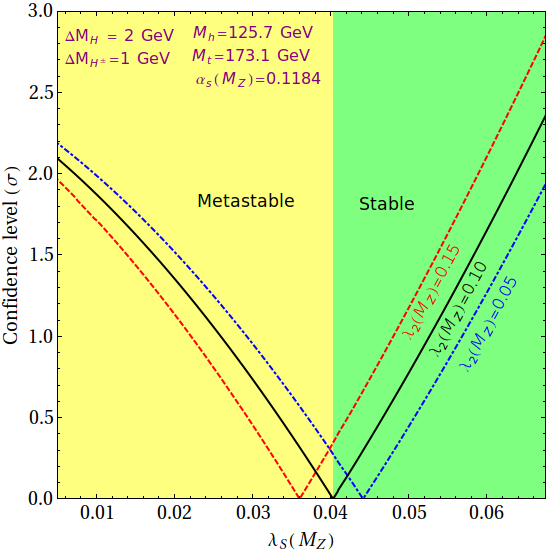}}
 \caption{\label{fig:confidenceIDM} \textit{ Dependence of confidence level at which EW vacuum stability is excluded (one-sided) or allowed on $\lambda_S(M_Z)$ and $\lambda_2(M_Z)$ in {\rm IDM}. Regions of absolute stability (green) and metastability (yellow) of EW vacuum are shown for $\lambda_2(M_Z)=0.1$. The positive slope of the line corresponds to the stable EW vacuum and negative slope corresponds to the metastability.     
 } }
 \end{center}
 \end{figure}

To study the impact of nonzero ID couplings, however, it is instructive to study the change in the confidence level ($\sigma$) at which EW stability is modified with respect to these couplings. As in Ref.~\cite{Khan:2014kba}, In Fig.~\ref{fig:confidenceIDM} ~$\sigma$ has been plotted against $\lambda_S(M_Z)$ for different values of $\lambda_2(M_Z)$. $M_A$ is varied along with $\lambda_S(M_Z)$ to keep DM relic density fixed at $\Omega h^2=0.1198$ throughout the plot. Note that changing $\lambda_2(M_Z)$ does not alter $\Omega h^2$. The masses of other ID particles  are determined using $\Delta M_{H^\pm}=1$~GeV and $\Delta M_H=2$~GeV. The parameter space considered does not yield too large DM-nucleon cross-section, inconsistent with XENON\,100 and LUX DM direct detection null results. For a specific value of $\lambda_2(M_Z)=0.1$, with the increase of $\lambda_S(M_Z)$, the confidence level at which EW is metastable (yellow region) gets reduced and becomes zero at $\lambda_S(M_Z)\simeq 0.04$. After this, EW vacuum enters in the stable phase (green). With further increases in $\lambda_S(M_Z)$, the confidence level at which EW is stable keeps increasing.
Two other values in the same plot is used to illustrate the role of $\lambda_2(M_Z)$.
The value of $\lambda_S(M_Z)$ at which the EW vacuum enters in the stable phase increases with decreases in $\lambda_2(M_Z)$, as expected. The yellow and green marked regions are not applicable when $\lambda_2(M_Z)=0.05, 0.15$.   
\subsection{Veltman's conditions in IDM}
\label{sec:veltman}
As the validity of the ID model have been extended till $\mpl$, it is interesting to explore whether Veltman's condition (VC) can be satisfied in this model at any scale on or before $\mpl$. It has been shown in Section~\ref{VCsINSM}, if one imposes VC in the SM at $\mpl$, then the top mass measurement $M_t=173.1\pm 0.6$~GeV implies $M_h\approx 135\pm2.5$~GeV, which is excluded at more than 5$\sigma$.  

Veltman's condition implies that the quadratic divergences in the radiative corrections to the Higgs mass can be handled if the coefficient multiplying the divergence somehow vanishes~\cite{Veltman:1980mj, Hamada:2012bp}. VC includes the contributions from the infrared degrees of freedom of the theory and does not carry any special information about the ultraviolet divergences. In SM, it suggests the combination 
\begp
\allowdisplaybreaks \beq
6 \lambda_1+\frac{9}{4} g_2^2+\frac{3}{4} g_1^2 -12 y_t^2 = 0\nn\, .
\eeq
\eegp
Due to the large negative contribution from the term containing the top Yukawa coupling, it is not possible to satisfy VC till $\mpl$ given the experimental measurements of $M_t$ and $M_h$ within the context of SM.

In the ID model, as more scalars are added a possibility opens up to satisfy VC, as their contributions can offset the large negative contribution from the top quark. 

The above VC for the SM associated with $m_{11}$ is promoted in the ID model to~\cite{Barbieri:2006dq, Chakraborty:2014oma}
\begp
\allowdisplaybreaks \beq
6 \lambda_1+ 2\lambda_3+\lambda_4+\frac{9}{4} g_2^2+\frac{3}{4} g_1^2 -12 y_t^2  = 0.
\eeq
\eegp
If $2\lambda_3+\lambda_4$ is positive, then it has been checked with the RG improved coupling constants, and it is possible to satisfy the above VC at a scale before $\mpl$. 

However, in the ID model $m_{22}$ also receives quadratically divergent radiative corrections. The corresponding VC reads as
\begp
\allowdisplaybreaks \beq
6 \lambda_2+ 2\lambda_3+\lambda_4+\frac{9}{4} g_2^2+\frac{3}{4} g_1^2 = 0\, .
\eeq
\eegp
Note that it lacks the Yukawa contribution as the unbroken $Z_2$ forbids fermionic interactions of the inert doublet $\Phi_2$. As $2\lambda_3+\lambda_4$ is already positive, this VC can be satisfied if $\lambda_2$ is negative. But a negative $\lambda_2$ renders the potential unbounded from below as evident from the earlier discussions. Note that amongst our RG improved coupling constants $\lambda_1$ can be driven to negative values at high scales. But this makes the required cancellations for VCs even worse. Hence, it is not possible to satisfy Veltman's conditions in a scenario where only the ID model reigns the entire energy regime up to the $\mpl$.

\section{Triplet ($Y=0$) scalar extension of the SM}
\label{sec:ITM}
In this case, the SM is extended by an additional hyperchargeless ${ SU(2)_L}$ scalar triplet. This triplet is odd under a discrete $Z_2$ symmetry, whereas, the standard model fields are even. Also this symmetry prohibits the inert triplet to acquire a vacuum expectation value.

The tree-level scalar potential of eqn.~\ref{ScalarpotHTM} with  additional $Z_2$-symmetry, can be written\footnote{Here $T$ is denoted as the scalar triplet, which is equivalent to $\widetilde{\Phi}$ of eqn.~\ref{TripPhiY0} and the $\lambda_i$'s are equivalent to $\widetilde{\lambda}_i$ of the eqn.~\ref{ScalarpotHTM}.} as, 
\begp
\allowdisplaybreaks \bea
V_0(\Phi,T) &=& \mu_1^2 |\Phi|^2 + \lambda_1 |\Phi|^4+ \frac{\mu_2^2}{2} ~tr|T|^2 + \frac{\lambda_2}{4} ~tr|T|^4+ \frac{\lambda_3}{2} |\Phi|^2 tr|T|^2   \, ,
\label{ScalarpotHTM1}
\eea
\eegp
where the SM Higgs doublet $\Phi$ and the inert scalar triplet $T$ are given by,
\begp
\allowdisplaybreaks \beq
	\Phi ~=~ \left( \begin{array}{c} G^+ \\ \frac{1}{\sqrt{2}}\left(v+h+i G^0\right) \end{array} \right),
	\qquad
	T ~=~ \left(
\begin{array}{cc}
   H/\sqrt{2}  &   -H^+ \\
     -H^- &  -H/\sqrt{2} 
\end{array}\right),
\eeq
\eegp
$T$ contains a neutral scalar $H$ and a pair of charged scalar fields $H^\pm$. The $Z_2$ symmetry prohibits these particles to decay entirely to SM particles. The neutral particle $H$ can then serve as a DM candidate. 

After electroweak symmetry breaking the scalar potential can be written as,
\begp
\allowdisplaybreaks \bea
V(h, H,H^\pm) &=&  \frac{1}{4} \left[ 2 \mu_1^2 (h+v)^2 + \lambda_1 (h+v)^4 +2 \mu_2^2 (H^2+2 H^+ H^-) \right. \nn \\
&& \left. + \lambda_2 (H^2 + 2 H^+ H^-)^2  + \lambda_3 (h+v)^2 (H^2+2 H^+ H^-) \right].\nn
\eea
\eegp
Masses of these scalars are given by,
\begp
\allowdisplaybreaks \begin{align}
	M_{h}^2 &= \mu_1^2 + 3 \lambda_1 v^2,\nn \\
	M_{H}^2 &= \mu_2^2 +  \frac{\lambda_3}{2} v^2, \nn \\
	M_{H^\pm}^2 &= \mu_2^2 + \frac{\lambda_3}{2}  v^2  \,\nn .
	\label{MassTrip}
\end{align}
\eegp
At the tree level both the neutral and charged particle masses are the same. If we include one-loop radiative correction then the charged particles become slightly heavier~\cite{Cirelli:2009uv,Cirelli:2005uq} than neutral ones.
\begp
\allowdisplaybreaks \beq
\Delta M=(M_{H^\pm}-M_{H})_{1\text{-}loop}=\frac{\alpha M_{H}}{4\pi s_W^2}\Big[f(\frac{M_W}{M_{H}}) -c_W^2 f(\frac{M_Z}{M_{H}})\Big]\nn
\label{massdifftrip}
\eeq
\eegp
where,
\begp
\allowdisplaybreaks \beq
f(x)=-\frac{x}{4}\Big[ 2 x^3 ~ {\rm log}(x)+(x^2-4)^{\frac{3}{2}}~ {\rm log}\Big[ \frac{1}{2}(x^2-2-x\sqrt{x^2-4}) \Big]\Big].\nn
\eeq
\eegp
One-loop effective Higgs potential in $\MS$ scheme in the Landau gauge is given by   
\begp
\allowdisplaybreaks \beq
V_1^{{\rm SM}+{\rm IT}}(h)= V_1^{\rm SM}(h) + V_1^{\rm IT}(h)
\eeq
\eegp
$V_1^{\rm SM}(h)$ can be found in eqn.~\ref{V1loop}. The additional contribution to the one-loop effective potential due to the inert triplet is given by~\cite{Forshaw:2003kh},
\begp
\allowdisplaybreaks \beq
V_1^{\rm IT}(h)= \sum_{j=H,H^+,H^-} \frac{1}{64 \pi^2} M_j^4(h) \left[ \ln\left(\frac{M_j^2(h)}{\mu^2(t)} \right)- \frac{3}{2} \right] 
\eeq
\eegp
where, 
\begp
\allowdisplaybreaks \beq
M_j^2(h)=\frac{1}{2} \,\lambda_{j}(t) \, h^2(t)+\mu_2^2(t) 
\eeq
\eegp
with $\lambda_{H,H^\pm}(t)= \lambda_{3}(t)$.
In the present work, in the Higgs effective potential, SM contributions are taken at two-loop level and the IT scalar contributions are considered at one-loop only. 

Similarly for large field value $h\gg v$, the Higgs effective potential can be approximated as
\begp
\allowdisplaybreaks \beq
V_{\rm eff}^{{\rm SM}+{\rm IT}}(h) \simeq \lambda_{\rm 1,eff}(h) \frac{h^4}{4}\, ,
\label{efflamIT}
\eeq
\eegp
with
\begp
\allowdisplaybreaks \beq
\lambda_{\rm 1,eff}(h) = \lambda_{\rm 1,eff}^{\rm SM}(h) +\lambda_{\rm 1,eff}^{\rm IT}(h)\, ,
\label{lameffIT}
\eeq
\eegp
where $\lambda_{\rm eff}^{\rm SM}(\phi)$ same as in eqn.~\ref{eq:effqurtic} and,
\begp
\allowdisplaybreaks \bea
 \lambda_{\rm 1,eff}^{\rm IT}(h)&=& e^{4\Gamma(h)} \left[\frac{3 \lambda_3^2}{256 \pi^2}  \left(\ln\left(\frac{\lambda_3}{2}\right)-\frac{3}{2}\right ) \right]\, . \label{efflamtriplet}
\eea
\eegp
 \begin{figure}[h!]
 \begin{center}
 \includegraphics[width=2.7in,height=2.7in, angle=0]{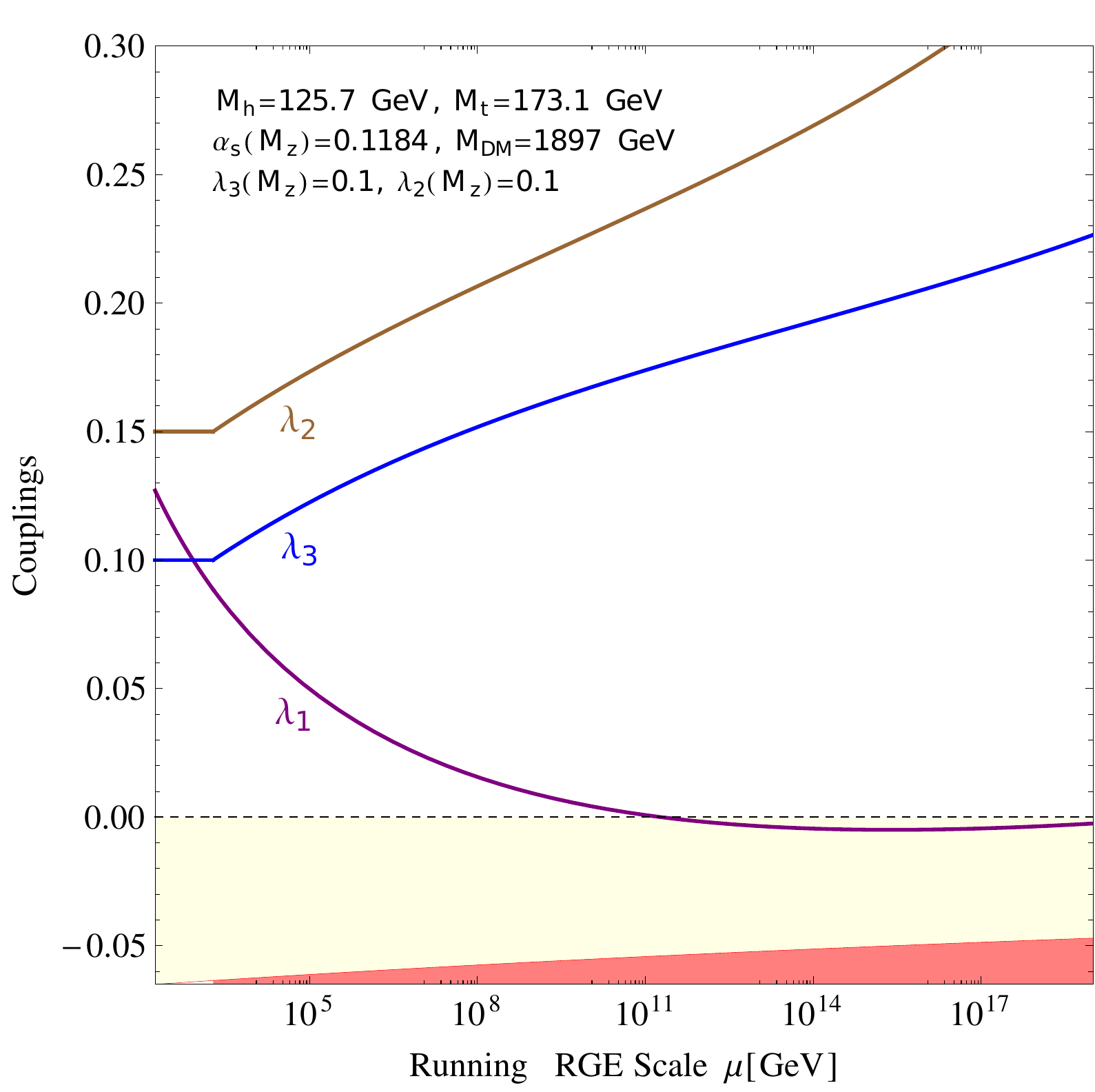}
 \caption{\label{fig:SMInert} \textit{{\rm ITM} RG evolution of the couplings $\lambda_1, \lambda_2, \lambda_3$ for the set of parameters in Table~\ref{table1IT}. } }
 \end{center}
 \end{figure}
\setlength\tablinesep{2pt}
\begin{table}[h!]
\begin{center}
    \begin{tabular}{| c | c | c | c | c | c | c | c |}
    \hline
      & $\lambda_1$ & $\lambda_2$ &  $\lambda_3$\\
\hline
   ~$M_{t}$  ~&~ 0.127054 ~&~ 0.10 ~&~ 0.10~\\
            \hline
   $\mpl$ & $-$0.00339962 & 0.267706 & 0.206306\\
              \hline
    \end{tabular}
    \caption{A set of values of all ITM coupling constants at  $M_t$ and $\mpl$ for $M_{DM}=1897$ GeV.}
    \label{table1IT}
\end{center}
\end{table}
To compute the RG evolution of all the couplings in these model, we first calculate all couplings with threshold corrections at $M_t$ and using the RG-eqns.~\ref{betal_HTM}$-$\ref{betal_3HTM}, one can obtain the couplings $\lambda_{1,2,3}$ at $\mpl$.
For a specific set of values of $\lambda_i$ at $M_t=173.1$~GeV, their values at $\mpl = 1.2 \times 10^{19}$~GeV for $M_h=125.7$~GeV and $\alpha_s\left(M_Z\right)=0.1184$ have been provided in Table~\ref{table1IT}. In Fig.~\ref{fig:SMInert} we explicitly show running of the scalar couplings $(\lambda_i)$ for this set of parameters. It has been found that for the specific choice of parameters, $\lambda_1$ assumes a small negative value leading to a metastable EW vacuum as discussed in the previous sections.
\subsection{Constraints on ITM}
The absolute stability of the EW vacuum of the scalar potential (eqn.~\ref{ScalarpotHTM1}), unitarity of scattering matrix, perturbativity and the EW precision measurements puts a stringent bound on the ITM parameter space which had been shown in Chapter~\ref{chap:EWSBextended}.

Constrains from LHC diphoton signal strength and dark matter will be discussed in the following subsections.
\subsubsection{Bounds from LHC diphoton signal strength}
The Higgs to diphoton signal strength $\mu_{\gamma\gamma}$ can be defined similar to eqn.~\ref{mugagalow}. If the IT particles have masses greater than $M_h/2$, i.e., $\Gamma\left(h\rightarrow \rm IT,IT\right)=0$ then,
\begp
\allowdisplaybreaks \beq 
 \mu_{\gamma\gamma}= \frac{\Gamma(h\rightarrow \gamma\gamma)_{\rm IT}}{\Gamma(h\rightarrow \gamma\gamma)_{\rm SM}}\, .
 \label{mugagahighIT}
\eeq
\eegp
In ITM, the additional contributions to $\mu_{\gamma\gamma}$ at one-loop due to the $H^\pm$ can be found in eqn.~\ref{hgaga}, where $\mu_{hH^+H^-}= \lambda_3 v$ stands for the coupling constant of $hH^+H^-$ vertex.
 One can see that a positive $\lambda_3$ leads to a destructive interference between SM and IT contributions in eqn.~\ref{hgaga} and {\it vice versa}. Hence, for IT particles heavier than $M_h/2$, $\mu_{\gamma\gamma}<1$ ($\mu_{\gamma\gamma}>1$) when $\lambda_3$ is positive (negative).
One can see from the eqn.~\ref{hgaga}, the contribution to the Higgs diphoton channel is proportional to $\frac{\lambda_3}{M_{H^\pm}^2}$.
It has been seen, if the charged scalar mass is greater than $300$ GeV then the contributions of inert triplet model to the diphoton signal is negligible.
\subsubsection{Constraints from dark matter relic density}
\begin{figure}[h!]
\begin{center}
{
\includegraphics[width=2.7in,height=2.7in, angle=0]{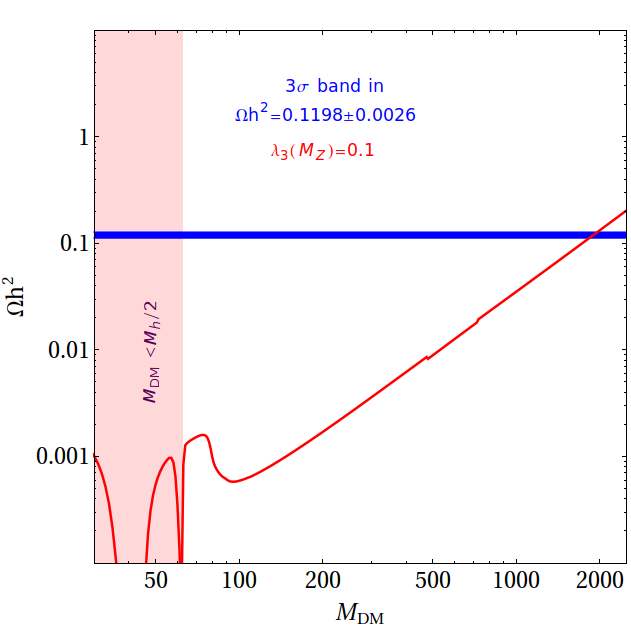}}
\caption{ \textit{Dark matter relic density $\Omega h^2$ as a function of the dark matter mass $M_{DM}(\equiv M_H)$ for the portal coupling: $\lambda_3(M_Z)=0.10$ (red) in {\rm ITM}. The thin blue band corresponds the relic density, $\Omega h^2=0.1198 \pm 0.0026$ (3$\sigma$).}}
\label{fig:relictriplet}
\end{center}
\end{figure}
In Fig.~\ref{fig:relictriplet}, we have plotted the relic density as a function of dark matter mass for ITM. Here in this plot, the Higgs portal coupling $\lambda_3(M_Z)=0.10$ has been used. The light red band is excluded from the Higgs invisible decay width data from the LHC. It has been seen that from eqn.~\ref{massdifftrip}, the mass difference between the neutral and charged particles is $\sim$150 MeV~\cite{Cirelli:2009uv,Cirelli:2005uq}. The co-annihilation cross-section of the dark matter $H$ with charged ($H^\pm$) particles are very large. It has been seen that for 500 GeV, the total cross-section is $\left\langle \sigma v \right\rangle \sim 10^{-25}$ ${\rm cm^3 s^{-1}}$ and so the relic density becomes $\sim 0.01$ which corresponds to under-abundance. For dark matter mass greater than $1.8$ TeV, one can get relic density of the dark matter in the right ballpark. 

\subsection{Tunneling Probability and Metastability in ITM}\label{TUNHTM}
\label{sec:metastabilityIT}
 \begin{figure}[h!]
 \begin{center}
\subfigure[]{
 \includegraphics[width=2.7in,height=2.7in, angle=0]{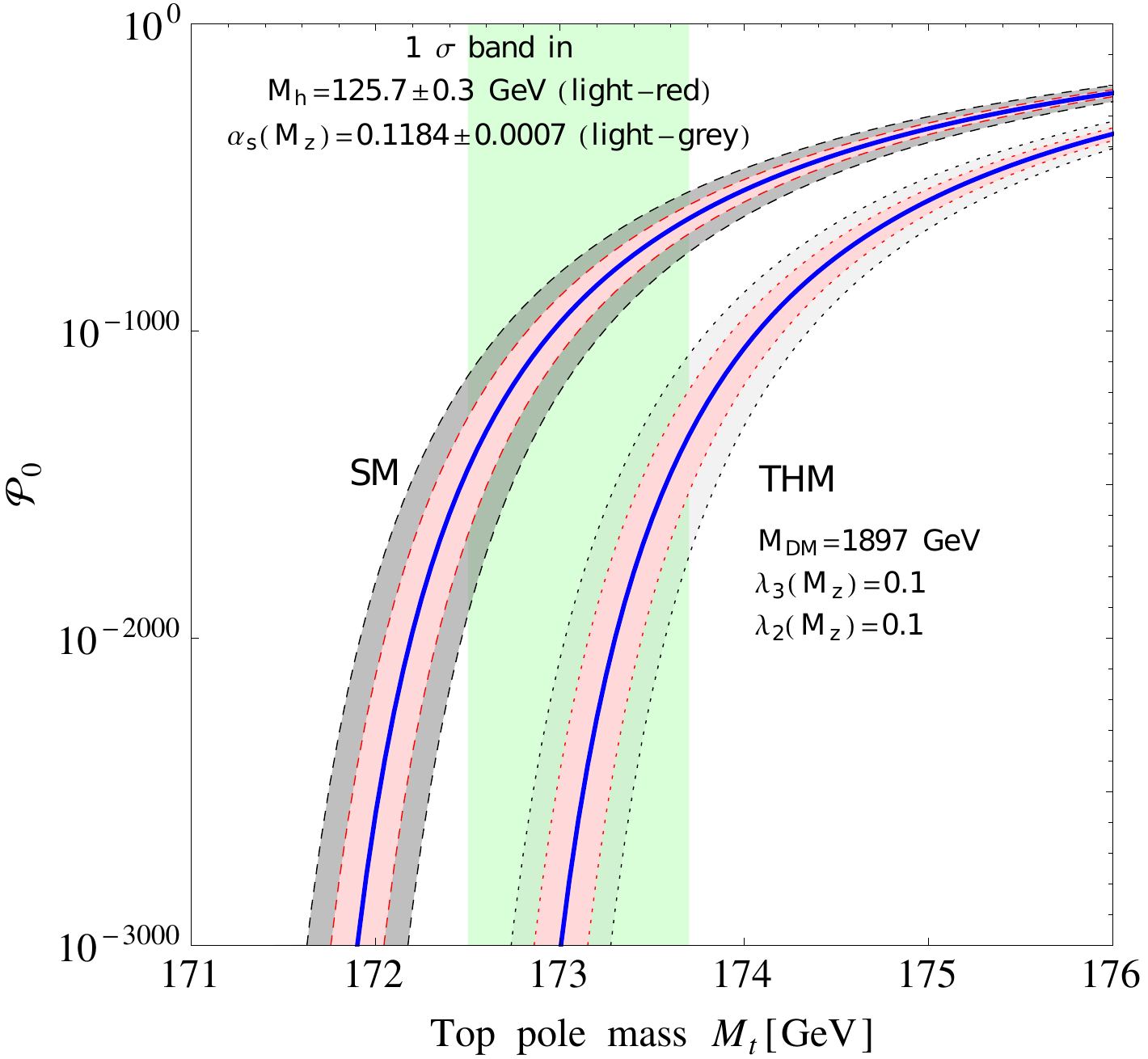}}
 \hskip 15pt
 \subfigure[]{
 \includegraphics[width=2.7in,height=2.7in, angle=0]{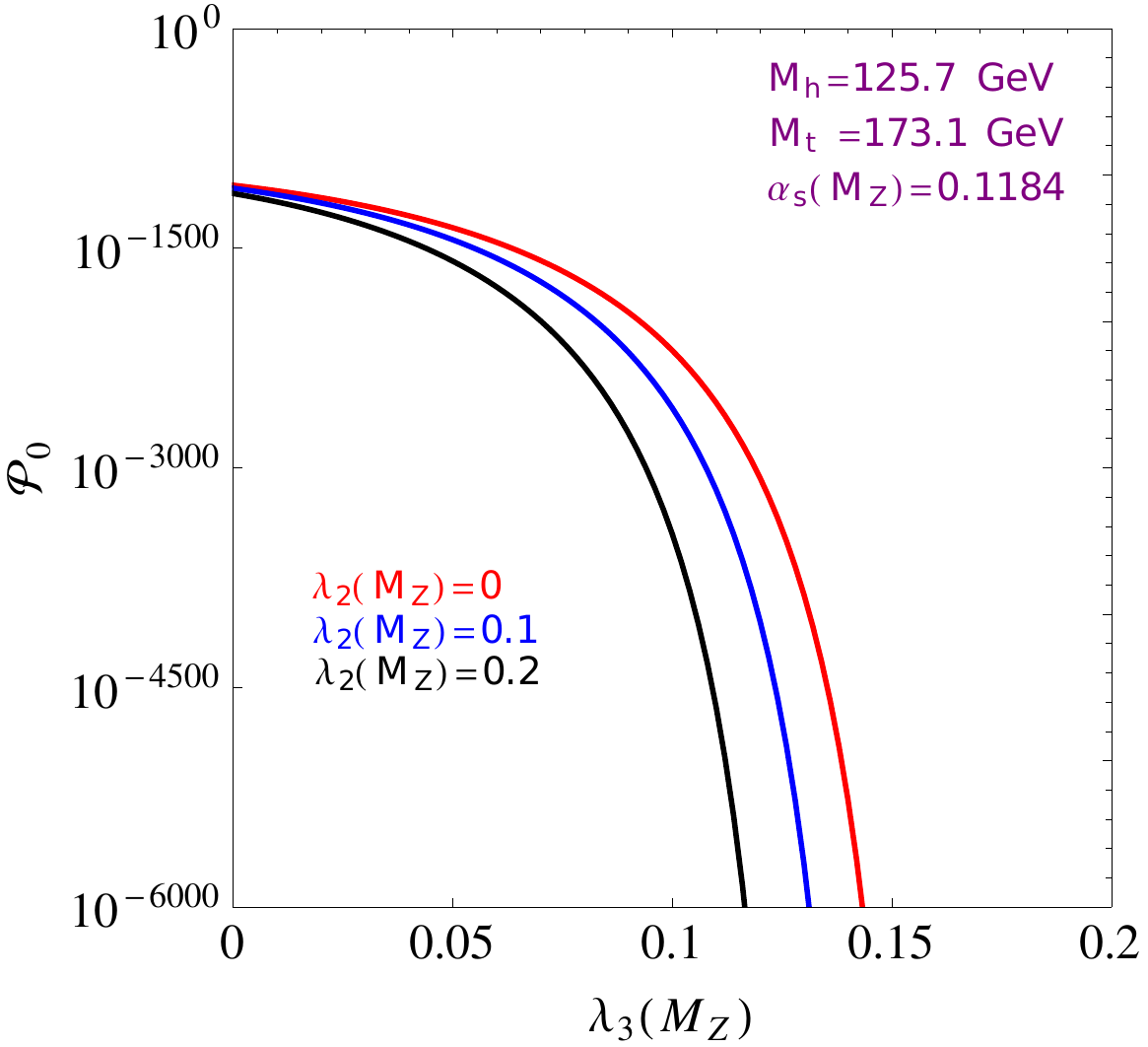}}
 \caption{\label{fig:TunIT} \textit{ (a) Tunneling probability ${\cal P}_0$ dependence on  $M_t$. The left band (between dashed lines) corresponds to {\rm SM}. The right one (between dotted lines) is for $IT$ model for DM mass $M_H=1897$~GeV. Dark matter constraints are respected for these specific choices of parameters. Light-green band stands for $M_t$ at $\pm 1\sigma$. (b) ${\cal P}_0$ is plotted against Higgs dark matter coupling $\lambda_S(M_Z)$ for different values of $\lambda_2(M_Z)$.}}
 \end{center}
 \end{figure}
The tunneling probability ${\cal P}_0$ (see eqn.~\ref{prob}) is computed by putting the minimum value of $\lambda_{1,\rm eff}$ of eqn.~\ref{lameffIT} in eqn.~\ref{Action}. In Fig.~\ref{fig:TunIT}$\left(a\right)$, we have plotted  ${\cal P}_0$ as a function of  $M_t$. The right band corresponds to the tunneling probability for our benchmark point as in Table~\ref{table1IT}. For comparison, we plot ${\cal P}_0$ for SM as the left band in Fig.~\ref{fig:TunIT}$(a)$. 1$\sigma$ error bands in $\alpha_s$ and $M_h$ are also shown. We plot ${\cal P}_0$ as a function of  $\lambda_3(M_Z)$ in Fig.~\ref{fig:TunIT}$\left(b\right)$ for different choices of $\lambda_2(M_Z)$, assuming  $M_h=125.7$~GeV, $M_t=173.1$~GeV and $\alpha_s=0.1184$. Here DM mass $M_{DM}$ is also varied with $\lambda_3$ to get  $\Omega h^2=0.1198$.

The modified vacuum stability conditions are, 
\begin{itemize}
\item
If $0>\lambda_1(\Lambda_B)>\lambda_{\rm 1,min}(\Lambda_B)$, then the vacuum is metastable. 
\item
If $\lambda_1(\Lambda_B)<\lambda_{\rm 1,min}(\Lambda_B)$, then the vacuum is unstable. 
\item
If $\lambda_2<0$, the potential is unbounded from below along the $H$ and $H^\pm$-direction. 
\item
If $\lambda_3(\Lambda_{\rm I})<0$, the potential is unbounded from below along a direction in between $H$ and $h$ also $H^\pm$ and $h$.
\end{itemize}
In the above, $\Lambda_{\rm I}$ represents any energy scale for which $\lambda_1$ is negative. 
\subsection{Phase diagrams in ITM}
 \begin{figure}[h!]
 \begin{center}
 \subfigure[]{\includegraphics[width=2.7in,height=2.7in, angle=0]{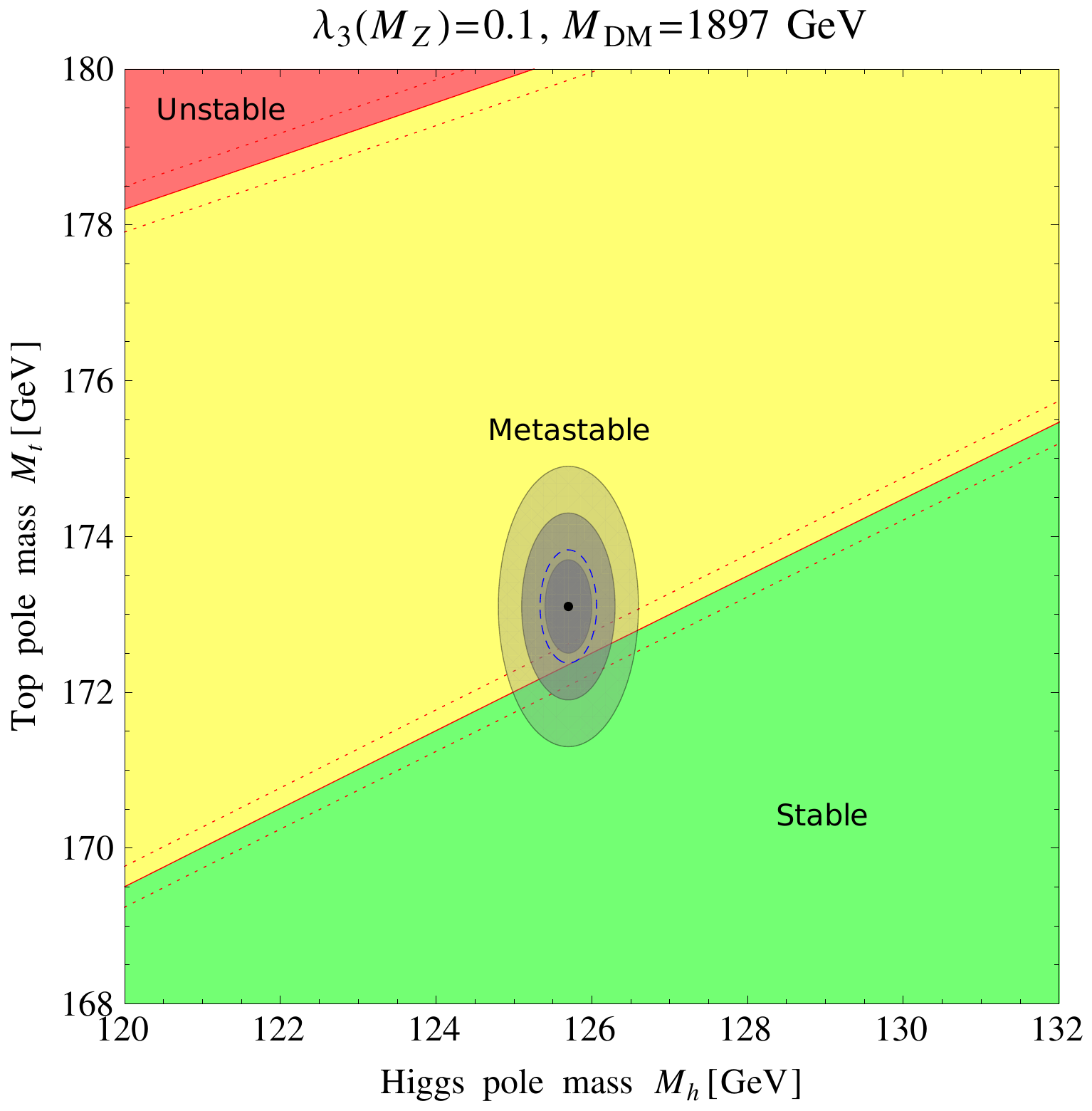}}
 \hskip 15pt
  \subfigure[]{
 \includegraphics[width=2.7in,height=2.7in, angle=0]{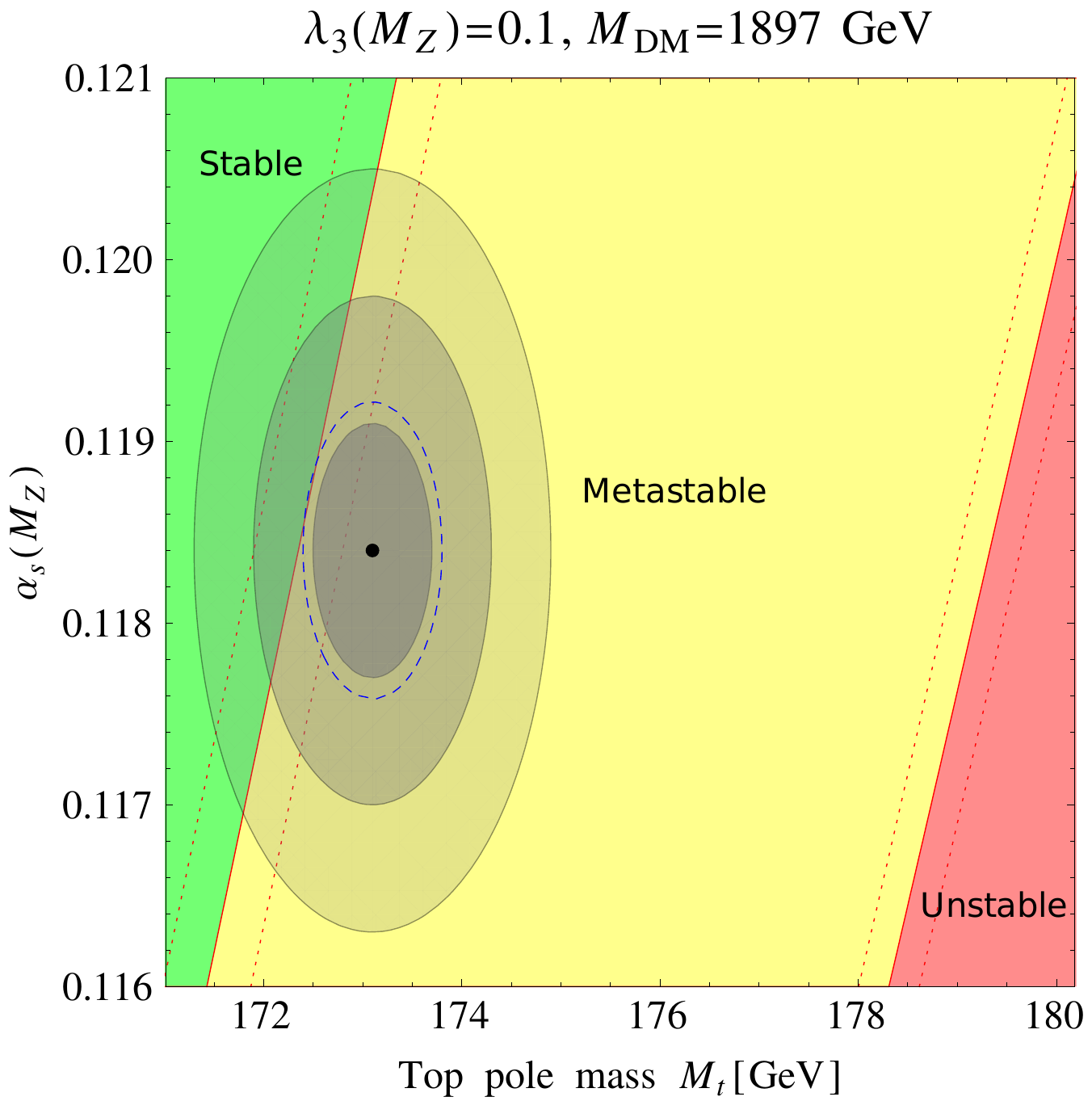}}
 \caption{\label{fig:Mt_MhIT} \textit{Phase diagrams in (a) $M_h - M_t$ plane and (b) $M_t-\alpha_s(M_Z)$ plane {\rm ITM}. Regions of 
absolute stability~(green), metastability~(yellow), instability~(red) of the EW vacuum are also marked. The gray zones represent error ellipses at $1$, $2$ and  $3\sigma$. The three boundary lines (dotted, solid and dotted red) correspond to $\alpha_s(M_Z)=0.1184 \pm 0.0007$.
} }
 \end{center}
 \end{figure} 
To explain the impact of inert triplet scalars to uplift the EW vacuum metastability, in Fig.~\ref{fig:Mt_MhIT} phase diagram in $M_t-M_h$ and $\alpha_s(M_Z)-M_t$ plane has been presented for the benchmark points $M_{DM}=1897$ GeV, $\lambda	_2(M_Z)=0.10$ and $\lambda_3(M_Z)=0.10$. For this points one can see from the phase diagram in Fig.~\ref{fig:Mt_MhIT} that the stability electroweak is excluded at 1.2$\sigma$ (one-sided).
 \begin{figure}[h!]
 \begin{center}{
 \includegraphics[width=2.7in,height=2.7in, angle=0]{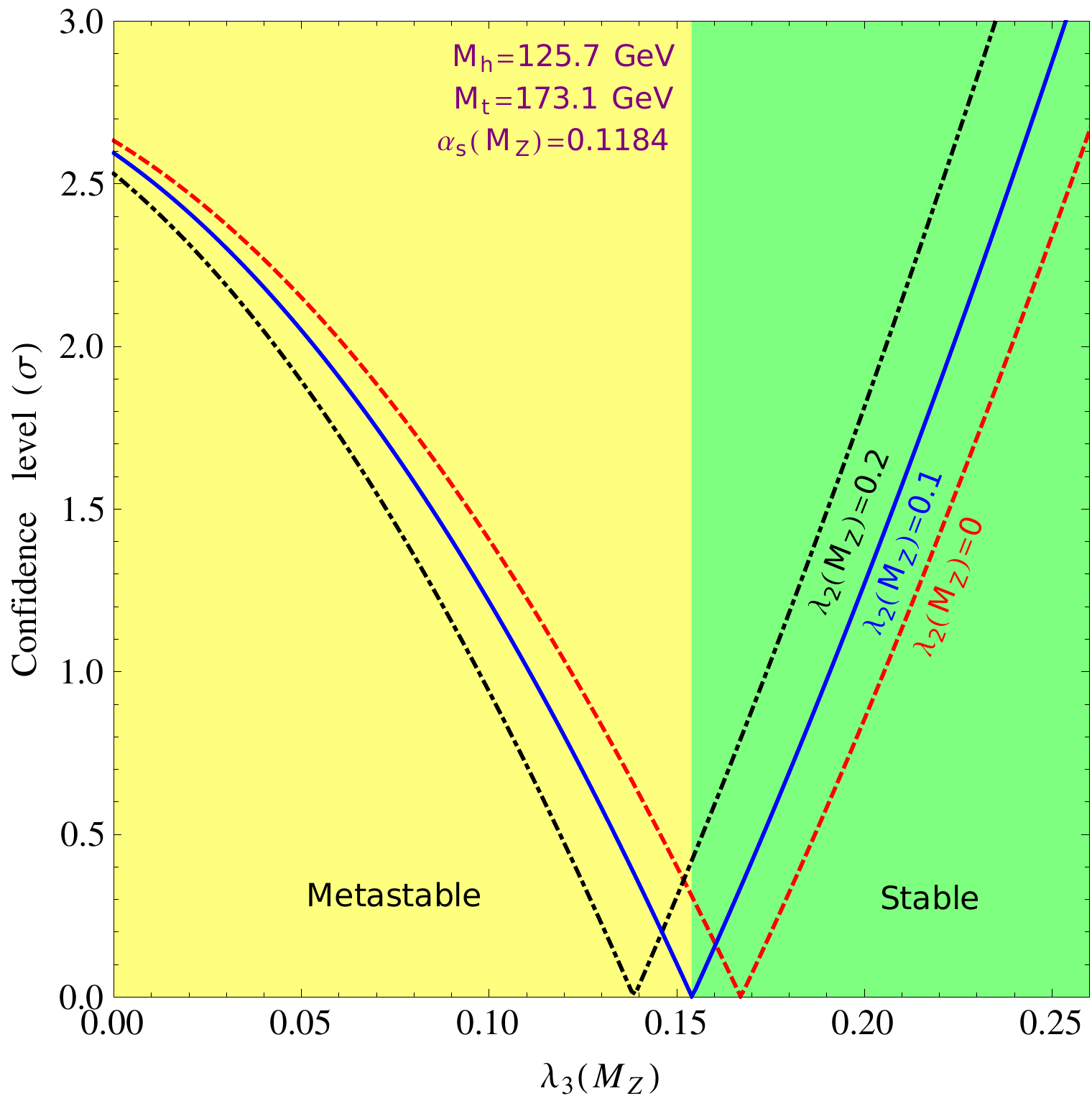}}
 \caption{\label{fig:confidenceIT} \textit{Dependence of confidence level  at which EW vacuum stability is excluded (one-sided) or allowed on $\lambda_3(M_Z)$ and $\lambda_2(M_Z)$ in {\rm ITM}. Regions of absolute stability (green) and metastability (yellow) of EW vacuum are shown for $\lambda_2(M_Z)=0.1$.
 } }
 \end{center}
 \end{figure}

The confidence level vs $\lambda_3(M_Z)$ diagram like Fig.~\ref{fig:kappaconfidence} of SM+$S$ and Fig.~\ref{fig:confidenceIDM} of IDM is presented in Fig.~\ref{fig:confidenceIT} for the inert triplet model.
If the ITM is valid up to the Planck scale, which may saturate the relic abundance of the dark matter of the Universe then this phase diagram becomes important to realize where the present EW vacuum is residing. Along the line (black, blue and red) in the Fig.~\ref{fig:confidenceIT} the dark matter mass change in such a way that the relic density $\Omega h^2 = 0.1198$ remains same.
One can see that with the increase of $\lambda_{2,3}(M_Z)$, the EW vacuum approaches the stability. In this model, the electroweak vacuum becomes absolutely stable after $\lambda_3(M_Z)= 0.154$ for $\lambda_2(M_Z)\approx 0.10$ (see blue line in the Fig.~\ref{fig:confidenceIT}).
This phase diagram has been presented for central values of $M_h$, $M_t$ and $\alpha_s(M_Z)$. However, if we increase the top mass or decrease the Higgs mass or decrease $\alpha_s(M_Z)$ within experimental uncertainties then 
the size of the region corresponding to the metastability of the EW vacuum increases.
With a maximum top mass $M_t=174.9$ GeV and a minimum $M_h=124.8$ GeV and a minimum $\alpha_s(M_Z)=0.1163$, allowed at 3$\sigma$, the EW vacuum the Higgs potential becomes absolutely stable for the dark matter mass more than 1912 GeV with $\lambda_{3}(M_Z)$ greater than $0.31$ for fixed $\lambda_2(M_Z)=0.1$.

In Fig.~\ref{fig:Rgammagamma}, we have shown the valid parameter spaces in $\lambda_3(M_Z)-M_H$ plane for central value of $M_t$, $M_h$ and $\alpha(M_Z)$. Here the lower (red) region are excluded, as the scalar potential becomes unbounded from below along the direction in between $H^\pm$ and $h$.
In this region the effective Higgs quartic coupling is negative and form a local minima along the Higgs $h$ direction near the Planck scale and at the same time $\lambda_3$ remains negative up to the Planck scale. It has been found that the parameter spaces with negative $\lambda_3(M_Z)$ is also allowed from the metastability.
In this case $\lambda_3$ becomes positive at the scale $\Lambda_{B}$ of global minimum and remains positive up to the Planck scale. The green region implies that the EW vacuum is absolutely stable. In the upper red region the unitary bounds are violated. In the plot, the right-side of the black dotted line are viable from $\mu_{\gamma\gamma}$ at 1$\sigma$.
 \begin{figure}[h!]
 \begin{center}
 \includegraphics[width=3.0in,height=2.7in, angle=0]{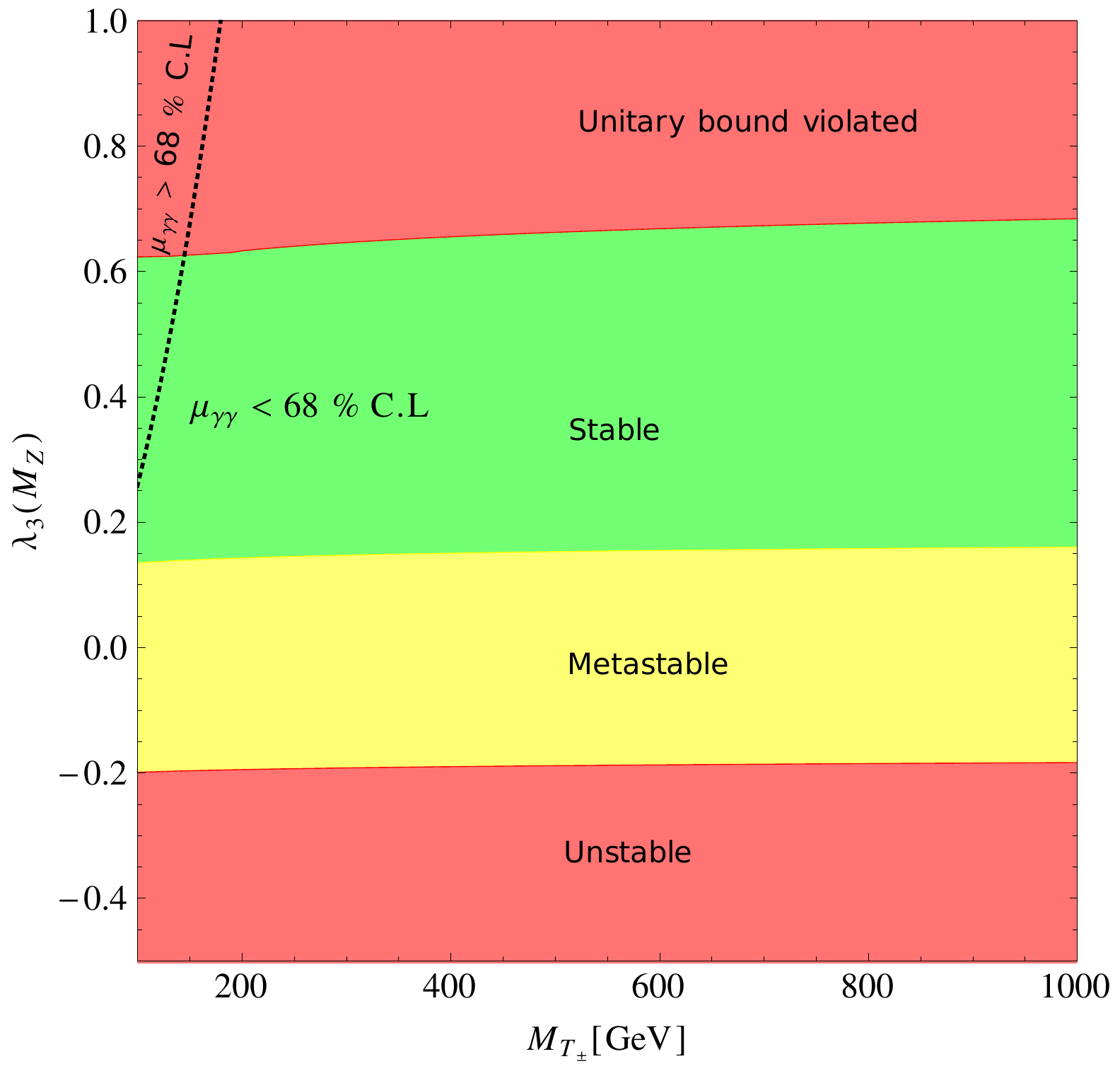}
 \caption{\label{fig:Rgammagamma} \textit{Phase diagram in $\lambda_3(M_Z)-M_H$ plane in {ITM}. Right side of the black-dotted line is allowed from the signal strength ratio of $\mu_{\gamma\gamma}$ within 68$\%$ confidence level. and the left side is excluded at 1$\sigma$. In the metastable region, the Higgs portal coupling $\lambda_3(M_Z)$ is negative, however, beyond the scale $\Lambda_{B}$ it is greater than zero.} }
 \end{center}
 \end{figure}

\section{Summary}
According to the standard model of particle physics with the present measured values of the SM parameters, the electroweak vacuum is lying right in between stability and instability, as if it is ready to tunnel into a regime of absolute instability.
Nevertheless, the transition time required for this is safely beyond the present lifetime of the Universe. Still, the question is, what prompts such a near-criticality? In this chapter, we did not try to find an answer to this question. But we explored the validity of this question in the context of the different models containing extra scalar field(s) that also offers a solution to the dark matter puzzle.

Near-criticality is best explained with the help of phase
diagrams which were used in the Chapter~\ref{chap:MetaSM} for the SM.
A similar endeavor has been made in this chapter for various kinds of extended scalar sectors namely SM+$S$, inert doublet and inert triplet models.
The next-to-next-to-leading order corrections have been included in the SM calculations. The
effects due to the extra scalar(s) were incorporated up to two-loop.

In this chapter, various kinds of dark matter models have been chosen to illustrate changes in EW vacuum stability as apart from neutrino masses, the presence of dark matter in the Universe is the most striking signature of new physics beyond the standard model of particle physics.

In these models, for some specific choice of parameter space, the scalar field(s) can rescue the EW vacuum from metastability, making it absolutely stable, so that $\lambda$ never turns negative.
But as the DM direct detection experiments or collider searches are yet to confirm the exact nature of the DM candidate, the parameter space allowed by DM relic density constraints, dictated by the cosmic microwave background radiation experiments such as Planck or WMAP, have been considered.
It has been checked that in these considerations, the related DM-nucleon cross-sections are beyond the present sensitivity of direct-detection experiments such as XENON100 and LUX.

The addition of scalar multiplet does not help in realizing the asymptotic safety scenario of gravity. But it is possible to have two degenerate vacua at somewhat lower energy (shown for the SM+$S$ model), which depends on the parameter space under consideration.

In short, near-criticality of EW vacuum indicates the presence of new dynamics other than the SM at a very high energy.
In this chapter, new scalars physics have been introduced at EW scale to demonstrate the influence of such scalars in shaping the minimum (if any) of the potential lying close to $\mpl$.

\chapter{Summary and Conclusions}
\label{chap:Conclusions}
\linespread{0.1}
\pagestyle{headings}
\noindent\rule{15cm}{1.5pt} 

The Higgs signal strength data is consistent with the theoretical predictions of the SM with small uncertainties. The LHC is yet to find any convincing signal suggesting the existence of any new physics beyond the standard model of particle physics. There are several well-motivated possibilities for an enlarged scalar sector that go far beyond the minimalistic one doublet scenario of the standard model. Small uncertainty in Higgs signal data at LHC does allow physics beyond the SM. Other experimental evidences point towards the existence of dark matter, which so far could have escaped detection in colliders and DM direct detection experiments.

In this thesis, several extended scalar sectors of the SM have been discussed.
The SM has been extended with either a $SU(2)$ singlet or doublet or triplet with different hypercharges.
It has been considered that the extra scalar fields transform under same standard model gauge group.

This work has two aspects. In the first case, it has been considered that both the SM doublet and the extra scalar field are responsible for the electroweak symmetry breaking, i.e., both the neutral $CP$-even component of SM doublet and extra scalar fields are getting vacuum expectation values.
In other words, the minima of the combined scalar potential form on the plane containing the $CP$-even components. These components mix and form a Higgs-like particle and extra heavy scalar particles.
The different extended scalar sectors contain different kinds of scalar particles such as charged, or neutral $CP$-even and $CP$-odd scalar(s).
These scalars can couple to the vector bosons. In this point of view, various kinds of vector boson scattering processes have been calculated for the
three models such as $(i)$ type-II two Higgs doublet model, $(ii)$ Higgs triplet model with a hyperchargeless scalar triplet, and $(iii)$ Higgs triplet model with a
scalar triplet with hypercharge $Y = 2$.
The exact expressions of longitudinal polarization vectors have been used to determine the scattering cross-sections.
Generally, the idea of vector boson scattering has been used for a better understanding of EWSB.
A generic expression of the amplitude of the vector boson scattering has been given for these models.
From this analysis, it is revealed that away from resonance the cross-sections are not significantly different from that of the SM. This is quite expected because only those parameter spaces have been chosen which satisfy all the existing constraints.
One can differentiate these three models from the standard model and between one another, looking for the resonance peaks in different modes of vector boson scattering. In future, if we observe any signature of these new scalars in the future collider experiments, then this study will help in revealing the characteristics of the extended scalar sector.

If the standard model is valid up to the Planck scale, the present measurements on the masses of the top quark and Higgs indicate the presence of a deeper minimum of the scalar potential at a very high energy scale, threatening the stability of the present electroweak vacuum. State of the art NNLO calculations performed to evaluate the probability that the present EW vacuum will tunnel into the deeper vacuum lying close to $\mpl$ suggest that the present EW vacuum is metastable at $\sim$3$\sigma$. As a part of this thesis, this has been reproduced. The lack of stability might be the artifact of the incompleteness of the SM.

It is important to look into the problem of EW stability in a scenario which addresses the issue of DM as well. In this case, at a time an extended scalar sector has been taken which provides a viable dark matter candidate. In the presence of these new scalar sectors, the detailed study of the Higgs potential has been discussed.
Assuming these models valid up to the $\mpl$ scale and allowing the metastability of the electroweak vacuum, new viable parameter spaces have been found.
As new extra scalar sector introduces a few new parameters and fields, the study of the parameter space is quite involved when one considers radiatively improved scalar potentials containing SM NNLO corrections and two-loop new sector contributions.
Inclusion of these NNLO corrections is mandatory to reproduce the correct confidence level at which EW vacuum is metastable in the SM.
In these models, the formation of the extra minimum near the $\mpl$ scale have been described. The detailed calculation of transition probability from the EW minimum to the new minimum at the Planck scale has also been shown.
In the presence of new scalar sectors, the conditions for absolute stability of the EW vacuum have been reviewed in this work. If one allows the metastable EW vacuum then new metastability conditions arise which have also been shown for these extended scalar sectors of the SM.

As extra scalar sectors can provide a viable dark matter candidate which may fulfill the relic abundance of the dark matter in the Universe, in this context, it is instructive to explore whether these extra scalars can also extend the lifetime of the Universe. This study will help in estimating the mean lifetime of the EW vacuum, especially if it still remains in the metastable state in models with extended scalar sectors.

\appendix
\graphicspath{{AppendixA/}}
\begin{appendices}
\chapter{ Analytical expressions for vector boson scattering \label{sec:wwscatdetails}}
\pagestyle{headings}
\pagenumbering{arabic}

\setcounter{equation}{0}  
\section{Amplitudes for different modes of $V_L V_L$ scattering}
\label{feynmanamp}

Let $p_1,~p_2$ are the four momenta of initial state gauge bosons and $k_1,~k_2$ are that for the final state gauge bosons. 
$\epsilon_\mu(p)$ be the polarization four vector of a gauge bosons $V(\equiv W^\pm,~Z$) with four momentum $p$. It can be written as, $\epsilon_\mu(p)\equiv\{\frac{|{\bold p}|}{M_V},\frac{E_V}{M_V}\hat{p}\}$, where $E_V=\sqrt{|{\bold p}|^2+M_V^2}$ is the energy of the gauge boson. Here, $M_V$ is the mass of $V$. We use the shorthand notations $\epsilon_1 \equiv \epsilon(p_1)$, $\epsilon_2 \equiv \epsilon(p_2)$, $ \epsilon_3 \equiv \epsilon(k_1) $, $ \epsilon_4 \equiv \epsilon(k_2) $, $c_W \equiv \cos{\theta_W}$ and $s_W \equiv \sin{\theta_W}$ and $x\equiv\cos\theta$, where $\theta$ is the scattering angle.
Mandelstam variables are defined as: $s=(p_1+p_2)^2$; $t=(p_1-k_1)^2$; $u=(p_1-k_2)^2$. 
\subsection{$W_L^+(p_1)~+~ W_L^-(p_2) \rightarrow W_L^+(k_1)~+~W_L^-(k_2)$}
Scattering amplitudes in terms of longitudinal polarization vectors and four momenta: 
\begp
\allowdisplaybreaks \bea
(a)&{\cal M}_{\rm p}&\hspace{-2ex} =~~ g_2^2 \lbrace 2 (\epsilon_1.\epsilon_3)( \epsilon_2.\epsilon_4)- (\epsilon_1.\epsilon_2)( \epsilon_3.\epsilon_4)-(\epsilon_1.\epsilon_4)( \epsilon_2.\epsilon_3)\rbrace.\nn\\
(b)&{\cal M}^{\gamma+{Z}}_{s} &\hspace{-2ex} =~- g_2^2 \left( \frac{s_W^2}{s} + \frac{c_W^2}{s-M_Z^2}\right)~\lbrace (p_1- p_2)^\mu (\epsilon_1.\epsilon_2)+2 (p_2.\epsilon_1)\epsilon_2^\mu -2 (p_1.\epsilon_2)\epsilon_1^\mu \rbrace\nn \\&&\hspace{4.6cm} \lbrace (k_2 - k_1)_\mu (\epsilon_3.\epsilon_4)-2 (k_2.\epsilon_3)\epsilon_{4\mu} -2 (k_1.\epsilon_4)\epsilon_{3\mu} \rbrace.\nn\\
(c)&{\cal M}^{\gamma+{Z}}_{t} &\hspace{-2ex} =~- g_2^2 \left( \frac{s_W^2}{t} + \frac{c_W^2}{t-M_Z^2}\right)~\lbrace (p_1 + k_1)^\mu (\epsilon_1.\epsilon_3)-2 (k_1.\epsilon_1)\epsilon_3^\mu -2 (p_1.\epsilon_3)\epsilon_1^\mu \rbrace\nn \\&&\hspace{4.6cm} \lbrace (p_2 + k_2)_\mu (\epsilon_2.\epsilon_4)-2 (k_2.\epsilon_2)\epsilon_{4\mu} -2 (p_2.\epsilon_4)\epsilon_{2\mu} \rbrace.\nn\\
 (d)&{\cal M}^{S}_{s} &\hspace{-2ex} =~ - \frac{(C~g_2 M_W)^2}{s-M_S^2} (\epsilon_1.\epsilon_2)(\epsilon_3.\epsilon_4).\nn\\
(e)&{\cal M}^{S}_{t} &\hspace{-2ex} =~  - \frac{(C~g_2 M_W)^2}{t-M_S^2} (\epsilon_1.\epsilon_3)(\epsilon_2.\epsilon_4).\nn\\
(f)&{\cal M}^{H^{++}}_{u} &\hspace{-2ex} =~ - \frac{(\hat{C}~g_2 M_W)^2}{u-M^2_{H^{++}}} (\epsilon_1.\epsilon_4)(\epsilon_2.\epsilon_3).\nn
\eea
\eegp
 \begin{figure}[h!]
 \begin{center}
 {
 \includegraphics[width=15cm,height=8cm, angle=0]{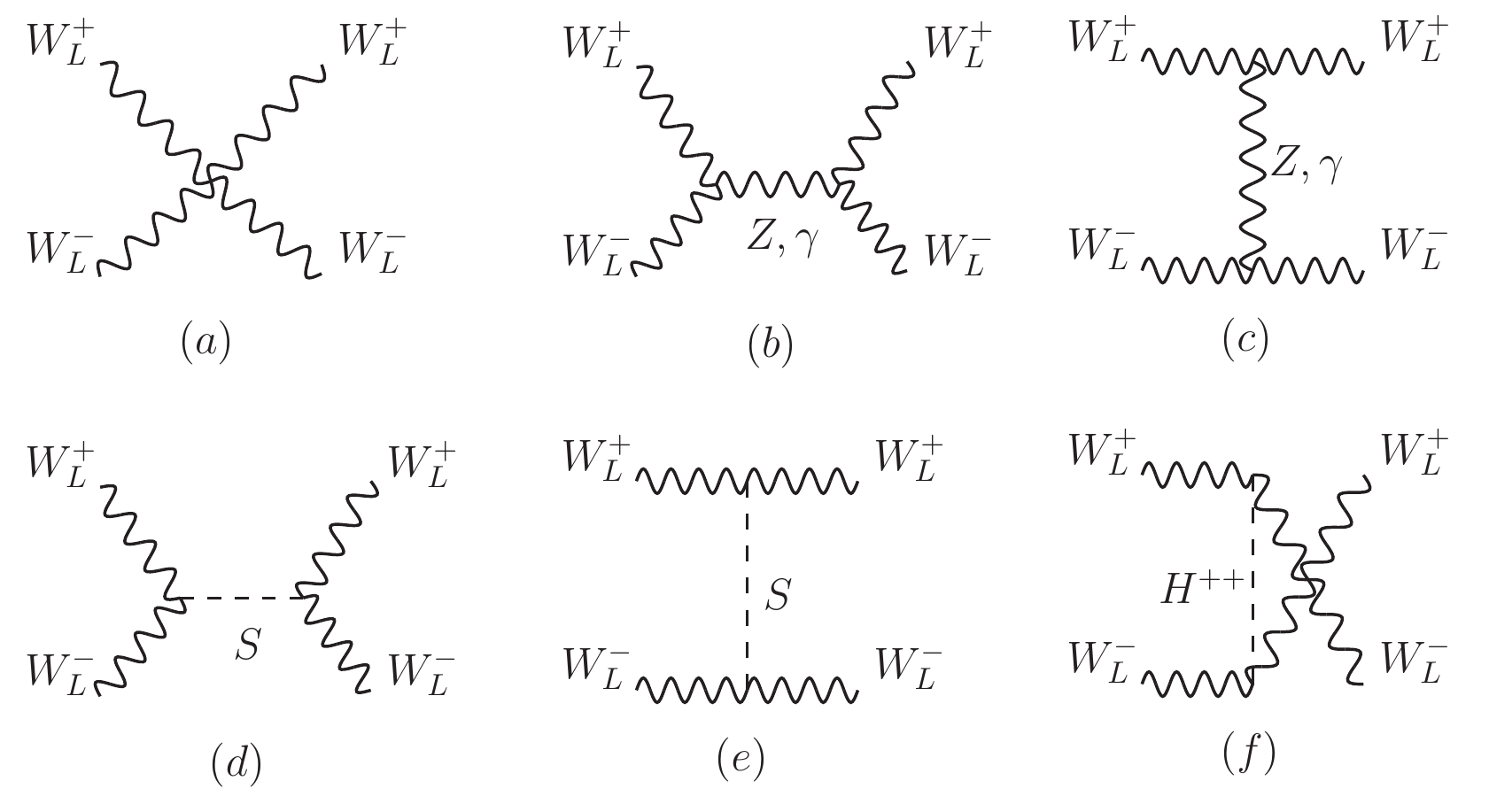}
 \caption{\label{fig:WpWmWpWm} \textit{Generic Feynman diagrams for $W_L^+(p_1)~+~ W_L^-(p_2) \rightarrow W_L^+(k_1)~+~W_L^-(k_2)$ scattering} }
 }
 \end{center}
 \end{figure}

Scattering amplitudes in terms of center of momentum energy and scattering angle:
\begp
\allowdisplaybreaks \bea
(a)& {\cal M}_{\rm p}&\hspace{-2ex} =~~\frac{E_{CM}^2 g_2^2}{16 M_{W}^4} \big\{8 M_{W}^2 (1-3 x)+E_{CM}^2 \left(-3+6 x+x^2\right)\big\}.\nn\\
(b)& {\cal M}^{\gamma+{Z}}_{s} &\hspace{-2ex} =~ -\frac{g_2^2}{4 M_{W}^4} \left( \frac{s_W^2}{s} + \frac{c_W^2}{s-M_Z^2}\right) \big( E_{CM}^6-12 E_{CM}^2 M_{W}^4-16 M_{W}^6\big) x.\nn\\
(c)&{\cal M}^{\gamma+{Z}}_{t} &\hspace{-2ex} =~ -\frac{g_2^2}{32M_{W}^4} \left( \frac{s_W^2}{t} + \frac{c_W^2}{t-M_Z^2}\right)\big\{-64 M_{W}^6 (1+x)+E_{CM}^6 (-1+x)^2 (3+x)\nn \\&&\hspace{0.35cm}+16 E_{CM}^2 M_{W}^4 \left(1-7 x+10 x^2\right)-4 E_{CM}^4 M_{W}^2 \left(3-13 x+9 x^2+x^3\right)\big\}.\nn\\
(d)& {\cal M}^{S}_{s} &\hspace{-2ex} =~  - \frac{(C~g_2 M_W)^2}{s-M_S^2}~\frac{\left(E_{CM}^2-2 M_{W}^2\right)^2}{4 M_W^4}.\nn\\
(e)& {\cal M}^{S}_{t} &\hspace{-2ex} =~  - \frac{(C~g_2 M_W)^2}{t-M_S^2}~\frac{\{4 M_W^2+E_{CM}^2 (-1+x)\}^2}{16 M_W^4}.\nn\\
(f)&{\cal M}^{H^{++}}_{u} &\hspace{-2ex} =~  - \frac{(\hat{C}~g_2 M_W)^2}{u-M^2_{H^{++}}}~\frac{\{-4 M_W^2+E_{CM}^2 (1+x)\}^2}{16 M_W^4}.\nn
\eea
\eegp
Here, $E_{CM}=\sqrt{s}$ is the center of momentum energy. 
\subsection{$W_L^+(p_1)~+~ W_L^+(p_2) \rightarrow W_L^+(k_1)~+~W_L^+(k_2)$}
\vspace{0cm}
 \begin{figure}[h!]
 \begin{center}
 {
 \includegraphics[width=15cm,height=8cm, angle=0]{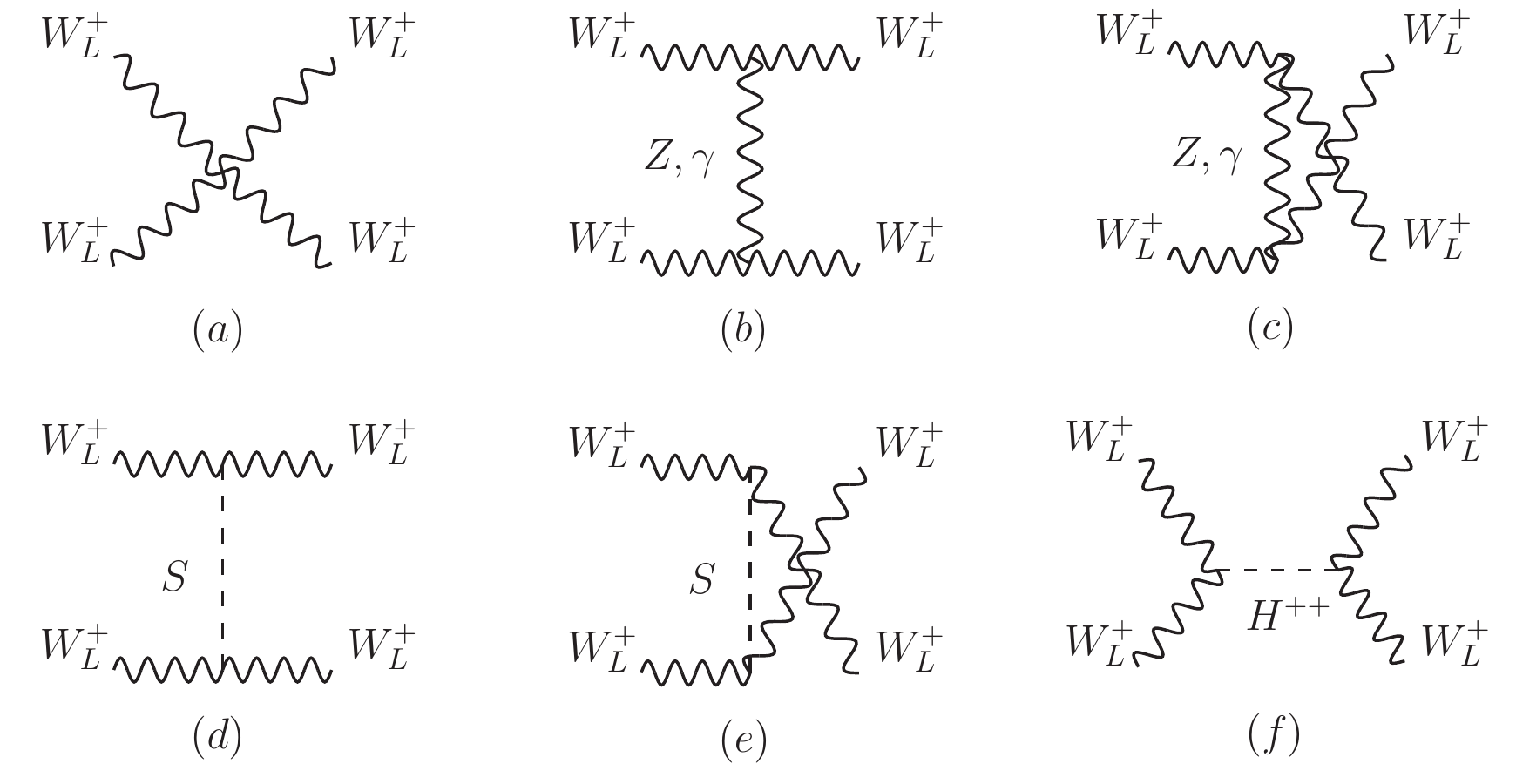}
 \caption{\label{fig:WpWpWpWp} \textit{Generic Feynman diagrams for $W_L^+(p_1)~+~ W_L^+(p_2) \rightarrow W_L^+(k_1)~+~W_L^+(k_2)$ scattering. } }
 }
 \end{center}
 \end{figure}

Scattering amplitudes in terms of longitudinal polarization vectors and four momenta: 
\begp
\allowdisplaybreaks \bea
(a)& {\cal M}_{\rm p}&\hspace{-2ex} =~~ g_2^2 \lbrace 2 (\epsilon_1.\epsilon_2)( \epsilon_3.\epsilon_4)- (\epsilon_1.\epsilon_3)( \epsilon_2.\epsilon_4)-(\epsilon_1.\epsilon_4)( \epsilon_2.\epsilon_3)\rbrace.\nn\\
(b)& {\cal M}^{\gamma+{Z}}_{t} &\hspace{-2ex} =~~  g_2^2 \left( \frac{s_W^2}{t} + \frac{c_W^2}{t-M_Z^2}\right)~\lbrace (p_1 + k_1)^\mu (\epsilon_1.\epsilon_3)-2 (k_1.\epsilon_1)\epsilon_3^\mu -2 (p_1.\epsilon_3)\epsilon_1^\mu \rbrace\nn \\&&\hspace{4.25cm} \lbrace (p_2 + k_2)_\mu (\epsilon_2.\epsilon_4)-2 (k_2.\epsilon_2)\epsilon_{4\mu} -2 (p_2.\epsilon_4)\epsilon_{2\mu} \rbrace.\nn\\
(c)& {\cal M}^{\gamma+{Z}}_{u} &\hspace{-2ex} =~~  g_2^2 \left( \frac{s_W^2}{u} + \frac{c_W^2}{u-M_Z^2}\right)~\lbrace (p_1 + k_2)^\mu (\epsilon_1.\epsilon_4)-2 (k_2.\epsilon_1)\epsilon_4^\mu -2 (p_1.\epsilon_4)\epsilon_1^\mu \rbrace\nn \\&&\hspace{4.25cm} \lbrace (p_2 + k_1)_\mu (\epsilon_2.\epsilon_3)-2 (k_1.\epsilon_2)\epsilon_{3\mu} -2 (p_2.\epsilon_3)\epsilon_{2\mu} \rbrace.\nn\\
(d)& {\cal M}^{S}_{t} &\hspace{-2ex} =~- \frac{(C~g_2 M_W)^2}{t-M_h^2} (\epsilon_1.\epsilon_3)(\epsilon_2.\epsilon_4).\nn\\
(e)& {\cal M}^{S}_{u} &\hspace{-2ex} =~- \frac{(C~g_2 M_W)^2}{u-M_h^2}~ (\epsilon_1.\epsilon_4)(\epsilon_2.\epsilon_3).\nn\\
(f)& {\cal M}^{H^{++}}_{s} &\hspace{-2ex} =~- \frac{(\hat{C}~g_2 M_W)^2}{s-M_{H^{++}}^2}~ (\epsilon_1.\epsilon_2)(\epsilon_3.\epsilon_4).\nn
\eea 
\eegp
Scattering amplitudes in terms of center of momentum energy and scattering angle:
\begp
\allowdisplaybreaks \bea
(a)& {\cal M}_{\rm p} &\hspace{-2ex} =~ -\bigg(\frac{g^2}{8 M_{W}^4}\bigg) \big\{ 8 E_{CM}^2 M_{W}^2+E_{CM}^4 \left(-3+x^2\right)\big\}.\nn\\
(b)& {\cal M}^{\gamma+{Z}}_{t}&\hspace{-2ex} =~~ \frac{g_2^2}{32M_{W}^4}\left( \frac{s_W^2}{t} + \frac{c_W^2}{t-M_Z^2}\right) \big\{-64 M_{W}^6 (1+x)+E_{CM}^6 (-1+x)^2 (3+x)\nn\\&&\hspace{0.35cm}+16 E_{CM}^2 M_{W}^4 \left(1-7 x+10 x^2\right)-4 E_{CM}^4 M_{W}^2 \left(3-13 x+9 x^2+x^3\right)\big\}.\nn\\ 
(c)& {\cal M}^{\gamma+{Z}}_{u} &\hspace{-2ex} =~~ \frac{g_2^2}{32M_{W}^4} \left( \frac{s_W^2}{u} + \frac{c_W^2}{u-M_Z^2}\right)\big\{ 64 M_{W}^6 (-1+x)-E_{CM}^6 (-3+x) (1+x)^2\nn\\&&\hspace{0.35cm}+16 E_{CM}^2 M_{W}^4 \left(1+7 x+10 x^2\right)+4 E_{CM}^4 M_{W}^2 \left(-3-13 x-9 x^2+x^3\right)\big\}.\nn\\
(d)& {\cal M}^{h}_{t} &\hspace{-2ex} =~~  - \frac{(C~g_2 M_W)^2}{t-M_h^2}~\frac{\{4 M_{W}^2+E_{CM}^2 (-1+x)\}^2}{16 M_W^4}.\nn\\
(e)& {\cal M}^{h}_{u} &\hspace{-2ex} =~~    - \frac{(C~g_2 M_W)^2}{u-M_h^2}~\frac{\{-4 M_{W}^2+E_{CM}^2 (1+x)\}^2}{16 M_W^4}.\nn\\
(f)& {\cal M}^{H^{++}}_{s} &\hspace{-2ex} =~~   - \frac{(\hat{C}~g_2 M_W)^2}{s-M_{H^{++}}^2}~ \frac{\left(E_{CM}^2-2 M_W^2\right)^2}{4 M_W^4}.\nn
\eea
\eegp
\subsection{$W_L^+(p_1)~+~ W_L^-(p_2) \rightarrow Z_L(k_1)~+~Z_L(k_2)$}
Scattering amplitudes in terms of longitudinal polarization vectors and four momenta: 
\begp
\allowdisplaybreaks \bea
(a)& {\cal M}_{\rm p} &\hspace{-2ex} =~ -g_2^2 c_W^2 \lbrace 2 (\epsilon_1.\epsilon_2)( \epsilon_3.\epsilon_4)- (\epsilon_1.\epsilon_3)( \epsilon_2.\epsilon_4)-(\epsilon_1.\epsilon_4)( \epsilon_2.\epsilon_3)\rbrace.\nn\\
(b)& {\cal M}^{W}_{t} &\hspace{-2ex} =~  -\frac{g_2^2 c_W^2}{t-M_W^2}~\bigg[\lbrace (p_1 + k_1)^\mu (\epsilon_1.\epsilon_3)-2 (k_1.\epsilon_1)\epsilon_3^\mu -2 (p_1.\epsilon_3)\epsilon_1^\mu \rbrace\nn \\&&\hspace{2.4cm} \lbrace (p_2 + k_2)_\mu (\epsilon_2.\epsilon_4)-2 (k_2.\epsilon_2)\epsilon_{4\mu} -2 (p_2.\epsilon_4)\epsilon_{2\mu} \rbrace\nn\\&&\hspace{2.7cm}+ \frac{(M_W^2-M_Z^2)^2}{M_W^2}(\epsilon_1.\epsilon_3)(\epsilon_2.\epsilon_4) \bigg].\nn\\
(c)& {\cal M}^{W}_{u} &\hspace{-2ex} =~  -\frac{g_2^2 c_W^2}{u-M_W^2}~\bigg[\lbrace (p_1 + k_2)^\mu (\epsilon_1.\epsilon_4)-2 (k_2.\epsilon_1)\epsilon_4^\mu -2 (p_1.\epsilon_4)\epsilon_1^\mu \rbrace\nn \\&&\hspace{2.4cm} \lbrace (p_2 + k_1)_\mu (\epsilon_2.\epsilon_3)-2 (k_1.\epsilon_2)\epsilon_{3\mu} -2 (p_2.\epsilon_3)\epsilon_{2\mu} \rbrace\nn\\&&\hspace{2.7cm}+ \frac{(M_W^2-M_Z^2)^2}{M_W^2}(\epsilon_1.\epsilon_4)(\epsilon_2.\epsilon_3)\bigg].\nn\\
(d)& {\cal M}^{S}_{s} &\hspace{-2ex} =~  -\frac{(C~g_2 M_W) (C^{\prime}~\frac{g_2 M_Z}{c_W})}{s-M_S^2} (\epsilon_1.\epsilon_2)(\epsilon_3.\epsilon_4).\nn\\
(e)& {\cal M}^{H^+}_{t} &\hspace{-2ex} =~ -\frac{(\widetilde{C}~\frac{g_2 M_Z}{c_W})^2}{t-M_{H^+}^2} (\epsilon_1.\epsilon_3)(\epsilon_2.\epsilon_4).\nn\\
(f)& {\cal M}^{H^+}_{u} &\hspace{-2ex} =~  -\frac{(\widetilde{C}~\frac{g_2 M_Z}{c_W})^2}{u-M_{H^+}^2} (\epsilon_1.\epsilon_4)(\epsilon_2.\epsilon_3).\nn
\eea
\eegp
 \begin{figure}[h!]
 \begin{center}
 {
 \includegraphics[width=16cm,height=7.2cm, angle=0]{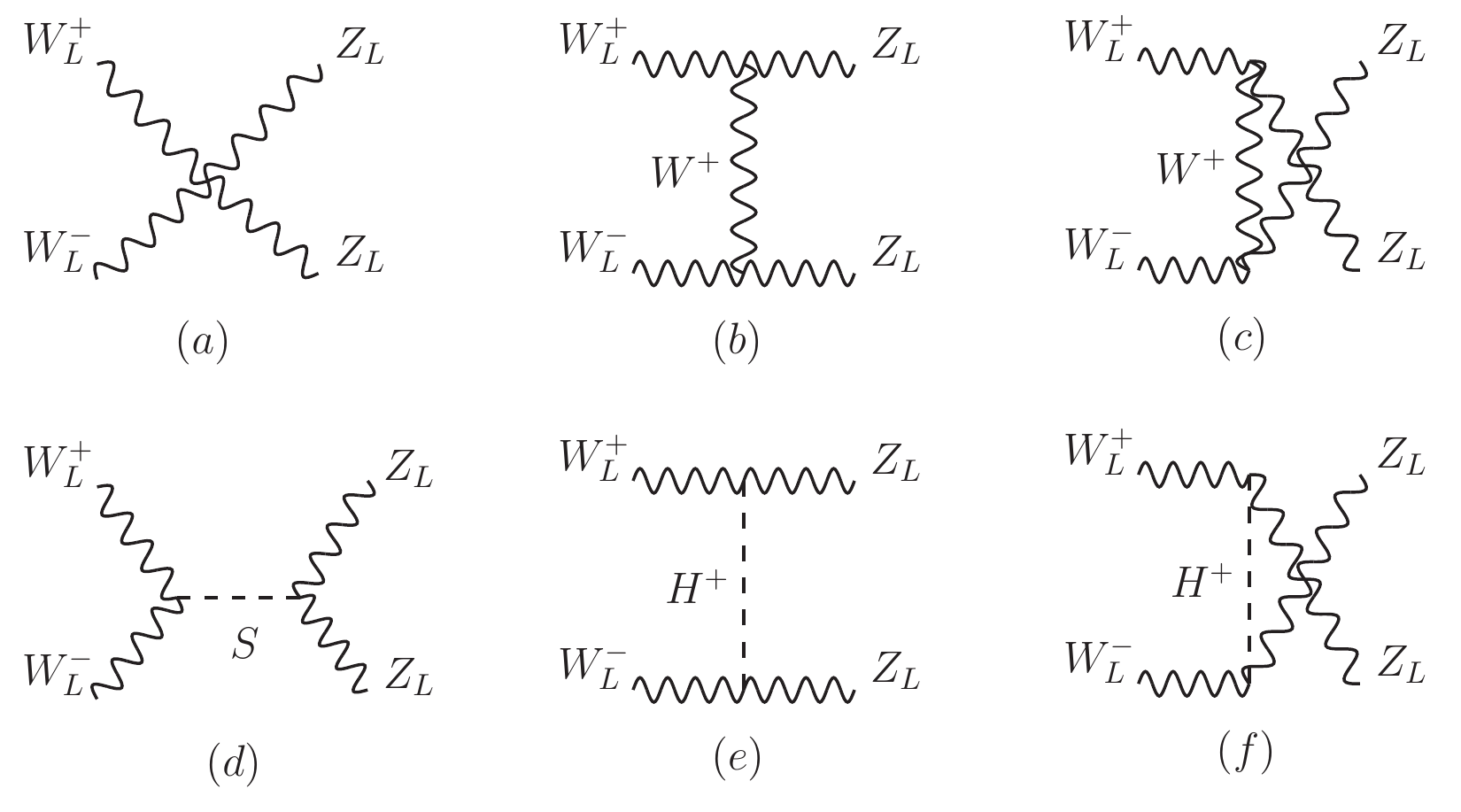}
 \caption{\label{fig:WpWmZZ} \textit{Generic Feynman diagrams for $W_L^+(p_1)~+~ W_L^-(p_2) \rightarrow Z_L(k_1)~+~Z_L(k_2)$ scattering.} }
 }
 \end{center}
 \end{figure}

Scattering amplitudes in terms of center of momentum energy and scattering angle:
\begp
\allowdisplaybreaks \bea
(a)& {\cal M}_{\rm p} &\hspace{-2ex} =~~  \frac{E_{CM}^2 g_2^2 c_W^2}{8M_{W}^2 M_{Z}^2}\big\{4 \left(M_{W}^2+M_{Z}^2\right)+E_{CM}^2 \left(-3+x^2\right)\big\}.\nn\\
(b)& {\cal M}^{W}_{t} &\hspace{-2ex} =~ -\frac{ c_W^2 g_2^2}{32M_{W}^4 M_{Z}^2(t-M_W^2)} \big\{  3 E_{CM}^6 M_{W}^2-4 E_{CM}^4 M_{W}^4-10 E_{CM}^4 M_{W}^2 M_{Z}^2\nn\\
&&\hspace{1cm}+8 E_{CM}^2 M_{W}^4 M_{Z}^2+2 E_{CM}^4 M_{Z}^4+16 E_{CM}^2 M_{W}^2 M_{Z}^4-96 M_{W}^4 M_{Z}^4\nn\\
&&\hspace{1cm}-8 E_{CM}^2 M_{Z}^6+32 M_{W}^2 M_{Z}^6+x (E_{CM}^4 M_{W}^2\text{  }x^2-5 E_{CM}^4 M_{W}^2\nn\\
&&\hspace{1cm}+12 E_{CM}^2 M_{W}^4+24 E_{CM}^2 M_{W}^2 M_{Z}^2+16 M_{W}^4 M_{Z}^2\nn\\
&&\hspace{1cm}-4 E_{CM}^2 M_{Z}^4)\sqrt{(-E_{CM}^2+4 M_{W}^2) (-E_{CM}^2+4 M_{Z}^2)}+E_{CM}^6 M_{W}^2 x^2\nn\\
&&\hspace{1cm}-16 E_{CM}^4 M_{W}^4 x^2+32 E_{CM}^2 M_{W}^6 x^2-22 E_{CM}^4 M_{W}^2 M_{Z}^2 x^2\nn\\
&&\hspace{1cm}+96 E_{CM}^2 M_{W}^4 M_{Z}^2 x^2+2 E_{CM}^4 M_{Z}^4 x^2+32 E_{CM}^2 M_{W}^2 M_{Z}^4 x^2 \big\}.\nn\\
(c)&{\cal M}^{W}_{u} &\hspace{-2ex} =~  -\frac{ c_W^2 g_2^2}{32M_{W}^4 M_{Z}^2(u-M_W^2)} \big\{  3 E_{CM}^6 M_{W}^2-4 E_{CM}^4 M_{W}^4-10 E_{CM}^4 M_{W}^2 M_{Z}^2\nn\\
&&\hspace{1cm}+8 E_{CM}^2 M_{W}^4 M_{Z}^2+2 E_{CM}^4 M_{Z}^4+16 E_{CM}^2 M_{W}^2 M_{Z}^4-96 M_{W}^4 M_{Z}^4\nn\\
&&\hspace{1cm}-8 E_{CM}^2 M_{Z}^6+32 M_{W}^2 M_{Z}^6- x (E_{CM}^4 M_{W}^2\text{  }x^2-5 E_{CM}^4 M_{W}^2\nn\\
&&\hspace{1cm}+12 E_{CM}^2 M_{W}^4+24 E_{CM}^2 M_{W}^2 M_{Z}^2+16 M_{W}^4 M_{Z}^2\nn\\
&&\hspace{1cm}-4 E_{CM}^2 M_{Z}^4)\sqrt{(-E_{CM}^2+4 M_{W}^2) (-E_{CM}^2+4 M_{Z}^2)}+E_{CM}^6 M_{W}^2 x^2\nn\\
&&\hspace{1cm}-16 E_{CM}^4 M_{W}^4 x^2+32 E_{CM}^2 M_{W}^6 x^2-22 E_{CM}^4 M_{W}^2 M_{Z}^2 x^2\nn\\
&&\hspace{1cm}+96 E_{CM}^2 M_{W}^4 M_{Z}^2 x^2+2 E_{CM}^4 M_{Z}^4 x^2+32 E_{CM}^2 M_{W}^2 M_{Z}^4 x^2 \big\}.\nn\\
(d)& {\cal M}^{S}_{s} &\hspace{-2ex} =~  -\frac{(C~g_2 M_W) (C^{\prime}~\frac{g_2 M_Z}{c_W})}{s-M_S^2}~ \frac{\left(E_{CM}^2-2 M_W^2\right) \left(E_{CM}^2-2 M_Z^2\right)}{4 M_W^2 M_Z^2}.\nn\\
(e)& {\cal M}^{H^+}_{t} &\hspace{-2ex} =~  -\frac{(\widetilde{C}~\frac{g_2 M_Z}{c_W})^2}{t-M_{H^+}^2}~ \frac{\big \{\sqrt{\left(E_{CM}^2-4 M_W^2\right) \left(E_{CM}^2-4 M_Z^2\right)}-E_{CM}^2 x\big \}^2}{16 M_W^2 M_Z^2}.\nn\\
(f)& {\cal M}^{H^+}_{u} &\hspace{-2ex} =~  -\frac{(\widetilde{C}~\frac{g_2 M_Z}{c_W})^2}{u-M_{H^+}^2} ~\frac{\big \{ \sqrt{\left(E_{CM}^2-4 M_W^2\right) \left(E_{CM}^2-4 M_Z^2\right)}+E_{CM}^2 x\big\}^2}{16 M_W^2 M_Z^2}.\nn
\eea
\eegp
\subsection{$W_L^+(p_1)~+~ Z_L(p_2) \rightarrow W_L^+(k_1)~+~Z_L(k_2)$}
\vspace{0cm}
Scattering amplitudes in terms of longitudinal polarization vectors and four momenta:
\begp
\allowdisplaybreaks \bea
(a)& {\cal M}_{\rm p} &\hspace{-2ex} =~ -c_W^2 g_2^2 \{ 2 (\epsilon_1.\epsilon_3)(\epsilon_2.\epsilon_4)-(\epsilon_1.\epsilon_4)(\epsilon_2.\epsilon_3)-(\epsilon_1.\epsilon_2)(\epsilon_3.\epsilon_4)\}.\nn\\
(b)& {\cal M}^{W}_{s} &\hspace{-2ex} =~~  \frac{g_2^2 c_W^2}{s-M_W^2}~\bigg[\lbrace (p_1- p_2)^\mu (\epsilon_1.\epsilon_2)+2 (p_2.\epsilon_1)\epsilon_2^\mu -2 (p_1.\epsilon_2)\epsilon_1^\mu \rbrace\nn \\&&\hspace{2.1cm} \lbrace (k_2 - k_1)_\mu (\epsilon_3.\epsilon_4)-2 (k_2.\epsilon_3)\epsilon_{4\mu} -2 (k_1.\epsilon_4)\epsilon_{3\mu} \rbrace\nn \\&&\hspace{2.3cm}-\frac{(M_W^2-M_Z^2)^2}{M_W^2}(\epsilon_1.\epsilon_2)(\epsilon_3.\epsilon_4)\bigg].\nn\\
(c)& {\cal M}^{W}_{u} &\hspace{-2ex} =~ -\frac{g_2^2 c_W^2}{u-M_W^2}~\bigg[\lbrace (p_1 + k_2)^\mu (\epsilon_1.\epsilon_4)-2 (k_2.\epsilon_1)\epsilon_4^\mu -2 (p_1.\epsilon_4)\epsilon_1^\mu \rbrace\nn \\&&\hspace{2.1cm} \lbrace (p_2 + k_1)_\mu (\epsilon_2.\epsilon_3)-2 (k_1.\epsilon_2)\epsilon_{3\mu} -2 (p_2.\epsilon_3)\epsilon_{2\mu} \rbrace\nn\\&&\hspace{2.3cm}- \frac{(M_W^2-M_Z^2)^2}{M_W^2}(\epsilon_1.\epsilon_4)(\epsilon_2.\epsilon_3)\bigg].\nn\\
(d)& {\cal M}^{S}_{t} &\hspace{-2ex} =~  -\frac{(C~g_2 M_W) (C^{\prime}~\frac{g_2 M_Z}{c_W})}{t-M_S^2} (\epsilon_1.\epsilon_3)(\epsilon_2.\epsilon_4).\nn\\
(e)& {\cal M}^{H^+}_{s} &\hspace{-2ex} =~ -\frac{(\widetilde{C}~\frac{g_2 M_Z}{c_W})^2}{s-M_{H^+}^2} (\epsilon_1.\epsilon_2)(\epsilon_3.\epsilon_4).\nn\\
(f)&{\cal M}^{H^+}_{u} &\hspace{-2ex} =~ -\frac{(\widetilde{C}~\frac{g_2 M_Z}{c_W})^2}{u-M_{H^+}^2} (\epsilon_1.\epsilon_4)(\epsilon_2.\epsilon_3).\nn
\eea
\eegp
 \begin{figure}[h!]
 \begin{center}
 {
 \includegraphics[width=15cm,height=8cm, angle=0]{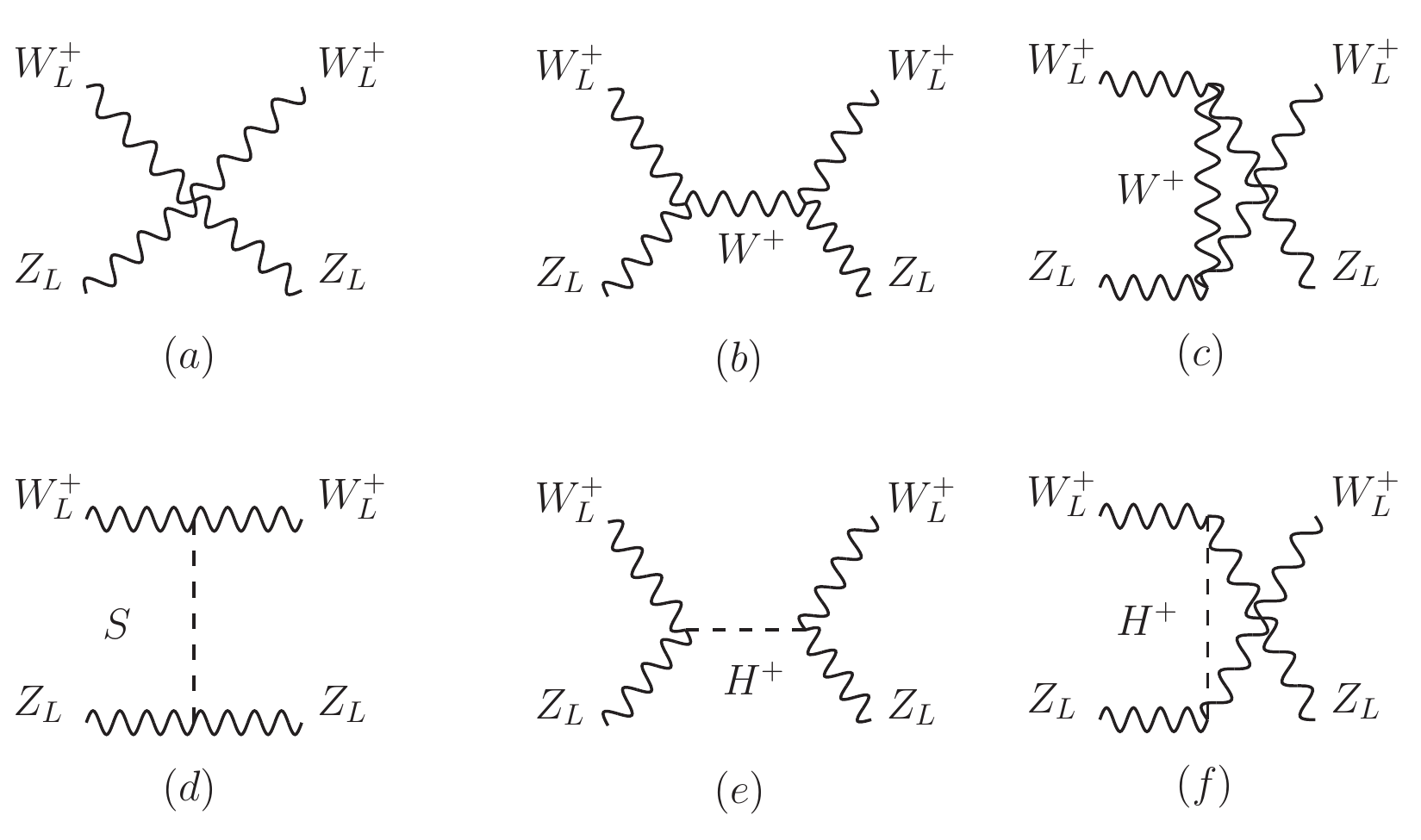}
 \caption{\label{fig:WpZWpZ} \textit{Generic Feynman diagrams for $W_L^+(p_1)~+~ Z_L(p_2) \rightarrow W_L^+(k_1)~+~Z_L(k_2)$ scattering.} }
 }
 \end{center}
 \end{figure}

Scattering amplitudes in terms of center of momentum energy and scattering angle:
\begp
\allowdisplaybreaks \bea
(a)& {\cal M}_{\rm p} &\hspace{-2ex} =~ -\frac{c_W^2 g^2}{16 E_{CM}^4 M_{W}^2 M_{Z}^2} \big\{4 E_{CM}^2 \left(M_{W}^2-M_{Z}^2\right)^2 \left(M_{W}^2+M_{Z}^2\right) (-1+x)\nn\\
&&\hspace{4cm}+\left(M_{W}^2-M_{Z}^2\right)^4 (-1+x)^2\nn\\
&&\hspace{3cm}+4 E_{CM}^6 \left(M_{W}^2+M_{Z}^2\right) (1+3 x)+E_{CM}^8 \left(-3-6 x+x^2\right)\nn\\
&&\hspace{4cm}-2 E_{CM}^4 \left(M_{W}^2-M_{Z}^2\right)^2 \left(-1+4 x+x^2\right)\big\}.\nn\\
(b)& {\cal M}^{W}_{s} &\hspace{-2ex} =~  -\frac{c_W^2 g^2}{4 E_{CM}^2 M_{W}^4 M_{Z}^2(s-M_W^2)} \big[ E_{CM}^6 \left(M_{W}^2-M_{Z}^2\right)^2+\left(M_{W}^5-M_{W} M_{Z}^4\right)^2 (-1+x)\nn\\
&&\hspace{4cm}+E_{CM}^8 M_{W}^2 x-E_{CM}^4 \big\{2 M_{Z}^6+4 M_{W}^4 M_{Z}^2 (-1+2 x)\nn\\
&&\hspace{4cm}+M_{W}^2 M_{Z}^4 (-1+2 x)+M_{W}^6 (3+2 x)\big\}\nn\\
&&\hspace{4cm}+E_{CM}^2 \big\{M_{W}^2+M_{Z}^2)
 (3 M_{W}^6+M_{W}^2 M_{Z}^4+M_{Z}^6\nn\\
&&\hspace{4cm}-M_{W}^4 M_{Z}^2 (5+8 x)\big\}\big].\nn\\
(c)& {\cal M}^{W}_{u} &\hspace{-2ex} =~~ \frac{M_{Z}^2c_W^2 g^2}{32 E_{CM}^6 M_{W}^4 (u-M_W^2)}  \big[\big\{ M_{W}^2 (M_{W}^2-M_{Z}^2)^6 (-1+x)^3\nn\\
&&\hspace{2.5cm}+E_{CM}^{12} M_{W}^2 (-3+x) (1+x)^2\nn\\
&&\hspace{2.5cm}-2 E_{CM}^2 (M_{W}^2-M_{Z}^2)^4 (-1+x)^2 (-M_{Z}^4+M_{W}^4 x+\nn\\
&&\hspace{2.5cm}+M_{W}^2 M_{Z}^2 (3+x)\big\} +2 E_{CM}^{10} \big\{M_{Z}^4 (1+x)^2\nn\\
&&\hspace{2.5cm}+M_{W}^2 M_{Z}^2 (1+9 x+7 x^2-x^3)+M_{W}^4 (4+15 x+10 x^2-x^3)\big\}\nn\\
&&\hspace{2.5cm}+4 E_{CM}^6\big\{-M_{Z}^8 (-3+x^2)-M_{W}^6 M_{Z}^2 (-9+9 x+7 x^2+x^3)\nn\\
&&\hspace{2.5cm}+M_{W}^8 (-2+x+10 x^2+x^3)+M_{W}^2 M_{Z}^6 (-9+x+15 x^2+x^3)\nn\\
&&\hspace{2.5cm}-M_{W}^4 M_{Z}^4 (-15+9 x+17 x^2+x^3)\big\}\nn\\
&&\hspace{2.5cm}-E_{CM}^4 (M_{W}^2-M_{Z}^2)^2 \big\{-8 M_{Z}^6 (-1+x)\nn\\
&&\hspace{2.5cm}+M_{W}^6 (-7-5 x+11 x^2+x^3)\nn\\
&&\hspace{2.5cm}+M_{W}^2 M_{Z}^4 (-23+11 x+11 x^2+x^3)\nn\\
&&\hspace{2.5cm}-2 M_{W}^4 M_{Z}^2 (-3-25 x+27 x^2+x^3)\big\}-E_{CM}^8 \big\{8 M_{Z}^6 (1+x)\nn\\
&&\hspace{2.5cm}+2 M_{W}^4 M_{Z}^2 (9+25 x+31 x^2-x^3)\nn\\
&&\hspace{2.5cm}+M_{W}^2 M_{Z}^4 (-13+19 x+49 x^2+x^3)\nn\\
&&\hspace{2.5cm}+M_{W}^6 (3+35 x+49 x^2+x^3)\big\}\big].\nn\\
(d)& {\cal M}^{S}_{t} &\hspace{-2ex} =~ -\frac{(C~g_2 M_W) (C^{\prime}~\frac{g_2 M_Z}{c_W})}{t-M_S^2} ~\bigg(\frac{1}{16 E_{CM}^4 M_W^2 M_Z^2}\bigg)~\big[\big\{E_{CM}^4 (-1+x)\nn\\
&&\hspace{2.5cm}+(M_W^2-M_Z^2)^2 (-1+x)\nn\\
&&\hspace{2.5cm}+2 E_{CM}^2 (-M_Z^2 (-1+x)+M_W^2 (1+x))\big\} \big\{E_{CM}^4 (-1+x)\nn\\
&&\hspace{2.5cm}+(M_W^2-M_Z^2)^2 (-1+x)+2 E_{CM}^2 (-M_W^2 (-1+x)\nn\\
&&\hspace{2.5cm}+M_Z^2 (1+x))\big\}\big].\nn\\
(e)& {\cal M}^{H^+}_{u} &\hspace{-2ex} =~ -\frac{(\widetilde{C}~\frac{g_2 M_Z}{c_W})^2}{u-M_{H^+}^2} ~\bigg(\frac{1}{16 E_{CM}^4 M_W^2 M_Z^2}\bigg) \big\{ 2 E_{CM}^2 (M_W^2+M_Z^2)\nn\\
&&\hspace{0.35cm}+(M_W^2-M_Z^2)^2 (-1+x)-E_{CM}^4 (1+x)\big\}^2.\nn\\
(f)& {\cal M}^{H^+}_{s} &\hspace{-2ex} =~~  -\frac{(\widetilde{C}~\frac{g_2 M_Z}{c_W})^2}{s-M_{H^+}^2} ~\frac{(-E_{CM}^2+M_W^2+M_Z^2)^2}{4 M_W^2 M_Z^2}.\nn
\eea
\eegp
\subsection{$Z_L(p_1)~+~ Z_L(p_2) \rightarrow Z_L(k_1)~+~Z_L(k_2)$}
Scattering amplitudes in terms of longitudinal polarization vectors and four momenta: 
\begp
\allowdisplaybreaks \bea
(a)& {\cal M}^{S}_{s} &\hspace{-2ex} =~~ -\frac{(C^{\prime}~\frac{g_2 M_Z}{c_W})^2}{(s-M_S^2)}(\epsilon_1.\epsilon_2)(\epsilon_3.\epsilon_4).\nn\\
(b)& {\cal M}^{S}_{t} &\hspace{-2ex} =~~ -\frac{(C^{\prime}~\frac{g_2 M_Z}{c_W})^2}{(t-M_S^2)}(\epsilon_1.\epsilon_3)(\epsilon_2.\epsilon_4).\nn\\
(c)& {\cal M}^{S}_{u} &\hspace{-2ex} =~~ -\frac{(C^{\prime}~\frac{g_2 M_Z}{c_W})^2}{(u-M_S^2)}(\epsilon_1.\epsilon_4)(\epsilon_2.\epsilon_3).\nn
\eea
\eegp
 \begin{figure}[h!]
 \begin{center}
 {
 \includegraphics[width=15cm,height=4cm, angle=0]{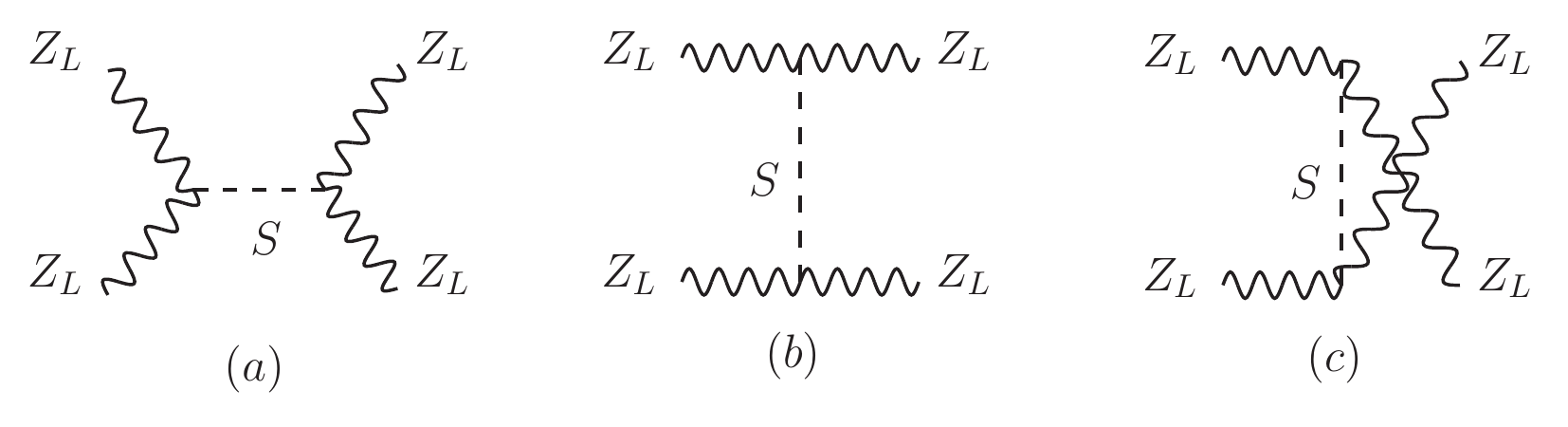}
 \caption{\label{fig:ZZZZ} \textit{Generic Feynman diagrams for $Z_L(p_1)~+~ Z_L(p_2) \rightarrow Z_L(k_1)~+~Z_L(k_2)$ scattering. } }
 }
 \end{center}
 \end{figure}

Scattering amplitudes in terms of center of momentum energy and scattering angle:
\begp
\allowdisplaybreaks \bea
(a)& {\cal M}^{S}_{s} &\hspace{-2ex} =~~ -\frac{(C^{\prime}~\frac{g_2 M_Z}{c_W})^2}{(s-M_S^2)}~\frac{(E_{CM}^2-2 M_Z^2)^2}{4 M_Z^4}.\nn\\
(b)& {\cal M}^{S}_{t} &\hspace{-2ex} =~~ -\frac{(C^{\prime}~\frac{g_2 M_Z}{c_W})^2}{(t-M_S^2)}~\frac{\big\{4 M_Z^2+E_{CM}^2 (-1+x)\big\}^2}{16 M_Z^4}.\nn\\
(c)& {\cal M}^{S}_{u} &\hspace{-2ex} =~~ -\frac{(C^{\prime}~\frac{g_2 M_Z}{c_W})^2}{(u-M_S^2)}~\frac{\big\{-4 M_Z^2+E_{CM}^2 (1+x)\big\}^2}{16 M_Z^4}.\nn
\eea
\eegp
\section{Required Feynman rules for $VV$ scattering}\label{fyenrules}
The Feynman rules for the different vertices with the assumption that all momenta and fields are incoming.

\vspace*{0.3cm}
\vspace{0.1cm}\begin{minipage}{50mm}
  {\includegraphics[]{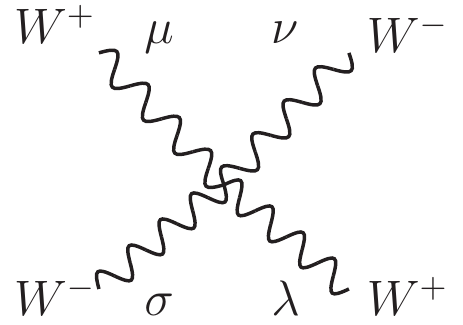}}
\end{minipage}
\begin{minipage}{10cm}
 $\displaystyle  : i g_2^2 (2 g^{\mu\lambda} g^{\sigma\nu}-g^{\mu\sigma} g^{\nu\lambda}-g^{\sigma\lambda}g^{\mu\nu}).$
\end{minipage}
\vspace{-12.8ex}\begin{equation}\end{equation}\vspace{4ex}

\vspace{0.4cm}
\begin{minipage}{50mm}
  {\includegraphics[]{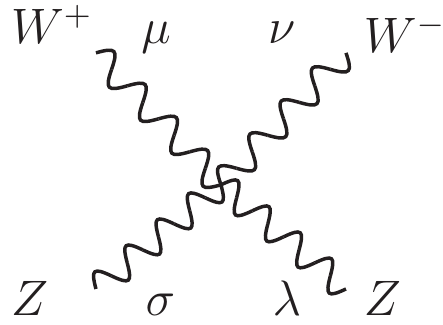}}
\end{minipage}
\begin{minipage}{10cm}
 $\displaystyle  : - i g_2^2 c_W^2(2 g^{\mu\nu} g^{\sigma\lambda}-g^{\nu\sigma} g^{\mu\lambda}-g^{\nu\lambda}g^{\mu\sigma}).$
\end{minipage}
\vspace{-12.8ex}\begin{equation}\end{equation}\vspace{4ex}

\vspace{0.4cm}
\begin{minipage}{50mm}
  {\includegraphics[]{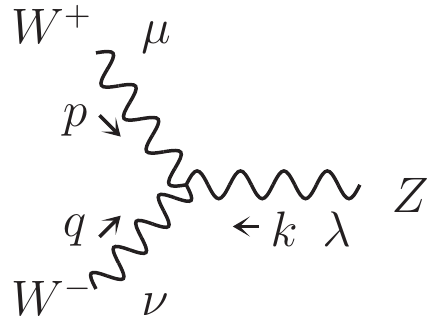}}
\end{minipage}
\begin{minipage}{10cm}
  $\displaystyle  : i g_2c_W \lbrace(p-q)^\lambda~g^{\mu\nu}+(q-k)^\mu~g^{\lambda\nu}+(k-p)^\nu~g^{\mu\lambda}\rbrace.$
\end{minipage}
\vspace{-9.8ex}\begin{equation}\end{equation}\vspace{4ex}

\vspace*{0.2cm}
\begin{minipage}{50mm}
  {\includegraphics[]{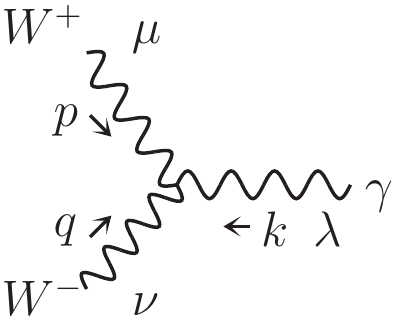}}
\end{minipage}
\begin{minipage}{10cm}
  \hspace{0cm} $\displaystyle  : i g_2s_W \lbrace(p-q)^\lambda~g^{\mu\nu}+(q-k)^\mu~g^{\lambda\nu}+(k-p)^\nu~g^{\mu\lambda}\rbrace.$
\end{minipage}
\vspace{-8.8ex}\begin{equation}\end{equation}\vspace{4ex}
\newpage	

\hspace{0.2cm}\begin{minipage}{50mm}
  {\includegraphics[]{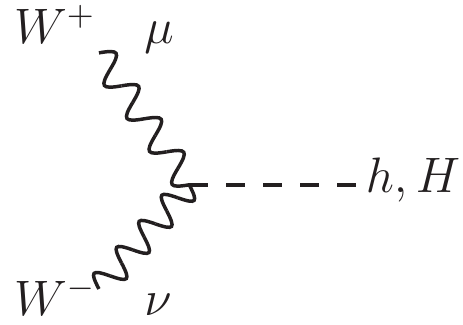}}
\end{minipage}
\begin{minipage}{10cm}\vspace{3ex}
  $\displaystyle  ~~~: {i g_2}{M_W} g_{\mu\nu} C$, where $C$ is given by:
\end{minipage}
\vspace*{0.4cm}
\begp
\allowdisplaybreaks
\begin{equation}
\begin{aligned}
  ~~~~&  {\rm SM}:  &
  &\left\{\begin{array}{l}{\rm for}~~h,~~C = 1,\\
      {\rm for}~~H,~~C = 0,\end{array}\right.\\
  & {\rm 2HDM}:  &
  &\left\{\begin{array}{l}{\rm for}~~h,~~C = \sin(\beta-\alpha),\\
      {\rm for}~~H,~~C = \cos(\beta-\alpha),\end{array}\right.\\
  & Y=0{\rm ~~HTM}:  &
  &\left\{\begin{array}{l}{\rm for}~~h,~~C = (\cos\widetilde{\beta}\cos{\gamma}+2\sin\widetilde{\beta}\sin{\gamma}),\\
      {\rm for}~~H,~~C = (-\cos\widetilde{\beta}\sin{\gamma}+2\sin\widetilde{\beta}\cos{\gamma}),\end{array}\right.\\
  & Y=2{\rm ~~HTM}:  &
  &\left\{\begin{array}{l}{\rm for}~~h,~~C = (\cos{\beta'}\cos{\gamma'}+\sqrt{2}\sin{\beta'}\sin{\gamma'}),\\
      {\rm for}~~H,~~C = (-\cos{\beta'}\sin{\gamma'}+\sqrt{2}\sin{\beta'}\cos{\gamma'}).\end{array}\right.
\end{aligned}
\end{equation}
\eegp
\vspace*{0.0cm}
\begin{minipage}{50mm}
  {\includegraphics[]{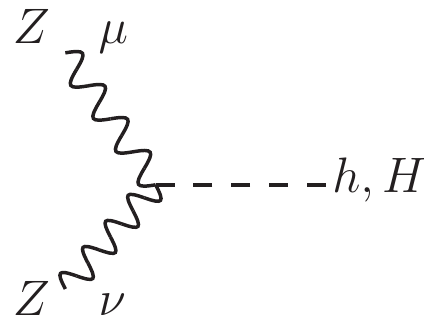}}
\end{minipage}
\begin{minipage}{10cm}\vspace{3ex}
  $\displaystyle  ~~~: \frac{i g_2 M_Z}{c_W} g_{\mu\nu} C'$, where $C'$ is given by:
\end{minipage}

\vspace*{0.4cm}
\begp
\allowdisplaybreaks
\begin{equation}
\begin{aligned}
  &  {\rm SM}:  &
  &\left\{\begin{array}{l}{\rm for}~~h,~~C' = 1,\\
      {\rm for}~~H,~~C' = 0,\end{array}\right.\\
  & {\rm 2HDM}:  &
  &\left\{\begin{array}{l}{\rm for}~~h,~~C' = \sin(\beta-\alpha),\\
      {\rm for}~~H,~~C' = \cos(\beta-\alpha),\end{array}\right.\\
  & Y=0{\rm ~~HTM}:  &
  &\left\{\begin{array}{l}{\rm for}~~h,~~C' = \cos\gamma,\\
      {\rm for}~~H,~~C' = -\sin{\gamma},\end{array}\right.\\
  & Y=2{\rm ~~HTM}:  &
  &\left\{\begin{array}{l}{\rm for}~~h,~~C' = (\cos{\delta'}\cos{\gamma'}+2\sin{\delta'}\sin{\gamma'}),\\
      {\rm for}~~H,~~C' = (-\cos{\delta'}\sin{\gamma'}+2\sin{\delta'}\cos{\gamma'}).\end{array}\right.
\end{aligned}
\end{equation}
\eegp

\vspace*{0.4cm}
\begin{minipage}{50mm}
  {\includegraphics[]{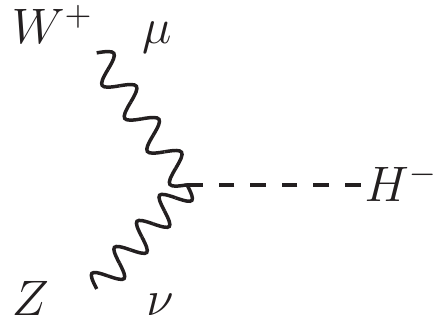}}
\end{minipage}
\begin{minipage}{10cm}\vspace{3ex}
$\displaystyle ~~~: \frac{i g_2 M_Z}{c_W} g_{\mu\nu} \widetilde{C}$, where $\widetilde{C}$ is given by:
\end{minipage}
\vspace*{0.5cm}
\begp
\allowdisplaybreaks
\begin{equation}
 \begin{aligned}
  ~~~~~&{\rm SM}:  \hspace{3cm} \widetilde{C}   =0,\\ 
  &{\rm 2HDM}: \hspace{2.4cm} \widetilde{C}   =0,\\
  &Y=0{\rm ~~HTM}: \hspace{1.3cm} \widetilde{C}   = \sin\widetilde{\beta}\cos\widetilde{\beta}\frac{M_W}{M_Z},\\
  &Y=2{\rm ~~HTM}: \hspace{1.3cm} \widetilde{C}   = c_W\bigg\lbrace\sin{\beta'}\cos{\delta'}s_W^2-\frac{(1+s_W^2)}{\sqrt{2}}\cos{\beta'}\sin{\delta'} \bigg\rbrace.  
 \end{aligned}
\end{equation}
\eegp

\begin{minipage}{50mm}
  {\includegraphics[]{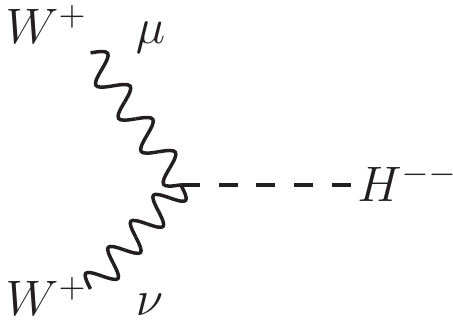}}
\end{minipage}
\begin{minipage}{10cm}\vspace{3ex}
$\displaystyle ~~~: {i g_2 M_W}g_{\mu\nu} \hat{C}$, where $\hat{C}$ is given by:
\end{minipage}

\vspace*{0.5cm}
\begp
\allowdisplaybreaks
\begin{equation}
 \begin{aligned}
 \hspace{-2cm} &{\rm SM}: \hspace{3cm} \hat{C}   =0,\\ 
  &{\rm 2HDM}: \hspace{2.4cm} \hat{C}   =0,\\
  &Y=0{\rm ~~HTM}: \hspace{1.3cm}\hat{C}   = 0,\\
  &Y=2{\rm ~~HTM}: \hspace{1.3cm} \hat{C}   = 2\sin{\beta'}.\qquad \qquad\qquad\qquad\qquad\qquad ~~~~~  
 \end{aligned}
\end{equation}
\eegp
\section{Restoration of unitarity in extended scalar sectors}
\label{highenergy}
Generally when $E_{CM}>>M_i$ ($i\equiv W,Z,h,H,H^+,H^{++}$), one can express $V_LV_L\rightarrow V_LV_L$ scattering amplitude as,
\begp
\allowdisplaybreaks \bea
{\cal M}&=& A_4~E_{CM}^4 + A_2~E_{CM}^2 + A_{0} + A_{-2}~E_{CM}^{-2}+....
\label{energyseries}
\eea
\eegp
If $A_{4,2}\neq 0$ then the scattering cross-section will increase with $E_{CM}$. The ${\rm SU(2)_L\times U(1)}_Y$ gauge symmetry implies that $A_{4}=0$ from the gauge mediated and cross diagrams. We need the scalar particles in the model so $A_2$ becomes exactly zero for $E_{CM}>>M_i$ and theory become unitarized, i.e., the cross-section will decrease with energy. 

The gauge and scalar contributions to $A_2$ and $A_0$ are denoted as,
    \begp
\allowdisplaybreaks  
\bea
A_2 &=& A_{2,g} + \sum_{S} A_{2,S},~~~~~~~S=h,H,H^+,H^{++}~,
\eea\eegp
and
\begp
\allowdisplaybreaks \bea
A_{0}&=& A_{0,g} + A_{0,S}.
\eea
\eegp

The expressions for $A_{2,g},~A_{2,S},~A_{0,g}$ and $A_{0,S}$ of different vector bosons scattering processes are presented for different models of extended scalar sector.

1. $W_L^+ W_L^- \rightarrow W_L^+ W_L^-$
\begin{table}[h!]
\begin{center}
\resizebox{15cm}{!}{
    \begin{tabular}{ | c | c |  c | c | c |}
         \hline
      & ${\rm SM}$ & ${\rm 2HDM}$ & ${\rm HTM}(Y=0)$ & ${\rm HTM}(Y=2)$ \\
    \hline
     $A_{2,g}$ & $\frac{g_2^2(4 M_W^2-3 c_W^2 M_Z^2)(1+x)}{2 M_W^4}$& $\frac{g_2^2(4 M_W^2-3 c_W^2 M_Z^2)(1+x)}{2 M_W^4}$& $\frac{g_2^2(4 M_W^2-3 c_W^2 M_Z^2)(1+x)}{2 M_W^4}$& $\frac{g_2^2(4 M_W^2-3 c_W^2 M_Z^2)(1+x)}{2 M_W^4}$  \\
     \hline
     $A_{2,S}({S\equiv h,H})$ & $-\frac{g_2^2}{2 M_W^2}~C^2~(1+x)$ & $-\frac{g_2^2}{2 M_W^2}~C^2~(1+x)$ & $-\frac{g_2^2}{2 M_W^2}~C^2~(1+x)$ &  $-\frac{g_2^2}{2 M_W^2}~C^2~(1+x)$ \\
     \hline
     $A_{2,S}(S\equiv H^+)$ & 0 & 0 & 0& 0  \\
     \hline
      $A_{2,S}(S\equiv H^{++})$ & 0 & 0 & 0 &  $\frac{g_2^2}{2 M_W^2}~\hat{C}^2~(1+x)$ \\
     \hline
     \end{tabular}
}
\end{center}
\caption{ \textit{ $A_{2,g}$ and $A_{2,S}$ for $W_L^+ W_L^- \rightarrow W_L^+ W_L^-$ process. $C$ and $\hat{C}$ can be found in Appendix \ref{fyenrules}.} }
    \label{tablea}
\end{table}

\begp
\allowdisplaybreaks \bea
A_{0,g} &=& \frac{g_2^2}{4 M_W^4 (-1+x)} \big\{-c_W^2 M_Z^4 (3+x^2)-2 c_W^2 M_W^2 M_Z^2 (-3-6 x+x^2)\nn\\&&\hspace{6cm}+4 M_W^4 (-1-4 x+x^2)\big\}\\
A_{0,S} &=&-\frac{1}{4 M_W^4}\bigg[({g_2 M_W}\hat{C})^2 \big\{M_{H^{++}}^2-2 M_W^2 (-1+x)\big\}\nn\\&&\hspace{3cm}+2 \sum_{S=h,H} (g_2 {M_W} C)^2 \big\{{M_S}^2+M_W^2 (-1+x)\big\}\bigg]
\eea
\eegp
\newpage
2. $W_L^+ W_L^+ \rightarrow W_L^+ W_L^+$
\begin{table}[h!]
\begin{center}
\resizebox{15cm}{!}{
    \begin{tabular}{ | c | c |  c | c | c |}
         \hline
      & ${\rm SM}$ & ${\rm 2HDM}$ & ${\rm HTM}(Y=0)$ & ${\rm HTM}(Y=2)$ \\
    \hline
     $A_{2,g}$ & $\frac{g_2^2(3 c_W^2 M_Z^2-4 M_W^2)}{ M_W^4}$& $\frac{g_2^2(3 c_W^2 M_Z^2-4 M_W^2)}{ M_W^4}$& $\frac{g_2^2(3 c_W^2 M_Z^2-4 M_W^2)}{ M_W^4}$& $\frac{g_2^2(3 c_W^2 M_Z^2-4 M_W^2)}{ M_W^4}$  \\
     \hline
     $A_{2,S}(S\equiv h,H)$ & $\frac{g_2^2}{M_W^2}~C^2$ & $\frac{g_2^2}{M_W^2}~C^2$ & $\frac{g_2^2}{M_W^2}~C^2$ &  $\frac{g_2^2}{M_W^2}~C^2$ \\
     \hline
     $A_{2,S}(S\equiv H^+)$ & 0 & 0 & 0& 0  \\
     \hline
      $A_{2,S}(S\equiv H^{++})$ & 0 & 0 & 0 &  $-\frac{g_2^2}{M_W^2}~\hat{C}^2$ \\
     \hline
     \end{tabular}
}
\end{center}
\caption{ \textit{$A_{2,g}$ and $A_{2,S}$ for $W_L^+ W_L^+ \rightarrow W_L^+ W_L^+$ process. $C$ and $\hat{C}$ can be found in Appendix \ref{fyenrules}.} }
    \label{tableb}
\end{table}
\begp
\allowdisplaybreaks \bea
A_{0,g} &=& \frac{{g_2}^2}{2 M_W^4 (-1+x^2)} \big\{c_W^2 M_Z^4 (3+x^2)+4 M_W^4 (1+3 x^2)\nn\\&&\hspace{3cm}-2 c_W^2 M_W^2 M_Z^2 (3+5 x^2)\big\}\\
A_{0,S} &=&-\frac{1}{4 M_W^4}\bigg\{({g_2 M_W}\hat{C})^2 (M_{H^{++}}^2-4 M_W^2)\nn\\&&\hspace{3cm}+2 \sum_{S=h,H} (g_2 {M_W} C)^2 (M_S^2+2 M_W^2)\bigg\}
\eea
\eegp
3. $W_L^+ W_L^- \rightarrow Z_L Z_L$
\begin{table}[h!]
\begin{center}
\resizebox{15cm}{!}{
    \begin{tabular}{ | c | c |  c | c | c |}
         \hline
      & ${\rm SM}$ & ${\rm 2HDM}$ & ${\rm HTM}(Y=0)$ & ${\rm HTM}(Y=2)$ \\
    \hline
     $A_{2,g}$ & $\frac{g_2^2 c_W^2 M_Z^2}{ M_W^4}$& $\frac{g_2^2 c_W^2 M_Z^2}{ M_W^4}$& $\frac{g_2^2 c_W^2 M_Z^2}{ M_W^4}$& $\frac{g_2^2 c_W^2 M_Z^2}{ M_W^4}$  \\
     \hline
     $A_{2,S}(S\equiv h,H)$ & $-\frac{g_2^2}{c_W M_W M_Z}~C C^{\prime}$ & $-\frac{g_2^2}{c_W M_W M_Z}~C C^{\prime}$ & $-\frac{g_2^2}{c_W M_W M_Z}~C C^{\prime}$ &  $-\frac{g_2^2}{c_W M_W M_Z}~C C^{\prime}$ \\
     \hline
     $A_{2,S}(S\equiv H^+)$ & 0 & 0 & $\frac{g_2^2}{c_W^2 M_W^2}~\tilde{C}^2$& $\frac{g_2^2}{c_W^2 M_W^2}~\tilde{C}^2$  \\
     \hline
      $A_{2,S}(S\equiv H^{++})$ & 0 & 0 & 0 &  0 \\
     \hline
     \end{tabular}
}
\end{center}
    \caption{ \textit{$A_{2,g}$ and $A_{2,S}$ for $W_L^+ W_L^- \rightarrow Z_L Z_L$ process. $C,C^{\prime}$ and $\tilde{C}$ can be found in Appendix \ref{fyenrules}.} }
    \label{tablec}
\end{table}

\begp
\allowdisplaybreaks \bea
A_{0,g} &=& -\frac{c_W^2 g_2^2 M_Z^2 }{2 M_W^4 (-1+x^2)}\big\{ M_Z^2 (-1+x^2)+2 M_W^2 (1+x^2)\big\}\\
A_{0,S} &=&\frac{1}{4 M_W^2 M_Z^2}\bigg[-2 \left(\frac{g_2 M_Z}{c_W}\widetilde{C}\right)^2 (M_{H^+}^2+M_W^2+M_Z^2)\nn\\&&\hspace{1cm}+\sum_{S=h,H} (g_2 M_W C) \left(\frac{g_2 M_Z}{c_W} C'\right) \big\{-{M_S}^2+2 (M_W^2+M_Z^2)\big\}\bigg]
\eea
\eegp

4. $W_L^+ Z_L \rightarrow W_L^+ Z_L$
\begin{table}[h!]
\begin{center}
   \resizebox{15cm}{!}{ \begin{tabular}{ | c | c |  c | c | c |}
         \hline
      & ${\rm SM}$ & ${\rm 2HDM}$ & ${\rm HTM}(Y=0)$ & ${\rm HTM}(Y=2)$ \\
    \hline
     $A_{2,g}$ & $-\frac{g_2^2 c_W^2 M_Z^2(1-x)}{8 M_W^4}$& $-\frac{g_2^2 c_W^2 M_Z^2(1-x)}{8 M_W^4}$& $-\frac{g_2^2 c_W^2 M_Z^2(1-x)}{8 M_W^4}$&$-\frac{g_2^2 c_W^2 M_Z^2(1-x)}{8 M_W^4}$ \\
     \hline
     $A_{2,S}({S\equiv h,H})$ & $\frac{g_2^2(1-x)}{8 c_W M_W M_Z}~C C^{\prime}$ & $\frac{g_2^2(1-x)}{8 c_W M_W M_Z}~C C^{\prime}$ & $\frac{g_2^2(1-x)}{8 c_W M_W M_Z}~C C^{\prime}$ &  $\frac{g_2^2(1-x)}{8 c_W M_W M_Z}~C C^{\prime}$ \\
     \hline
     $A_{2,S}(S\equiv H^+)$ & 0 & 0 & $-\frac{g_2^2}{8 c_W^2 M_W^2}(1-x)~\tilde{C}^2$& $\frac{g_2^2}{8 c_W^2 M_W^2}(1-x)~\tilde{C}^2$  \\
     \hline
      $A_{2,S}(S\equiv H^{++})$ & 0 & 0 & 0 &  0 \\
     \hline
     \end{tabular}
     }
    \caption{ \textit{$A_{2,g}$ and $A_{2,S}$ for $W_L^+ Z_L \rightarrow W_L^+ Z_L$ process. $C,C^{\prime}$ and $\tilde{C}$ can be found in Appendix \ref{fyenrules}.} }
    \label{tabled}
\end{center}
\end{table}

\begp
\allowdisplaybreaks \bea
A_{0,g} &=& \frac{c_W^2 {g_2}^2 M_Z^2 }{4 M_W^4 (1+x)}\big\{2 M_W^2 (-1+x)+M_Z^2 (1+x)^2\big\}\\
A_{0,S} &=&-\frac{1}{4 M_W^2 M_Z^2}\bigg[\left(\frac{g_2 M_Z}{c_W}\widetilde{C}\right)^2 \big\{2 M_{H^+}^2-(M_W^2+M_Z^2) (1+x)\big\}\nn\\&&\hspace{0.1cm}+\sum_{S=h,H} (g_2 M_W C) \left(\frac{g_2 M_Z}{c_W} C'\right) \big\{(M_S^2+(M_W^2+M_Z^2) (1+x)\big\}\bigg]
\eea
\eegp

5. $Z_L Z_L \rightarrow Z_L Z_L$

There are no gauge contributions in this process so $A_{2,V}=0$ and the contributions to ${A_{0}}$ are given as follows:
\begp
\allowdisplaybreaks \bea
A_{0,V} &=& 0,~~~~~({\rm no~gauge~contributions}),\\
A_{0,S} &=&-\frac{3}{4M_Z^4} \sum_{S=h,H} \left(	\frac{ g_2 M_Z}{c_W} C'\right)^2 M_S^2.
\eea
\eegp
\chapter{Effective Higgs quartic coupling and $RG$-equations for SM\label{sec:Higgsp}}
\pagestyle{headings}
\pagenumbering{arabic}
\section{Effective Higgs quartic coupling for SM}
The Higgs effective quartic coupling including one-loop and two-loop radiative corrections\cite{Buttazzo:2013uya}.
\begp
\allowdisplaybreaks \bea
\lambda_{\rm eff}^{\rm SM}(\phi) &=& e^{4\Gamma(\phi)}[ \lambda(\mu=\phi) + \lambda_{\rm eff}^{(1)}(\mu=\phi) +  \lambda_{\rm eff}^{(2)}(\mu=\phi)]
\label{eq:effqurtic}
\eea
\eegp
Here,
\begp
\allowdisplaybreaks \bea
(16 \pi^2)~\lambda_{\rm eff}^{(1)} &=& \frac{3}{8} g_2^4 ~\bigg(\ln\frac{g_2^2}{4}-\frac{5}{6}+2 \Gamma \bigg)+\frac{3}{16} (g_1^2+g_2^2)^2 ~ \bigg(\ln\frac{g_1^2+g_2^2}{4}-\frac{5}{6}+2 \Gamma \bigg)\nonumber\\ 
&& -3 y_t^4 ~ \bigg(\ln\frac{y_t^2}{2}-\frac{3}{2}+2 \Gamma \bigg)+ 3 \lambda^2 ~( 4  \ln\lambda-6 + 3\ln3 + 8 \Gamma)
\eea
\eegp
and
\begp
\allowdisplaybreaks \bea
(16 \pi^2)^2 \lambda_{\rm eff}^{(2)}&=&\frac{1}{576} g_1^4 g_2^2 \Bigg\lbrace 4359+218 \pi ^2-36  \bigg(2 \Gamma+\ln\frac{g_2^2}{4} \bigg)-153  \bigg(2 \Gamma+\ln\frac{g_2^2}{4} \bigg)^2\nonumber\\
&&-4080  \bigg(2 \Gamma+\ln\frac{g_1^2+g_2^2}{4} \bigg)+306  \bigg(2 \Gamma+\ln\frac{g_2^2}{4} \bigg)  \bigg(2 \Gamma+\ln\frac{g_1^2+g_2^2}{4} \bigg)\nonumber\\
&&+924  \bigg(2 \Gamma+\ln\frac{g_1^2+g_2^2}{4} \bigg)^2+132  \bigg(2 \Gamma+\ln\frac{g_1^2+g_2^2}{4} \bigg)  \bigg(2 \Gamma+\ln\frac{y_t^2}{2} \bigg)\nonumber\\
&&-66  \bigg(2 \Gamma+\ln\frac{y_t^2}{2} \bigg)^2\Bigg\rbrace +\frac{1}{192} g_1^2 g_2^4 \Bigg\lbrace 817+46 \pi ^2+213  \bigg(2 \Gamma+\ln\frac{g_2^2}{4} \bigg)^2 \nonumber\\
&&  -6  \bigg(2 \Gamma+\ln\frac{g_2^2}{4} \bigg) (50+
53  \bigg(2 \Gamma+\ln\frac{g_1^2+g_2^2}{4} \bigg)\nonumber\\
&&+4  \bigg(2 \Gamma+\ln\frac{g_1^2+g_2^2}{4} \bigg)  \bigg(-91+57 (2 \Gamma+\ln\frac{g_1^2+g_2^2}{4} \bigg)\nonumber\\
&& \hspace{-2cm}+12  \bigg(2 \Gamma+\ln\frac{g_1^2+g_2^2}{4} \bigg)  \bigg(2 \Gamma+\ln\frac{y_t^2}{2} \bigg)-6  \bigg(2 \Gamma+\ln\frac{y_t^2}{2} \bigg)^2\Bigg\rbrace \nonumber\\
&&\hspace{-2cm}+ 8 g_3^2 y_t^4\Bigg\lbrace 9-8  \bigg(2 \Gamma+\ln\frac{y_t^2}{2} \bigg)+3  \bigg(2 \Gamma+\ln\frac{y_t^2}{2} \bigg)^2\Bigg\rbrace
\nonumber\\
 &&\hspace{-2cm}+\frac{1}{8} g_1^2 g_2^2 y_t^2 \Bigg\lbrace-57+44  \bigg(2 \Gamma+\ln\frac{g_1^2+g_2^2}{4} \bigg)+4  \bigg(2 \Gamma+\ln\frac{y_t^2}{2} \bigg)\nonumber\\
&&\hspace{-2cm}-18  \bigg(2 \Gamma+\ln\frac{g_1^2+g_2^2}{4} \bigg)  \bigg(2 \Gamma+\ln\frac{y_t^2}{2} \bigg)+9  \bigg(2 \Gamma+\ln\frac{y_t^2}{2} \bigg)^2\Bigg\rbrace\nonumber\\
&&\hspace{-2cm}+\frac{1}{48} g_1^4 y_t^2 \Bigg\lbrace 189-28  \bigg(2 \Gamma+\ln\frac{g_1^2+g_2^2}{4} \bigg)-68  \bigg(2 \Gamma+\ln\frac{y_t^2}{2} \bigg)\nonumber\\
&&\hspace{-2cm}-54  \bigg(2 \Gamma+\ln\frac{g_1^2+g_2^2}{4} \bigg)  \bigg(2 \Gamma+\ln\frac{y_t^2}{2} \bigg)+27  \bigg(2 \Gamma+\ln\frac{y_t^2}{2} \bigg)^2\Bigg\rbrace\nonumber\\
&&\hspace{-2cm}+\frac{1}{576} g_1^6 \Bigg\lbrace 2883+206 \pi ^2-9  \bigg(2 \Gamma+\ln\frac{g_2^2}{4} \bigg)^2 +708  \bigg(2 \Gamma+\ln\frac{g_1^2+g_2^2}{4} \bigg)^2\nonumber\\
&&\hspace{-2cm}-102  \bigg(2 \Gamma+\ln\frac{y_t^2}{2} \bigg)^2+6  \bigg(2 \Gamma+\ln\frac{g_1^2+g_2^2}{4} \bigg)  \bigg(-470+3  \bigg(2 \Gamma+\ln\frac{g_2^2}{4} \bigg)\nonumber\\
&&\hspace{-2cm}+34  \bigg(2 \Gamma+\ln\frac{y_t^2}{2} \bigg) \bigg)\Bigg\rbrace+\frac{1}{6} y_t^6 \Bigg\lbrace-9 g_2^2  \bigg(1-\ln\frac{g_2^2}{4}+\ln\frac{y_t^2}{2} \bigg)+4 g_1^2  \bigg(9-8  \bigg(2 \Gamma\nonumber
\\ &&\hspace{-2cm}+\ln\frac{y_t^2}{2} \bigg)+3  \bigg(2 \Gamma+\ln\frac{y_t^2}{2} \bigg)^2 \bigg)\Bigg\rbrace+\frac{1}{192} g_2^6 \Bigg\lbrace-2067+90 \pi ^2+1264  \bigg(2 \Gamma\nonumber\\
&&\hspace{-2cm}+\ln\frac{g_2^2}{4} \bigg)+69  \bigg(2 \Gamma+\ln\frac{g_2^2}{4} \bigg)^2+632  \bigg(2 \Gamma+\ln\frac{g_1^2+g_2^2}{4} \bigg)-414  \bigg(2 \Gamma+\ln\frac{g_2^2}{4} \bigg)  \bigg(2 \Gamma \nonumber\\
&&\hspace{-2cm}+\ln\frac{g_1^2+g_2^2}{4} \bigg)+ 156  \bigg(2 \Gamma+\ln\frac{g_1^2+g_2^2}{4} \bigg)^2+36  \bigg(2 \Gamma+\ln\frac{g_1^2+g_2^2}{4} \bigg)  \bigg(2 \Gamma+\ln\frac{y_t^2}{2} \bigg)\nonumber\\
&&\hspace{-2cm}54  \bigg(2 \Gamma+\ln\frac{y_t^2}{2} \bigg)^2-144  \bigg(-\ln\frac{g_2^2}{4}+\ln\frac{y_t^2}{2} \bigg)  \bigg(2 \Gamma+\ln\frac{g_2^2 y_t^2}{4} \bigg)\Bigg\rbrace\nonumber\\
 &&\hspace{-2cm}+\frac{1}{2} y_t^6 \Bigg\lbrace -69-\pi ^2+48  \bigg(2 \Gamma+\ln\frac{y_t^2}{2} \bigg)-6  \bigg(2 \Gamma+\ln\frac{g_2^2}{4} \bigg)  \bigg(2 \Gamma+\ln\frac{y_t^2}{2} \bigg)\nonumber\\
&&\hspace{-2cm}-3  \bigg(2 \Gamma+\ln\frac{y_t^2}{2} \bigg)^2-6  \bigg(-\ln\frac{g_2^2}{4}+\ln\frac{y_t^2}{2} \bigg)  \bigg(2 \Gamma+\ln\frac{g_2^2 y_t^2}{4} \bigg)\Bigg\rbrace\nonumber\\
&&\hspace{-2cm}+\frac{3}{16} g_2^4 y_t^2 \Bigg\lbrace15+2 \pi ^2+8  \bigg(2 \Gamma+\ln\frac{g_2^2}{4} \bigg)+4  \bigg(2 \Gamma+\ln\frac{g_1^2+g_2^2}{4} \bigg)-12  \bigg(2 \Gamma+\ln\frac{y_t^2}{2} \bigg)\nonumber\\
\qquad &&\hspace{-2cm}-6  \bigg(2 \Gamma+\ln\frac{g_1^2+g_2^2}{4} \bigg)  \bigg(2 \Gamma+\ln\frac{y_t^2}{2} \bigg)-3  \bigg(2 \Gamma+\ln\frac{y_t^2}{2} \bigg)^2\nonumber\\
&&\hspace{-2cm}+12  \bigg(-\ln\frac{g_2^2}{4}+\ln\frac{y_t^2}{2} \bigg)  \bigg(2 \Gamma+\ln\frac{g_2^2 y_t^2}{4} \bigg)\Bigg\rbrace +\frac{3}{4}  \bigg(g_2^6-3 g_2^4 y_t^2+4 y_t^6 \bigg)~ \text{Polylog}_2\left[\frac{g_2^2}{2 y_t^2}\right]\nonumber\\
&&\hspace{-2cm}+\frac{1}{64} g_2^2  \bigg(g_1^4+18 g_1^2 g_2^2-51 g_2^4-\frac{48 g_2^6}{g_1^2+g_2^2} \bigg) \sqrt{-\frac{4 (g_1^2+g_2^2)}{g_2^2}+\frac{(g_1^2+g_2^2)^2}{g_2^4}} \Bigg\lbrace\frac{\pi ^2}{3}\nonumber\\
 &&\hspace{-2cm}-\ln\left[\frac{g_1^2+g_2^2}{g_2^2}\right]^2+2 \ln\left[\frac{1}{2}-\frac{1}{2} \sqrt{-3+\frac{4 g_1^2}{g_1^2+g_2^2}}\right]^2\nonumber\\
 &&\hspace{-2cm}-4 \text{Polylog}_2\left[\frac{1}{2}-\frac{1}{2} \sqrt{-3+\frac{4 g_1^2}{g_1^2+g_2^2}}\right]\Bigg\rbrace +\frac{1}{96} \sqrt{(g_1^2+g_2^2) (g_1^2+g_2^2-8 y_t^2)} ~ \bigg(17 g_1^4\nonumber\\
 &&\hspace{-2cm}-6 g_1^2 g_2^2+9 g_2^4+2  \bigg(7 g_1^2-73 g_2^2+\frac{64 g_2^2}{g_1^2+g_2^2} \bigg) y_t^2 \bigg)\Bigg\lbrace\frac{\pi ^2}{3}-\ln\left[\frac{g_1^2+g_2^2}{2 y_t^2}\right]^2\nonumber\\
&&\hspace{-2cm}+2 \ln\left[\frac{1}{2}  \bigg(1-\sqrt{1-\frac{8 y_t^2}{g_1^2+g_2^2}} \bigg)\right]^2-4 \text{Polylog}_2\left[\frac{1}{2}  \bigg(1-\sqrt{1-\frac{8 y_t^2}{g_1^2+g_2^2}} \bigg)\right]\Bigg\rbrace,
\label{efflamSM}~~~\\
{\rm where,}&&\nn\\
\Gamma(\phi)&=&\int_{M_t}^{\phi} \gamma(\mu)\,d\,\ln\mu \, .\nn
\eea
\eegp
Anomalous $\gamma(\mu)$ function of the Higgs field takes care of  its wave function renormalization.
\begp
\allowdisplaybreaks \bea
\gamma &=& \frac{1}{16 \pi ^2}\Bigg\lbrace\frac{3 g_1^2}{4}+\frac{9 g_2^2}{4}-3 y_t^2\Bigg\rbrace + \frac{1}{(16 \pi^2)^2}\Bigg\lbrace -\frac{431 g_1^4}{96}-\frac{9 g_1^2 g_2^2}{16}+\frac{271 g_2^4}{32}-6 \lambda^2\nonumber\\
&&+y_t^2  \bigg(-\frac{85 g_1^2}{24}-\frac{45 g_2^2}{8}-20 g_3^2+\frac{27 y_t^2}{4} \bigg)\Bigg\rbrace + \frac{1}{(16 \pi^2)^3}\Bigg\lbrace -\frac{1315 g_1^6}{54}+\frac{193 g_1^4 g_2^2}{36}+\frac{181 g_1^2 g_2^4}{60}\nonumber\\
&&-\frac{15851 g_2^6}{100}+\frac{419 g_1^4 g_3^2}{36}+\frac{2857 g_2^4 g_3^2}{100}+\frac{107 g_1^4 \lambda}{36}+\frac{119}{20} g_1^2 g_2^2 \lambda+\frac{223 g_2^4 \lambda}{25}\nonumber\\
&&-15 g_1^2 \lambda^2-45 g_2^2 \lambda_1^2+36 \lambda^3+\frac{1499 g_1^4 y_t^2}{36}-\frac{1321}{60} g_1^2 g_2^2 y_t^2+\frac{117 g_2^4 y_t^2}{5}\nonumber\\
&&-\frac{291}{20} g_1^2 g_3^2 y_t^2-\frac{757}{100} g_2^2 g_3^2 y_t^2-\frac{4462 g_3^4 y_t^2}{25}+\frac{135 \lambda^2 y_t^2}{2}+\frac{471 g_1^2 y_t^4}{20}+\frac{4011 g_2^2 y_t^4}{100}\nonumber\\
&&+\frac{1581 g_3^2 y_t^4}{20}-45 \lambda y_t^4-\frac{6013 y_t^6}{13}\Bigg\rbrace\nn
\label{anogama}
\eea
\eegp
\section{Standard Model $\beta$-functions}
The beta functions of the coupling parameters $\chi$ ($\chi \equiv  g_{1}$, $g_{2}$, $g_{3}$, $y_f$, $\lambda$ and $m^2$) for the SM are defined as, 
\beq
\beta_{\chi}= \frac{\partial\chi}{\partial \ln \mu}\,\label{betadef}. 
\eeq 

Beta function of the standard model coupling constants and the mass term up to three loop are presented here for completeness~\cite{Mihaila:2012fm,  Chetyrkin:2012rz, Zoller:2012cv, Chetyrkin:2013wya, Zoller:2013mra, Buttazzo:2013uya},

\begp
\allowdisplaybreaks \bea
\beta_{g_1} &=& g_1^3\Bigg[\frac{1}{16 \pi ^2}\Big( \frac{41}{6}\Big)+ \frac{1}{(16 \pi^2)^2} \Bigg\lbrace \frac{1}{18}  \big(199 g_1^2+81 g_2^2+264 g_3^2-51 y_t^2 \big) \Bigg\rbrace \nonumber\\
&&+ \frac{1}{(16 \pi^2)^3} \Bigg\lbrace -\frac{388613 g_1^4}{5184}+\frac{1315 g_2^4}{64}+99 g_3^4-3 \lambda^2-\frac{29 g_3^2 y_t^2}{3}\nonumber\\
&&+\frac{315 y_t^4}{16}+\frac{1}{864} g_1^2 (1845 g_2^2-4384 g_3^2 + 1296 \lambda-8481 y_t^2) \nonumber\\
&&+ g_2^2  \bigg(-g_3^2+\frac{3 \lambda}{2}-\frac{785 y_t^2}{32} \bigg) \Bigg\rbrace \Bigg]
\label{betag1}\\
\beta_{g_2} &=& g_2^3\Bigg[\frac{1}{16 \pi ^2}\Big(  -\frac{19}{6} \Big) + \frac{1}{(16 \pi^2)^2} \Bigg\lbrace  \frac{1}{6}  \big(9 g_1^2+35 g_2^2+72 g_3^2-9 y_t^2 \big)  \Bigg\rbrace\nonumber\\
&& + \frac{1}{(16 \pi^2)^3} \Bigg\lbrace  -\frac{5597 g_1^4}{576}+\frac{324953 g_2^4}{1728}+81 g_3^4-3 \lambda^2-7 g_3^2 y_t^2\nonumber\\
&&+\frac{147 y_t^4}{16}+\frac{1}{96} g_1^2 (873 g_2^2-32 g_3^2+48 \lambda-593 y_t^2)\nonumber\\
&&+g_2^2  \big(39 g_3^2+\frac{3 \lambda}{2}-\frac{729 y_t^2}{32} \big)  \Bigg\rbrace \Bigg]
\label{betag2}\\
\beta_{g_3} &=& g_3^3\Bigg[\frac{1}{16 \pi ^2}\Big( -7 \Big) + \frac{1}{(16 \pi^2)^2} \Bigg\lbrace  \frac{11 g_1^2}{6}+\frac{9 g_2^2}{2}-2 (13 g_3^2+y_t^2)  \Bigg\rbrace\nonumber\\
&& + \frac{1}{(16 \pi^2)^3} \Bigg\lbrace  -\frac{2615 g_1^4}{216}+\frac{109 g_2^4}{8}+\frac{65 g_3^4}{2}-40 g_3^2 y_t^2+15 y_t^4\nonumber\\
&&+g_2^2   \bigg(21 g_3^2-\frac{93 y_t^2}{8} \bigg)-\frac{1}{72} g_1^2  \big(9 g_2^2-616 g_3^2+303 y_t^2 \big)  \Bigg\rbrace \Bigg]
\label{betag3}
\eea
\eegp
\begp
\allowdisplaybreaks \bea
\beta_{y_t} &=& y_t \Bigg[ \frac{1}{16 \pi ^2} \Big\lbrace \frac{1}{12}\big(-17 g_1^2 - 27 g_2^2 - 96 g_3^2 + 54 yt^2 \big)\Bigg\rbrace + \frac{1}{(16 \pi^2)^2} \Big\lbrace  \frac{1187 g_1^4}{216}-\frac{23 g_2^4}{4}\nonumber\\&&+g_1^2  \bigg(-\frac{3 g_2^2}{4}+\frac{19 g_3^2}{9}+\frac{131 y_t^2}{16} \bigg)+g_2^2  \bigg(9 g_3^2+\frac{225 y_t^2}{16} \bigg)\nonumber\\&&-6 (18 g_3^4-\lambda^2-6 g_3^2 y_t^2+2 \lambda y_t^2+2 y_t^4)  \Bigg\rbrace \nonumber\\&&+\frac{1}{(16 \pi^2)^3}\frac{1}{24} \Big\lbrace  24 g_3^2 (16 \lambda y_t^2-157 y_t^4)\nonumber\\&&+4 g_3^4 y_t^2 (3827-1368 \zeta(3))+16 g_3^6 (-2083+960 \zeta(3))\nonumber\\&&+9 \big(-96 \lambda^3+10 \lambda^2 y_t^2+528 \lambda y_t^4+y_t^6 (113+36 \zeta(3))\big)  \Big\rbrace \Bigg]
\label{betayt}
\eea
\eegp
\begp
\allowdisplaybreaks \bea
\beta_{\lambda} & = & \Bigg[\frac{1}{16 \pi ^2}\Bigg\lbrace  \frac{3}{8} (2 g_2^4+(g_1^2+g_2^2)^2)+24 \lambda^2-6 y_t^4-3 \lambda (g_1^2+3 g_2^2-4 y_t^2) \Bigg\rbrace \nonumber\\
&&+ \frac{1}{(16 \pi^2)^2}\frac{1}{48} \Bigg\lbrace -379 g_1^6-559 g_1^4 g_2^2-289 g_1^2 g_2^4+915 g_2^6\nonumber\\
&&+48 \lambda \bigg(\frac{629 g_1^4}{24}-\frac{73 g_2^4}{8}+108 g_2^2 \lambda-312 \lambda^2+g_1^2  \bigg(\frac{39 g_2^2}{4}+36 \lambda \bigg)\bigg)\nonumber\\
&&-4 \big(57 g_1^4-2 g_1^2 (63 g_2^2+85 \lambda)+3 (9 g_2^4-90 g_2^2 \lambda+64 \lambda (-5 g_3^2+9 \lambda))\big) y_t^2\nonumber\\
&&-16 (8 g_1^2+96 g_3^2+9 \lambda) y_t^4+1440 y_t^6 \Bigg\rbrace +   \frac{1}{(16 \pi^2)^3} \frac{1}{12} \Bigg\lbrace  20952 \lambda^3 y_t^2\nonumber\\
&&+288 \lambda^4 (299+168 \zeta(3))\nonumber\\
&&-y_t^4 \big(g_3^4 (2128-768 \zeta(3))+48 g_3^2 y_t^2 (19-120 \zeta(3))+9 y_t^4 (533+96 \zeta(3))\big)\nonumber\\
&&+108 \lambda^2 y_t^2 \big(16 g_3^2 (-17+16 \zeta(3))+y_t^2 (191+168 \zeta(3))\big)\nonumber\\&& +\lambda y_t^2  \big(27 y_t^4 (13-176 \zeta(3))-32 g_3^4 (-311+36 \zeta(3))\nonumber\\&&-24 g_3^2 y_t^2 (-895+1296 \zeta(3)) \big) \Bigg\rbrace+0.0000133607 g_3^6 y_t^4\Bigg]
\label{betalam}
\eea
\eegp
\begp
\allowdisplaybreaks \bea
\beta_{m^2}&=& m^2 \Bigg[ \frac{1}{(16\pi^2)} \bigg\lbrace 6 \lambda +3  y_t^2 
-\frac{9 g_2^2}{4} -\frac{3 g_1^2 }{4} \bigg\rbrace
 + \frac{1}{(16\pi^2)^2} \bigg\lbrace \lambda  (-30 \lambda -36 y_t^2
 +36 g_2^2+12 g_1^2 )\nn \\
	&& +y_t^2 \bigg(-\frac{27 y_t^2}{4} +20 g_3^2
+\frac{45 g_2^2}{8}  +\frac{85 g_1^2}{24}  \bigg)+
-\frac{145}{32} g_2^4 +\frac{557}{96} g_1^4+\frac{15 g_2^2 g_1^2}{16} \bigg\rbrace\nn \\ &&
 + \frac{1}{(16\pi^2)^3} \bigg\lbrace\lambda^2  (1026 \lambda +148.5 y_t^2 -192.822 g_2^2-64.273 g_1^2 )
 +\lambda y_t^2  (347.394  y_t^2\nonumber \\ &&+80.385 g_3^2
 -318.591 g_2^2-99.498 g_1^2)
 +\lambda  ( -64.5145 g_2^4-182.79 g_1^4\nonumber \\ &&-63.0385 g_2^2 g_1^2)+ y_t^4  ( 154.405  y_t^2
 -209.24 g_3^2-3.82928 g_2^2-12.5128 g_1^2 )
 \nonumber \\ &&+ y_t^2  ( 178.484 g_3^4 -102.627 g_2^4-77.0028 g_1^4 +7.572 g_3^2 g_2^2
 +14.545 g_3^2 g_1^2\nonumber \\ &&+19.1167 g_2^2 g_1^2 )
 +g_2^4  ( -28.572  g_3^2+301.724 g_2^2+16.552  g_1^2 )+g_1^4  ( -11.642 g_3^2 \nonumber \\ &&+27.161 g_2^2 + 38.786 g_1^2 ) \bigg\rbrace \Bigg]
 \label{masterm}
\eea
\eegp

\section{$\beta$-functions for singlet scalar extended SM}
As a discrete $Z_2$ symmetry have been imposed on the extra singlet scalar, $odd$ number of scalars do not couple with the standard model particles. So the  $\beta$-functions of $g_1, g_2, g_3$ and Yukawa couplings remain unchanged. Only the Higgs quartic coupling $\beta_{\lambda}$ gets modified. At one loop, we have to add $\frac{k^2/2}{16 \pi^2}$ with $\beta_{\lambda}$ of eqn.~\ref{betadef}. The $\beta$-functions of $\lambda_S$ and $\kappa$ are given by~\cite{Haba:2013lga, Clark:2009dc, Davoudiasl:2004be,Khan:2014kba},
\begp
\allowdisplaybreaks \begin{align}
  &\beta_\kappa
    =\left\{
      \begin{array}{ll}
       0 & \mbox{for }\mu<M_S\\
       \frac{\kappa}{16\pi^2}\left[4\kappa+12\lambda+6y^2-\frac{3}{2}(g'{}^2+3g^2)+\lambda_S\right] & \mbox{for }\mu\geq M_S
      \end{array}
     \right. \, ,   \\
  &\beta_{\lambda_S}
    =\left\{
      \begin{array}{ll}
       0 & \mbox{for }\mu<M_S\\
       \frac{1}{16\pi^2}\left[3\lambda_S^2+12\kappa^2\right] & \mbox{for }\mu\geq M_S
      \end{array}
     \right. \, .
     \label{eq:betasinglet} 
 \end{align}
 \eegp

\section{$\beta$-functions for inert doublet extended SM}
As in inert doublet model pseudoscalar $A$ has been taken as a lightest particle, the expressions of $\beta$-functions at one loop are given as follows~\cite{Khan:2015ipa},

For $\mu<M_A$,
\begp
\allowdisplaybreaks \beq
\beta_{\lambda_1} = \beta_{\lambda}^{SM} ~~~{\rm and}~~~ \beta_{\lambda_{2,3,4,5}} =0,\label{betal_IDMM}
\eeq
\eegp
and for $\mu>M_A$
\begp
\allowdisplaybreaks \bea
\beta_{\lambda_1} &=& \frac{1}{16\pi^2}\big[24 \lambda_1^2 + 2 \lambda_3^2 + 2 \lambda_3 		\lambda_4 +  \lambda_4^2 +  \lambda_5^2 \nn \\
	&&+\frac{3}{8} \left( 3 g_2^4 + g_1^4 + 2 g_2^2 g_1^2 \right) - 3 \lambda_1 \left( 3 g_2^2 + g_1^2 \right) \nn \\	
	&&+ 4 \lambda_1 \left( y_\tau^2 + 3 y_b^2 + 3 y_t^2 \right) - 2 \left( y_\tau^4 + 3 y_b^4 + 3 y_t^4 \right)\big], \label{betal_1}\\
\beta_{\lambda_2} &=& 24 \lambda_2^2 + 2 \lambda_3^2 + 2 \lambda_3 \lambda_4 +  \lambda_4^2 +  \lambda_5^2 \nn \\
	&&+\frac{3}{8} \left( 3 g_2^4 + g_1^4 + 2 g_2^2 g_1^2 \right) - 3 \lambda_2 \left( 3 g^2 + g_1^2 \right)\big] , \\	
\beta_{\lambda_3} &=& \frac{1}{16\pi^2}\big[4\left( \lambda_1 + \lambda_2 \right) \left( 3 \lambda_3 +  \lambda_4 \right) + 4 \lambda_3^2 + 2 \lambda_4^2 + 2 \lambda_5^2 \nn \\
	&&+	\frac{3}{4} \left( 3 g_2^4 + g_1^4 - 2 g_2^2 g_1^2 \right) - 3 \lambda_3 \left( 3 g_2^2 + g_1^2 \right) \nn \\
	&&+2 \lambda_3 \left( y_\tau^2 + 3 y_t^2 + 3 y_b^2 \right)\big] , \\	
\beta_{\lambda_4} &=& \frac{1}{16\pi^2}\big[ 4 \lambda_4 \left( \lambda_1 + \lambda_2 + 2 \lambda_3 +  \lambda_4 \right) + 8 \lambda_5^2 \nn\\
	&&+ 3 g_2^2 g_1^2 - 3 {\lambda_4} \left( 3 g_2^2 + g_1^2 \right)\nn \\
	&&+ 2 \lambda_4 \left( y_\tau^2 + 3 y_t^2 + 3 y_b^2 \right) \big],\\   
\beta_{\lambda_5} &=& \frac{1}{16\pi^2}\big[ 4 \lambda_5 \left( \lambda_1 + \lambda_2 + 2 \lambda_3 + 3 \lambda_4 \right) \nn\\
	&& - 3 \lambda_5 \left( 3 g_2^2 + g_1^2 \right)  \nn\\
	&& +2 \lambda_5 \left( y_\tau^2 + 3 y_t^2 + 3 y_b^2 \right)\big].\label{betal_5}
\eea
\eegp
In this case the symbol $\lambda_1$ is the same as Higgs quartic coupling $\lambda$.
Let us note the $y_t$ dependence of these expressions. While $\beta_{\lambda_1}$ is dominated by the $y_t^4$ term,  $\beta_{\lambda_2}$ does not depend on $y_t$. The $y_t$ dependence of other $\beta_{\lambda_i}$s are softened by the corresponding $\lambda_i$ multiplying the $y_t^2$ terms. Two-loop RGEs used in this work have been generated using {\tt SARAH}~\cite{Staub:2013tta}.
\section{$\beta$-functions for inert triplet ($Y=0$) extended SM}
Similarly for $\beta$-functions for inert triplet model at one-loop given as follows~\cite{Forshaw:2003kh}:

For $\mu<M_H$,
\begp
\allowdisplaybreaks \beq
\beta_{\lambda_1} = \beta_{\lambda}^{SM} ~~~{\rm and}~~~ \beta_{\lambda_{2,3}} =0,\label{betal_HTM}
\eeq
\eegp

and for $\mu>M_H$
\begp
\allowdisplaybreaks \bea
\beta_{\lambda_1} &=& \frac{1}{16\pi^2}\big[ 24 \lambda_1^2 + \frac{3}{2} \lambda_3^2 +\frac{3}{8} \left( 3 g_2^4 + g_1^4 + 2 g_2^2 g_1^2 \right) - 3 \lambda_1 \left( 3 g_2^2 + g_1^2 \right) \nn \label{betal_1HTM}\\	
	&&+ 4 \lambda_1 \left( y_\tau^2 + 3 y_b^2 + 3 y_t^2 \right) - 2 \left( y_\tau^4 + 3 y_b^4 + 3 y_t^4 \right)\big], \\
\beta_{\lambda_2} &=& \frac{1}{16\pi^2}\big[ 22 \lambda_2^2 + 2 \lambda_3^2 + 12 g_2^4 -24 g_2^2 \lambda_2 \big], \label{betal_2HTM}\\	
\beta_{\lambda_3} &=& \frac{1}{16\pi^2}\big[6 g_2^4 + 12 \lambda_1 \lambda_3 + 10 \lambda_2 \lambda_3 + 4 \lambda_3^2 + \lambda_3 (6 y_t^2 -\frac{33}{2} g_2^2 -\frac{3}{2} g_1^2 )\big].\label{betal_3HTM}
\eea
\eegp
In this work two-loop RGEs are used for the extra scalar sectors.
\end{appendices}
%

\end{document}